\def\mearth{{\rm\,M_\oplus}}
\def\rearth{{\rm\,R_\oplus}}
\def\msun{{\rm\,M_\odot}}

\def\norm#1{{\left\| #1 \right\|}}

\documentclass[graybox]{svmult}

\usepackage{type1cm}        
\usepackage{makeidx}         
\usepackage{graphicx}        
\usepackage{multicol}        
\usepackage[bottom]{footmisc}

\usepackage{newtxtext}       %
\usepackage{newtxmath}       

\makeindex             

\begin{document}

\title*{Planet formation: key mechanisms and global models}
\author{Sean N. Raymond and Alessandro Morbidelli}
\institute{Sean N. Raymond \at Laboratoire d'Astrophysique de Bordeaux, CNRS and Universit{\'e} de Bordeaux, Pessac, France\\ \email{rayray.sean@gmail.com}
\and Alessandro Morbidelli \at Laboratoire Lagrange, Observatoire de la Cote d'Azur, Nice, France\\
\email{morby@oca.eu}}
%
%
\maketitle

\vskip -0.7in
\abstract{\\Models of planet formation are built on underlying physical processes. In order to make sense of the origin of the planets we must first understand the origin of their building blocks.
\vskip .1in
\noindent This review comes in two parts. The first part presents a detailed description of six key mechanisms of planet formation:
\begin{itemize}
\item The structure and evolution of protoplanetary disks
\item The formation of planetesimals
\item Accretion of protoplanets
\item Orbital migration of growing planets
\item Gas accretion and giant planet migration
\item Resonance trapping during planet migration
\end{itemize}
While this is not a comprehensive list, it includes processes for which our understanding has changed in recent years or for which key uncertainties remain.  
\vskip .1in
\noindent The second part of this review shows how global models are built out of planet formation processes. We present global models to explain different populations of known planetary systems, including close-in small/low-mass planets (i.e., {\em super-Earths}), giant exoplanets, and the Solar System's planets. We discuss the different sources of water on rocky exoplanets, and use cosmochemical measurements to constrain the origin of Earth's water. We point out the successes and failings of different models and how they may be falsified.   
\vskip .1in
\noindent Finally, we lay out a path for the future trajectory of planet formation studies.
}

\section{Observational constraints on planet formation models} 
\label{sec:intro}

If planet building is akin to cooking, then a review of planet formation is a cookbook. Planetary systems -- like dishes -- come in many shapes and sizes. Just as one cooking method cannot produce all foods, a single growth history cannot explain all planets. While the diversity of dishes reflects a range of cooking techniques and tools, they are all drawn from a common set of cooking methods. Likewise, the diversity of planetary systems can be explained by different combinations of processes drawn from a common set of physical mechanisms. Our goal in this review is first to describe the key processes of planet formation and then to show how they may be combined to generate global models, or recipes, for different types of planetary systems.

\begin{figure}[t]
\includegraphics[width=11cm]{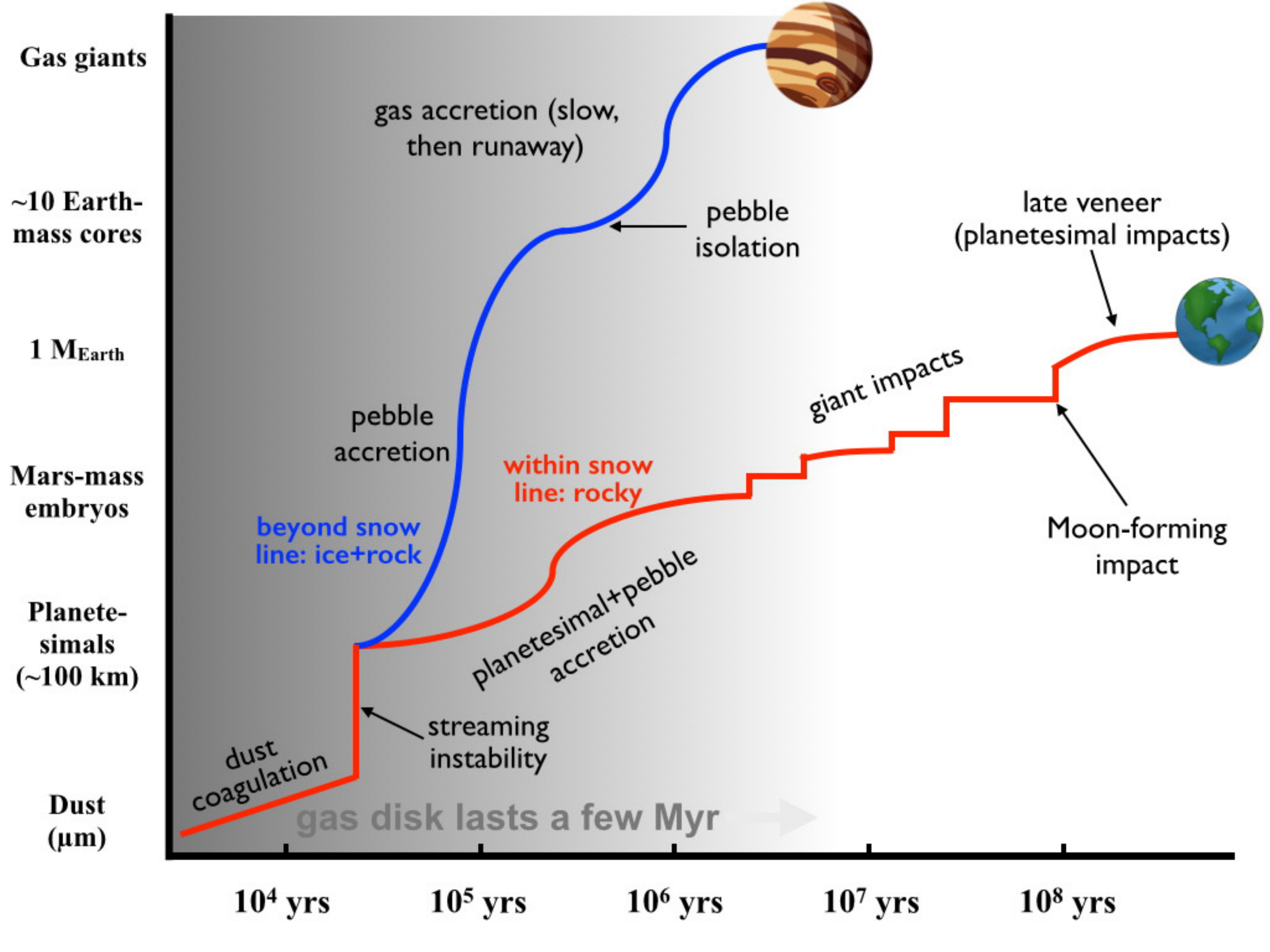}
\caption{Schematic view of some of the processes involved in forming Jupiter and Earth. This diagram is designed to present a broad view of the relevant mechanisms but still does not show a number of important  effects. For instance, we know from the age distribution of primitive meteorites that planetesimals in the Solar System formed in many generations, not all at the same time.  In addition, this diagram does not depict the large-scale migration thought to be ubiquitous among any planets more massive than roughly an Earth-mass (see discussion in text).  Adapted from \cite{meech19}.  }
\label{fig:growth_cartoon}       
\end{figure}

To illustrate the processes involved, Fig.~\ref{fig:growth_cartoon} shows a cartoon picture of our current vision for the growth of Earth and Jupiter.  Both planets are thought to have formed from planetesimals in different parts of the Solar System.  In our current understanding, the growth tracks of these planets diverge during the pebble accretion process, which is likely to be much more efficient past the snow line~\cite{lambrechts14,morby15}. There exists a much larger diversity of planets than just Jupiter and Earth, and many vital processes are not included in the Figure, yet it serves to illustrate how divergent formation pathways can contribute to planetary diversity. 

We start this review by summarizing the key constraints on planet formation models. Constraints come from Solar System measurements (e.g., meteorites), observations of other planetary systems (e.g., exoplanets and protoplanetary disks), as well as laboratory measurements (e.g., to measure the sticking properties of small grains).

\paragraph{Solar System Constraints}
Centuries of human observation have generated a census of the Solar System, albeit one that is still not 100\% complete. The most important constraints for planet formation include our system's orbital architecture as well as compositional and timing information gleaned from in-situ measurements. An important but challenging exercise is to distill the multitude of existing constraints into just a few large-scale factors to which resolution-limited models can be compared. 

The central Solar System constraints are:
\begin{itemize}

\item {\bf The masses and orbits of the terrestrial planets.}\footnote{The terms ``terrestrial'' and ``rocky'' planet are interchangeable: the Solar System community generally uses the term {\em terrestrial} and the exoplanet community uses {\em rocky}. We use both terminologies in this review to represent planets with solid surfaces that are dominated (by mass) by rock and iron.} The key quantities include their number, their absolute masses and mass ratios, and their low-eccentricity, low-inclination orbits. These have been quantified in studies that attempted to match their orbital distribution. For example the normalized angular momentum deficit $AMD$ is defined as\cite{laskar97,chambers01}:
\begin{equation} 
AMD = \frac{\sum_{j} m_j \sqrt{a_j} \left(1 - cos(i_j) \sqrt{1-e_j^2}\right)} {\sum_j m_j \sqrt{a_j}}, 
\end{equation}
\noindent where $a_j$, $e_j$, $i_j$, and $m_j$ correspond to planet $j$'s semimajor axis, eccentricity, orbital inclination, and mass. The Solar System's terrestrial planets have an $AMD$ of 0.0018.  

The radial mass concentration statistic $RMC$ (called $S_c$ by \cite{chambers01}) is a measure of the radial mass profile of the planets.  It is defined as:
\begin{equation} 
RMC = max \left(\frac{\sum m_j}{\sum m_j [log_{10}(a/a_j)]^2} \right).
\end{equation}
The function in brackets is calculated sweeping $a$ across all radii, and the $RMC$ represents the maximum.  For a one-planet system $RMC$ is infinite.  The $RMC$ is higher when the planets' masses are concentrated in narrow radial zones (as is the case in the terrestrial planets, with two large central planets and two small exterior ones).  The $RMC$ becomes smaller for systems that are more spread out and systems in which all planets have similar masses. The Solar System's terrestrial planets' $RMC$ is 89.9.  

Confronting distributions of simulated planets with these empirical statistics (as well as other ones) has become a powerful and commonly-used discriminant of terrestrial planet formation models \cite{chambers01,obrien06,raymond06b,raymond09c,raymond14,clement18,lykawka19}.  

\item {\bf The masses and orbits of the giant planets.} As for the terrestrial planets, the number (two gas giant, two ice giant), masses and orbits of the giant planets are the central constraints. The orbital spacing of the planets is also important, for instance the fact that no pair of giant planets is located in mean motion resonance.  An important, overarching factor is simply that the Solar System's giant planets are located far from the Sun, well exterior to the orbits of the terrestrial planets. 

\item {\bf The orbital and compositional structure of the asteroid belt.}  While spread over a huge area the asteroid belt contains only $\sim 4.5 \times 10^{-4} \mearth$ in total mass \cite{krasinsky02,kuchynka13,demeo13}, orders of magnitude less than would be inferred from models of planet-forming disks such as the very simplistic minimum-mass solar nebula model \cite{weidenschilling77,hayashi81}.  The orbits of the asteroids are excited, with eccentricities that are roughly evenly distribution from zero to 0.3 and inclinations evenly spread from zero to more than $20^\circ$ (a rough stability limit given the orbits of the planets).  While there are a number of compositional groups within the belt, the general trend is that the inner main belt is dominated by S-types and the outer main belt by C-types~\cite{gradie82,demeo13,demeo14}.  S-type asteroids are associated with ordinary chondrites, which are quite dry (with water contents less than 0.1\% by mass), and C-types are linked with carbonaceous chondrites, some of which (CI, CM meteorites) contain $\sim 10\%$ water by mass~\cite{robert77,kerridge85,alexander18}.
 
\item {\bf The cosmochemically-constrained growth histories of rocky bodies in the inner Solar System.}  Isotopic chronometers have been used to constrain the accretion timescales of different solid bodies in the Solar System.  Ages are generally measured with respect to CAIs (Calcium and Aluminum-rich Inclusions), mm-sized inclusions in chondritic meteorites that are dated to be 4.568 Gyr old \cite{bouvier10}.  Cosmochemical measurements indicate that chondrules, which are similar in size to CAIs, started to form at roughly the same time \cite{connelly08,nyquist09}.  Age dating of iron meteorites suggests that differentiated bodies -- large planetesimals or planetary embryos -- were formed in the inner Solar System within 1 Myr of CAIs \cite{halliday06,kruijer14,schiller15}.  Isotopic analyses of Martian meteorites show that Mars was fully formed within 5-10 Myr after CAIs \cite{nimmo07,dauphas11}, whereas similar analyses of Earth rocks suggest that Earth's accretion did not finish until much later, roughly 100 Myr after CAIs \cite{touboul07,kleine09}.  

There is evidence that two populations of isotopically-distinct chondritic meteorites -- the so-called carbonaceous and non-carbonaceous meteorites -- have similar age distributions \cite{kruijer17}.  Given that chondrules are expected to undergo very fast radial drift within the disk \cite{weidenschilling77b,lambrechts12}, this suggests that the two populations were kept apart and radially segregated, perhaps by the early growth of Jupiter's core \cite{kruijer17}.

\end{itemize}

\paragraph{Constraints from Observations of Planet-forming disks around other stars}
Gas-dominated protoplanetary disks are the birthplaces of planets. Disks' structure and evolution plays a central role in numerous processes such as how dust drifts \cite{birnstiel16}, where planetesimals form \cite{drazkowska16,drazkowska17}, and what direction and how fast planets migrate \cite{bitsch15}.

\begin{figure}[t]
\centering
\includegraphics[width=5.1cm]{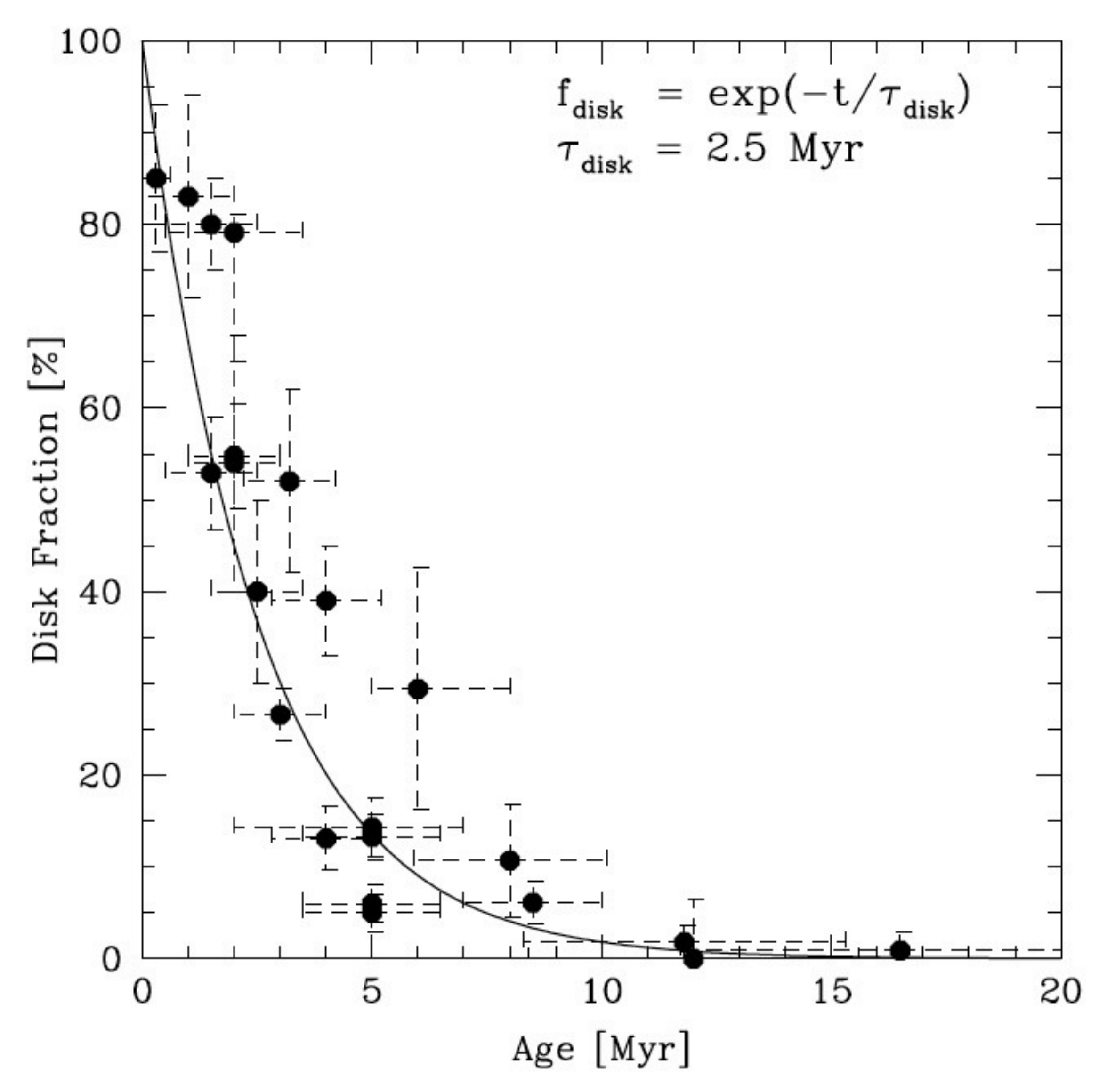}
\includegraphics[width=5.9cm]{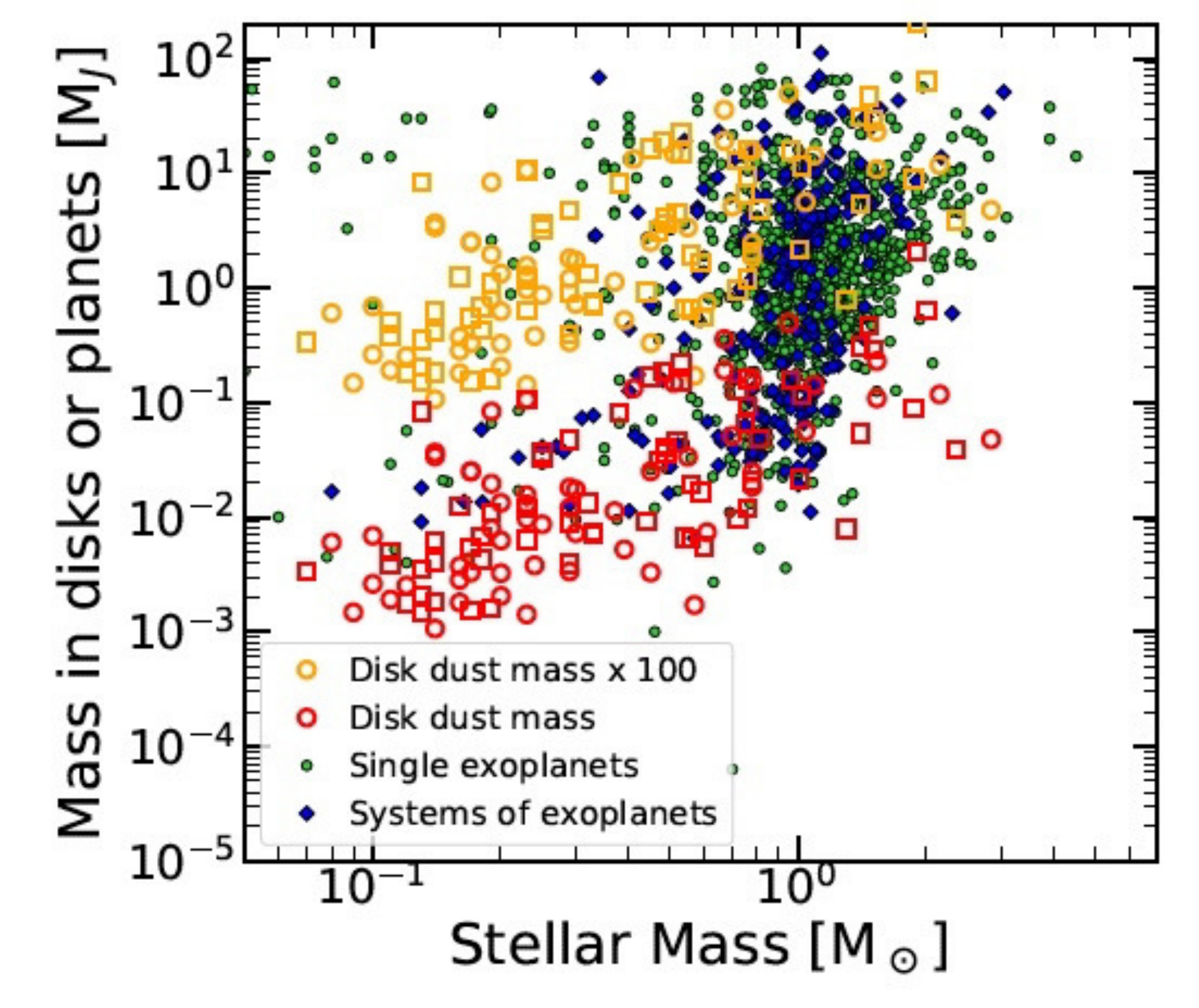}

\caption{Two observational constraints on planet-forming disks.  {\bf Left:} The fraction of stars that have detectable disks in clusters of different ages.  This suggests that the typical gaseous planet-forming disk only lasts a few Myr \cite{haisch01,briceno01,hillenbrand08}. From \cite{mamajek09}. {\bf Right:} A comparison between inferred disk masses and the mass in planets in different systems, as a function of host star mass.  The dust mass (red) is measured using sub-mm observations (and making the assumption that the emission is optically thin), and the gas mass is inferred by imposing a 100:1 gas to dust ratio. There is considerable tension, as the population of disks does not appear massive enough to act as the precursors of the population of known planets. The solution to this problem is not immediately obvious.  Perhaps disk masses are systematically underestimated~\cite{greaves11}, or perhaps disks are continuously re-supplied with material from within their birth clusters via Bondi-Hoyle accretion \cite{throop08,moeckel09}. From \cite{manara18}.}
\label{fig:disk_constraints}       
\end{figure}

We briefly summarize the main observational constraints from protoplanetary disks for planet formation models (see also dedicated reviews \cite{williams11,armitage11,alexander14}):
\begin{itemize}
\item {\bf Disk lifetime}. In young clusters virtually all stars have detectable hot dust, which is used as a tracer for the presence of gaseous disks \cite{haisch01,briceno01}.  However, in old clusters very few stars have detectable disks.  Analyses of a large number of clusters of different ages indicate that the typical timescale for disks to dissipate is a few Myr \cite{haisch01,briceno01,hillenbrand08,mamajek09,pecaut16}. Fig.~\ref{fig:disk_constraints} shows this trend, with the fraction of stars with disks decreasing as a function of cluster age.  It is worth noting that observational biases are at play, as the selection of stars that are members of clusters can affect the interpreted disk dissipation timescale \cite{pfalzner14}.
\item {\bf Disk masses}.  Most masses of protoplanetary disks are measured using sub-mm observations of the outer parts of the disk in which the emission is thought to be optically thin \cite{williams11}.  Disk masses are commonly found to be roughly equivalent to 1\% of the stellar mass, albeit with a 1-2 order of magnitude spread \cite{eisner03,andrews05,andrews07a,andrews10,williams11}.  It has recently been pointed out that there is tension between the inferred disk masses and the masses of exoplanet systems, as a large fraction of disks do not appear to contain enough mass to produce exoplanet systems \cite{manara18,mulders18}, even assuming a very high efficiency of planet formation (see Fig.~\ref{fig:disk_constraints}). 
\item {\bf Disk structure and evolution}.  ALMA observations suggest that disks are typically 10-100 au in scale \cite{barenfeld17}, similar to the expected dimensions of the Sun's protoplanetary disk \cite{hayashi81,gomes04,kretke12}. Sub-mm observations at different radii indicate that the surface density of dust $\Sigma$ in the outer parts of disks follows a roughly $\Sigma \sim r^{-1}$ radial surface density slope \cite{mundy00,looney03,andrews07b,andrews09}, consistent with simple models for accretion disks. Many disks observed with ALMA show ringed substructure~\cite{alma15,andrews16,andrews18}. Disks are thought to evolve by accreting onto their host stars, and the accretion rate itself has been measured to vary as a function of time; indeed, the accretion rate is often used as a proxy for disk age \cite{hartmann98,armitage11}.  As disks age, they evaporate from the inside-out by radiation from the central star \cite{alexander06,owen11,alexander14} and, depending on the stellar environment, may also evaporate from the outside-in due to external irradiation \cite{hollenbach94,adams04}.
\item {\bf Dust around older stars.}  Older stars with no more gas disks often have observable dust, called {\em debris disks} (recent reviews: \cite{matthews14,hughes18}).  Roughly 20\% of Sun-like stars are found to have dust at mid-infrared wavelengths~\cite{bryden06,trilling08,montesinos16}. This dust is thought to be associated with the slow collisional evolution of outer planetesimal belts akin to our Kuiper belt but generally containing much more mass~\cite{wyatt08,krivov10}.  The occurrence rate of dust is observed to decrease with the stellar age~\cite{meyer08,carpenter09}. More [less] massive stars have significantly higher [lower] occurrence of debris disks~\cite{su06,lestrade12}. There is no clear observed correlation between debris disks and planets \cite{moromartin07,marshall14,moromartin15}.  A significant fraction of old stars have been found to have warm or hot {\em exo-zodiacal} dust \cite{absil13,ertel14,kral17}. The origin of this dust remains mysterious as there is no clear correlation between the presence of cold and hot dust \cite{ertel14}.
\end{itemize}

\begin{figure*}
\centering
\includegraphics[width=10cm]{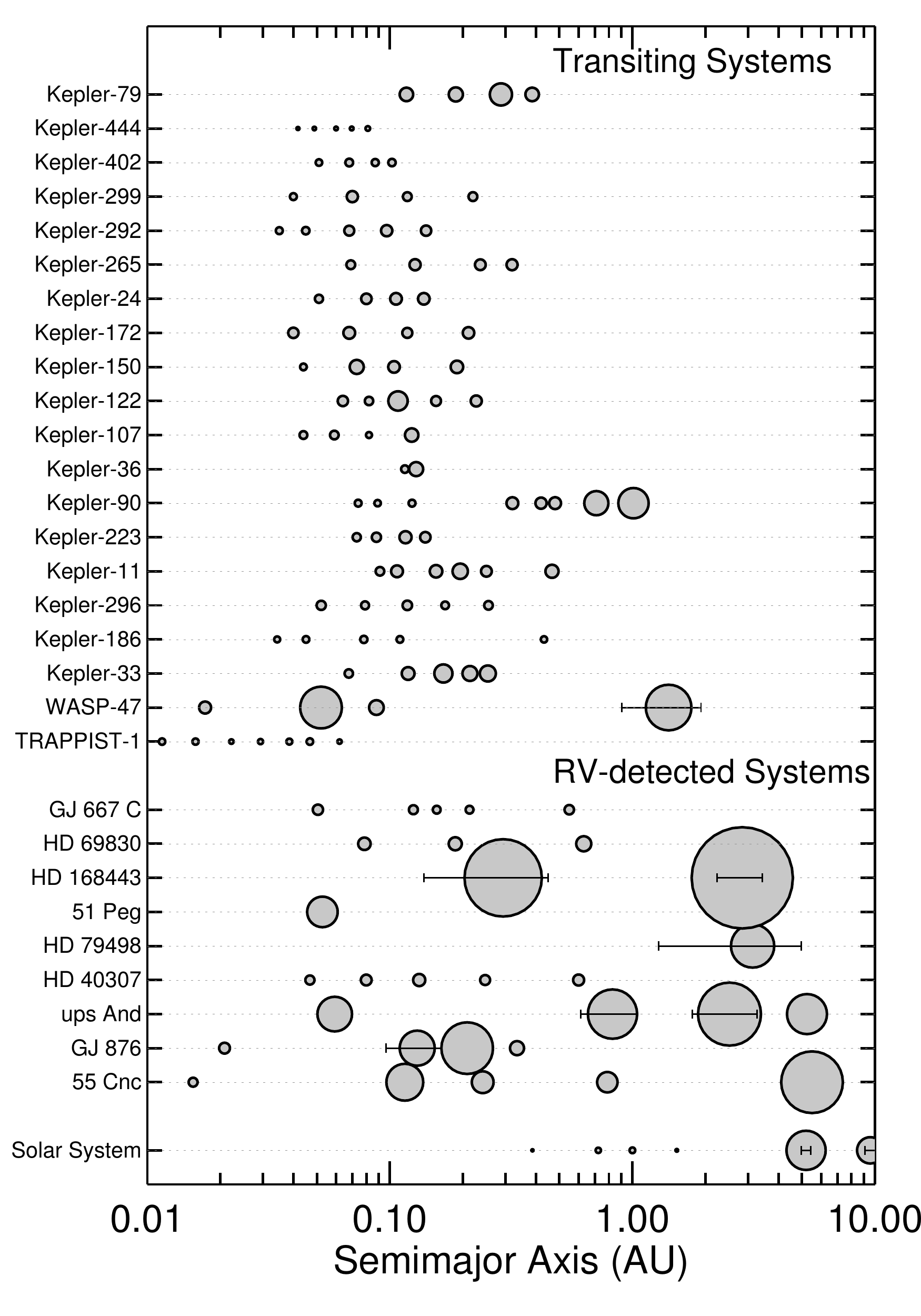}
\caption{A sample of exoplanet systems selected by hand to illustrate their diversity (from \cite{raymond18d}). The systems at the top were discovered by the transit method and the bottom systems by radial velocity (RV). Of course, some planetsare detected in both transit and RV  (e.g. 55 Cnc e; ~\cite{demory11}).  A planet's size is proportional to its actual size (but is not to scale on the x-axis). For RV planets without transit detections we used the $M sin i \propto R^{2.06}$ scaling derived by \cite{lissauer11b}. For giant planets ($M > 50 \mearth$) on eccentric orbits ($e > 0.1$; also for Jupiter and Saturn), the horizontal error bar represents the planet's pericenter to apocenter orbital excursion. The central stars vary in mass and luminosity; e.g., TRAPPIST-1 is an ultracool dwarf star with mass of only $0.08 \msun$~\cite{gillon17}. A handful of systems have $\sim$Earth-sized planets in their star's habitable zones, such as Kepler-186~\cite{quintana14}, TRAPPIST-1~\cite{gillon17}, and GJ 667 C~\cite{anglada13}. Some planetary systems -- for example, 55 Cancri~\cite{fischer08} -- are found in multiple star systems.}
\label{fig:interesting}       
\end{figure*}

\paragraph{Constraints from Extra-Solar Planets}
With a catalog of thousands of known exoplanets, the constraints from planets around other stars are extremely rich and constantly being improved. Figure~\ref{fig:interesting} shows the orbital architecture of a (non-representative) selection of known exoplanet systems. While there exist biases in the detection methods used to find exoplanets~\cite{winn15}, their sheer number form the basis of a statistical framework with which to confront planet formation theories.

We can grossly summarize the exoplanet constraints as follows:
\begin{itemize}
\item {\bf Occurrence and demographics.}  Over the past few decades it has been shown using multiple techniques that exoplanets are essentially ubiquitous~\cite{mayor11,cassan12,batalha13}. Despite the observational biases, a huge diversity of planetary systems has been discovered.  Yet when drawing analogies with the Solar System, it is worth noting that, if our Sun were to have been observed with present-day technology Jupiter is the only planet that could have been detected~\cite{morbyraymond16,raymond18d}. This makes the Solar System unusual at roughly the1\% level.  In addition, the Solar System is borderline unusual in {\em not} containing any close-in low-mass planets~\cite{martin15,mulders18}. For the purposes of this review we focus on two categories of planets: gas giants and close-in low-mass planets, made up of high-density `super-Earths' and puffy `mini-Neptunes'.
\item {\bf Gas giant planets: occurrence and orbital distribution.}  Radial velocity surveys have found giant planets to exist around roughly 10\% of Sun-like stars~\cite{cumming08,mayor11}.  Roughly one percent of Sun-like stars have {\em hot Jupiters} on very short-period orbits \cite{howard10,wright11}, very few have {\em warm Jupiters} with orbital radii of up to 0.5-1 au \cite{butler06,udry07b}, and the occurrence of giant planets increases strongly and plateaus between 1 to several au, and there are hints that it decreases again farther out~\cite{mayor11,fernandes19}; see Fig.~\ref{fig:giant_freq}. Direct imaging surveys have found a dearth of giant planets on wide-period orbits, although only massive young planets tend to be detectable~\cite{bowler18}. Microlensing surveys find a similar overall abundance of gas giants as radial velocity surveys and have shown that ice giant-mass planets appear to be far more common than their gas giant counterparts~\cite{gould10,suzuki16b}.  Giant planet occurrence has also been shown to be a strong function of stellar metallicity, with higher metallicity stars hosting many more giant planets~\cite{gonzalez97,santos01,laws03,fischer05,dawson13}.
\item {\bf Close-in low-mass planets: occurrence and orbital distribution.}  Perhaps the most striking exoplanet discovery of the past decade was the amazing abundance of close-in small planets.  Planets between roughly Earth and Neptune in size or mass with orbital periods shorter than 100 days have been shown to exist around roughly 30-50\% of all main sequence stars~\cite{mayor11,howard12,fressin13,dong13,petigura13,winn15}. Both the masses and radii have been measured for a subset of planets~\cite{marcy14} and analyses have shown that the smaller planets tend to have high densities and the larger ones have low densities, which has been interpreted as a transition between rocky `super-Earths' and gas-rich `mini-Neptunes' with a transition size or mass of roughly $1.5-2 \rearth$ or $\sim 3-5 \mearth$~\cite{weiss13,weiss14,rogers15,wolfgang16,chen17}.  For the purposes of this review we generally lump together all close-in planets smaller than Neptune and call them {\em super-Earths} for simplicity.  The super-Earth population has a number of intriguing characteristics that constrain planet formation models.  While they span a range of sizes, within a given system super-Earths tens to have very similar sizes~\cite{millholland17,weiss18}.  Their period ratios form a broad distribution and do not cluster at mean motion resonances\cite{lissauer11b,fabrycky14}.  Finally, in the Kepler survey the majority of super-Earth systems only contain a single super-Earth~\cite{batalha13,rowe14}, which contrasts with the high-multiplicity rate found in radial velocity surveys~\cite{mayor11}.  
\end{itemize}

\paragraph{Outline of this review}

The rest of this chapter is structured as follows.  

In Section 2 we will describe six essential mechanisms of planet formation.  These are:
\begin{itemize}
\item The structure and evolution of protoplanetary disks (Section 2.1)
\item The formation of planetesimals (Section 2.2)
\item Accretion of protoplanets (Section 2.3)
\item Orbital migration of growing planets (Section 2.4)
\item Gas accretion and giant planet migration (Section 2.5)
\item Resonance trapping during planet migration (Section 2.6)
\end{itemize}
This list does not include every process related to planet formation.  These processes have been selected because they are both important and are areas of active study.  

Next we will build global models of planet formation from these processes (Section 3).  We will first focus on models to match the intriguing population of close-in small/low-mass planets: the {\em super-Earths} (Section 3.1).  Next we will turn our attention to the population of giant planets, using the population of known giant exoplanets to guide our thinking about the formation of our Solar System's giant planets (Section 3.2).  Next we will turn our attention to matching the Solar System itself (Section 3.3), starting from the classical model of terrestrial planet formation (Section 3.3.1) and then discussing newer ideas: the Grand Tack (Section 3.3.2), Low-mass Asteroid belt (Section 3.3.3) and Early Instability (Section 3.3.4) models. Then we will discuss the different sources of water on rocky exoplanets, and use cosmochemical measurements to constrain the origin of Earth's water (Section 3.4).  Finally, in Section 4 we will lay out a path for the future trajectory of planet formation studies.

\section{Key processes in planet formation} 

In this Section we review basic properties that affect the disk's structure, planetesimals and planet formation as well as dynamical evolution. These processes build the skeleton of our understanding of how planetary systems are formed. As pieces of a puzzle, they will then be put together to develop models on the origin of the different observed structures of planetary systems in Sec.~\ref{sec:global}.

\subsection{Protoplanetary disks: structure and evolution}

Planet formation takes place in gas-dominated disks around young stars. These disks were inferred by Laplace~\cite{Laplace} from the near-perfect coplanarity of the orbits of the planets of the Solar System and of angular momentum conservation during the process of contraction of gas towards the central star. Disks are now routinely observed (imaged directly or deduced through the infrared excess in the spectral energy distribution) around young stars. The largest among protoplanetary disks are now resolved by the ALMA mm-interferometer (\cite{andrews18}). Here we briefly review the viscous-disk model and the wind-dominated model. For more in depth reading we recommend \cite{armitage11} and \cite{turner14}.

\subsubsection{Viscous-disk model(s)}
\label{viscous-disk}

The simplest model of a protoplanetary disk is a donut of gas and dust in rotation around the central star  evolving under the effect of its internal viscosity.  This is hereafter dubbed the {\em viscous-disk model}. Because of Keplerian shear, different rings of the disk rotate with different angular velocities, depending on their distance from the star. Consequently, friction is exerted between adjacent rings. The inner ring, rotating faster, tends to accelerate the outer ring (i.e. it exerts a positive torque) and the outer ring tends to decelerate the inner ring (i.e. exerting a negative torque of equal absolute strength). It can be demonstrated (see for instance \cite{hahn09}) that such a torque is
\begin{equation}
T=3\pi\Sigma\nu r^2 \Omega \ ,
\label{torque}
\end{equation}
where $\Sigma$ is the surface density of the disk at the bounday between the two rings, $\nu$ is the viscosity,  and $\Omega$ is the rotational frequency at the distance $r$ from the star.

A fundamental assumption of a viscous-disk model is that it is in steady state, which means the the mass flow of gas $\dot{M}$ is the same at any distance $r$. Under this assumption it can be demonstrated that the gas flows inwards with a radial speed
\begin{equation}
v_r=-{{3}\over{2}} {{\nu}\over{r}}
\label{v_r}
\end{equation}
and that the product $\nu\Sigma$ is independent of $r$. That is, the radial dependence of $\Sigma$ is the inverse of the radial dependence of $\nu$. Of course the steady-state assumption is valid only in an infinite disk. In a more realistic disk with a finite size, this assumption is good only in the inner part of the disk, whereas the outer part expands into the vacuum under the effect of the viscous torque \cite{lyndenbell74}.

If viscosity rules the radial structure of the disk, pressure rules the vertical structure. At steady state, the disk has to be in hydrostatic equilibrium, which means that the vertical component of the gravitational force exerted by the star has to be equal and opposite to the pressure force, i.e.:
\begin{equation}
{{GM_*}\over{r^3}}z=-{{1}\over{\rho}}{{{\rm d}P}\over{{\rm d}z}} \ ,
\label{vertical}
\end{equation}
where $M_*$ is the mass of the star, $z$ is the height over the disk's midplane, $\rho$ is the volume density of the disk and $P$ is its pressure. Using the perfect gas law $P={\cal R}/ \rho T/\mu$ (where $\cal{R}$ is the gas constant, $\mu$ is the molecular weight of the gas and $T$ is the temperature) and assuming that the gas is vertically isothermal (i.e. $T$ is a function of $r$ only), equation (\ref{vertical} gives the solution:
\begin{equation}
\rho(z)=\rho(0)\exp\left(-{{z^2}\over{2H^2}}\right) \ ,
\label{rho}
\end{equation}
where $H=\sqrt{{\cal R} r^3 T/\mu}$ is called the {\it pressure scale-height} of the disk (the gas extends to several scale-heights, with exponentially vanishing density).

We now need to compute $T(r)$. The simplest way is to assume that the disk is solely heated by the radiation from the central star (passive disk assumption, see \cite{chiang97}). Most of the disk is opaque to radiation, so the star can illuminate and deposit heat only on the surface layer of the disk, here defined as the layer where the integrated optical depth along a stellar ray reaches unity. For simplicity we assume that the stellar radiation hits a hard surface, whose height over the midplane is proportional to the pressure scale-height $H$. Then, the energy deposited on this surface between $r$ and $r+\delta r$ from the star is:
\begin{equation}
E_+^{Irr}= \left({{L_*}\over{4\pi r^2}}\right) (2\pi r) (r\delta h)\ ,
\label{Eplus}
\end{equation}
where $\delta h$ is the change in aspect ratio $H/r$ over the range $\delta r$, namely ${\rm d}(H/r)/{\rm d}r \delta r$; the parentheses have been put in (\ref{Eplus}) to regroup the terms corresponding respectively to (i) stellar brightness $L_*$ at distance $r$, (ii) circumference of the ring and (iii) projection of $H(r+\delta r)- H(r)$ on the direction orthogonal to the stellar ray hitting the surface. On the other had, the same surface will cool by black-body radiation in space at a rate
\begin{equation}
E_-^{Irr}=2\pi r \delta r  \sigma T^4
\label{Eminus}
\end{equation}
where $\sigma$ is Boltzman's constant. Equating (\ref{Eplus}) and (\ref{Eminus}) and remembering the definition of $H$ as a function of $r$ and $T$ leads to
\begin{equation}
T(r)\propto r^{-3/7} \ {\rm and} \quad H/r\propto r^{2/7} \ .
\label{passive}
\end{equation}
The positive exponent in the dependence of the aspect ratio $H/r$ on $r$ implies that the disk is {\it flared}. Notice that neither quantities in Eq.~\ref{passive} depend on disk's surface density, opacity or viscosity.

However, because we are dealing with a viscous disk, we cannot neglect the heat released by viscous dissipation, i.e. in the friction between adjacent rings rotating at different speeds. Over a radial width $\delta r$, this friction dissipates energy at a rate (\cite{armitage11}
\begin{equation}
E_+^{Visc}={9\over 8} \Sigma\nu \Omega^2 2\pi r \delta r \ .
\label{Eplusnu}
\end{equation}
This heat is dissipated mostly close to the midplane, where the disk's volume density is highest. This changes the cooling with respect to Eq.~\ref{Eminus}. The energy cannot be freely irradiated in space; it has first to be transported from the midplane through the disk, which is opaque to radiation, to the ``surface'' boundary with the optically thin layer. Thus the cooling term in Eq.~\ref{Eminus} has to be divided by $\kappa\Sigma$, where $\kappa$ is the disk's opacity. Again by balancing heating and cooling and the definition of $H$ we find:
\begin{equation}
H/r\propto (\dot{M}^2\kappa/\nu r)^{1/8} \ ,
\label{active}
\end{equation}
where we have used that $\dot{M}=2\pi r v_r \Sigma= -3\pi\nu\Sigma$.

To know the actual radial dependence of this expression, we need to know the radial dependences of $\kappa$ and $\nu$ (remember that $\dot{M}$ is assumed to be independent of $r$). The opacity depends on temperature, hence on $r$ in a complicated manner, with abrupt transitions when the main chemical species (notably water) condense \cite{bell94}. Let's ignore this for the moment. Concerning $\nu(r)$, Shakura and  Sunyaev \cite{shakura73} proposed from dimensional analysis that the viscosity is proportional to the square of the characteristic length of the system and is inversely proportional to the characteristic timescale. At a distance $r$ from the star, the characteristic length of a disk is $H(r)$ and the characteristic timescale is the inverse of the orbital frequency $\Omega$. Thus they postulated $\nu=\alpha H^2\Omega$, where $\alpha$ is an unknown coefficient of proportionality. If one adopts this prescription for the viscosity, the viscous-disk model is qualified as an {\it $\alpha$-disk model}. Injecting this definition of $\nu$ into Eq.~\ref{active} one obtains
\begin{equation}
H/r\propto \left({{\kappa \dot{M}^2}\over{\alpha}}\right)^{1/10} r^{1/20}\ .
\label{finalh}
\end{equation}
This result implies that the aspect ratio of a viscously heated disk is basically independent of $r$ (and $T\propto 1/r$), in sharp contrast with the aspect ratio of a passive disk. Because the disk is both heated by viscosity and illuminated by the star, its aspect ratio at each $r$ will be the maximum between Eq.~\ref{finalh} and Eq.~\ref{passive}: it will be flat in the inner part and flared in the outer part. Because Eq.~\ref{finalh} depends on opacity, accretion rate and $\alpha$, the transition from the flat disk and the flared disk will depend on these quantities. In particular, given that $\dot{M}$ decreases with time as the disk is consumed by accretion of gas onto the star \cite{hartmann98}, this transition moves towards the star as time progresses \cite{bitsch15}. The effects of non-constant opacity introduce wiggles of $H/r$ over this general trend \cite{bitsch15}.

The viscous disk model is simple and neat, but its limitation is in the understanding of the origin of the disk's viscosity. The molecular viscosity of the gas is by orders of magnitude insufficient to deliver the observed accretion rate $\dot{M}$ onto the central star \cite{hartmann98} given a reasonable disk's density, comparable to that of the Minimum Mass Solar Nebula model (MMSN: \cite{weidenschilling77}, \cite{hayashi81}). It was thought that the main source of viscosity is turbulence and that turbulence was generated by the magneto-rotational instability (see \cite{balbus98}). But this instability requires a relatively high ionization of the gas, which is prevented when grains condense in abundance, at a temperature below $\sim 1,000$K \cite{desch15}. Thus, only the very inner part of the disk is expected to be turbulent and have a high viscosity. Beyond the condensation line of silicates, the viscosity should be much lower. Remembering that $\nu\Sigma$ has to be constant with radius, the drop of $\nu$ at the silicate line implies an abrupt increase of $\Sigma$. As we will see, this property has an important role in the drift of dust and the migration of planets. It was expected that near the surface of the disk, where the gas is optically thin and radiation from the star can efficiently penetrate, enough inoization may be produced to sustain the Magneto Rotational Instability (MRI) \cite{stone98}. However, these low-density regions are also prone to other effects of non-ideal magneto-hydrodynamics (MHD), like ambipolar diffusion and the Hall effect \cite{bai13}, which are expected to quench turbulence. Thus, turbulent viscosity does not seem to be large enough beyond the silicate condensation line to explain the stellar accretion rates that are observed. This has promoted an alternative model of disk structure and evolution, dominated by the existence of disk winds, as we review next.

\subsubsection{Wind-dominated disk models}

Unlike viscous disk models, which can be treated with simple analytic formulae, the emergence of disk winds and their effects are consequence of non-ideal MHD. Thus their mathematical treatment is complicated and the results can be unveiled only through numerical simulations. Thus, this Section will remain at a phenomenological level. For an in-depth study we recommend \cite{bai13,bai17,turner14}.

As we have seen above, the ionized regions of the disk have necessarily a low density. If the disk is crossed by a magnetic field, the ions atoms in these low-density regions can travel along the magnetic field lines without suffering collision with neutral molecules. This is the essence of {\it ambipolar diffusion}.

Consider a frame co-rotating with the disk at radius $r_0$ from the central star.  In this frame,  fluid parcels feel an effective potential combining the gravitational and centrifugal potentials.  If poloidal $(r,z)$ magnetic field  lines  act  as  rigid  wires  for  fluid  parcels (which happens as long as the poloidal flow is slower than the local Alfven speed)\footnote{The Alfven speed is defined as the ratio between the magnetic field intensity B and $\sqrt{\mu_0\rho}$, where $\mu_0$ is the permeability of vacuum and $\rho$ is the total mass density of the charged plasma particles. Apart from relativity effects, the Alfven speed is the phase speed of the Alfven wave, which is a magnetohydrodynamic wave in which ions oscillate in response to a restoring force provided by an effective tension on the magnetic field lines.},  then  a parcel initially at rest at $r=r_0$ can undergo a runaway if the field line to which it is attached is more inclined than a critical angle.  Along such a field line, the effective potential decreases with distance, leading to an acceleration of magnetocentrifugal origin.  This yields  the  inclination  angle  criterion $\theta >30^\circ$ for  the disk-surface poloidal field with respect to the vertical direction. Here  fluid  parcels  rotate  at constant  angular  velocity and so increase their specific angular momentum. Angular  momentum  is  thus  extracted  from  the disk  and transferred  to  the  ejected material.  As the disk loses angular momentum, some material has to be transferred towards the central star, driving the stellar accretion. The  efficiency  of this process  is  directly  connected  to  the  disk's  magnetic field strength, with a stronger field leading to faster accretion.

\begin{figure*}
\centering
\includegraphics[height=10cm]{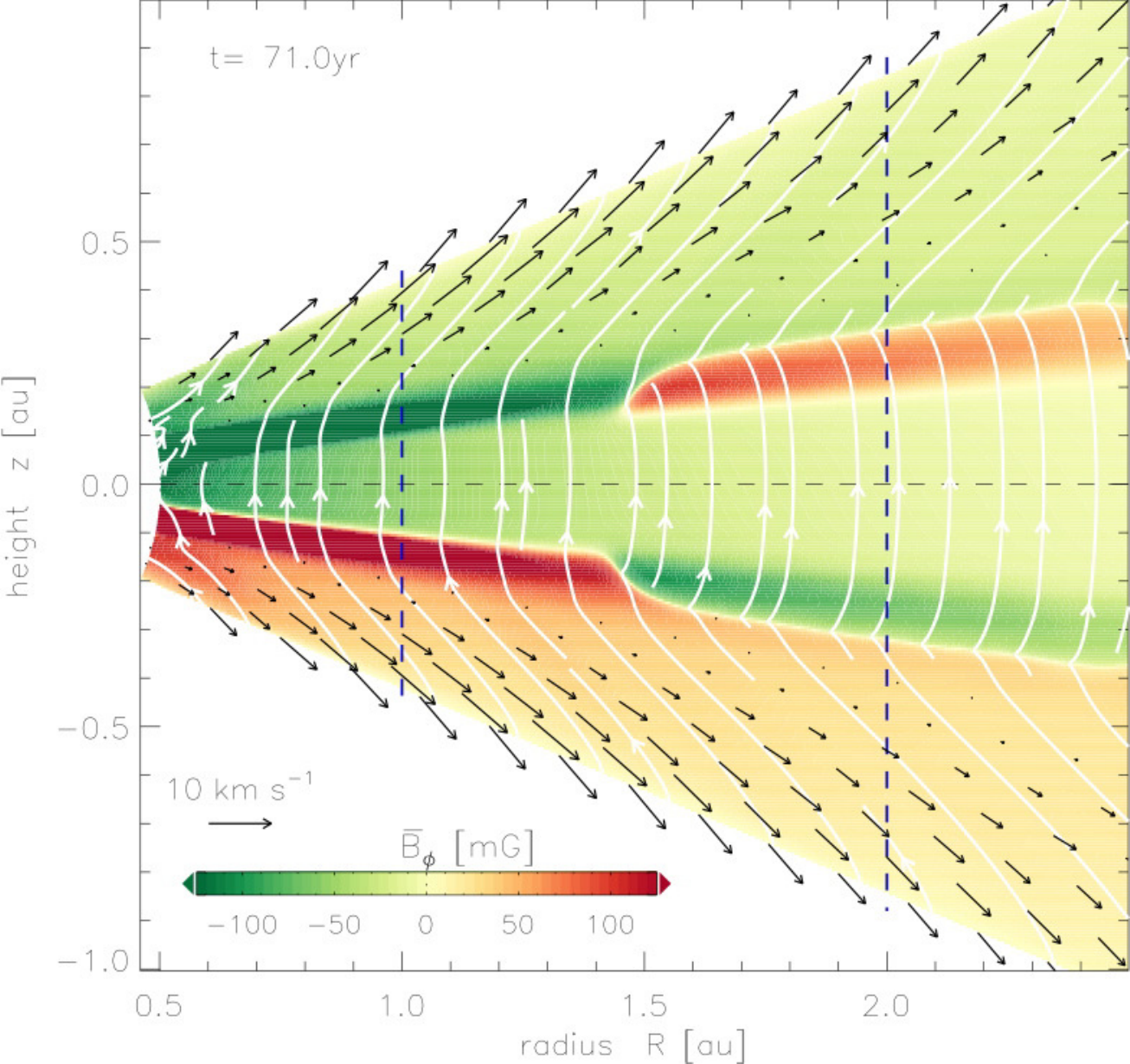} \includegraphics[height=10cm]{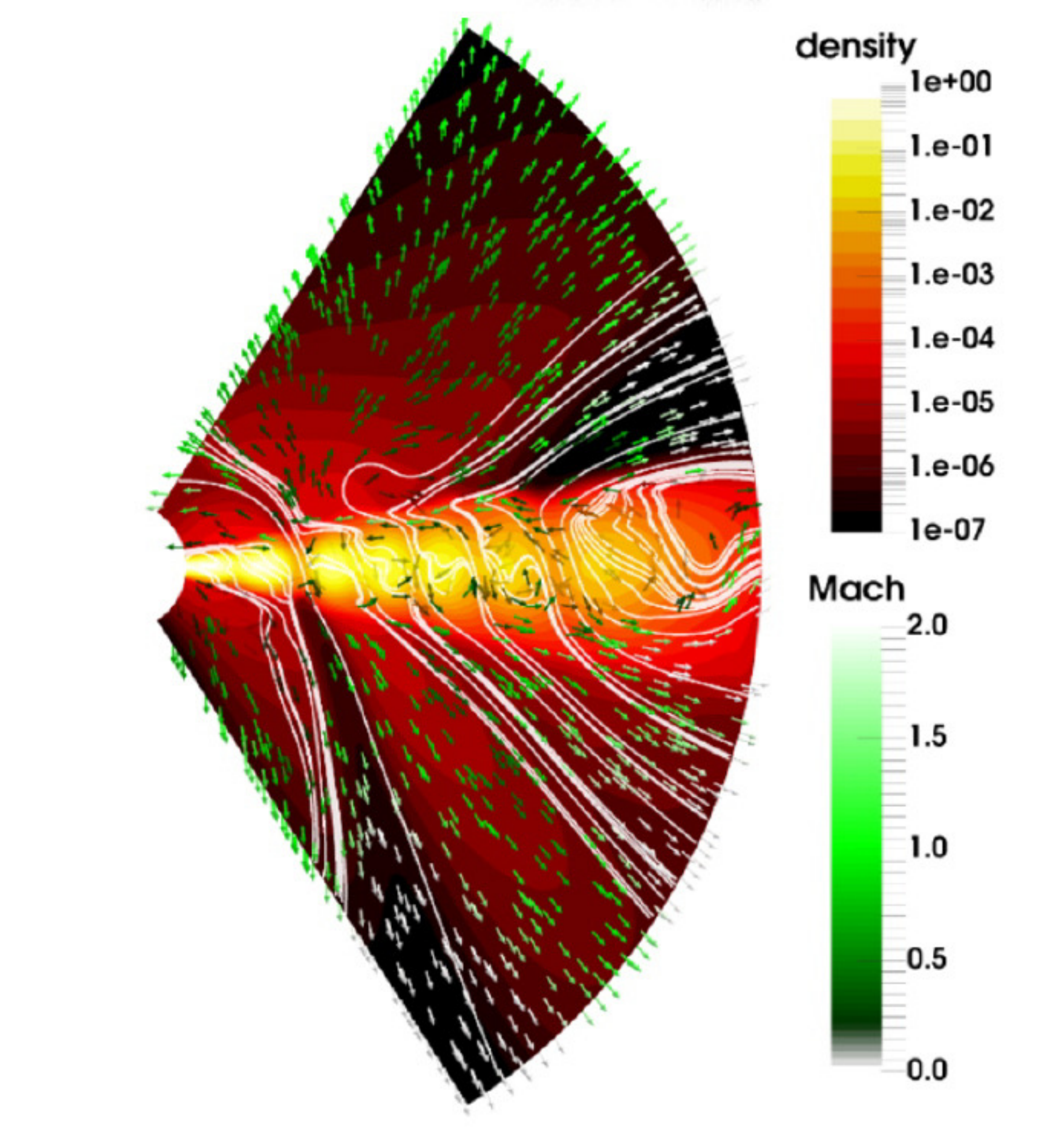}
\caption{Disk structure in two different wind-dominated disk global models. In the top panel from \cite{gressel15} shows the intensity and polarity of the magnetic field, the field lines and the velocity vectors of the wind. This disk model is symmetric relative to the mid-plane (anti-symmetric in polarity). The bottom panel from \cite{bethune17} shows the gas density, the magnetic field lines and the wind velocity vectors. This model has no symmetry relative to the midplane. Moreover, the disk is fragmented in rings by the accumulation of magnetic field lines in specific radial intervals.}
\label{winds}
\end{figure*}

In wind-dominated disk models the viscosity can be very low and the disk's density can be of the order of that of the MMSN. The observed accretion rate onto the central star is not due to viscosity (the small value of $\nu\Sigma$ provides only a minor contribution) but is provided by the radial, fast advection of a small amount of gas, typically at 3-4 pressure scale heights $H$ above the disk's midplane \cite{gressel15}. The global structure of the disk can be symmetric (Fig.~\ref{winds}a) or asymmetric (Fig.~\ref{winds}b) depending on simulation parameters and the inclusion of different physical effects (e.g. the Hall effect). The origin of the asymmetry is not fully understood. In some cases, the magnetic field lines can be concentrated in narrow radial bands that can fragment the disk into concentric rings (\cite{bethune17}; see Fig.~\ref{winds}b). This effect is intriguing in light of the recent ALMA observations of the ringed structure of protoplanetary disks \cite{alma15,andrews16,andrews18}.  It should be stressed, however, that there is currently no consensus on the origin of the observed rings. An alternative possibility is that they are the consequence of planet formation \cite{Zhang18} or of other disk instabilities \cite{tominaga19}. Understanding whether ring formation in protoplanetary disks is a prerequisite for, or a consequence of, planet formation is an essential goal of current research. 

\begin{figure}[t]
\centering
\includegraphics[width=12cm]{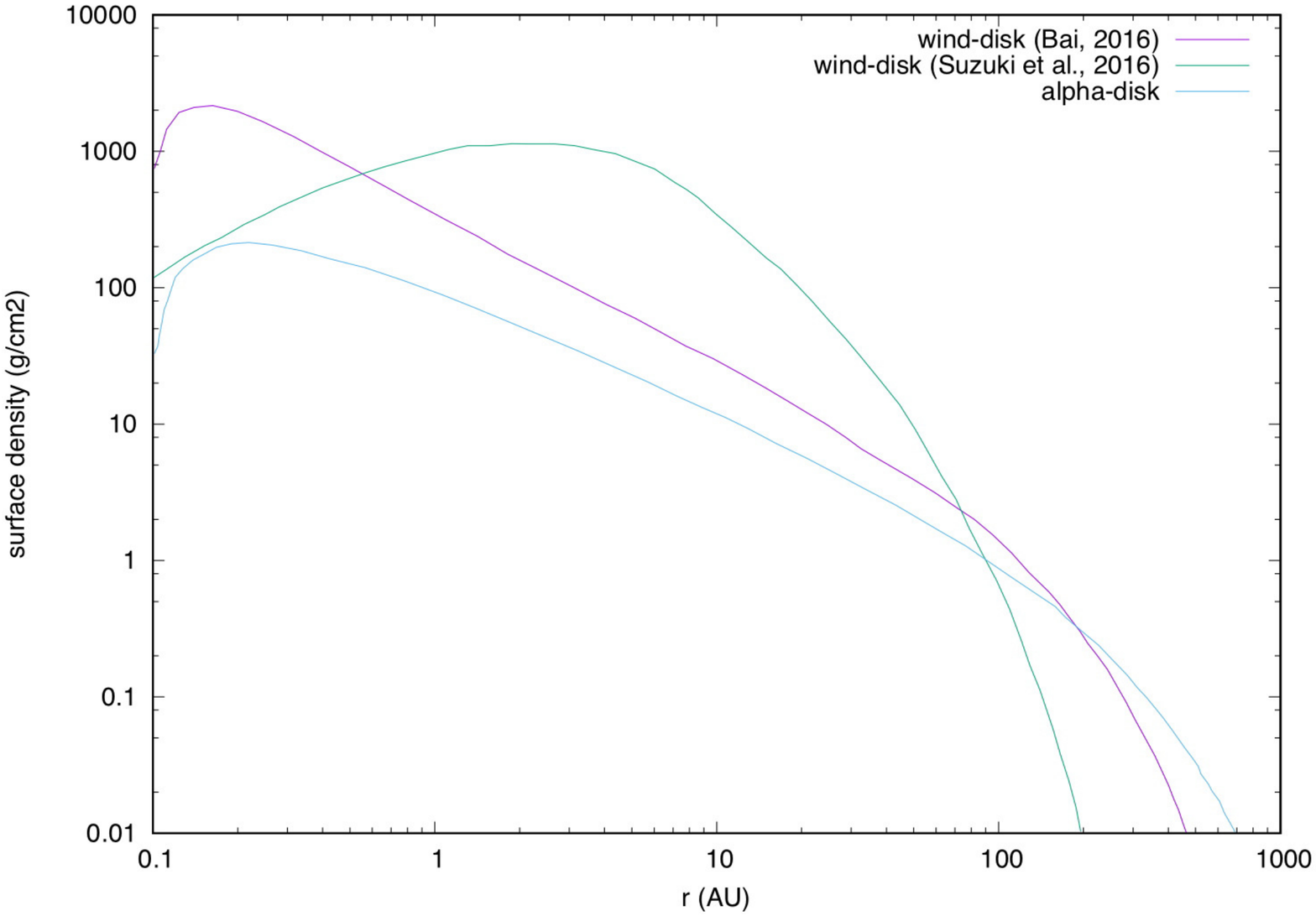}
\caption{Comparison between the radial surface density distribution of the wind-dominated disk models of (\cite{suzuki16}), (\cite{Bai16}) and an $\alpha$-disk model.}
\label{suzukibai}
\end{figure}

Depending on assumptions on the radial gradient of the magnetic field, and hence of the strength of the wind, the gas density of the disk may be partially depleted in its inner part \cite{suzuki16} or preserve a global power-law structure similar to that of viscous-disk models \cite{Bai16}  (see Fig.~\ref{suzukibai}). A positive surface density gradient as in \cite{suzuki16} has implication for the radial drift of dust and planetesimals \cite{ogihara18b} and for the migration of protoplanets \cite{ogihara18}. In addition in \cite{suzuki16} the maximum of the surface density (where dust and migrating planets tend to accumulate) moves outwards as time progresses and the disk evolves.

Disk winds do not generate an appreciable amount of heat. In wind-dominated models the disk temperature is close to that of a passive disk \cite{mori19,chambers19}, which has a snowline inward of 1 au \cite{bitsch15}. The deficit of water in inner solar system bodies (the terrestrial planets and the parent bodies of enstatite and ordinary chondrites) demonstrates that the protoplanetary solar disk inwards of 2-3 au was warm, at least initially \cite{morby16}. This implies that either the viscosity of the disk was quite high or another form of heating -- for instance the adiabatic compression of gas as it fell onto the disk from the interstellar medium -- was operating early on.  

\subsubsection{Dust dynamics}

The dynamics of dust particles is largely driven by gas drag. Any time that there is a difference in velocity between the gas and the particle, a drag force is exerted on the particle which tends to erase the velocity difference. The {\it friction time} $t_f$ is defined as the coefficient which relates the accelerations felt by the particle to the gas-particle velocity difference, namely:
\begin{equation}
\dot{\vec{v}}=-{1\over{t_f}} (\vec{v}-\vec{u}) \ ,
\end{equation}
where $\vec{v}$ is the particle velocity vector and $\vec{u}$ is the gas velocity vector, while $\dot{\vec{v}}$ is the particle's acceleration. The smaller a dust particle the shorter its $t_f$. In the Epstein regime, where the particle size is smaller than the mean free path of a gas molecule, $t_f$ is linearly proportional to the particle's size $R$. In the Stokes regime the particle size is larger than the mean free path of a gas molecule and  $t_f\propto \sqrt{R}$. It is convenient to introduce a dimensionless number, called the {\it Stokes number}, defined as $\tau_s=\Omega t_f$, which represents the ratio between the friction time and the orbital timescale.

The effects of gas drag are mainly the sedimentation of dust towards the midplane and its radial drift towards pressure maxima.

To describe an orbiting particle as it settles in a disk, cylindrical coordinates are the natural choice. The stellar gravitational force can be decomposed into a radial and a vertical component. The radial component is cancelled by the centrifugal force due to the orbital motion. The vertical component, $F_{g,z}= -m \Omega^2z$, where $m$ is the mass of the particle and $z$ its vertical coordinate, instead accelerates the particle towards the midplane, until its velocity $v_{settle}$ is such that the gas drag force $F_D=m v_{settle}/t_f$ cancels $F_{g,z}$. This sets $v_{settle}=\Omega^2z t_f$ and gives a settling time $T_{settle}=z/v_{settle}=1/(\Omega^2 t_f)=1/(\Omega\tau_s)$. Thus, for a particle with Stokes number $\tau_s=1$ the settling time is the orbital timescale. However, turbulence in the disk stirs up the particle layer, which therefore has a finite thickness. Assuming an $\alpha$-disk model, the scale height of the particle layer is \cite{chiang10}:
\begin{equation}
H_p={H\over{\sqrt{1+\tau_s/\alpha}}} \ .
\label{Hp}
\end{equation}

Dust particles undergo radial drift due to a small difference of their orbital velocity relative to the gas. The gas feels the gravity of the central star and its own pressure. The pressure radial gradient exerts a force $F_r=-(1/\rho) {\rm d}P/{\rm d}r$ which can oppose or enhance the gravitational force. As we saw above, $P\propto \rho T$, and $\rho\sim \Sigma/H$. Because $\Sigma$ and $T$ in general decrease with $r$, ${\rm d}P/{\rm d}r<0$. The pressure force opposes to the gravity force, diminishing it. Consequently, the gas parcels orbit the star at a speed that is slightly slower than the Keplerian speed at the same location. The difference between the Keplerian speed $v_k$ and the gas orbital speed is $\eta v_K$, where
\begin{equation}
\eta=-{1\over{2}}\left({H\over{r}}\right)^2{{{\rm d}\log P}\over{{\rm d}\log r}}
\label{eta}
\end{equation}

The radial velocity of a particle is then:
\begin{equation}
v_r= -2\eta v_K \tau_s/(\tau_s^2-1) + u_r/(\tau_s^2-1) \ ,
\label{drift}
\end{equation}
where $u_r$ is the radial component of the gas velocity. Except for extreme cases $u_r$ is very small and hence the second term in the right-hand side of Eq.~\ref{drift} is negligible relative to the first term. Consequently, the direction of the radial drift of dust depends on the sign of $\eta$, i.e., from Eq.~\ref{eta}, on the sign of the pressure gradient. If the gradient is negative as in most parts of the disk, the drift is inwards. But in the special regions where the pressure gradient is positive, the particle's drift is outwards. Consequently, dust tends to accumulate at pressure maxima in the disk. We have seen above that the MHD dynamics in the disk can create a sequence of rings and gaps, where the density is alternatively maximal and minimal (Fig.~\ref{winds}b). Each of these rings therefore features a pressure maximum along a circle. In absence of diffusive motion, the dust would form an infinitely thin ring at the pressure maximum. Turbulence produces diffusion of the dust particles in the radial direction, as it does in the vertical direction. Thus, as dust sedimentation produces a layer with thickness given by Eq.~\ref{Hp}, dust migration produces a ring with radial thickness $w_p=w/\sqrt{1+\tau_s/\alpha}$ around the pressure maximum, where $w$ is the width of the gas ring assuming that it has a Gaussian profile \cite{dullemond18}.

Consequently, observations of the dust distribution in protoplanetary disks can provide information on the turbulence in the disk. The fact that the width of the gaps in the disk of HL Tau appears to be independent of the azimuth despite the fact that the disk is viewed with an angle smaller than $90^\circ$, suggests that the vertical diffusion of dust is very limited such that $\alpha$ in Eq.~\ref{Hp} must be be $10^{-4}$ or less \cite{pinte16}. In contrast, the observation that dust is quite broadly distributed in each ring of the disks, suggest that $\alpha$ could be as large as $10^{-3}$, depending on the particles Stokes number $\tau_s$, which is not precisely known \cite{dullemond18}. These observations therefore suggest that turbulence in the disks is such that the vertical diffusion it produces is weaker than the radial diffusion. It is yet unclear which mechanism could generate turbulence with this property. 

\subsection{Planetesimal formation}

Dust particles orbiting within a disk often collide. If collisions are sufficiently gentle, they stick through electrostatic forces, forming larger particles \cite{blum08}. One could imagine that this process continues indefinitely, eventually forming macroscopic bodies called planetesimals. However, as we have seen above, particles drift through the disk at different speeds depending on their size (or Stokes number). Thus, there is a minimum speed at which particles of different sizes can collide. Particles of equal size also have a distribution of impact velocities due to turbulent diffusion.

\begin{figure}[t]
\centering
\includegraphics[width=12cm]{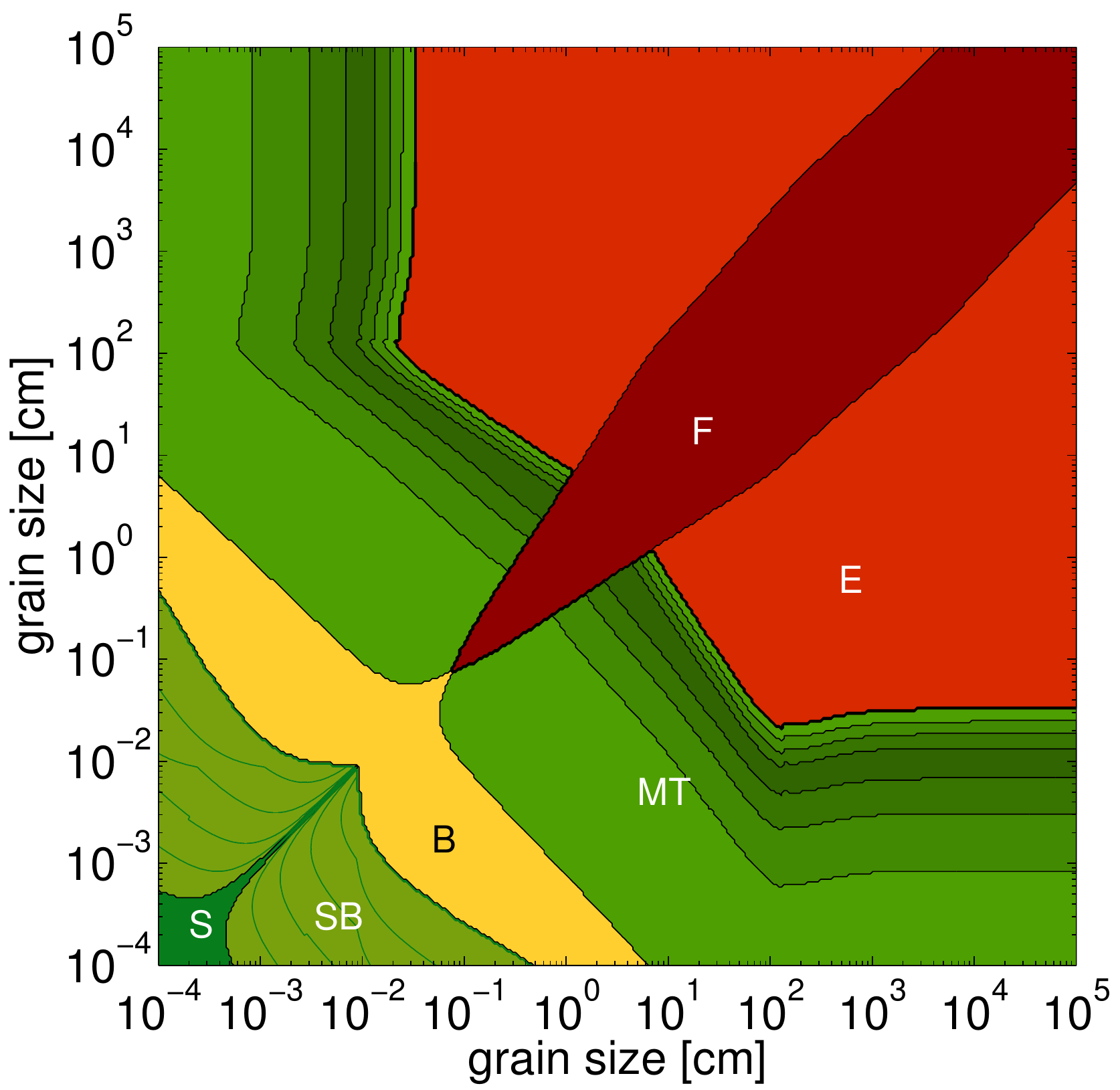}
\caption{Map of collisional outcome in the disk. The sizes of colliding particles are reported on the axes. The colours denote the result of each pair-wise collision. Green denotes growth, red denotes erosion and yellow denotes neither of the above (i.e. a bounce). The label S stands for sticking, SB for stick and bounce, B for bounce, MT for mass transfer, E for erosion and F for fragmentation. Form (\cite{windmark12}).  This map is computed for compact (silicate) particles, at 3 au.}
\label{Windmark-fig}
\end{figure}

Fig.~\ref{Windmark-fig} shows a map of the outcome of dust collisions within a simple disk model, from \cite{windmark12}. Using laboratory experiments on the fate of collisions as a function of particle sizes and mutual velocities, and considering a disk with turbulent diffusion $\alpha=10^{-3}$ and drift velocities as in a MMSN disk, Windmark et al. computed the growth/disruption maps for different heliocentric distances. The one from Fig.~\ref{Windmark-fig} is for a distance of 3 au. The figure shows that, in the inner part of the disk, particles cannot easily grow beyond a millimeter in size. A bouncing barrier prevents particles to grow beyond this limit. If a particle somehow managed to grow to $\sim 10$cm, its growth could potentially resume by accreting tiny particles. But as soon as particles of comparable sizes hit each other, erosion or catastrophic fragmentation occurs, thus preventing the formation of planetesimal-size objects.

The situation is no better in the outer parts of the disk.  In the colder regions, due to the lower velocities and the sticking effect of water ice, particles can grow to larger sizes.  But this size is nevertheless limited to a few centimeters due to the so-called {\it drift barrier} (i.e. large enough particles start drifting faster than they grow: \cite{birnstiel16}). It has been proposed that if particles are very porous, they could absorb better the collisional energy, thus continuing to grow without bouncing or breaking \cite{okuzumi12}. Very porous planetesimals could in principle form this way and their low densities would make them drift very slowly through the disk. But eventually these planetesimals would become compact under the effect of their own gravity and of the ram pressure of the flowing gas \cite{kataoka13}.  This formation mechanism for planetesimals is still not generally accepted in the community. At best, it could work only in the outer part of the disk, where icy monomers have the tendency to form very porous structures, but not in the inner part of the disk, dominated by silicate particles. Moreover, meteorites show that the interior structure of asteroids is made mostly of compact particles of 100 microns to a millimeter in size, called chondrules, which is not consistent with the porous formation mode.

A mechanism called the {\it streaming instability} \cite{youdin05} can bypass these growth bottlenecks to form planetesimals. Although originally found to be a linear instability (see \cite{jacquet11}), this instability raises even more powerful effects which can be qualitatively explained as follows. This instability arises from the speed difference between gas and solid particles. As the differential makes particles feel drag, the friction exerted from the particles back onto the gas accelerates the gas toward the local Keplerian speed. If there is a small overdensity of particles, the local gas is in a less sub-Keplerian rotation than elsewhere; this in turn reduces the local headwind on the particles, which therefore drift more slowly towards the star. Consequently, an isolated particle located farther away in the disc, feeling a stronger headwind and drifting faster towards the star, eventually joins this overdense region. This enhances the local density of particles and reduces further its radial drift. This process drives a positive feedback, i.e. an instability, whereby the local density of particles increases exponentially with time. 

\begin{figure*}
\centering
\includegraphics[width=12cm]{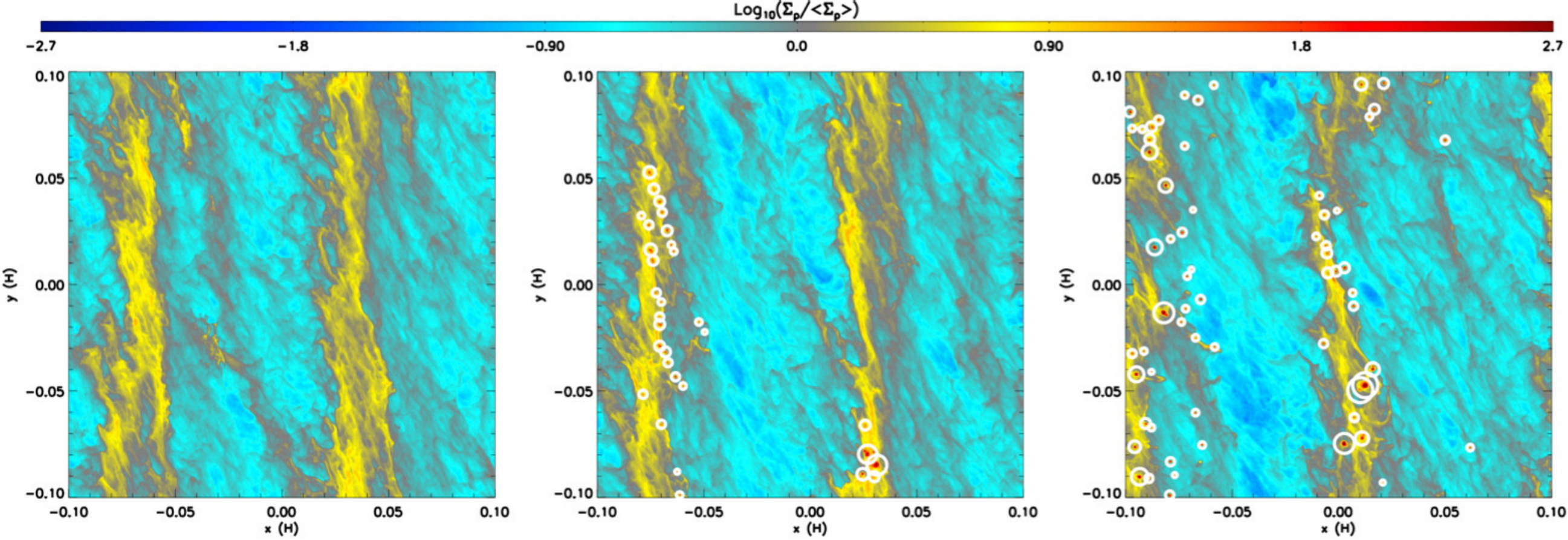}
\caption{Snapshots in time of a simulation of the streaming instability (from \cite{simon16}). The color scale shows the vertically integrated particle surface density normalized to the average particle surface density (log scale). Time increases from left to right. The left panel shows the clumping due to the streaming instability in the absence of self-gravity but right before self-gravity is activated ($t= 110\Omega^{−1}$). The middle panel corresponds to a point shortly after self-gravity was activated ($t= 112.5\Omega^{−1}$), and the right panel corresponds to a time in which most of the planetesimals have formed ($t= 117.6\Omega^{-1}$). In the middle and right panel, each planetesimal is marked via a circle of the size of the Hill sphere.}
\label{streaming}
\end{figure*}

Particle clumps generated by the streaming instability can become self-gravitating and contract to form planetesimals. Numerical simulations of the streaming instability process \cite{johansen15,simon16,simon17,schaefer17,abod19} show that planetesimals of a variety of sizes are produced, but those that carry most of the final total mass are those of $\sim 100$~km in size. This size is indeed prominent in the observed size-frequency distributions of both asteroids and Kuiper-belt objects. Thus, these models suggest that planetesimals form (at least preferentially) big, in stark contrast with the collisional coagulation model in which planetesimals would grow progressively from pair wise collisions.  If the amount of solid mass in small particles is large enough, even Ceres-size planetesimals can be directly produced from particle clumps (Fig.~\ref{streaming}). 

While Fig.~\ref{streaming} shows that the streaming instability can clearly form planetesimals, a concern arises from the initial conditions of such simulations.  Simulations find that quite large particles are needed for optimal concentration, corresponding to at least decimeters in size when applied to the asteroid belt. Chondrules -- a ubiquitous component of primitive meteorites -- typically have sizes from 0.1 to 1 mm but such small particles are hard to concentrate in vortices or through the streaming instability. High-resolution numerical simulations \cite{carrera15,yang17} show that chondrule-size particles can trigger the streaming instability only if the initial mass ratio between these particles and the gas is larger than about 4\%. The initial solid/gas ratio of the Solar System disk is thought to have been $\sim$1\%. At face value, planetesimals should not have formed as agglomerates of chondrules. A possibility is that future simulations with even-higher resolution and run on longer timescales will show that the instability can occur for a smaller solid/gas ratio, approaching the value measured in the Sun.

Certain locations within the disk may act as preferred sites of planetesimal formation.  Drifting particles may first accumulate at distinct radii in the disc where their radial speed is slowest and then, thanks to the locally enhanced particle/gas ratio, locally trigger the streaming instability. Two locations have been identified for this preliminary radial pile-up. One is in the vicinity of the snowline, where water transitions from vapor to solid form \cite{ida16b,armitage16,schoonenberg17}. The other is in the vicinity of 1 au \cite{drazkowska16}. These could be the two locations where planetesimals could form very early in the proto-planetary disk \cite{drazkowska18}. Elsewhere in the disk, the conditions for planetesimal formation via the streaming instability would only be met later on, when gas was substantially depleted by photo-evaporation from the central star, provided that the solids remained abundant (\cite{throop08} \cite{Carrera17}).

At least at the qualitative level, this picture is consistent with available data for the Solar System. The meteorite record reveals that some planetesimals formed very early (in the first few $10^5$y \cite{kleine09,kruijer14,schiller15}). Because of the large abundance of short-lived radioactive elements present at the early time \cite{grimm93,monteux18}, these first planetesimals melted and differentiated, and are today the parent bodies of iron meteorites. But a second population of planetesimals formed 2 to 4 My later \cite{villeneuve09}. These planetesimals did not melt and are the parent bodies of the primitive meteorites called the chondrites. We can speculate that differentiated planetesimals formed at the two particle pile-up locations mentioned above, whereas the undifferentiated planetesimals formed elsewhere, for instance in the asteroid belt while the gas density was declining. Yet these preferred locations were certainly themselves evolving in time \cite{drazkowska18}.

Strong support for the streaming instability model comes from Kuiper belt binaries. These binaries are typically made of objects of similar size and identical colors (see \cite{noll08}). It has been shown \cite{nesvorny10b} that the formation of a binary is the natural outcome of the gravitational collapse of the clump of pebbles formed in the streaming instability if the angular momentum of the clump is large. Simulations of this process can reproduce the typical semi-major axes, eccentricities and size ratios of the observed binaries. The color match between the two components is a natural consequence of the fact that both are made of the same material. This is a big strength of the model because such color identity cannot be explained in any capture or collisional scenario, given the observed intrinsic difference in colors between any random pair of Kuiper belt objects (KBOs - this statement holds even restricting the analysis to the cold population, which is the most homogenous component of the Kuiper belt population). Additional evidence for the formation of equal-size KBO binaries by streaming instability is provided by the spatial orientation of binary orbits. Observations \cite{noll08} show a broad distribution of binary inclinations with $\simeq$80\% of prograde orbits ($i_{\rm b}<90^\circ$) and $\simeq$20\% of retrograde orbits ($i_{\rm b}>90^\circ$). To explain these observations, Nesvorny et al. \cite{Nesvorny19} analyzed high-resolution simulations and determined the angular momentum vector of the gravitationally bound clumps produced by the streaming instability. Because the orientation of the angular momentum vector is approximately conserved during collapse, the distribution obtained from these simulations can be compared with known binary inclinations. The comparison shows that the model and observed distributions are indistinguishable. This clinches an argument in favor of the planetesimal formation by the streaming instability and binary formation by gravitational collapse. No other planetesimal formation mechanism has been able so far to reproduce the statistics of orbital plane orientations of the observed binaries. 

\subsection{Accretion of protoplanets}

Once planetesimals appear in the disk they continue to grow by mutual collisions. Gravity plays an important role by bending the trajectories of the colliding objects, which effectively increases their collisional cross-section by a factor
\begin{equation}
	F_g = 1+ V_{esc}^2/V_{rel}^2 \ ,
\label{focussing}
\end{equation}
where $V_{esc}$  is the mutual escape velocity  defined as  $V_{esc} = [2G (M_1+M_2)/ (R_1+R_2)]^{1/2}$ , $M_1, M_2, R_1, R_2$ are the masses and radii of the colliding bodies, $V_{rel}$ is their relative velocity before the encounter and $G$ is the gravitational constant. $F_g$ is called the gravitational focussing factor \cite{safronov72}.

The mass accretion rate of an object becomes 
\begin{equation}
	{{{\rm d}M}\over{{\rm d}t}} \propto R^2 F_g   \propto  M^{2/3} F_g
\label{accretion}
\end{equation}
where the bulk density of planetesimals is assumed to be independent of their mass, so that the planetesimal physical radius $R \propto M^{1/3}$. These equations imply two distinct growth modes called runaway and oligarchic growth.

\subsubsection{Runaway growth}
If one planetesimal (of mass $M$) grows quickly then its escape velocity $V_{esc}$ becomes much larger than its relative velocity $V_{rel}$ with respect to the rest of the planetesimal population. Then one can approximate $F_g$ as $V_{esc}^2/V_{rel}^2$. Notice that the approximation $R \propto M^{1/3}$  makes $V_{esc}^2 \propto  M^{2/3}$.

Substituting this expression into Eq.~\ref{accretion} leads to:
\begin{equation}
{{{\rm d}M}\over{{\rm d}t}} \propto {{M^{4/3}}\over{V_{rel}^2}} \ , 
\end{equation}
or, equivalently:
\begin{equation}
{1\over{M}}{{{\rm d}M}\over{{\rm d}t}} \propto{{M^{1/3}}\over{V_{rel}^2}} \ .
\end{equation}
This means that the relative mass-growth rate is a growing function of the body's mass. In other words, small initial differences in mass among planetesimals are rapidly magnified, in an exponential manner. This growth mode is called {\it  runaway growth} \cite{greenberg78,wetherill89,wetherill93,kokubo96,kokubo98}. 

Runaway growth occurs as long as there are objects in the disk for which $V_{esc} \gg  V_{rel}$. While $V_{esc}$ is a simple function of the largest planetesimals' masses, $V_{rel}$ is affected by other processes.  There are two dynamical damping effects that act to decrease the relative velocities of planetesimals. The first is gas drag. Gas drag not only causes the drift of bodies towards the central star, as seen above, but it also tends to circularize the orbits, thus reducing their relative velocities $V_{rel}$. Whereas orbital drift vanishes for planetesimals larger than about 1 km in size, eccentricity damping continues to influence bodies up to several tens of kilometers across. However, in a turbulent disk gas drag cannot damp $V_{rel}$ down to zero: in presence of turbulence the relative velocity evolves towards a size-dependent equilibrium value \cite{ida08}. The second damping effect is that of collisions. Particles bouncing off each other tend to acquire parallel velocity vectors, reducing their relative velocity to zero. For a given total mass of the planetesimal population, this effect has a strong dependence on the planetesimal size, roughly $1/r^4$ \cite{wetherill93}. 

Meanwhile, relative velocities are excited by the largest growing planetesimals by gravitational scattering, whose strength depends on those bodies' escape velocities. A planetesimal that experiences a near-miss with the largest body has its trajectory permanently perturbed and will have a relative velocity $V_{rel} \sim V_{esc}$ upon the next return. Thus, the planetesimals tend to acquire relative velocities of the order of the escape velocity from the most massive bodies, and when this happens runaway growth is shut off (see below)

To have an extended phase of runaway growth in a planetesimal disk, it is essential that the bulk of the solid mass is in small planetesimals, so that the damping effects are important. Because small planetesimals collide with each other frequently and either erode into small pieces or grow by coagulation, this condition may not hold for long. Moreover, if planetesimals really form with a preferential size of $\sim100$~km, as in the streaming instability scenario, the population of small planetesimals would have been insignificant and therefore runaway growth would have only lasted a short time if it happened at all.

\subsubsection{Oligarchic growth}
When the velocity dispersion of planetesimals becomes of the order of the escape velocity from the largest bodies, the gravitational focussing factor (Eq.~\ref{focussing})  becomes of order unity. Consequently the mass growth equation (Eq.~\ref{accretion}) becomes
\begin{equation}
{1\over{M}}{{{\rm d}M}\over{{\rm d}t}} \propto{{M^{-1/3}}\over{V_{rel}^2}} \ .
\end{equation}
In these conditions, the relative growth rate of the large bodies slows with the bodies' growth. Thus, the mass ratios among the large bodies tend to converge to unity.
 
In principle, one could expect that the small bodies also narrow down their mass difference with the large bodies. But in reality, the large value of  $V_{rel}$  prevents the small bodies from accreting each other. Small bodies only contribute to the growth of the large bodies (i.e. those whose escape velocity is of the order of $V_{rel}$ ). This phase is called {\it oligarchic growth} \cite{kokubo98,kokubo00}.

In practice, oligarchic growth leads to the formation of a group of objects of roughly equal masses, embedded in the disk of planetesimals. The mass gap between oligarchs and planetesimals is typically of a few orders of magnitude. Because of dynamical friction -- an equipartition of orbital excitation energy \cite{chandrasehkar43} -- planetesimals have orbits that are much more eccentric than the oligarchs. The orbital separation among the oligarchs is of the order of 5 to 10 mutual Hill radii $R_H$, where:
\begin{equation}
	 R_{H}= {{a_1+a_2}\over{2}} \left({{M_1+M_2}\over{3M_*}}\right)^{1/3} \ ,
         \label{HillRadius}
         \end{equation}
and $a_1, a_2$ are the semi-major axes of the orbits of the objects with masses $M_1, M_2$, and $M_*$ is the mass of the star.

\subsubsection{The need for an additional growth process} 

In the classic view of planet formation \cite{wetherill92,lissauer87,lissauer93} the processes of runaway growth and oligarchic growth convert most of the planetesimals mass into a few massive objects: the protoplanets (sometimes called {\em planetary embryos}). However, this picture does not survive close scrutiny. 

In the Solar system, two categories of protoplanets formed within the few Myr lifetime of the gas component of the protoplanetary disk (see Fig.~\ref{fig:disk_constraints}). In the outer system, a few planets of multiple Earth masses formed and were massive enough to be able to gravitationally capture a substantial mass of H and He from the disk and become the observed giant planets, from Jupiter to Neptune. In the inner disk, instead, the protoplanets only reached a mass of the order of the mass of Mars and eventually formed the terrestrial planets after the disappearance of the gas (see Section 3.3.1). Thus, the proptoplanets in the outer part of the disk were 10-100 times more massive of those in the inner disk. This huge mass ratio is even more surprising if one considers that the orbital periods, which set the natural clock for all dynamical processes including accretion, are ten times longer in the outer disk. 

The snowline represents a divide between the inner and the outer disk. The surface density of solid material is expected to increase beyond the snowline due to the availability of water ice \cite{hayashi81}. However, this density-increase is only of a factor of $\sim 2$ \cite{lodders03}, which is insufficient to explain the huge mass ratio between protoplanets in the outer and inner parts of the disc \cite{morby15}. 

In addition, whereas in the inner disk oligarchic growth can continue until most of the planetesimals have been accreted by protoplanets, the situation is much less favorable in the outer disc. There, when the protoplanets become sufficiently massive (about 1 Earth mass), they tend to scatter the planetesimals away, rather than accrete them. In doing this they clear their neighboring regions, which in turn limits their own growth \cite{levison10}.  In fact, scattering dominates over growth when the ratio $V_{esc}^2/2 V_{orb}^2>1$, where $V_{esc}$ is the escape velocity from the surface of the protoplanet and $V_{orb}$ is its orbital speed (so that $\sqrt{2} V_{orb}$ is the escape velocity from the stellar potential well from the orbit of the protoplanet). This ratio is much larger in the outer disc than in the inner disc because $V_{orb}^2 \propto 1/a$, where $a$ is the orbital semi major axis. 

Consequently, understanding the formation of the multi-Earth-mass cores of the giant planets and their huge mass ratio with the protoplanets in the inner Solar System is a major problem of the runaway/oligarchic growth models, and it has prompted the elaboration of a new planet growth paradigm, named {\it pebble accretion}. 

\subsubsection{Pebble accretion}

Let's take a step back to what seems to be most promising planetesimal formation model: that of self-gravitating clumps of small particles (hereafter called pebbles even though in the inner disc they are expected to be at most mm-size, so that grains would be a more appropriate term). Once a planetesimal forms it remains embedded in the disk of gas and pebbles and it can keep growing by accreting individual pebbles.  This process was first envisioned by Ormel and Klahr \cite{ormel10} and then studied in detailed by Lambrechts and Johansen \cite{lambrechts12,lambrechts14,johansen17}. To avoid confusion, we call below  the accreting body a {\it protoplanet} and we denote the accreted body as a pebble or a planetesimal, depending on whether it feels strong gas drag. 

\begin{figure*}
\centering
\includegraphics[width=12cm]{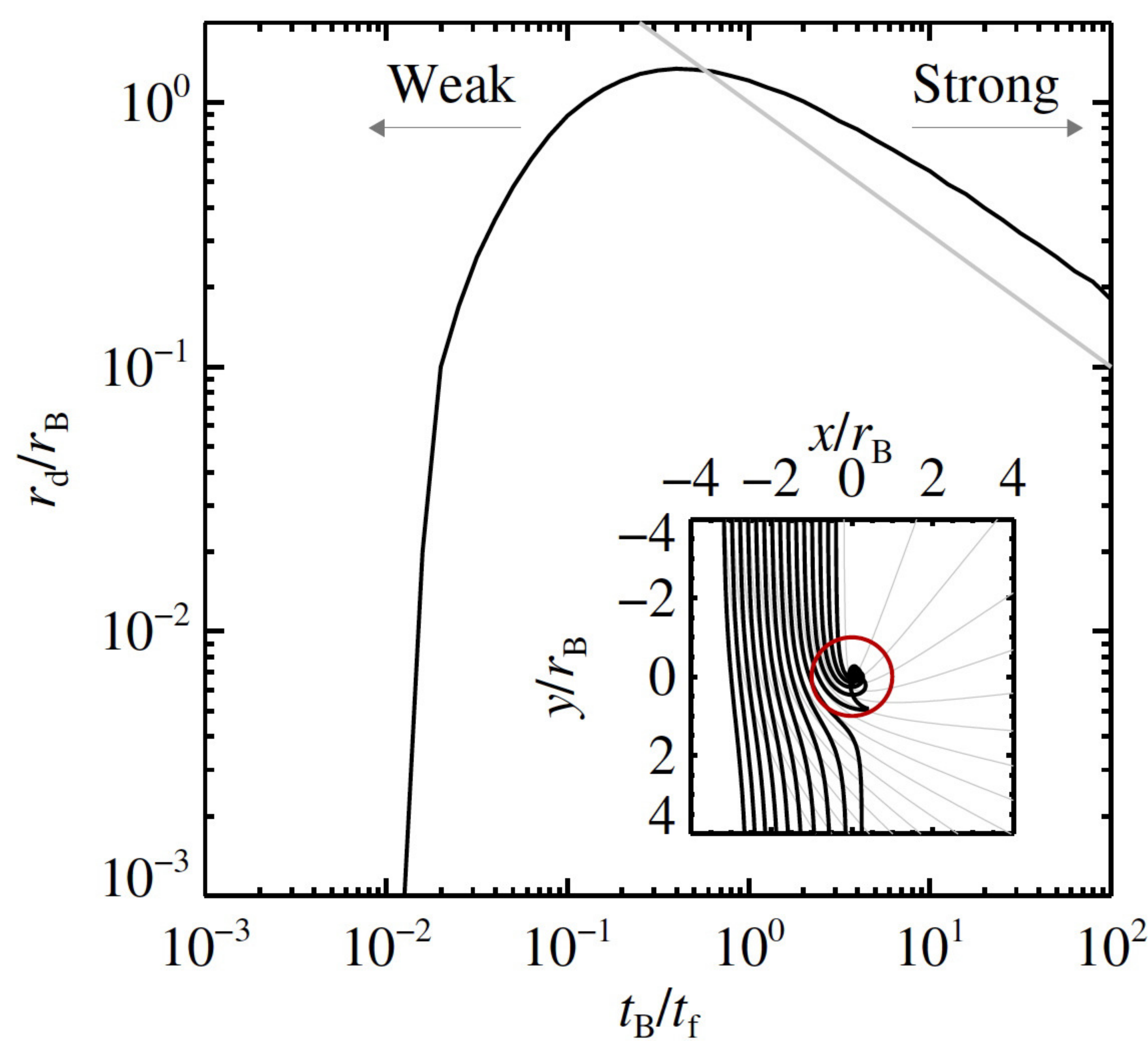}
\caption{Efficiency of pebble accretion. The outer plot shows the accretion radius $r_d$, normalized to the Bondi radius $r_B$, as a function of $t_B/t_f$, where $t_f$ is the friction time and $t_B$ is the time required to cross the Bondi radius at the encounter velocity $v_{rel}$. The smaller is the pebble the larger is $t_B/t_f$. The inset shows pebble trajectories (black curves) with $t_B/t_f=1$, which can be compared with those of objects with $t_B/t_f\to 0$ (grey curves). Clearly the accretion radius for the former is much larger. A circle of Bondi radius is plotted in red. From (\cite{lambrechts12}).}
\label{Lambrechts}
\end{figure*}

Pebble accretion is more efficient than planetesimal accretion for two reasons. First, the accretion cross-section for a protoplanet-pebble encounter is much larger than for a protoplanet-planetesimal encounter. As seen above, in a protoplanet-planetesimal encounter the accretion cross-section is $\pi R^2 F_g$, where $R$ is the physical size of the protoplanet and $F_g$ is the gravitational focussing factor. But in a protoplanet-pebble encounter it can be as large as $\pi r_d^2$, where $r_d$ is the distance at which the protoplanet can deflect the trajectories of the incoming objects. This is because, as soon as the pebble's trajectory starts to be deflected, its relative velocity with the gas increases and gas-drag becomes very strong. Thus, the pebble's trajectory spirals towards the protoplanet. This is shown in the inlet of Fig.~\ref{Lambrechts}, whereas the outer panel of the figure shows the value of $r_d$ as a function of the pebble's friction time, normalized to the Bondi radius $r_B=GM/v_{rel}^2$ ($v_{rel}$ being the velocity of the pebble relative to the protoplanet, typically of order $\eta v_K$).

The second reason that pebble accretion is more efficient than planetesimal accretion is that pebbles drift in the disk. Thus, the orbital neighborhood of the protoplanet cannot become empty. Even if the protoplanet accretes all the pebbles in its vicinity, the local population of pebbles is renewed by particles drifting inward from larger distances. This does not happen for planetesimals because their radial drift in the disk is negligible.
 
Provided that the mass-flux of pebbles through the disk is large enough, pebble accretion can grow protoplanets from about a Moon-mass up to multiple Earth-masses, i.e. to form the giant planets cores within the disc's lifetime \cite{lambrechts12,lambrechts14}. The large mass ratio between protoplanets in the outer vs. inner parts of the disc can be explained by remembering that icy pebbles can be relatively large (a few centimeters in size), whereas in the inner disc the pebble's size is limited to sub-millimetre by the bouncing silicate barrier (chondrule-size particles) and by taking into account that pebble accretion is more efficient for large pebbles than for chondrule-size particles \cite{morby15}.

For all these reasons, while some factors remain unknown (particularly the pebble flux and its evolution during the disk lifetime), pebble accretion is now considered to the dominant process of planet formation.

An important point is that pebble accretion cannot continue indefinitely. When a planet grows massive enough it starts opening a  gap in the disk. This eventually creates a pressure bump at the outer edge of the gap which stops the flux of pebbles. The mass at which this happens is called {\it pebble isolation mass} \cite{morby12,lambrechts14b} and depends on disk's viscosity and scale height \cite{bitsch18}. Once a planet reaches the pebble isolation mass it stops accreting pebbles. Given that it blocks the inward pebble flux, this means that all protoplanets on interior orbits are starved of pebbles, regardless of their masses. Turbulent diffusion can allow some pebbles to pass through the pressure bump \cite{weber18}, particularly the smallest ones, because the effects of diffusion are proprtional to $\sqrt{\alpha/\tau_s}$.  

\subsection{Orbital migration of planets}
\label{sect:migr}

Once a massive body forms in the disk, it perturbs the distribution of the gas in which it is embedded. We generically denote the perturbing body as a {\em planet}. In this Section we consider planets smaller than a few tens of Earth masses. The case of giant planets will be discussed in the next Section. 

\begin{figure}
\centering
\includegraphics[width=12cm]{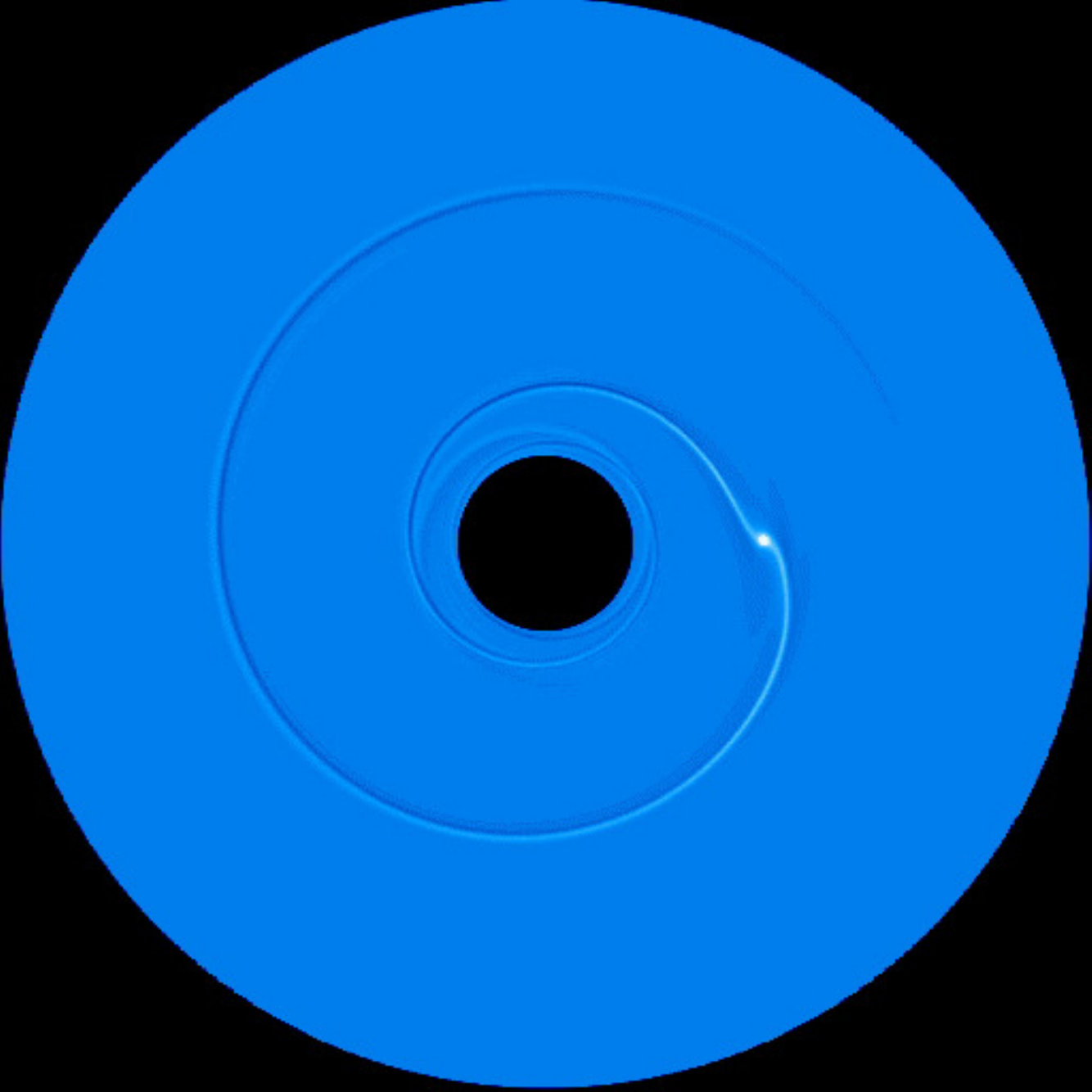}
\caption{The spiral density wave launched by a planet in the gas disk. The colour brightness is proportional to the gas surface density. Courtesy of F. Masset.}
\label{wake}
\end{figure}

Analytic and numerical studies have shown that a planet generates a spiral density wave in the disk, as shown in Fig.~\ref{wake}~\cite{goldreich79,goldreich80,lin79,ward86,tanaka02}. The exterior wave trails the planet. The gravitational attraction that the wave exerts on the planet produces a negative torque that slows the planet down. The interior wave leads the planet and exerts a positive torque. The net effect on the planet depends on the balance between these two torques of opposite signs. It was shown by \cite{ward86} that for axis-symmetric disks with any power-law radial density profiles, the negative torque exerted by the wave in the outer disk wins. This is because a power-law disk is in slight sub-keplerian rotation, so that the gravitational interaction of the planet with a disk's ring located at $a_p+\delta a$ ($a_p$ being the orbital radius of the planet) is stronger than with the ring located at $a_p-\delta a$, given that the relative velocity with the former is smaller. As a consequence of this imbalance, the planet must lose angular momentum and its orbit shrinks: the planet migrates towards the central star. This process is called {\it Type-I migration}. The planet migration speed is:
\begin{equation}
	{{{\rm d}a}\over{{\rm d}t}} \propto M_p \Sigma_g \left({a\over{H}}\right)^2\ ,
\label{eq.TypeI}
\end{equation}
where $a$ is the orbital radius of the planet (here assumed to be on a circular orbit), $M_p$ is its mass, $\Sigma_g$  is the surface density of the gas disk and $H$ is its height at the distance $a$ from the central star. A precise migration formula, function of the power-law index of the density and temperature radial profiles, can be found in Paardekooper et al. \cite{paardekooper10,paardekooper11}. The planet-disk interaction also damps the planet's orbital eccentricity and inclination if these are initially non-zero. These damping timescales are a factor $(H/a)^2$ smaller than the migration timescale \cite{tanaka04,cresswell07}.

Precise calculations show that an Earth-mass body at 1 au, in a Minimum Mass Solar Nebula ($\Sigma_g =1700$g/cm$^2$) with scale height $H/a$ = 5\%, migrates into the star in 200,000~y. For different planets or different disks, the migration time can be scaled using the relationship reported in Eq.~\ref{eq.TypeI}. So, Lunar- to Mars-mass protoplanets are only mildly affected by Type-I migration because their migration timescales exceed the few Myr lifetime of the gas disk. Conversely, for more massive planets, migration should be substantial and should bring them close to the star before that the disk disappears. 

Planet-disk interactions through the spiral density wave are only part of the story. An important interaction occurs along the planet's orbit due to fluid elements that are forced to do horseshoe-like librations in a frame corotating with the planet. Along these librations, as a fluid element passes from inside the planet's orbit to outside, it receives a positive angular momentum kick and exerts an equivalent but negative kick onto the planet. The opposite happens when a fluid element passes from outside of the planet's orbit to inside. It can be proven \cite{masset06} that, if the radial surface density gradient at the planet's location is proportional to $1/r^{3/2}$ (i.e. the {\it vortensity} of the disk is constant with radius) the positive and negative kicks cancel out perfectly, and there is no net effect on the planet. But for different radial profiles there is a net torque on the planet, named the {\it vortensity-driven corotation torque}\cite{paardekooper10,paardekooper11}. If the disk's profile is shallower than $1/r^{3/2}$ this corotation torque is positive and it slows down migration relative to the rate from Eq.~\ref{eq.TypeI}. Moreover, if the disk's radial surface density gradient is positive and sufficiently steep, the corotation torque (positive) can exceed the (negative) torque exerted by the wave and reverse migration~\cite{masset06}. This implies the existence of a location in the disk -- typically near the density maximum -- where migration stops, dubbed {\it planet trap} \cite{lyra10}. Positive surface density gradients could exist at the inner edge of the protoplanetary disk, where the disk is truncated by the stellar magnetic torque \cite{chang10}, or at transition from the MRI-active to the MRI-inactive parts of the disk~\cite{flock17,flock19} -- also very close to the central star -- or at the inner edge of each ring observed in MHD simulations (see Fig.~\ref{winds}b). Therefore, there can be several {\it planet traps} in the disk \cite{hasegawa12,baillie15}.    

The corotation region can also exert a positive torque on the planet in a region of the disk where the radial temperature gradient is steeper that $1/r$ \cite{paardekooper06}. This torque is called {\it entropy-driven corotation torque} \cite{paardekooper10,paardekooper11}. Steep temperature gradients exist behind the ``bumps'' of the disk's aspect ratio that are generated by opacity transitions \cite{bitsch15}. However, because the disk evolves over time towards a passive disk, with a temperature gradient shallower than $1/r$, the outward migration regions generated by the entropy-driven corotation torque exist only temporarily \cite{bitsch15}.  

Other torques can act on the planet and affect its migration in specific cases. If the viscosity of the disk is very small, {\it dynamical torques} are produced as a feedback of planet migration \cite{paardekooper14,pierens15,pierens16}. The feedback is negative, i.e. it acts to decelerate the migration, if the disk's surface density profile is shallower than $1/r^{3/2}$ and migration is inwards, or if the profile is steeper than $1/r^{3/2}$ and migration is outwards. In the opposite cases, the dynamical torque accelerates the migration. 

Low-viscosity disks are also prone to a number of instabilities generating vortices when submitted to the perturbation of a planet. As a result, the migration of the planet can become stochastic, due to the interaction with these variable density structures \cite{mcnally19}.

As it approaches the planet gas is compressed then decompressed so that its temperature first increases then decreases. Because hot gas loses energy by irradiation, the situation is not symmetric and the gas is colder (i.e. denser) after the conjunction with the planet than it was before conjunction. This generates a negative torque \cite{lega14}. On the other hand, if a planet is accreting solids, gravitational energy is released as heat. This source of heat modifies the density of the gas in the vicinity of the planet. In some conditions, this {\it heating torque} can exceed the previous effect, so that the net effect is positive and can even overcome the negative torque exerted by the wave \cite{benitez15}. This torque, however, also enhances the orbital eccentricity of the planet \cite{eklund17}, which in turn reduces its accretion rate. Thus some self-regulated regime can be achieved \cite{masset17}. 

Finally, even the steady-state dust distribution can be perturbed by the presence of the planet, acquiring asymmetries that can exert torques on the planet \cite{benitez18}.

To summarize, although the migration of a small-mass planet is typically inward and fast, there can be locations in the disk where migration is halted, as well as a number of temporary mechanisms that can reduce or enhance the migration rate. Therefore, the actual migration of a planet must be investigated in a case-by-case basis and requires a realistic modeling of the disk, given that its density and temperature gradients, opacity, viscosity and dust distribution play a key role. Unfortunately, so far our limited theoretical and observational knowledge of disks hampers our ability to model planet migration quantitatively.

\subsection{Gas accretion and giant planet migration}

A massive planet immersed in a gas disk can attract gas by gravity and build up an atmosphere. To distinguish between the solid part of the planet from its atmosphere, we will call the formed the {\it core}.

The closed set of equations that govern the distribution of gas in the atmosphere are:
\begin{equation}
{{{\rm d} P}\over{{\rm d}r}}={{G M(r) \rho(r)}\over{r}} \ ,
\label{hydrostatic}
\end{equation}
where $\rho(r)$ is the density of the gas at a distance $r$ from the center of the planet, which describes hydrostatic equilibrium (gravity balanced by the internal pressure gradient);
\begin{equation}
{{{\rm d} M}\over{{\rm d}r}}=4\pi r^2 \rho(r)\ ,
\label{mass}
\end{equation}
which describes the planet's mass-radius relationship $M(r)$ from $M(r_c)=M_c$, where $r_c$ is the radius of the core and $M_c$ is its mass;
\begin{equation}
{{{\rm d} T}\over{{\rm d}r}}=-{{3 \kappa L \rho}\over{64\pi\sigma r^2 T^3}} \ ,
\label{temp}
\end{equation}
where $\sigma$ is Boltzman's constant, $\kappa$ the gas opacity and $L\propto M_c \dot{M}_c/r_c$ is the luminosity of the core, due to the release of the gravitational energy delivered by the accretion of solids at a rate $\dot{M}_c$;
\begin{equation}
P=\frac{{\cal R}}{\mu\rho T},
\label{EOS}
\end{equation}
which is the equation of state, here for a perfect gas ($\cal{R}$ being the perfect gas constant and $\mu$ the molecular weight).

\begin{figure}[t]
\centering
\includegraphics[width=12cm]{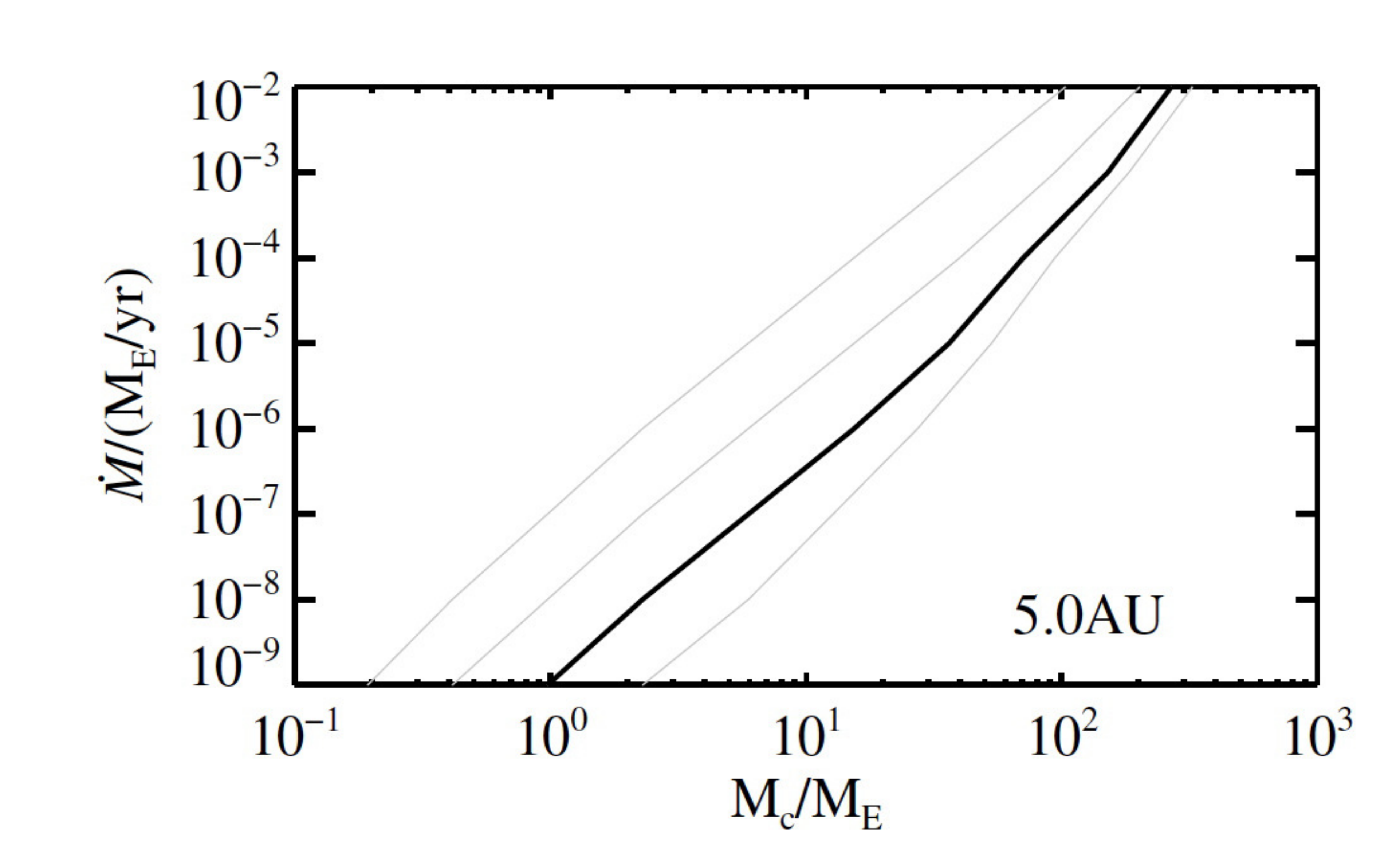}
\caption{The value of $M_{crit}$ (on the horizontal axis) as a function of core's accretion rate (on the vertical axis), for different values of the opacity -- increasing from left to right. The atmosphere can be in hydrostatic equilibrium only if $M_c<M_{crit}$. From \cite{lambrechts14}.}
\label{Mcrit}
\end{figure}

One can attempt to solve this set of equations using boundary condition $\rho(r_b)=\rho_0$ and $T(r_b)=T_0$, where $\rho_0$ and $T_0$ are the disk's values for gas density and temperature, respectively, and $r_b$ is the disk-planet boundary, typically the Bondi radius $r_b=2GM/c_s^2$ ($c_s$ being the sound speed). A solution exists only for $M_c<M_{crit}(\kappa,L)$ where $M_{crit}(\kappa,L)$ is a threshold value depending on opacity and luminosity (and disk's properties), as shown in Fig.~\ref{Mcrit} \cite{piso14,lambrechts14}.

$M_{crit}$ tends to zero as the core's accretion rate tends to zero. If $L=0$ no hydrostatic solution can exist. Recall from the previous Section that when a planet reaches the pebble isolation mass the accretion pebbles effectively stops \cite{morby12,bitsch18}. This drastically changes the value of $L$ and hence  $M_{crit}$. If the atmosphere of the planet was in hydrostatic equilibrium up to that point, it may be out of equilibrium. As a rule of thumb, when $M_c$ approaches $M_{crit}$ the mass of the atmosphere in hydrostatic equilibrium approaches that of the core.  This triggers runaway gas accretion\cite{lambrechts14}. 

\begin{figure}[t]
\centering
\includegraphics[width=12cm]{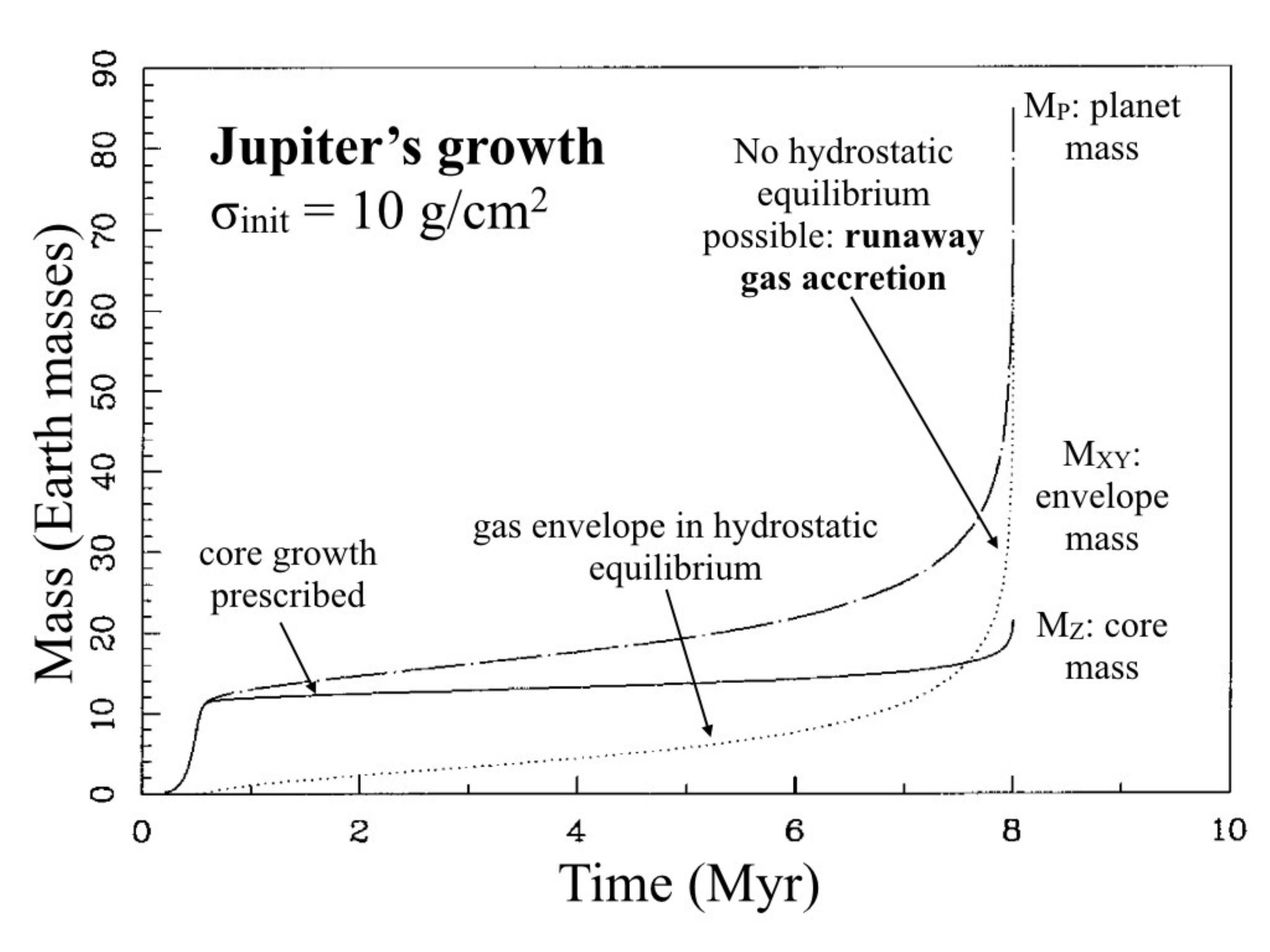}
\caption{Runaway growth of a giant planet near Jupiter's orbital radius. The solid curve shows the mass of the solid core, the dotted curve the mass of the atmosphere and the dashed curve the total mass of the planet. Here the accretion of solids onto the core is prescribed according to a now obsolete planetesimal accretion model. The approximate boundary in time between the hydrostatic regime and the runaway regime is at $\sim$7.5 Myr. While the runaway gas accretion regime is not in hydrostatic equilibrium it can be modeled as a series of equilibrium states.  Adapted from \cite{pollack96}.}
\label{runaway}
\end{figure}

When the atmosphere is no longer in hydrostatic equilibrium, it contracts under the effect of gravity. The compression of gas releases energy, so the atmosphere can only contract on the Kelvin-Helmoltz timescale, which is effectively the atmosphere's cooling timescale through irradiation. As the atmosphere contracts, new gas can be captured within the Bondi radius $r_b$. This increases the mass of the atmosphere and hence the gravity of the full planet. This triggers a positive feedback on the accretion rate, so that the mass of the planet's atmosphere increases exponentially with time (see Fig.~\ref{runaway}) \cite{pollack96,hubickyj05,lissauer09} .

A number of studies have modeled atmospheric accretion during the runaway phase \cite{ikoma00,coleman17,lambrechts19b} using different approaches. All have confirmed the runaway gas accretion with a mass-doubling timescales for Jupiter-mass planets of order $10^4$--$10^5$~y in a MMSN disk. This raises the question of why Saturn-mass exoplanets outnumber Jupiter-mass ones, which drastically outnumber super-Jupiters ~\cite{butler06,udry07b}. One possibility is that giant planets enter in their runaway phase late, as the disk is disappearing \cite{pollack96,bitsch15b}. Given that the mass doubling timescale in the runaway phase is so much shorter than the disk's lifetime, this appears to be a surprising coincidence. The other possibility is that the growth of a planet is limited by the ability of the disk to transport gas radially. We know from observations, however, that the gas accretion rate onto the central star is typically of the order of $10^{-8}M_\odot$/y \cite{hartmann98}, which means that a Jupiter-mass of gas passes through the orbit of a giant planet in only $10^5$y, again much shorter than the disk's lifetime. Perhaps the study of giant planet growth in low-viscosity disks dominated by winds can bring a solution to the problem, given the different geometry and mechanism of transport for the accreting gas relative to a classic, viscous disk.   

\subsubsection{Gap opening and Type-II migration}

As we have seen in Sect.~\ref{sect:migr} a planet embedded in a disk exerts a positive torque on the outer part of the disk and a negative torque on the inner part. The torque is proportional to the planet's mass. If the planet is small, its torque is easily overcome by the viscous torque that the annuli of the disk exert on each other. The global surface density profile of the disk is not changed and only the spiral density wave appears. But if the planet is massive enough the torque it exerts on the disk overwhelms the disk's viscous torque. In this case, the planet effectively pushes gas away from its orbit: outer gas outwards and inner gas inwards. A gap opens in the gas distribution around the orbit of the planet.

As gas is removed, the gap becomes deeper and wider. But as the gradients of the disk's surface density distribution become steeper at the edges of the gap, the disk's viscous torque is enhanced. In fact, the viscous torque acting  on  elementary  rings  can  be  computed  by  differentiation of Eq.~\ref{torque} to give:
\begin{equation} 
\delta T_\nu= -{3\over{2}}\nu\Omega\left[{r\over{\Sigma}}{{{\rm d}\Sigma}\over{{\rm d}r}}+{1\over{2}}\right](2\pi r \Sigma)\ .
\label{viscdiff}
\end{equation}
Once the density gradient becomes steep enough, the viscous torque balances the torque that the planet exerts on the same annulus of the disk \cite{varniere04}.

The depth of a gap -- defined as the ratio between the surface density of the disk perturbed by the planet $\Sigma_p$ and the original value $\Sigma_u$ -- is \cite{kanagawa18}:
\begin{equation}
{{\Sigma_p}\over{\Sigma_u}}={1\over{1+0.04K}}\ ,
\label{Kanagawa-gap}
\end{equation}
where
\begin{equation}
K=\left({M_p\over{M_*}}\right)^2 \left({H\over{r}}\right)^{-5}\alpha^{-1}
\end{equation}
and $M_p, M_*$ are the masses of the planet and the star, respectively, the aspect ratio of the disk $H/r$ is taken at the planet's location, and as usual $\alpha=\nu/(H^2\Omega)$. This formula holds up to $K\sim 10^4$ (see Fig. 11 in \cite{kanagawa18}). However it cannot hold indefinitely, in particular in the limit of low $\alpha$.  This is because if the density gradients at the edges of the gap become too steep, the rotational properties of the gas change so much under the effect of the pressure gradient that the specific angular momentum of the disk $r^2\Omega(r)$ is no-longer a growing function of $r$ \cite{kanagawa15}. When this happens the disk becomes Rayleigh unstable and develops local turbulence, in turn enhancing the local viscosity. This effectively limits the steepness of the gap ``walls'' and the depth of the gap. Because the pressure gradient is also proportional to $(H/r)^2$, this narrative implies that, in the limit of vanishing viscosity, the denser the disk the shallower the gap. Thus, a gap opening criterion --i.e. the minimal mass of a planet to cause a depletion of 90\% of the gas in the gap-- must depend not only on viscosity but also on the disk's aspect ratio. An often-quoted quoted criterion is the following:
\begin{equation}
{3\over{4}} {H\over{R_H}} +{50\over{qR_e}} < 1 \ ,
\label{crida-crit}
\end{equation}
where $q=M_p/M_*$, $R_H=(q/3)^{1/3}$ and $R_e=r_p^2\Omega_p /\nu$ \cite{crida06}.

The formation of a gap profoundly changes a planet's migration.  This migration mode has been dubbed {\it Type-II migration}. The gap must migrate along with the planet. In particular, as the planet moves inwards, the disk has to refill the portion of the gap ``left behind'' by the planet's radial motion. Because the radial velocity of the gas in a viscous unperturbed disk is $v_r=-(3/2) \nu/r$, this was the expected migration speed of the planet, independent of the planet's mass and disk's scale aspect ratio \cite{ward97}.

\begin{figure}[t]
\centering
\includegraphics[width=12cm]{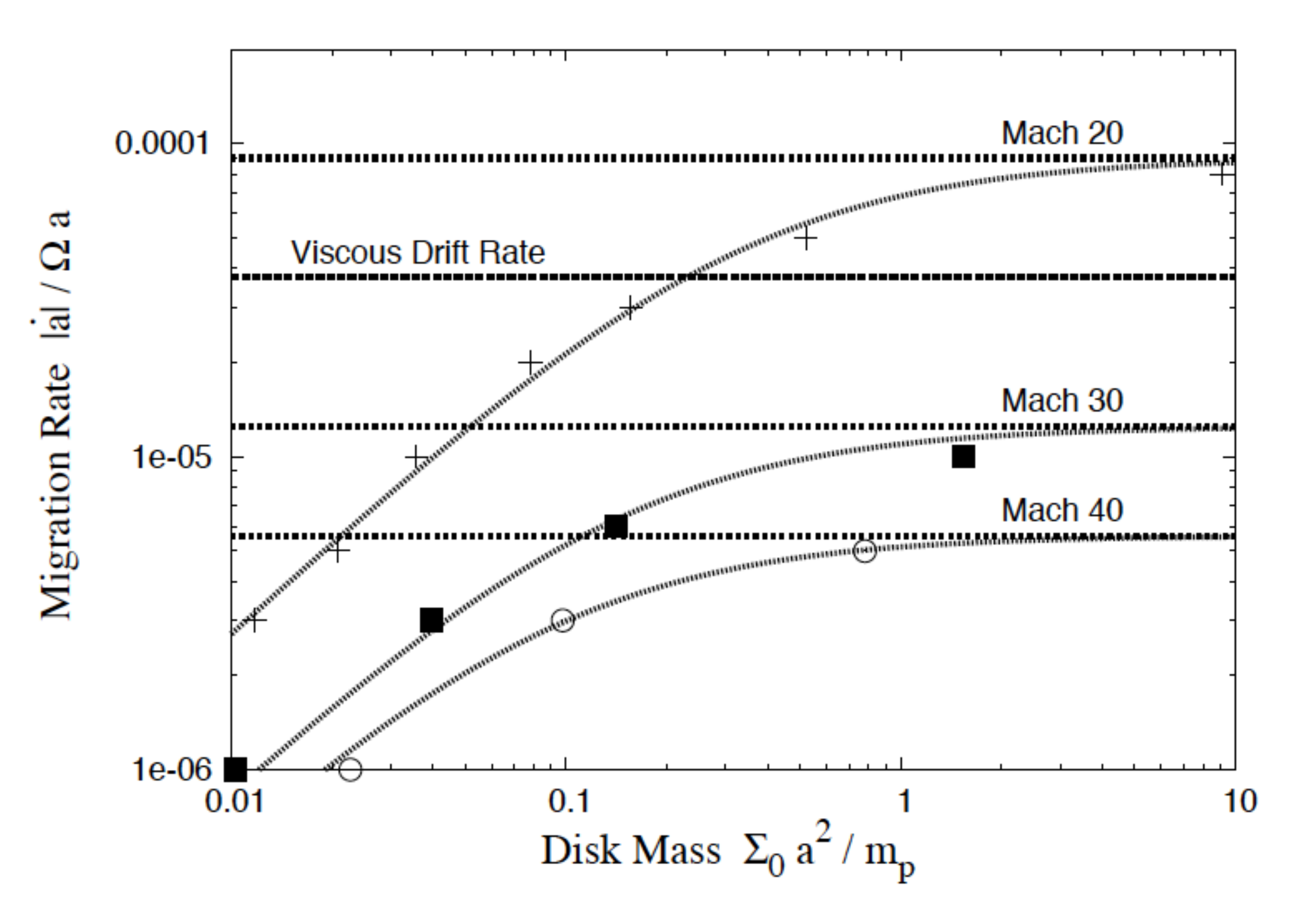}
\caption{The migration rate as a function of the ratio $r_p^2\Sigma/M_p$ for a giant planet in a disk with $\alpha=3\times 10^{-3}$, for three different disk's aspect ratios. Notice that the migration rate can be smaller or larger than the viscous drift rate of the gas, i.e. the idealized Type-II migration speed. From \cite{duffell14}.}
\label{TypeII}
\end{figure}

However, the planet's migration rate is not so simple \cite{duffell14,durmann15}. It depends on the ratio between the disk's density and the mass of the planet, exemplified by the dimensionless ratio $r_p^2\Sigma/M_p$ (see Fig.~\ref{TypeII}), and also on the disk's aspect ratio. Depending on these quantities, the radial velocity of the planet can be smaller or larger than the ``idealized'' Type-II migration speed of $-(3/2) \nu/r$.

It is easy to understand that a planet's migration speed is slower than the idealized speed for the case of a low disk mass. A light disk obviously cannot push a heavy planet. This is the inertial limit: the planet is an obstacle to the flow of the gas. But how can a planet migrate faster than the radial speed of the gas? In principle the gas should remain behind and the planet, losing contact from the outer disk, should feel a weaker negative torque, eventually slowing down its migration until it moves at the same speed of the gas. The explanation given in \cite{duffell14,durmann15} was that, as the planet migrates inwards, the gas from the inner disk can pass through the gap, refilling the left-behind part of the gap. In this way, the planet does not have to wait for the gas to drift-in at viscous speed. But the passage of gas through the gap is insignificant if the gap is significantly wider than the planet's horseshoe region \cite{robert18}, which is the case for massive planets in low-viscosity disks. In that case, as the planet moves inwards, the gap must be refilled from the outer disk. However, the steep density gradient at the edge of the gap enhances the viscous torque, as discussed above, so that the gas radial speed can be several times faster than $-(3/2) \nu/r$. This explains why the planet can exceed the idealized Type-II migration rate. Nevertheless, the migration rate of the planet must linearly proportional to the viscosity of the disk \cite{robert18}.

The dependence of the migration rate of giant planets on disk's viscosity opens the possibility that, in low-viscosity disks, the migration timescale can exceed the disk's lifetime. This would be convenient to explain why most giant planets are warm/cold Jupiters and not hot Jupiters \cite{butler06,udry07b}.\footnote{We define a hot Jupiter as a giant planet within 0.1 au from its star and warm/cold Jupiters as those that are beyond 0.5 au. Very few giant planets fall between 0.1 and 0.5 au so the exact values of these boundaries are not important. Debiased observations suggest that hot Jupiters are about 1/10th or less as abundant as warm/cold Jupiters \cite{fernandes19,wittenmyer16,wittenmyer20}.} However, recall from Sect.~\ref{viscous-disk} that low-viscosity disks cannot explain the accretion rates observed for the central stars. If viscosity is low, there must be some additional mechanism for the radial transport of gas, possibly induced by angular momentum removal in disk winds. The migration rate of giant planets in these kind of disks has not yet been studied.

\subsection{Resonance trapping during planet migration}

Numerical simulations show that multiple planets migrating in a disk have the tendency to lock into mutual mean motion resonance, where the period ratio is very close to the ratio of integer numbers~\cite{terquem07,pierens08,izidoro17}. Typically, these numbers differ by just one such that the period ratios are 2:1, 3:2, 4:3 etc. These are called {\em first-order} resonances.

To understand this propensity to form resonant chains we need to dive into the complex world of dynamical planet-planet interactions. To stay simple, we will consider the planar three body problem, where two planets orbit a star on coplanar orbits. This is the simplest example of dynamics of a planetary system, yet it already captures most of the complexities of real systems. The best mathematicians made huge efforts to find an analytic solution of this problem, until Poincar\'e \cite{poincare1892} demonstrated that this was impossible. A general analytic solution does not exist. The system can exhibit chaotic behavior \cite{henon64}. Yet, some description can be provided, such as for instance for the dynamics of resonant planets with small libration amplitude, which is that of interest to understand the formation of resonant chains. This is what we attempt to do in this Section.

To study the three body problem, the most effective approach is to use the Hamiltonian formalism. A system of first-order differential equations

\begin{equation}
{{{\rm d}\vec{x}}\over{{\rm d}t}}=\vec{f}(\vec{x},\vec{y})\ ,\quad
{{{\rm d}\vec{y}}\over{{\rm d}t}}=\vec{g}(\vec{x},\vec{y})
\label{eq}
\end{equation}
is said to be in Hamiltonian forms, if there exists a scalar function ${\cal H}(\vec{x},\vec{y})$ such as
\begin{equation}
f_i(\vec{x},\vec{y})={{\partial {\cal H}}\over{\partial y_i}}\ , \quad g_i(\vec{x},\vec{y})=-{{\partial {\cal H}}\over{\partial x_i}}\ ,
\label{Ham-eq}
\end{equation}
for each component $i=1,\ldots,N$ of the vectors $\vec{x}, \vec{y}, \vec{f}, \vec{g}$ The function $\cal{H}$ is called the {\it Hamiltonian} of the problem; $\vec{x}$ is called the vector of {\it coordinates} and $\vec{y}$ the vector of {\it momenta}. If $H$ is periodic on $\vec{x}$, the coordinates are also called {\it angles} and the momenta {\it actions}.

An important property of Hamiltonian dynamics is that ${\cal H}$ is constant over the flow $\vec{x}(t), \vec{y}(t)$ that is solution of the equations of motion. This means that, in the special case where ${\cal H}$ is a function of only one component of $\vec{x}$, namely ${\cal H}\equiv {\cal H}(x_1,\vec{y})$, the evolution of the system can be easily computed: $y_2,\ldots,y_N$ are constant of motion from (\ref{eq}) and (\ref{Ham-eq}); the motion of $x_1, y_1$ can be obtained from the level curve of ${\cal H}(x_1, y_1, y_2, \ldots, y_N)={\cal H}(\vec{x}(0),\vec{y}(0))$, which is a 1D curve in a 2D $x_1, y_1$ space (called {\it phase space}). In this case, the problem is integrable (i.e. the solution is provided with analytic functions of time).

In studying a problem written in Hamiltonian form, a typical goal is to find a transformation of variables $\vec{x}\to \vec{x}', \vec{y}\to \vec{y}'$ that transforms the Hamiltonian function into one that is independent of $x'_2,\ldots,x'_N$. However, only transformations that preserve the form of Hamilton equations (Eqns~\ref{eq} and~\ref{Ham-eq}), called {\it canonical transformations}, are allowed. There are many forms of canonical transformations. In this Section we will use only linear transformations $\vec{x}'=A\vec{x}, \vec{y}'=B\vec{y}$ where $A, B$ are $N\times N$  matrices. It can be proven that the transformation is canonical if and only if
\begin{equation}
B=\left[A^{-1}\right]^T \ .
\label{transform}
\end{equation}

In general, it is not possible to find a canonical transformation that makes the Hamiltonian independent of $N-1$ coordinates. Then, a goal in {\it perturbation theory} is to find a canonical transformation that turns ${\cal H}(\vec{x},\vec{y})$ into
\begin{equation}
{\cal H}'(\vec{x}',\vec{y}')={\cal H}'_0(x'_1,\vec{y}')+\epsilon {\cal H}'_1(\vec{x}',\vec{y'})\ .
\label{perturb}
\end{equation}
In this case ${\cal H}'_0$ is called the {\it integrable approximation} and ${\cal H}'_1$ the {\it perturbation}. The latter can be neglected if one is interested in the dynamics up to a time $t<1/\epsilon$. Depending on the goal in terms of accuracy, $\epsilon$ has to be sufficiently small.

After this broad and shallow introduction to Hamiltonian dynamics, let's turn to the planar three-body problem. The problem admits a Hamiltonian description, with Hamiltonian function:
\begin{equation}
{\cal H}=\sum_{j=1,2} {{\norm{\vec{p}_j}^2}\over{2\mu_j}} - {{G(M_*+m_j)\mu_j}\over{\norm{\vec{r}_j}}} + {{\vec{p}_1 . \vec{p}_2}\over{M_*}} - {{G m_1 m_2}\over{\norm{\Delta}}} \ ,
\label{3BP}
\end{equation}
where $\vec{r}_j$ is the {\it heliocentric} position vector of planet $j$ of reduced mass $\mu_j=m_jM_*/(m_j+M_*)$, $\vec{p}_j=m_j\vec{v}_j$ with $\vec{v}_j$ being the {\it barycentric} velocity vector (not a typo!: positions and velocities have to be taken in different reference frames \cite{poincare1897} if one wants a Hamiltonian description of the problem) and $\vec{\Delta}=\vec{r}_1-\vec{r}_2$.

Using the canonical Delaunay variables:
\begin{equation}
\begin{array}{rcr}
\Lambda_j=\mu_j\sqrt{G(M_*+m_j)a_j}\ , && \lambda_j= M_j+\varpi_j\nonumber\\
\Gamma_j=\Lambda_j(1-\sqrt{1-e_j^2}) \ , &&
\gamma_j=-\varpi_j\nonumber\\
\end{array}
\label{Delauney}
\end{equation}
where $a_j, e_j$ are the semi major axes and eccentricities, $\varpi_j$ are the perihelion longitudes, $M_j$ the mean anomalies and $G$ is the gravitational constant, the Hamiltonian (Eq.~\ref{3BP}) becomes:
\begin{equation}
{\cal H}=-G^2\sum_{j=1,2}{{\mu_j^3(M_*+m_j)^2}\over{2\Lambda_j^2}}+{\cal H}_1(\Lambda_{1,2},\Gamma_{1,2},\lambda_{1,2},\gamma_{1,2}) \ .
\label{3BP-del}
\end{equation}
The first term in the r.h.s. of (Eq.~\ref{3BP-del}), denoted ${\cal K}$ hereafter, taken alone, is an integrable Hamiltonian, but the flow that it describes is trivially that of two uncoupled Keplerian motions: the actions $\Lambda_{1,2}, \Gamma_{1,2}$ are constant, the longitude of perihelia $\gamma_{1,2}$ are constants, and only the mean longitudes $\lambda_{1,2}$ move with constant frequency $G^2\mu_j^3(M_*+m_j)^2/\Lambda_j^3=\sqrt{G(M_*+m_j)/a_j^3}$. So, this kind of integrable approximation of the full Hamiltonian is not sufficient for our purposes and we need to find a better one.

To this end, we expand ${\cal H}_1$ in power series of $\sqrt{\Gamma_j}\sim \sqrt{\Lambda_j/2} e_j$ and in Fourier series of the angles $\lambda_j,\gamma_j$. The general form is therefore
\begin{eqnarray}
{\cal H}_1=\sum_{l_1,l_2,k_1,k_2,j_1,j_2} && c_{l_1,l_2,k_1,k_2,j_1,j_2}(\Lambda_1,\Lambda_2) \Gamma_1^{j_1/2}\Gamma_2^{j_2/2} \cos(k_1\lambda_1+k_2\lambda_2+l_1\gamma_1,l_2\gamma_2)\cr +&& s_{l_1,l_2,k_1,k_2,j_1,j_2}(\Lambda_1,\Lambda_2) \Gamma_1^{j_1/2}\Gamma_2^{j_2/2} \sin(k_1\lambda_1+k_2\lambda_2+l_1\gamma_1,l_2\gamma_2) \cr
&&\hbox{} 
\label{expansion}
\end{eqnarray}
The so-called {\it D'Alembert rules} give us information on which terms of this series can have non-zero coefficients, namely:
\begin{itemize}
\item[-] only the $c$ coefficients can be non-zero, because the Hamiltonian must be invariant for a change of sign of all angles (measuring angles clockwise or counter-clockwise is arbitrary), so that the Fourier expansion can contain only $\cos$ terms;
\item[-] only the $c_{l_1,l_2,k_1,k_2,j_1,j_2}$ coefficients with $k_1+k_2-l_1-l_2=0$ can be non-zero, because the Hamiltonian has to be invariant by a rotation of the reference frame, namely increasing all angles by an arbitrary phase $\delta_0$ (remember that $\gamma=-\varpi$, so if all orbital angles are increased by $\delta_0$, $\gamma$ is decreased by $\delta_0$). 
\item[-] only the $c_{l_1,l_2,k_1,k_2,j_1,j_2}$ coefficients with $j_1=|l_1|+2n$ and $j_2=|l_2|+2i$ (with $n$ and $i$ non-negative integer numbers) can be non-zero. This is because the Hamiltonian is not singular for circular orbits (i.e. $\Gamma_1=0$ and/or $\Gamma_2=0$) so that it has become a polynomial expression in the canonical variables $p_1=\sqrt{2\Gamma_1}\cos\gamma_1, q_1=\sqrt{2\Gamma_1}\sin\gamma_1, p_2=\sqrt{2\Gamma_2}\cos\gamma_2, q_2=\sqrt{2\Gamma_2}\sin\gamma_2$.
\end{itemize}

If we are interested in two planets near a mean motion resonance where $P_2\sim k/(k-1) P_1$, where $P_1$ and $P_2$ are the orbital periods and $k$ is a positive integer number, the angle $k\lambda_2-(k-1)\lambda_1$ will have an almost null time-derivative (as one can see remembering that $\dot{\lambda}=2\pi/P$ and using the relationship between the orbital periods written above). Thus, it is a slow angle, whereas both $\lambda_1$ and $\lambda_2$ are fast angles, as it is their difference. To highlight this difference in timescales, let us define new angles:
\begin{equation}
\delta\lambda=\lambda_1-\lambda_2\ ,\quad \theta=k\lambda_2-(k-1)\lambda_1 \ .
\label{1trans-angles}
\end{equation}
Using the rule (\ref{transform}) this linear transformation of the angles can be made canonical by changing the actions as:
\begin{equation}
\Delta\lambda=k\Lambda_1+(k-1)\Lambda_2\ ,\quad \Theta=\Lambda_1+\Lambda_2\ ,
\label{1trans-actions}
\end{equation}
so that $(\Delta\lambda, \delta\lambda)$ and $(\Theta, \theta)$ are pairs of canonical action-angle variables. 

Because we are interested in the long-term evolution of the dynamics, we can average the Hamiltonian over $\delta\lambda$, which means that the averaged Hamiltonian will be independent of this angle and, consequently, $\Delta\lambda$ will be a constant of motion. Because the units of semi-major axis are arbitrary, one can always chose them so that $\Delta\lambda=1$. In other words, changing the values of $\Delta\lambda$ does not change the dynamics; it simply changes the unit of measure of the semi major axes.

Using the D'Alembert rules described above, the function in Eq.~\ref{expansion} takes the form
\begin{equation}
{\cal H}_1=\sum_{m,n,i>0,j>0} d_{m,n,i,j}(\Theta,\Delta\lambda)\Gamma_1^{|n-m|/2+j}\Gamma_2^{|m|/2+i}\cos[n(\theta+\gamma_1)+m(\gamma_2-\gamma_1)]\ ,
\label{averaged}
\end{equation}
where the coefficients $d_{m,n,i,j}$ come from the original coefficients $c_{l_1,l_2,k_1,k_2,j_1,j_2}$ through a simple index algebra that follows trivially from the redefinition of the angles (Eq.~\ref{1trans-angles}).

Because there are only two possible combinations of angles in the harmonics of Eq.~
\ref{averaged} it is convenient to identify each of them with a single angle, namely:
\begin{equation}
\psi_1=\theta+\gamma_1 \,\quad \delta\gamma=\gamma_2-\gamma_1\ , \quad \gamma_2'=\gamma_2
\label{new-ang}
\end{equation}
Again, using the rule (Eq.~\ref{transform}) this linear transformation of the angles can be made canonical by changing the actions as:
\begin{equation}
\Psi_1=\Theta\ ,\quad \Psi_2=\Theta-\Gamma_1\ ,\quad {\cal L}=\Gamma_1+\Gamma_2-\Theta
\label{new-act}
\end{equation}
Now, the Hamiltonian ${\cal K}+{\cal H}_1$ depends only on the angles $\psi_1$ and $\delta\gamma$, and ${\cal L}$ is a new constant of motion (related to the angular momentum of the system).

\begin{figure}[t]
\centering
\includegraphics[width=12cm]{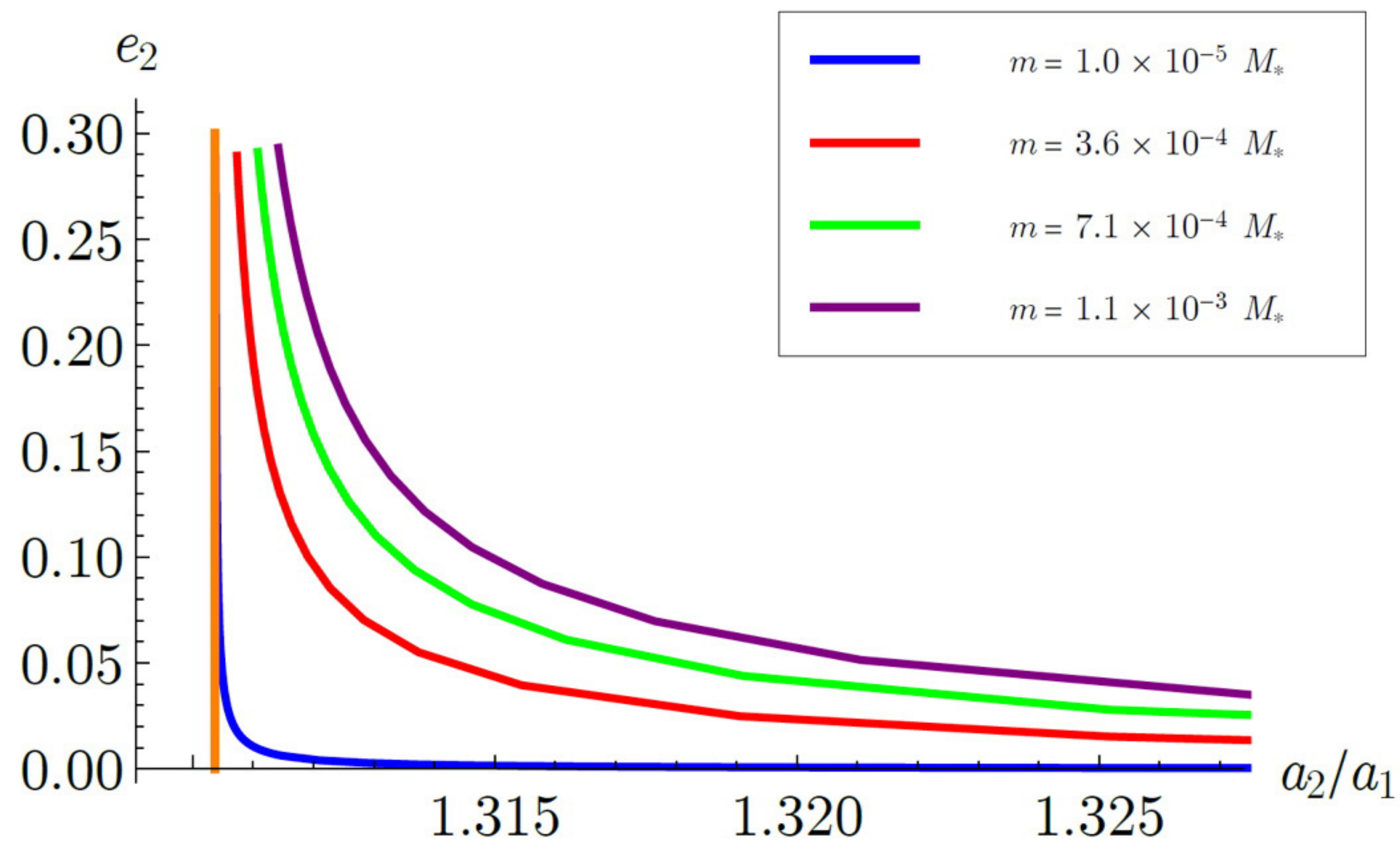}
\caption{The locus of equilibrium points for the 3:2 mean motion resonance in the plane $e_2$ vs. $a_2/a_1$, for different planetary masses (here assumed to be equal to each other) in different colors. The vertical orange line shows the location of the resonance in the Keplerian approximation (the same $a_2/a_1$ for any eccentricity because the orbital frequencies depends only on semi major axes). Adapted from Pichierri et al. \cite{pichierri18}.}
\label{Pichierri18}
\end{figure}

This Hamiltonian is still not integrable, because it depends on two angles. An integrable approximation (dependent on only one angle) can be obtained if one retains in Eq.~\ref{averaged} only the terms linear in the planets' eccentricities, i.e. proportional to $\sqrt{\Gamma_{1,2}}$, and doing some cumbersome change of variables \cite{sessin84,batygin13b}. In this case one can then trace global dynamical diagrams by plotting the level curves of the Hamiltonian (see for instance Fig. 3 in \cite{batygin13b}). However, even in the general non-integrable case one can look for the stable equilibrium points in $(\Psi_1,\psi_1,\Psi_2,\delta\gamma$) as a function of ${\cal L}$ (recall that $\Delta\lambda$ can be fixed to unity). The locus of equilibrium points, once transformed back into the original orbital elements,  describes a curve in $e_2,a_2/a_1$ or, equivalently, $e_1,a_2/a_1$, like that shown in Fig.~\ref{Pichierri18}. Note that on the curve $a_2/a_1\to \infty$ as $e_1\to 0, e_2\to 0$. This feature comes from the fact that, from Hamilton's equations $\dot{\gamma}_{1.2}=\partial{\cal H}/\partial{\Gamma}_{1.2}$ applied to (\ref{averaged}), one has  $\dot{\gamma}_{1,2}\propto \Gamma_{1,2}^{-1/2}$, i.e. $\dot{\gamma}_{1,2}\to \infty$ as $\Gamma_{1,2}\propto e^2_{1,2} \to 0$. Thus, to have the equilibrium $\dot{\psi_1}=0$ the value of $\dot{\theta}$ has to diverge, which means that $a_2/a_1$ has to diverge as well. 

This feature of the curve of equilibrium points is the key to understand resonant capture. If the planets are far from resonance (i.e. $a_2/a_1$ is much larger than the resonant ratio assuming Keplerian motion; a 3:2 resonance being located at $a_2/a_1 = 1.3103$ in Fig.~\ref{Pichierri18}), the protoplanetary disk exerts damping forces on their eccentricities, so that the planet's orbits are basically circular. This means, from the shape of the curve of equilibrium points, that the planets will be on the equilibrium. As migration proceeds and $a_2$ approaches $a_1$ (this is the case if the outer planet migrates faster towards to the star, which happens in Type-I migration if it is more massive or if the inner planet is blocked at a planet trap), the ratio $a_2/a_1$ decreases. If this happens slowly compared to the libration period around the equilibrium point, the dynamical evolution has to react {\it adiabatically} \cite{neishtadt84}. This means that the amplitude of libration around the equilibrium point has to be conserved. Because initially the planets have vanishing amplitude of libration (their eccentricities are basically null as those characterizing the equilibrium point), this means that the planets have to evolve from one equilibrium point to the other, i.e. they have to follow the curve traced in Fig.~\ref{Pichierri18}. As $a_2/a_1$ asymptotically approaches the Keplerian resonant value, the eccentricities of the two planets increase.

If the convergent migration is too fast, the adiabatic condition is broken. The amplitude of libration is not conserved. The planets can jump off resonance and  continue to approach each other more closely. But, because the libration frequency of a resonance $P_2\sim k/(k-1) P_1$ increases with $k$, eventually the planets will find a resonance with $k$ large enough and libration frequencies fast enough that the adiabatic condition is satisfied. Then, they will be trapped in resonance. In essence, the faster is the convergent migration, the higher the index $k$ of the resonance in which the planets will be trapped. But trapping always occurs, eventually. These arguments apply for each pair of neighboring planets in a multi-planet system. This is why the formation of configurations in resonant chains is a typical outcome of planet migration.

If the adiabatic condition is satisfied, is resonant trapping stable? Fig~\ref{Pichierri18} suggests that the eccentricities of the planets should grow indefinitely. But in reality the disk exerts eccentricity damping, so that the eccentricities grow until an equilibrium is established between the eccentricity damping from the disk and the resonant conversion of convergent migration into eccentricity pumping. A precise formula to compute this equilibrium in a variety of configurations (damping exerted on both planets or only on one, inner planet at a planet-trap or not etc.) can be found in the appendix of \cite{pichierri18}. The equilibrium eccentricity is typically of order $(H/r)^2$, where $(H/r)$ is the aspect ratio of the protoplanetary disk.

There is a complication. For the adiabatic principle to be applied, the dissipative forces have to act on the parameters of the Hamiltonian, not on the dynamical variables. This is the case for migration. The change in semi major axis ratio changes the otherwise constant of motion ${\cal L}$, i.e. a parameter of the Hamiltonian. But the eccentricity damping affects $\Gamma_{1,2}$, i.e. dynamical variables. Then, the equilibrium point can become a focus, which means that the dynamical evolution spirals around it. The spiral can be inward if the focus is stable (which means that any initial amplitude of libration would shrink to 0) or outwards if the focus is unstable (which means that the amplitude of libration grows indefinitely, even if it is initially arbitrarily small). This unstable evolution, first pointed out in \cite{goldreich14}, is called {\it overstability}).

\begin{figure}[t]
\centering
\includegraphics[width=5.6cm]{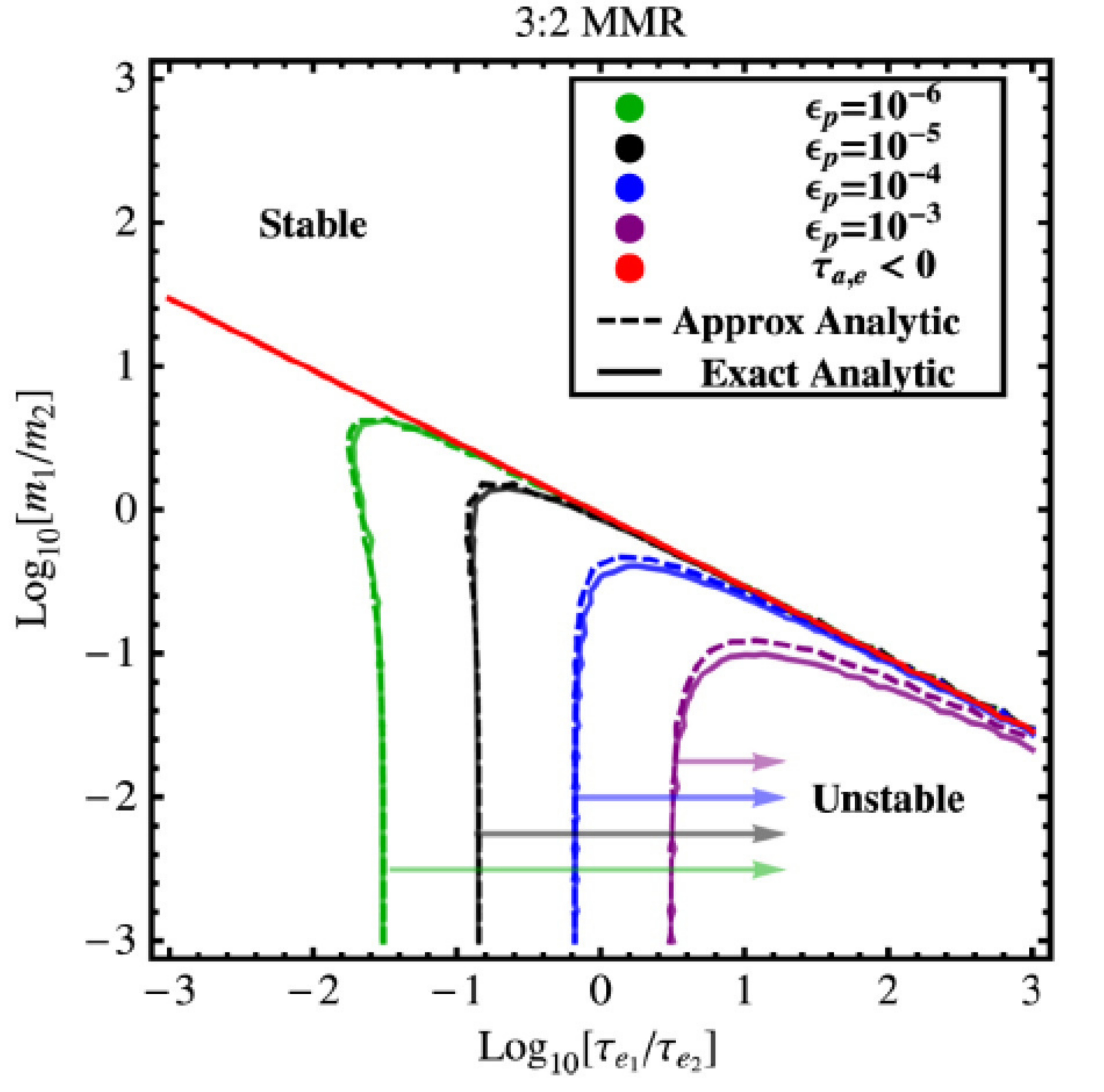}
\includegraphics[width=5.6cm]{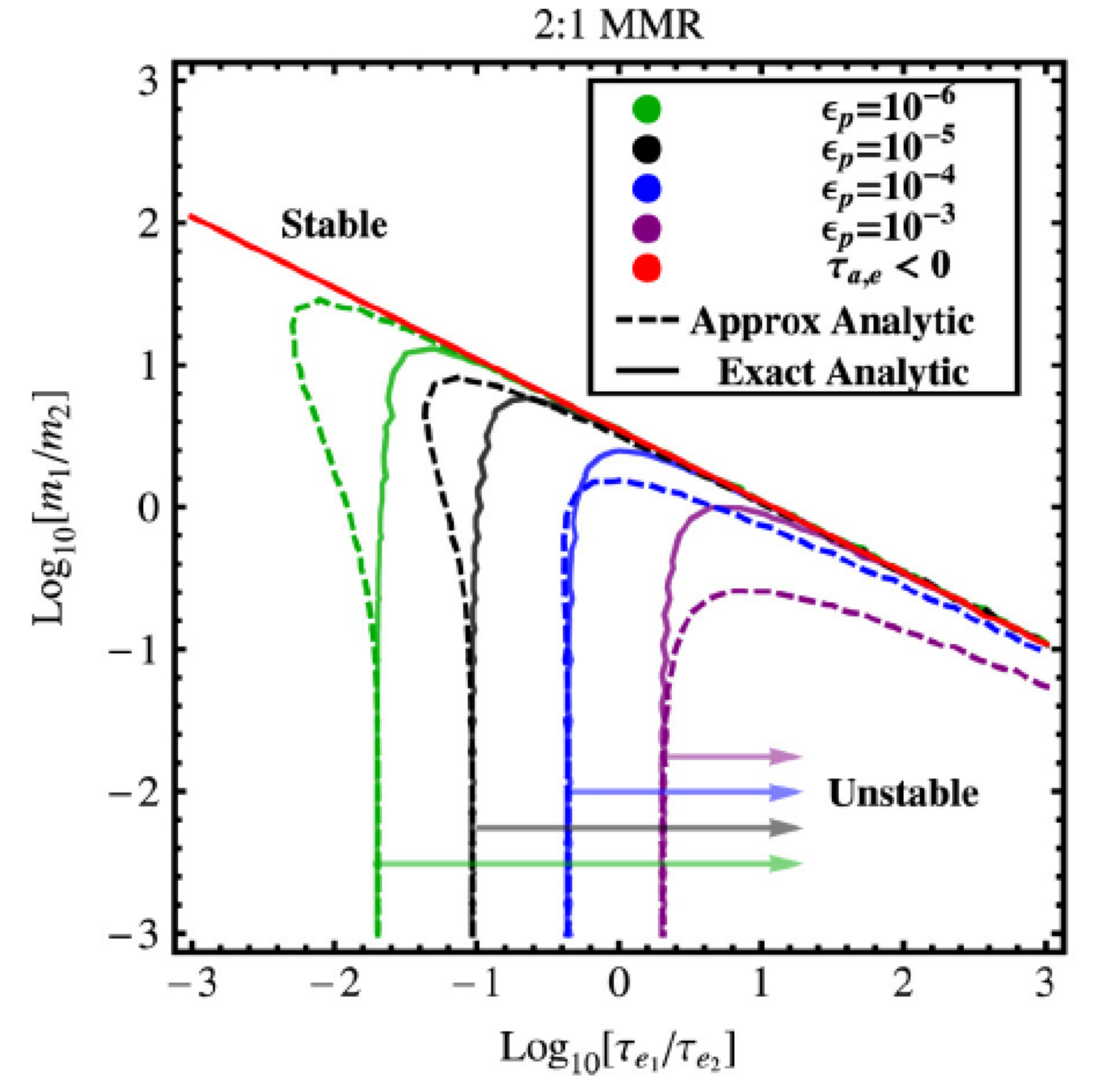}
\caption{Each plot shows the region of the parameter plane $\tau_{e_1}/\tau_{e_2}$ and $m_1/m_2$, where $\tau_{e_{1,2}}$ is the damping timescale of the eccentricity of planets 1 and 2, respectively, with masses $m_1$ and $m_2$, for different planetary masses $\epsilon_p=(m_1+m_2)/M_*$. The left plot is for the 3:2 resonance, the right plot for the 2:1 resonance. For a planet of mass $m$ at semi-major axis $a$ the eccentricity damping timescale is $\tau_e\propto (H/a)^4/(m\Sigma\sqrt{a})$, where $\Sigma$ is the surface density of the disk and $H/a$ its aspect ratio. Thus, if the disk has a constant aspect ratio and $\Sigma\propto 1/\sqrt{a}$ the system should place on the dashed diagonal line (i.e. $m_1/m_2=tau_{e_1}/tau_{e_2}$). This shows, for instance, that a 3:2 system is stable if $m_1<m_2$. If instead the inner planet is in a low-density region (i.e. a disk's cavity), $\tau_{e_1}\to\infty$ and therefore the system is stable for all mass ratios. Adapted from (\cite{deck15}). }
\label{overstability}
\end{figure}

Fig.~\ref{overstability} shows a summary of the most detailed investigation of the stability/overstability of a resonance in presence of eccentricity damping \cite{deck15}. If the eccentricity damping is proportional to the planetary masses (as in the case where both planets are embedded in the disk), for the 3:2 and higher-$k$ resonances the resonant configuration is stable whenever $m_1>m_2$. In the opposite case, the stability/overstability depends on the total mass ratio $\epsilon_p=(m_1+m_2)/M_*$. There is a limit value of the ratio $m_1/m_2$ below which the system is overstable, and this ratio decreases with increasing $\epsilon_p$ (Fig.~\ref{overstability}a). For the 2:1 resonance the situation is qualitatively similar, but the planets can be overstable even if $m_1/m_2>1$ if $\epsilon_p$ is small enough (Fig.~\ref{overstability}b).  On the other hand, if there is no eccentricity damping on the inner planet ($\tau_{e_1}\to \infty$, which happens if the inner planet has been pushed into a disk's cavity), the resonant configuration is always stable for any $m_1/m_2$ ratio. 

Even if resonant planets are not in the overstable regime, they may be unstable because of other processes. Due to their proximity and eccentric orbits they may approach  each other too much over their resonant trajectories and be destabilized by a close-encounter \cite{pichierri18}. There can also be subtle secondary resonances between a combination of the libration frequencies and the synodic frequency $\dot{\lambda}_1-\dot{\lambda}_2$. These secondary resonances cannot be described with the averaged Hamiltonian (Eq.~\ref{averaged}) because the terms in $\lambda_1-\lambda_2$ have been removed by the averaging procedure.~But they can be studied following a more precise and convoluted approach. As the number of planets in a resonant chain increases, the number of libration frequencies increases as well and therefore a richer set of combined frequencies is possible. This explains, at least at the qualitative level, why long resonant chains are more fragile than short chains, as observed in numerical simulations \cite{matsumoto12,cossou13}. The long-term evolution of multi-planet resonant chains remains nevertheless an active area of research in celestial mechanics. 

\section{Global models of planet formation} 
\label{sec:global}

Building global models of planet formation is akin to putting together a puzzle.  We have a vague picture of what the puzzle should look like (i.e., from exoplanet demographics) but observational biases cloud our view.  And the puzzle pieces -- the planet formation mechanisms -- often change in number and in shape.  The puzzle-builders must constantly have an eye both on the evolution of the big picture and on the set of viable puzzle pieces.  And one must not hesitate to discard a model when it no longer serves.

We will describe our current best global models for the origin of super-Earth systems (Section 3.1), giant planet systems (Section 3.2) and our own Solar System (Section 3.3).  Then we will look at how water may be delivered to rocky planets (Section 3.4).

\subsection{Origin of close-in super-Earths}

The key properties of super-Earth systems that must be matched by any formation model can be very simply summarized as follows:

\begin{itemize}
\item A large fraction (roughly one third to one half) of stars have close-in super-Earths (with periods shorter than100 days; \cite{fressin13,mulders18}), but many (perhaps most) of them do not.
\item Most super-Earth systems only have a single planet detected in transit, whereas a fraction of systems is found with many planets in transit~\cite{lissauer11b,fang12,tremaine12}.
\item In multi-planet systems, pairs of neighboring super-Earths are rarely found to be in mean motion resonance~\cite{lissauer11b,fabrycky14}.
\item The masses of super-Earths extend from Earth to Neptune, with a preference for a few $\mearth$~\cite{weiss13,marcy14,wolfgang16,chen17}.
\end{itemize}

Models for the formation of super-Earths were developed before they were even discovered \cite{raymond08a}.  While certain models have been refined in recent years, only a single new model has been developed.

{\it In-situ accretion} is the most intuitive and simplest model for super-Earth formation yet it has a fatal flaw. That model proposes that super-Earths simply accreted from local material very close to their stars in a similar fashion to the classical model of terrestrial planet formation in the Solar System (see Section 3.3.1 below).  In-situ accretion was proposed in 2008 \cite{raymond08a} and discarded because the masses implied in the innermost parts of disks seemed prohibitively large.  In 2013 this idea was revisited by Chiang and Laughlin \cite{chiang13}, who used the population of known super-Earths to generate a ``minimum-mass extrasolar nebula'' representing a possible precursor disk that would have formed the population of super-Earths.  The high masses inferred in inner disks conflicts with measurements, but those measurements are only of the outermost parts of disks \cite{williams11}.  While it is possible to imagine that inner disks can pile up material, there is a simply-understood timescale problem.  With very high densities in the inner disk the growth timescale for super-Earths is extremely fast \cite{lissauer07,raymond07b,raymond14b,inamdar15,schlichting14,ogihara15}.  In fact, the growth timescales are so fast and the requisite disks so massive that migration is simply unavoidable \cite{ogihara15}. Even aerodynamic drag is strong enough to cause rapid orbital drift \cite{inamdar15,grishin15}.  Thus, we cannot consider the planets to have formed in-situ because they must have migrated and their final orbits cannot represent their starting ones.  Nonetheless, it has been shown that if the right conditions were to arise, with the requisite amount of solid material close to the star, accretion should indeed produce planetary systems similar to the observed super-Earths~\cite{hansen12,hansen13,dawson15,dawson16,moriarty16,lee14,lee16,lee17}.

A number of the first processes to be explored for forming super-Earths relied on giant planets.  For example, migrating giant planets can shepherd material interior to their orbits and stimulate the growth of super-Earths~\cite{zhou05,fogg05,fogg07,raymond06c,mandell07}. Moving secular resonances driven by giant planet interactions can do the same \cite{zhou05}.  However, these models have been ruled out as the main formation pathways for super-Earths because most super-Earth systems do not appear to have an associated giant planet (although testing the correlation between super-Earths and outer gas giants is an active area of study \cite{bryan19,zhu18,barbato18}). 

\begin{figure}[t]
\centering
\includegraphics[width=12cm]{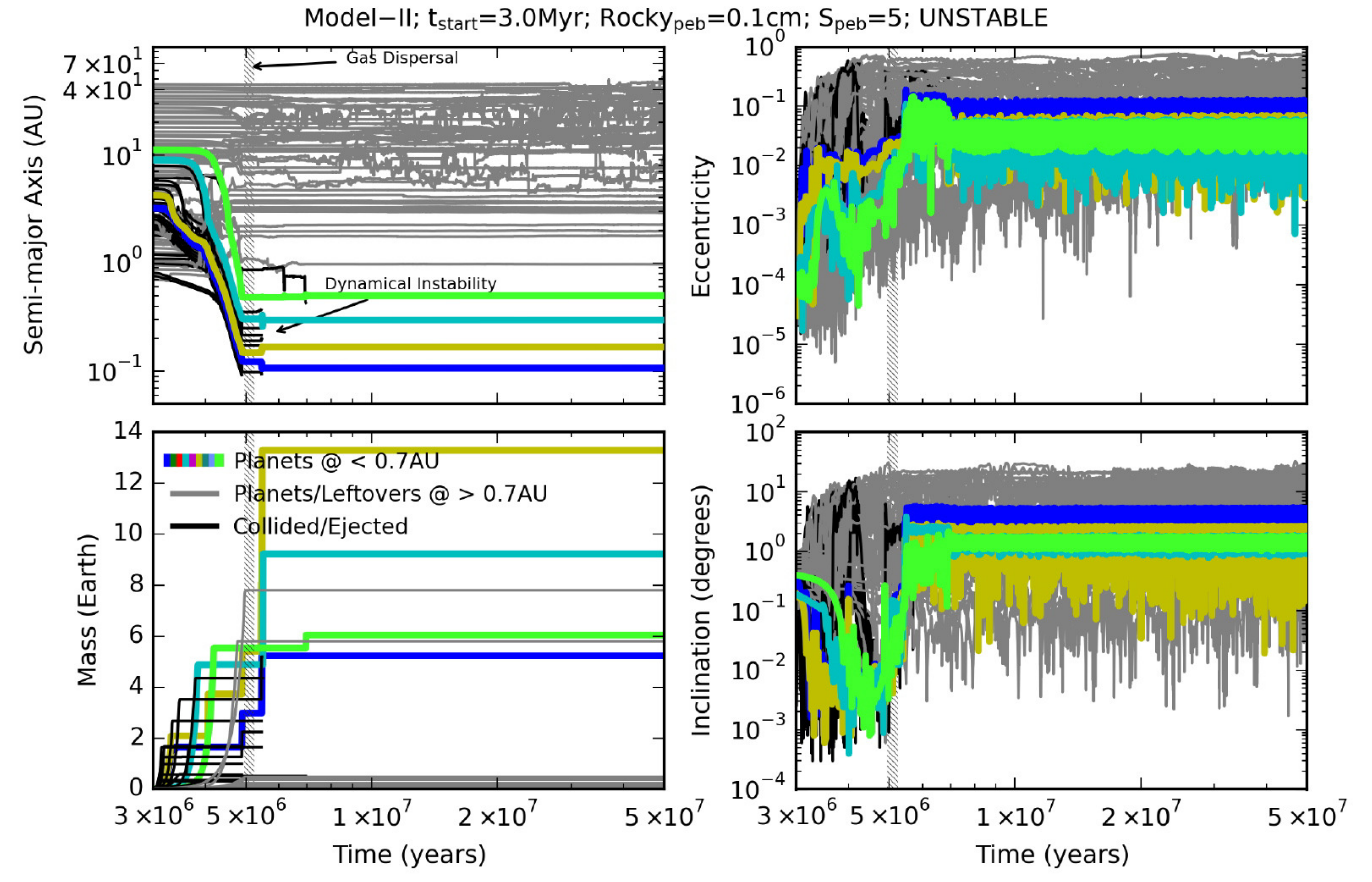}
\caption{Evolution of a simulation of the {\em breaking the chains} migration model for the origin of close-in super-Earths. The panels show the evolution of a population of $\sim$lunar-mass planetary embryos that grow by accreting pebbles, migrate inward and form a resonant chain anchored at the inner edge of the disk.  The resonant chain is destabilized shortly after the dissipation of the gaseous disk, leading to a late phase of scattering and collisions that spreads out the planets and increases their eccentricities and inclinations. From \cite{izidoro19}. }
\label{fig:btc_evol}       
\end{figure}

The migration model has proven quite successful model in reproducing the observed population of super-Earths\cite{terquem07,ogihara09,mcneil10,ida10,cossou14,izidoro17,izidoro19,raymond18b,carrera18}.  In that model, large planetary embryos grow throughout the disk and migrate inward, driven by the gaseous disk (Fig.~\ref{fig:btc_evol} shows an example of a simulation of the migration model).  It is natural to think that embryos would form first past the snow line, where pebble accretion is thought to be more efficient (see Section 2.3 and \cite{morby15}). However, it is also possible that in some disks large embryos can form very close to their stars.  This might occur if inward-drifting pebbles are concentrated at a pressure bump, perhaps associated with the inner edge of a dead zone~\cite{chatterjee14,chatterjee15}.\footnote{This idea forms the basis of a model of super-Earth formation sometimes called {\em inside-out planet formation}~\cite{boley13,chatterjee14,chatterjee15,hu16,hu17}. That model invokes the direct formation of super-Earths from pebbles at pressure bumps in the inner disk.}  While the migration rate and direction depends on the disk model  \cite{lyra10,kretke12,bitsch13,bitsch14}, inward migration is generally favored.  The inner edge of the disk creates a strong positive torque \cite{flock19} that acts to trap any inward-migrating embryo \cite{masset06}. This leads to a convergence of bodies near the disk's inner edge. Convergent migration leads to resonant trapping (see Section 2.6) and tends to produce planets in long chains of resonances. Of course, the observed super-Earths are not found in resonance.  However, most resonant chains become dynamically unstable as the gaseous disk dissipates.  This leads to a phase of giant collisions between embryos that is quite similar to that simulated by studies that ignored the migration phase and invoked a large population of embryos in a dissipating gaseous disk \cite{hansen12,hansen13,lee14,dawson15,dawson16}.  The instability phase leads to scattering among embryos, breaks the resonant chains, and causes systems to spread out and become dynamically excited.  A small fraction of resonant chains remain stable after the disk dissipates; these may represent iconic systems such as Trappist-1 \cite{gillon17,luger17} and Kepler-223 \cite{mills16}.  

Simulations of the migration model have shown that the surviving systems quantitatively match the population of observed super-Earths as long as more than 90\% of resonant chains become unstable~\cite{izidoro17,izidoro19}.  When run through a simulated observing pipeline, the significant dynamical excitation of the surviving systems implies large enough inclinations such that most viewing geometries can only see a single planet in transit.  This solves the so-called Kepler dichotomy problem~\cite{johansen12,fang12} and implies that all super-Earth systems are inherently multiple.  The period ratio distribution of simulated systems matches observations, again taking observational biases into account~\cite{izidoro17,izidoro19,mulders19}. 

Yet questions remain.  If super-Earth formation is as efficient as in simulations, why don't all stars have them?  One possibility is that when an outer gas giant planet forms, it blocks the inward migration of large embryos and those instead become ice giants \cite{izidoro15a,izidoro15b}. This implies an anti-correlation between the presence of outer gas giants and systems with many super-Earths, and it remains unclear if such an anti-correlation exists\cite{bryan19,zhu18,barbato18}. The fraction of stars with gas giants appears to be far less than the fraction of stars without super-Earths, which makes it difficult to imagine gas giants being the main cause. Is it possible, instead, that outer ice giants or super-Earths are the culprit? Probably not. While their occurrence rate is high \cite{gould10,suzuki16b}, outer ice giants cannot efficiently block the inward migration of other planets and are generally too low-mass to block the pebble flux \cite{morby12,lambrechts14,bitsch18}. Perhaps, instead, many disks are subdivided into radial zones \cite{johansen09}. Pebbles trapped within a given zone may not be able to drift past the zone boundary such that the pebble flux in certain regions of the disk would remain too low for planets to grow fast enough for long-range migration \cite{lambrechts19}. Such a scenario would also be compatible with the ringed structures seen in many ALMA disks~\cite{andrews18}.

The compositions of super-Earths may also provide a constraint for formation models.  It naively seems that the migration model should produce very volatile-rich super-Earths because the main source of mass is beyond the snow line, where the efficiency of pebble accretion is higher \cite{morby15,izidoro19,bitsch19,bitsch19b}.  This need not be true in 100\% of cases, as inward-migrating icy embryos can in some cases stimulate the growth of inner, purely rocky planets~\cite{raymond18b}.  However, most super-Earths in the migration model should be ice-rich \cite{izidoro19}.  It is unclear whether this is consistent with the observed distribution of bulk densities of super-Earths.  While it has been claimed that most super-Earths appear to be ``rocky'' \cite{owen13,owen17,lopez17,jin18}, measurement uncertainties preclude any clear determination of the compositions of super-Earths \cite{dorn15,dorn18}.  In addition, ``rocky'' planets may in some context include water contents up to $\sim$20\% by mass~\cite{gupta19}.  For context, that is similar to the approximate water contents of comet 67P \cite{paetzold19}, Pluto, and the most water-rich meteorites known (thought to originate from the outer Solar System \cite{alexander19a,alexander19b}).  Finally, short-lived radionuclides such as Al-26 may efficiently dry out some super-Earths \cite{lichtenberg19}. The rocky vs. icy nature of super-Earths remains an important outstanding issue.

The fact that super-Earths seem to ubiquitously have large radii consistent with atmospheres of a few percent H/He by mass~\cite{weiss14,wolfgang16,fulton17,fulton18} confirm that these planets formed during the gaseous disk phase and certainly by a process quite different than the formation of our own terrestrial planets. Formation models are starting to be coupled to models of atmospheric accretion and loss \cite{lee14,inamdar15,ginzburg16,lee16,carrera18}. However, given the complexities in these processes \cite{lambrechts17} this remains an ongoing challenge.

\subsection{Giant exoplanets: formation and dynamics}

Drawing from numerous radial velocity, transit and microlensing surveys for exoplanets, the essential constraints on giant planet formation are:
\begin{itemize}
\item Gas giants exist around roughly 10\% of Sun-like stars \cite{cumming08,mayor11,winn15,foremanmackey16,suzuki16b,fernandes19} and are more/less common around more/less massive stars \cite{lovis07,johnson07,clanton14,clanton16}.
\item Most gas giants are on relatively wide orbits past 0.5-1 au \cite{butler06,udry07b,mayor11,rowan16,wittenmyer16,wittenmyer20}.
\item Gas giant exoplanets tend to have much higher eccentricities than Jupiter and Saturn, following a broad eccentricity distribution with a median of $\sim 0.25$ \cite{butler06,udry07b,wright09,bonomo17}.
\item There is a strong correlation between a star's metallicity and the probability that it hosts a gas giant planet \cite{gonzalez97,santos01,fischer05}, especially among hot Jupiters and gas giants with eccentric orbits \cite{dawson13}.
\end{itemize}

\begin{figure}[t]
\centering
\includegraphics[width=5.6cm]{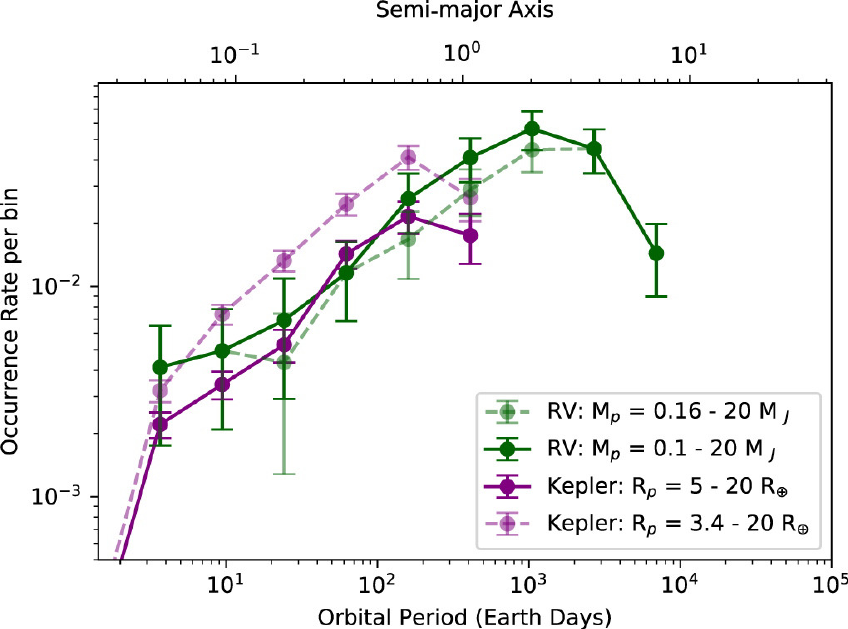}
\includegraphics[width=5.6cm]{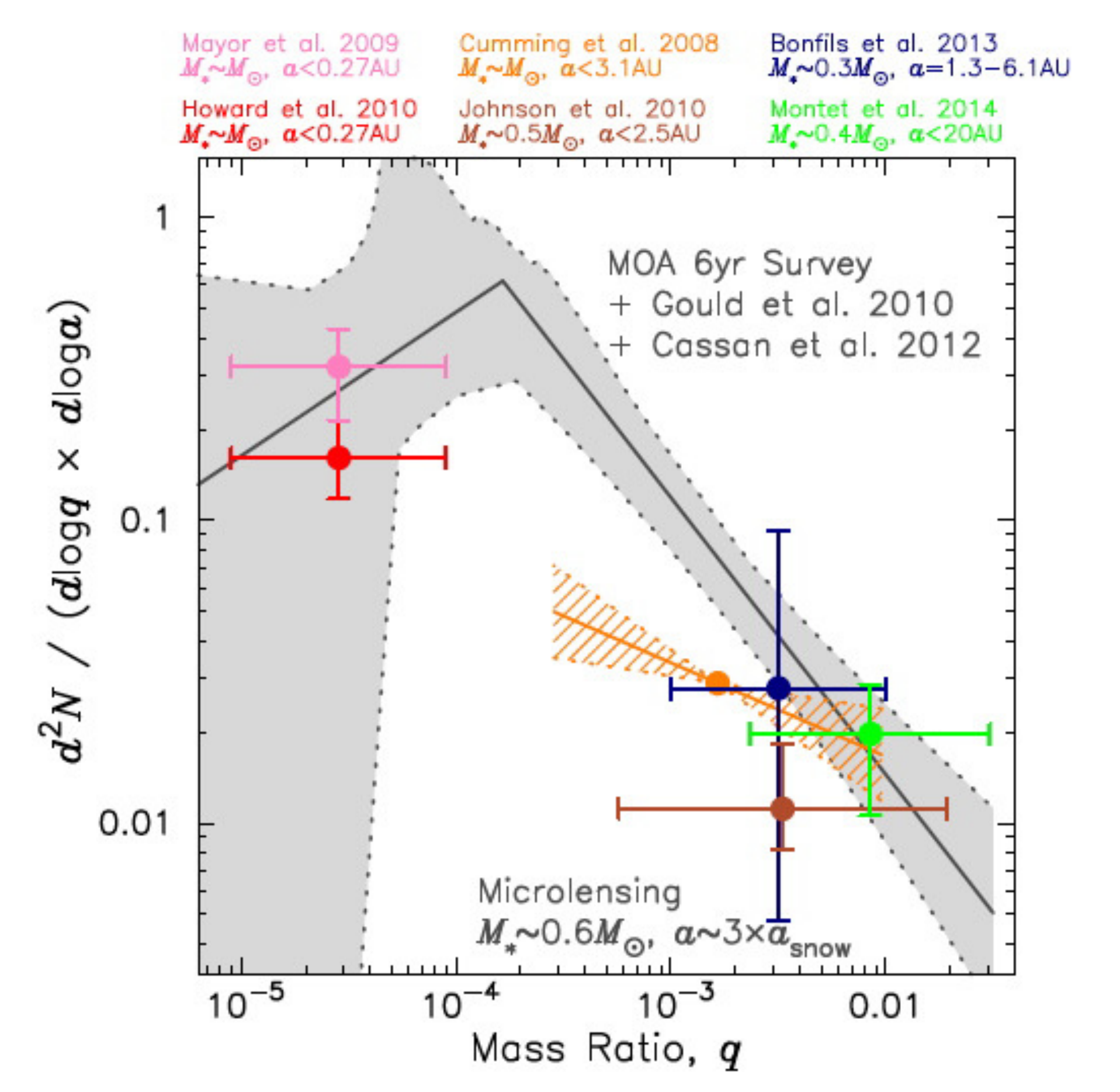}
\caption{Mass and orbital radius distributions of giant exoplanets.  {\bf Left:} The frequency of giant planets as a function of orbital distance from radial velocity surveys (green) and the Kepler transit statistics (purple).  From \cite{fernandes19}. {\bf Right:} The occurrence rate of giant planets as a function of planet-to-star mass ratio $q$, from \cite{suzuki16b}. The different symbols correspond to different radial velocity studies whereas the black line and grey regions are the results from microlensing surveys. }
\label{fig:giant_freq}       
\end{figure}

There exist two categories of formation models for giant planets \cite{boley09,helled14}.  The first is a top-down, collapse scenario in which a localized instability in a protoplanetary disk can lead to the direct formation of one or more giant planets.  The second is the bottom-up, core accretion scenario (which we argue below should really be called the {\em core-migration-accretion} scenario).  In recent years the core accretion model has become the dominant one, yet it is plausible that some systems may be explained by the disk instability model.

\paragraph {Disk Instability} 
The disk instability scenario \cite{boss97,boss98,boss00,mayer02,mayer07} invokes a localized region of the disk that becomes gravitationally unstable.  This is usually quantified with the Toomre $Q$ criterion \cite{toomre64}:
\begin{equation}
\frac{c_s \, \Omega}{\pi \, G \, \Sigma} < Q_{crit} \approx  1,
\end{equation}
\noindent where $c_s$ is the sound speed, $\Omega$ is the local orbital frequency, $G$ is the gravitational constant, and $\Sigma$ is the local surface density of the disk. Simulations show that in disks that have $Q \lesssim 1$, fragmentation does indeed occur and Jupiter-mass clumps form quickly \cite{boss98,mayer07,boley10}.  To become true gas giants, these clumps must cool quickly to avoid being sheared apart by the Keplerian flow.

It is generally thought that only the outermost parts of massive disks are able to attain the criteria needed for instability \cite{meru11,armitage11,kimura12}.  The very distant, massive planets such as those of the HR8799 system \cite{marois08,marois10} may perhaps be explained by disk instability, as other mechanism struggle to form such massive planets so far away from their stars. However, it is not clear whether disk instability can produce Jupiter-like planets (although the giant planet orbiting the low-mass star GJ3512 is a very good candidate -- \cite{morales19}). Planets or clumps that form very far out would also migrate inward rapidly \cite{baruteau14} and it is unclear whether they would survive. In the {\em tidal downsizing model}, gravitational instability forms clumps in the outer parts of disks that migrate inward but are often disrupted to act as the seeds of much smaller planets \cite{nayakshin15}. 

\paragraph {Core-{\em Migration}-Accretion} 

The bottom-up, core accretion scenario forms the basis of the current paradigm of giant planet formation. In its standard form the core-accretion model essentially represents a combination of processes summarized in Sections 2.3 (growth of protoplanets) and 2.5 (gas accretion).  As we shall see, another important process (orbital migration -- described in Section 2.4) must be included.

Since its inception the canonical picture has suffered setbacks related to core growth, migration, and gas accretion (see Section 2.3).  Early models that invoked planetesimals as the building blocks of large cores could not explain the rapid growth of the $\sim10 \mearth$ cores needed to trigger rapid gas accretion~\cite{levison01,thommes03,rafikov04,chambers06,levison10}. Even in the most optimistic scenarios planetesimal accretion was simply too slow. In recent years it has been shown that pebble accretion is far more efficient and can indeed -- given that there is a sufficiently massive and drawn-out supply of pebbles \cite{johansen17,bitsch19} -- explain the rapid growth of giant planet cores~\cite{ormel10,lambrechts12,lambrechts14}.  It has also been shown that gas accretion is actually not halted, and only modestly slowed down, by the generation of a gap in the disk \cite{dangelo03,uribe13}.

\begin{figure}[t]
\centering
\includegraphics[width=12cm]{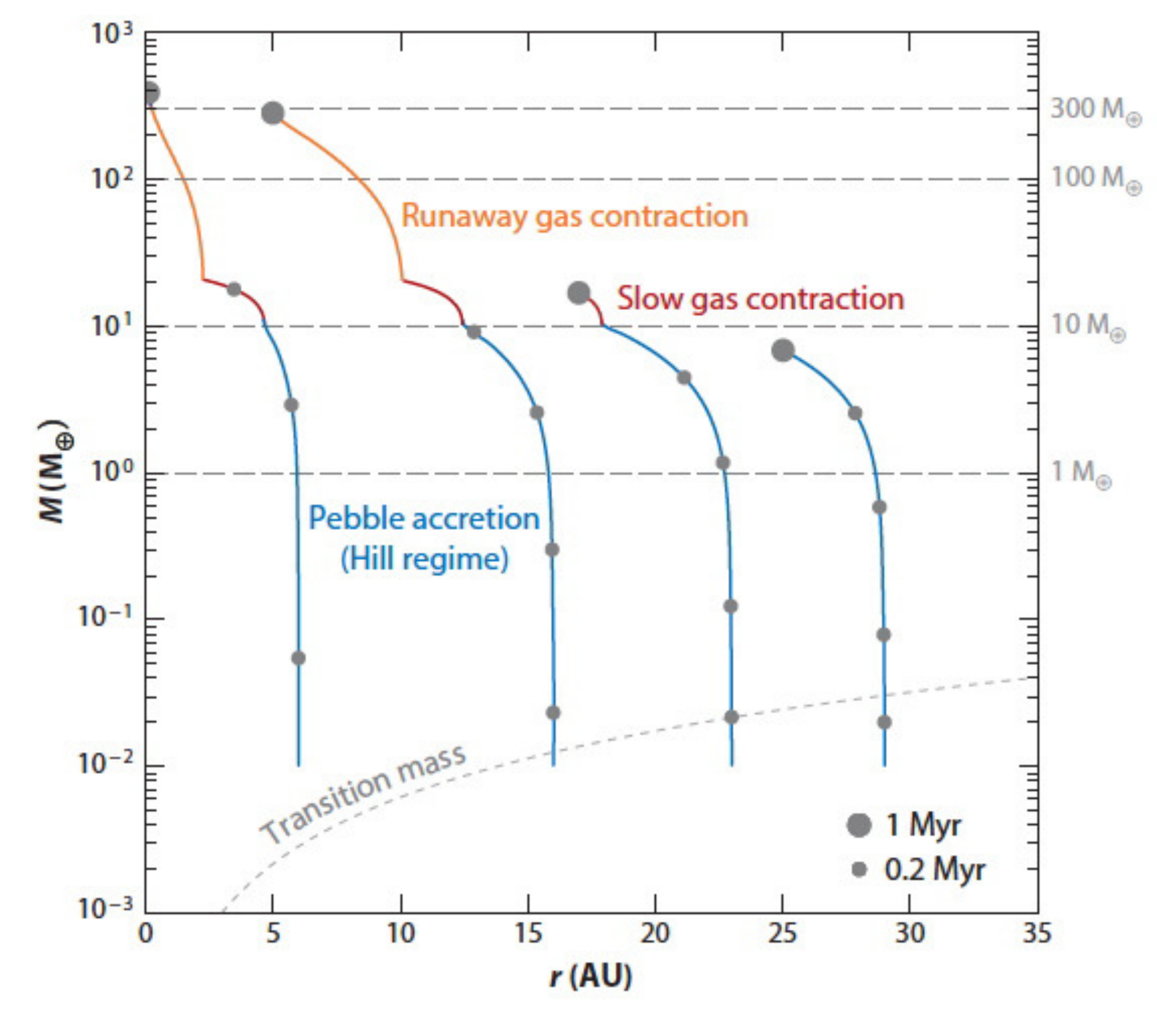}
\caption{Growth tracks for four different types of planets from simulations that include pebble accretion and migration and gas accretion.  The leftmost planet grows fast enough to migrate a long distance and become a hot Jupiter.  The second planet from the left starts as a core at $\sim15 au$ and ends up as a Jupiter-like planet at 5 au. The two planets on the right are ice giant analogs that never undergo rapid gas accretion.  In this case the pebble isolation mass -- the mass above which the pebble flux is blocked and pebble accretion shut off \cite{lambrechts14b,bitsch18} -- was fixed at $10 \mearth$. Each small dot denotes a time interval of 0.2 Myr along the growth tracks and each large dot a time interval of 1 Myr. From \cite{johansen17}.}
\label{fig:growthtracks}       
\end{figure}

Migration remains a giant issue for the core accretion model, so much so that the model itself could plausibly be renamed the core-{\em migration}-accretion model.  It is not so much a problem as an added dimension.  The mass scale at which gas accretion becomes an important phenomenon is similar to that at which migration becomes important.  Thus, planets that are undergoing gas accretion must necessarily be migrating at the same time.  Growth tracks of planets must include both radial and mass growth. 

Figure~\ref{fig:growthtracks} shows the growth tracks of four different simulated planets.  Starting from $\sim$lunar-mass cores, each planet's growth is initially determined simply by the flux of pebbles across its orbit \cite{lambrechts14,ida16}.  Once each planet reaches several Earth masses it starts to migrate inward. At the same time it starts to slowly accrete gas.  Above the pebble isolation mass (defined in Sec 2.3) pebble accretion is stopped but migration and gas accretion continue. The two inner planets in the simulation grew fast enough to undergo rapid gas accretion and become gas giants, whereas the two outer planets did not accrete enough gas before the disk dispersed and thus ended up as ice giants.  

Fig.~\ref{fig:growthtracks} leads to a naive-sounding but surprisingly profound question: {\em Why is Jupiter at 5 au?} In the model from Fig.~\ref{fig:growthtracks}, any core that started within roughly 10 au ended up as a hot Jupiter.  Likewise, in order to finish at 5 au a core needed to start at 15-20 au (see also \cite{bitsch15,ndugu18}).  How, then did our own Jupiter avoid this fate? There are currently four possible solutions.  First, perhaps Jupiter's core simply did form at 15-20 au \cite{bitsch15b}.  While this is hard to rule out, it seems unlikely because most material that originated interior to Jupiter's orbit would have remained interior to Jupiter's final orbit \cite{raymond17,pirani19}. This would therefore require that only a few Earth masses of material formed in the entire Solar System interior to 15-20 au.  Second, perhaps Jupiter's migration was much slower than that shown in Fig.~\ref{fig:growthtracks}.  That could be the case if the disk's viscosity was much lower than generally assumed, which is consistent with new disk models (see Section 2.1).  Third, perhaps Jupiter's migration was halted because the inner gas disk was evaporated away by energetic radiation from the active young Sun \cite{alexander12b}.  Indeed, disks are thought to disperse by being photo-evaporated by their central stars \cite{alexander14}, and this may be a consequence.  Fourth, perhaps Jupiter's migration was held back by Saturn.  Hydrodynamical simulations show that, while a Jupiter-mass planet on its own migrates inward, the Jupiter-Saturn pair can avoid rapid inward migration and sometimes even migrate outward \cite{masset01,morby07,pierens08,zhang10,pierens11,pierens14}.  This forms the basis of the Grand Tack model of Solar System formation, which we will discuss in detail in Section 3.3.2.  However, that model's potential fatal flaw is that avoiding inward migration requires that Jupiter and Saturn maintain a specific mass ratio of roughly 3-to-1 \cite{masset01} and it is uncertain whether that ratio can be maintained in the face of gas accretion.

Despite these uncertainties, the core accretion model can match a number of features of the known exoplanet population~\cite{ida04,mordasini09,ndugu18}, including the observed giant planet-metallicity correlation~\cite{fischer05} and the much higher abundance of Neptune-mass compared with Jupiter-mass planets~\cite{butler06,udry07b,suzuki16b}.

\paragraph {Unstable Giant Planet Systems and the Exoplanet Eccentricity Distribution} 

Gas giant exoplanets are often found on non-circular orbits. This is surprising because, as we saw in Section 2, planets form in disks and so on orbits similar to the disks' streamlines, which are all circular and coplanar. The broad eccentricity of giant planets was for years a subject of broad speculation (for a historical discussion and a comprehensive list of proposed mechanisms, see \cite{ford08}).  The most likely culprit is a mechanism often referred to as {\em planet-planet scattering} \cite{rasio96,weidenschilling96,lin97}.  The planet-planet scattering model proposes that the giant exoplanets formed in systems with many planets and that those that we see are the survivors of dynamical instabilities.  Instabilities lead to orbit crossing, followed by a phase of gravitational scattering that usually concludes when one or more planets are ejected from the system entirely, typically leaving the surviving planets with eccentric orbits \cite{marzari02,adams03,moorhead05,chatterjee08,juric08,nagasawa08,raymond08b,raymond09a,raymond10,raymond11,raymond12,beauge12}.

Whereas instabilities among systems of rocky planets tend to lead to collisions, among systems of giant planets they lead to scattering and ejection.  This can be understood by simply considering the conditions required for a planet to be able to give a strong enough gravitational kick to eject another object.  The Safronov number $\Theta$ is simply the ratio of the escape speed from a planet's surface to the escape speed from the star at that orbital radius.  It is defined as:
\begin{equation}
\Theta^2 = \left( \frac{G \, m_p}{R_p} \right) \left( \frac{a_p}{G \, M_\star} \right) = \frac{m_p}{M_\star} \frac{a_p}{R_p},
\end{equation}
\noindent where $m_p$ and $M_\star$ are the planet and star mass, respectively, $a_p$ is the planet's orbital radius, and $R_p$ its physical radius. When $\Theta < 1$ collisions are (statistically-speaking) favored over ejection, and when $\Theta > 1$ ejection is favored. Scattering and ejection are therefore favored for massive planets far from their stars.  It is interesting to note that $\Theta$ is defined in the same way as the gravitational focussing factor $F_G$ from Eq.~\ref{focussing} in Section 2.2, which described the gravitational cross section of a growing protoplanet such that $F_G = 1 + \left( V_{esc}/V_{rel}\right)^2$. Here, in the context of scattering, the dynamics are the same but in reverse. While the gravitational focussing factor $F_G$ describes from how far away can a planet accrete another object, $\Theta$ is a measure of how far a planet can launch another object.  

The planet-planet scattering model matches the observed eccentricity distribution of giant exoplanets \cite{adams03,juric08,chatterjee08,raymond09a,raymond10}.  The only requirement is that a large fraction of systems have undergone instabilities. In simulations performed to date, at least 75\% -- and probably more like 90-95\% -- of giant planets systems must have undergone planet-planet scattering in their past \cite{juric08,raymond10,raymond11}.  Scattering can even explain the highest-eccentricity giant exoplanets \cite{carrera19}.

There is a anti-correlation between giant exoplanet masses and eccentricities.  More massive planets are observed to have higher eccentricities \cite{jones06,ribas07,wright08}.  This is the opposite of what one would expect from planet-planet scattering. Given that scattering is a process of equipartition of energy, one would naturally expect the low-mass planets to end up with high eccentricities and the high-mass planets with low-eccentricities.  This is indeed what is seen in simulations of unstable systems starting with a dispersion of different planet masses.  The solution to this conundrum may be quite simply that massive planets form in systems with many, roughly equal-mass planets \cite{raymond10,ida13}.  In that case, the most massive planets do indeed end up on the most eccentric orbits, as observed.

\paragraph{A {\em Breaking the Chains} Scenario for Giant Planets}

\begin{figure}[t]
\includegraphics[width=12cm]{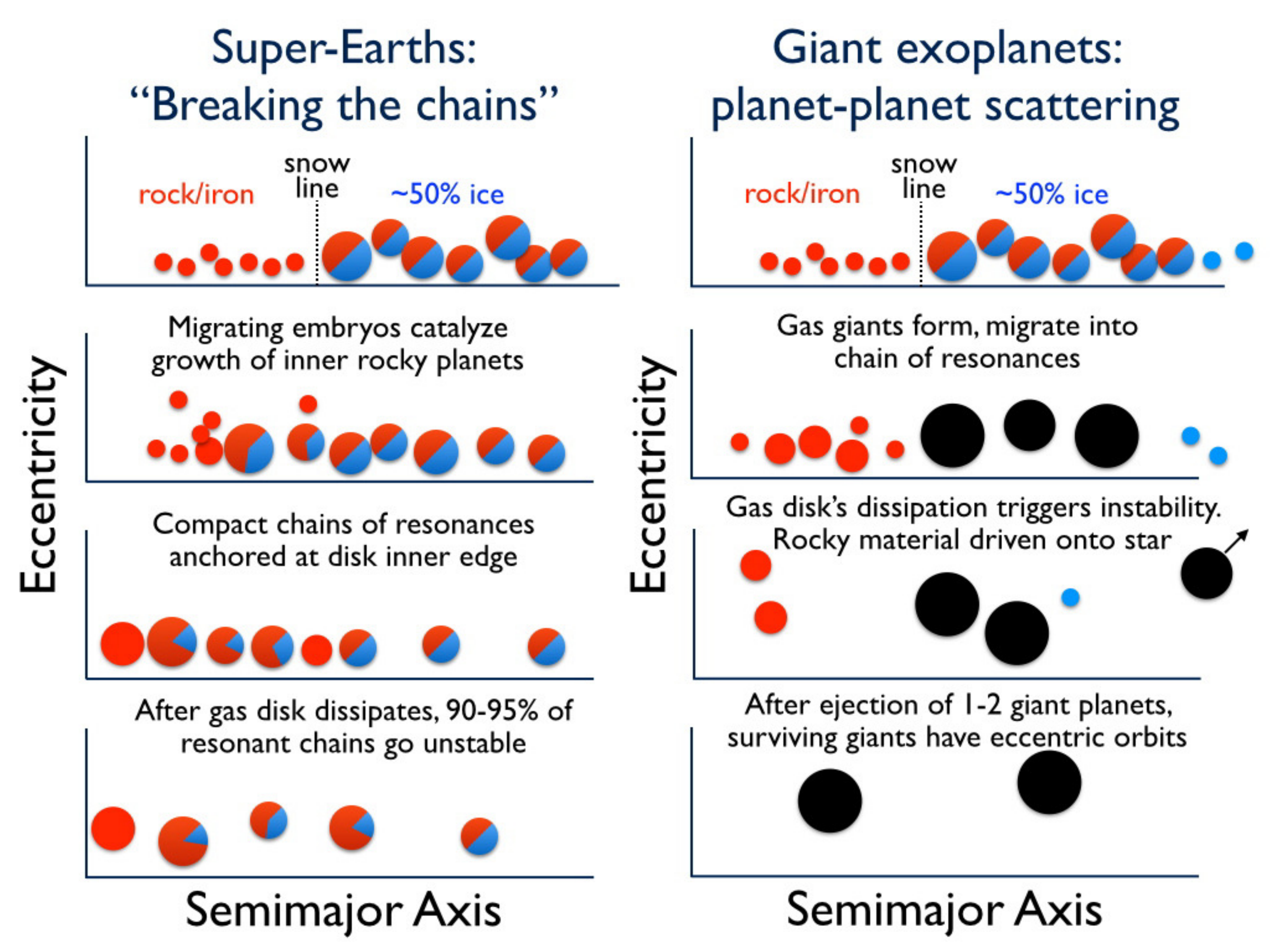}
\caption{Cartoon comparison of the evolution of super-Earth and giant planets systems. The same general processes could very well causing both populations to follow {\em breaking the chains}-like pathways. From \cite{raymond18d}. }
\label{fig:exoplanet_cartoon}       
\end{figure}

The evolution of giant planet systems may well parallel that of super-Earths.  As we saw in Section 3.1, the prevailing model for super-Earth formation -- called {\em breaking the chains} \cite{izidoro17} -- involves migration into resonant chains followed by instability.  Could this same evolutionary pathway apply to giant planet systems (Fig.~\ref{fig:exoplanet_cartoon})?

The answer appears to be yes.  Giant planets may indeed follow a {\em breaking the chains}-style evolution that is similar, but not identical, to that of super-Earth systems.  

Survival of resonances after the gaseous disk phase may be somewhat more common for giant planet systems. Only a few dozen giant planets are known to be found in resonance (e.g., the GJ876 resonant chain of giant planets \cite{rivera10}). Yet it is possible that many more resonant pairs of giant planets are hiding in plain site. The radial velocity signatures of resonant planets can mimic those of a single eccentric planet \cite{anglada10} and it is possible that up to 25\% of the current sample of eccentric planets are actually pairs of resonant planets \cite{boisvert18}. The PDS 70 protoplanetary disk may host a pair of young giant planets in resonance \cite{bae19}, and other disk signatures may require multi-resonant planets to explain them \cite{dodsonrobinson11}.  When multiple giant planets form within a given disk it may thus slow and limit migration. This contrasts with the case of super-Earths, which migrate all the way to the disk's inner edge.

Instabilities appear to be ubiquitous among both super-Earths and giant exoplanets. As discussed above, the eccentricities of giant exoplanets are easily matched if most systems are survivors of instability.  Instabilities may be triggered by the dispersal of the gaseous disk \cite{matsumura10}, chaotic diffusion within the giant planets' orbits \cite{marzari02,batygin15b}, or external perturbations, e.g. from wide binary stars \cite{kaib13}.

\paragraph{Connection with Rocky Planets and Debris Disks}

\begin{figure}[t]
\centering
\includegraphics[width=12cm]{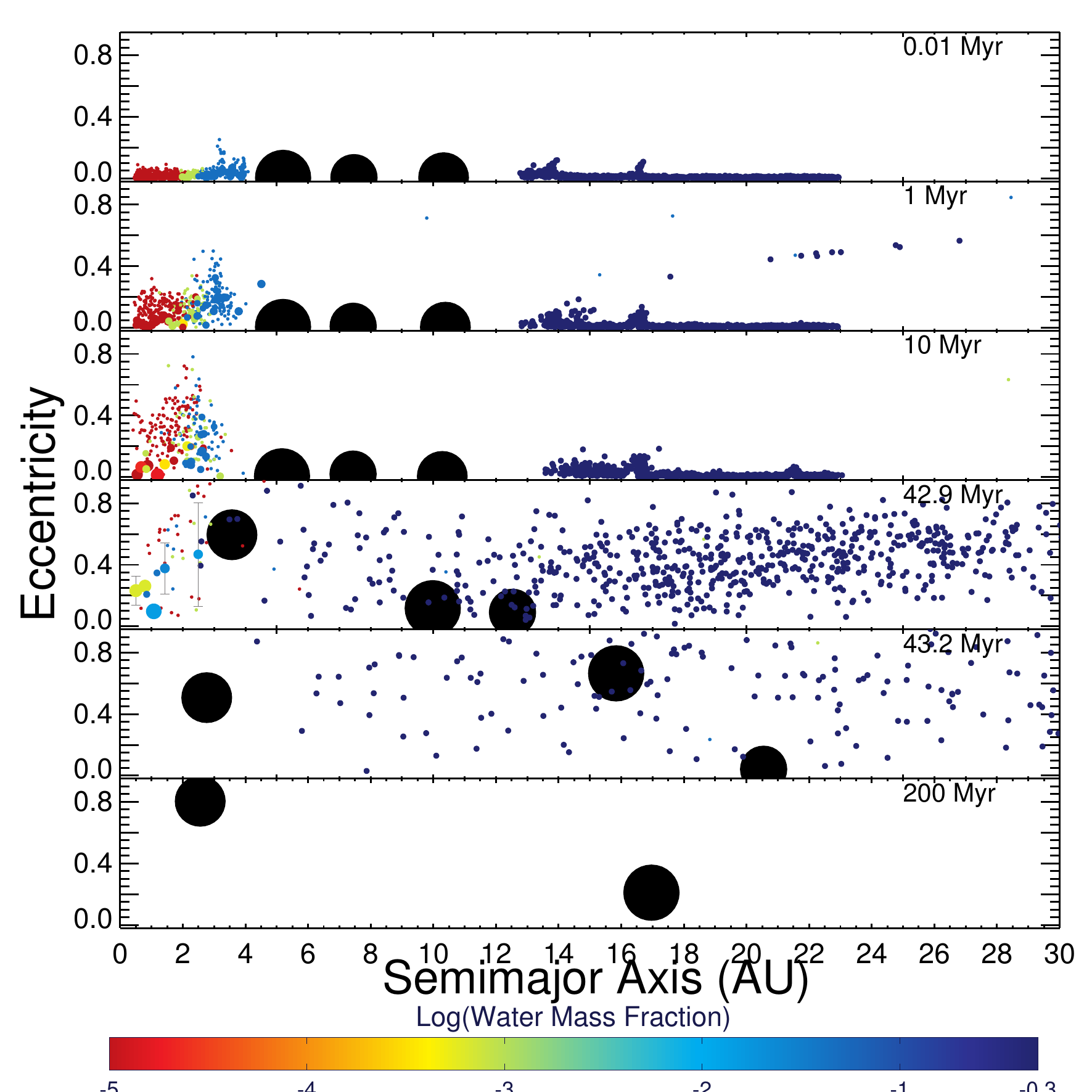}
\caption{Evolution of a simulation in which a system of gas giant went unstable, resulting in the destruction of both an inner system of growing terrestrial planets (that were driven into collisions with the central star) and an outer disk of ice-rich planetesimals (that was ejected into interstellar space; see \cite{raymond18,raymond18c}.  From \cite{raymond12}.}
\label{fig:unstab}       
\end{figure}

When giant planets go unstable they affect their entire system.  Given their large masses and the high eccentricities they reach during the scattering phase, giant planets can wreak havoc on their inner and outer systems.  Giant planet scattering systematically disrupts inner rocky planet systems \cite{veras05,veras06,matsumura13,carrera16} or their building blocks \cite{raymond11,raymond12}, usually by driving inner bodies onto such eccentric orbits that they collide with their host stars (see Fig.~\ref{fig:unstab}).  Scattering also tends to destroy outer planetesimals disks by ejecting planetesimals into interstellar space \cite{raymond11,raymond12,raymond13,marzari14}. Outer planetesimal belts -- when they survive and contain enough mass to self-excite to a moderate degree -- evolve collisionally to produce cold dust observed as debris disks \cite{wyatt08,krivov10,matthews14,hughes18}, .  

Raymond et al~\cite{raymond11,raymond12} proposed that, by influencing both the inner and outer parts of their systems, giant planets induce a correlation between rocky inner planets and debris disks.  Hints of a correlation have been found \cite{wyatt12} but more data are needed.  It has also been suggested that giant planets -- especially those on very eccentric orbits -- should anti-correlate with the presence of debris disks.  While there may be a tendancy for debris disks to be less dense in systems with eccentric giant planets\cite{bryden09}, no strong correlation or anti-correlation between giant planets and debris disks has been observed to date \cite{moromartin07,moromartin15}.

Nonetheless, connections between planetary system architecture and the presence and characteristics of dust remains an active area of study.

\subsection{Solar System formation}

The standard timeline of Solar System formation proceeds as follows.  Time zero is generally taken as the formation of Calcium- and Aluminum-rich Inclusions, or CAIs.  CAIs are the oldest known inclusions in chondritic meteorites aged to be 4.568 Gyr old\cite{amelin02,bouvier10,connelly12}.  Radioactive ages of iron meteorites suggest that their parent bodies -- several hundred to a thousand-km scale planetary embryos -- were fully formed within 1 Myr after CAIs~\cite{halliday06,kruijer14,schiller15}.  There are two isotopically distinct types of meteorites: the so-called carbonaceous and non-carbonaceous groups~\cite{warren11}.  There is a broad distribution in the ages of both types of meteorites, whose overlap indicate that these very different types of meteorites accreted simultaneously~\cite{amelin02,warren11,kruijer17}.  Chondrules -- the primitive, mm-scale building blocks of chondritic meteorites -- are found at the size scale at which objects drift rapidly through the disk due to aerodynamic drag~\cite{weidenschilling77b,lambrechts12,johansen15}.  The separation of the two different types of chondritic meteorites is interpreted as indicating that two isotopically-distinct reservoirs were kept spatially segregated, perhaps by the rapid growth of a large planetary embryo to the pebble isolation mass~\cite{budde16,kruijer17}.  This may thus constrain the timing of the growth of Jupiter's core to have been very rapid, reaching the critical mass to block the pebble flux (of $\sim 10-20 \mearth$; \cite{bitsch18}) within 1-1.5 Myr after CAIs \cite{kruijer17}.  

The gaseous disk probably lasted for roughly 5 Myr.  Gaseous disks around other stars are observed to dissipate on a few Myr timescale (see Fig.~\ref{fig:disk_constraints}).  The oldest chondritic meteorites are the CB chondrites, which formed roughly 5 Myr after CAIs albeit perhaps in the absence of gas \cite{kita05,krot05,johnson16}.  Given that gas is thought to be needed for planetesimal formation \cite{johansen14}, this implies 5 Myr as an upper limit on the gas disk lifetime (at least in some regions).  

Hafnium-Tungsten (Hf/W) measurements of Martian meteorites indicate that Mars' formation was basically finished within 5-10 Myr \cite{nimmo07,dauphas11}, meaning that it grew very little after the disk had dispersed.  In contrast, Hf/W measurements indicate that Earth's last differentiation event -- generally thought to have been the Moon-forming impact -- did not take place until $\sim$100 Myr after CAIs \cite{touboul07,kleine09}. However, uncertainties in the degree of equilibration of the Hf/W isotopic system during giant impacts make it hard to determine a more accurate timeframe \cite{fischer18}.  

Highly-siderophile (``iron-loving'') elements should in principle be sequestered in the core during differentiation events.  Thus, any highly-siderophile elements in the terrestrial mantle and crust should in principle have been delivered {\em after} the Moon-forming impact \cite{kimura74}.  Simulations show that there is a clear anti-correlation between the timing of the last giant impact on Earth and the mass in planetesimals delivered after the last giant impact \cite{jacobson14}.  The total amount of highly-siderophile elements can therefore constrain the timing of the impact. Assuming a chondritic composition, Earth accreted the last $\sim 0.5\%$ of its mass as a veneer after the Moon-forming impact \cite{morbywood15}.  This implies that the Moon-forming impact took place roughly 100 Myr after CAIs \cite{jacobson14}, consistent with the Hf/W chronometer.

The Solar System's giant planets are thought to have undergone an instability.  The so-called {\em Nice model} showed that the giant planets' current orbital configuration \cite{tsiganis05,nesvorny12} -- as well as the orbital properties of Jupiter's Trojans \cite{morby05,nesvorny13}, the irregular satellites \cite{nesvorny14}, and the Kuiper belt \cite{levison08,nesvorny15} -- could be explained by a dynamical instability in the giant planets' orbits \cite{nesvorny18b}.  The instability is generally thought to have been triggered by interactions between the giant planets and a remnant planetesimal disk (essentially the primordial Kuiper belt \cite{gomes05,morby07,levison11,quarles19}), although a self-driven instability is also possible \cite{raymond10,ribeiro20}.  The instability was first proposed to explain the so-called terminal lunar cataclysm, i.e. an abrupt increase in the flux of projectiles hitting the Moon roughly 500 Myr after CAIs \cite{tera74,gomes05,morby12}.  However, newer analyses suggest that no terminal lunar cataclysm ever took place \cite{chapman07,boehnke16,zellner17,morby18,hartmann19}, and that the instability may have taken place anytime in the first $\sim$100 Myr after CAIs \cite{morby18,nesvorny18,mojzsis19}.  

\begin{figure}[t]
\centering
\includegraphics[width=12cm]{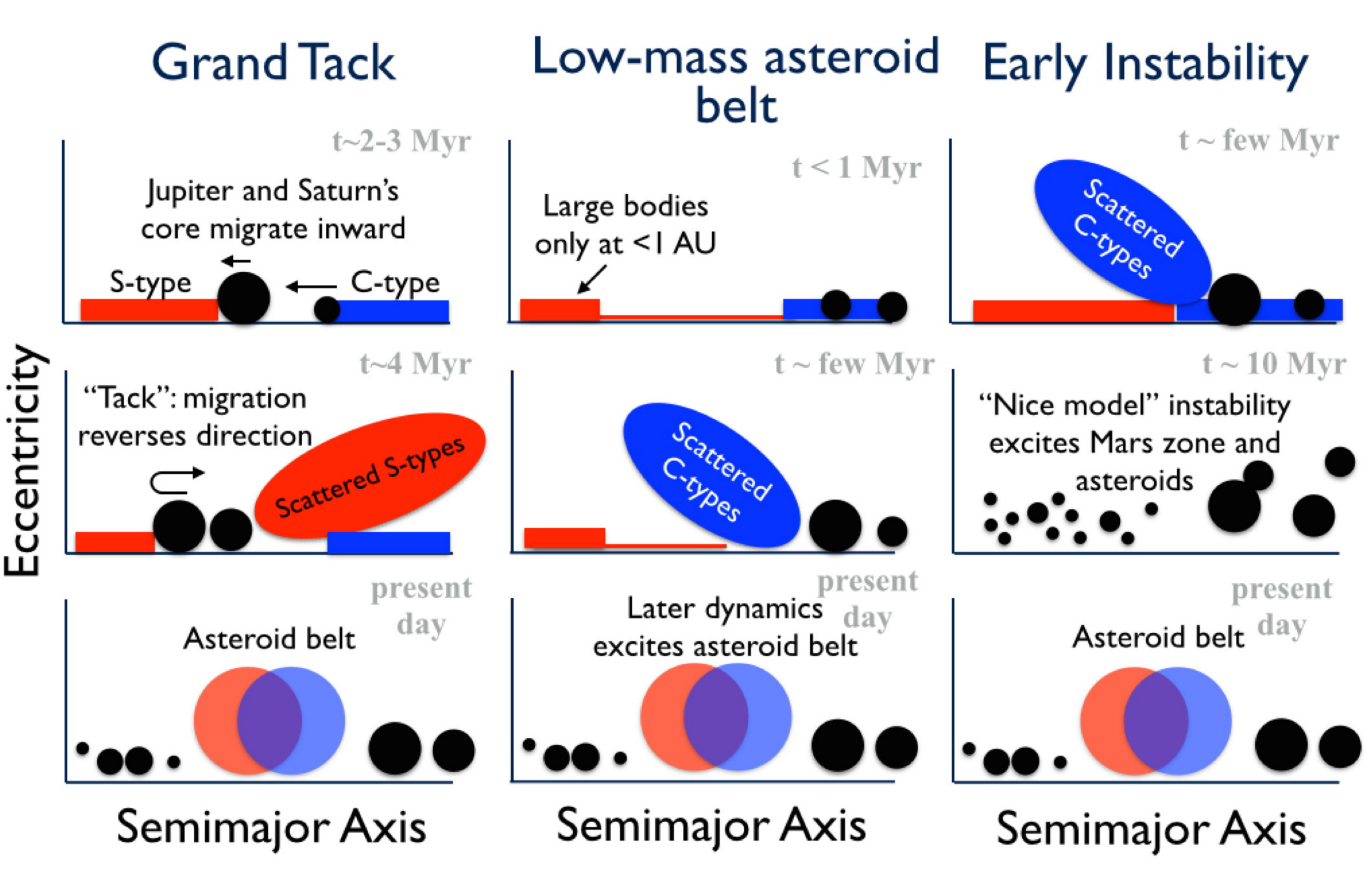}
\caption{Cartoon comparison between the global evolution of three plausible models for Solar System formation described in Sections 3.3.2-3.3.4. From \cite{raymond18d}. }
\label{fig:comparison}       
\end{figure}

Over the past decade, global models of Solar System formation have been revolutionized (see Fig~\ref{fig:comparison}).  Decades-old models that assumed local growth of the planets (i.e., the so-called {\em classical model}~\cite{weidenschilling77,wetherill78,wetherill92,chambers98,chambers01,raymond04,obrien06,raymond09c,raymond14} have been supplanted with models that explain the distribution of the planets and small body belts by invoking processes such as long-range migration of the giant planets (the {\em Grand Tack} model \cite{walsh11,raymond14c,brasser16}), non-uniform planetesimal formation within the disk (the {\em Low-mass Asteroid belt} model \cite{hansen09,drazkowska16,izidoro15c,raymond17b}) and an early instability among the giant planets' orbits (the {\em Early Instability} model \cite{clement18,clement19a,clement19b}).  Pebble accretion has been proposed to play an important role during the terrestrial planets' growth \cite{levison15b}, the importance of which may be constrained by isotopic measurements of different types of meteorites as well as Earth samples \cite{warren11,kruijer17,schiller18,budde19}.  

Here we first describe the classical model and then the other competing models.  We then compare the predictions of different models.

\subsubsection{The Classical model}

The classical model of terrestrial planet formation makes two dramatically simplifying assumptions.  First, it assumes that the planets formed roughly in place. This implies that one can reconstruct the approximate mass distribution of the protoplanetary disk by simply spreading the planets' masses out in concentric annuli (the so-called `minimum-mass solar nebula' model; \cite{weidenschilling77,hayashi81,davis05}).  Second, it assumes that giant- and terrestrial planet formation can be treated separately.  Thus, one can study the influence of the giant planets' orbits on terrestrial planet formation after the gas disk had dispersed without the need to consider their earlier effects.

\begin{figure}[t]
\centering
\includegraphics[width=12cm]{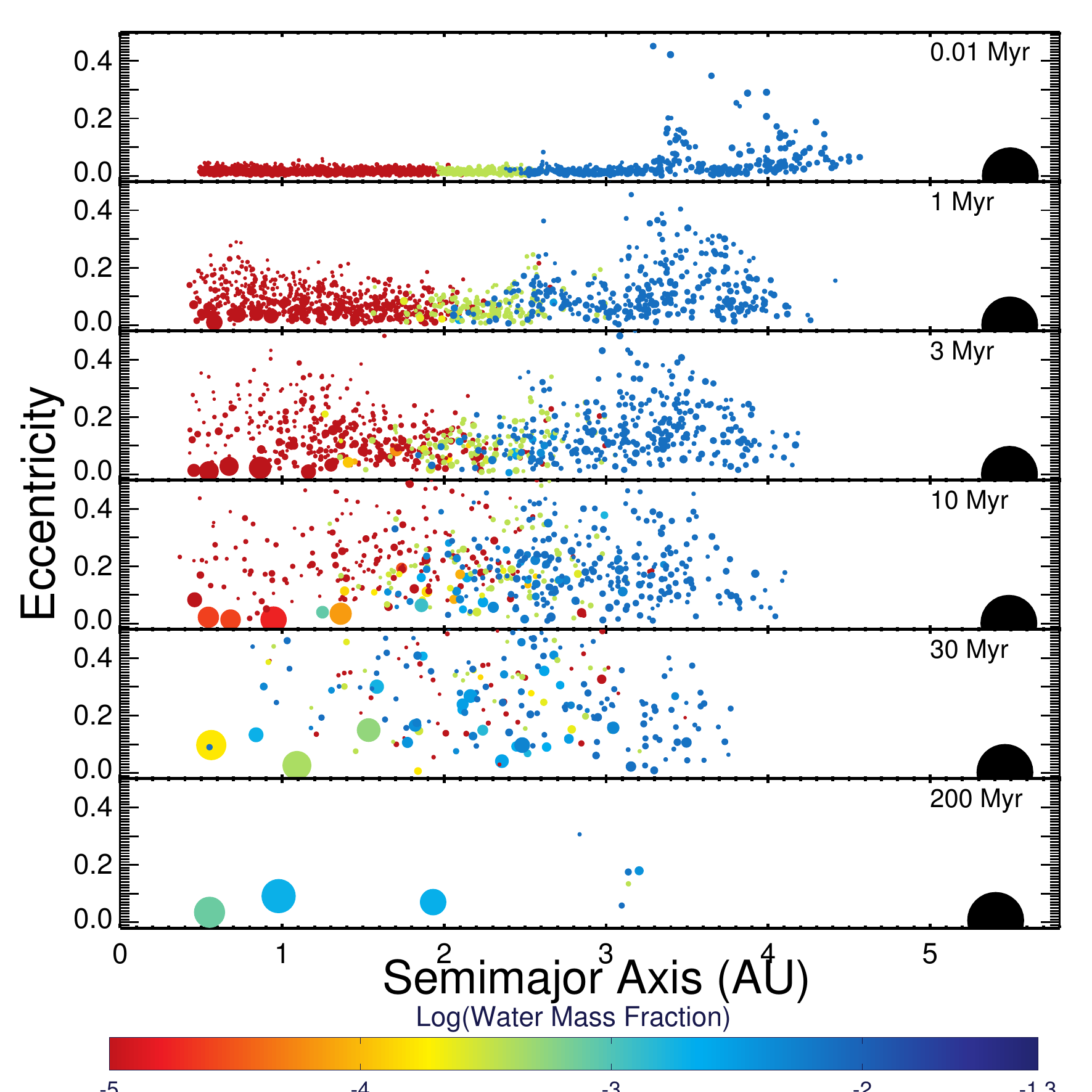}
\caption{Simulation of the classical model of terrestrial planet formation from \cite{raymond06b}. The size of each planetary embryo scales with its mass $m$ as $m^{1/3}$.  Colors represent the water contents, initially calibrated to match the water contents of current orbital radii of different populations of primitive asteroids \cite{gradie82,raymond04}. Jupiter is the large black circle.  }
\label{fig:classical}     
\end{figure}

Figure~\ref{fig:classical} shows a simulation of the classical model of terrestrial planet formation (from \cite{raymond06b}).  In this case Jupiter is included on a supposed pre-instability orbital configuration, on a near-circular orbit. The terrestrial disk is initially composed of roughly 2000 planetary embryos with masses between that of Ceres and the Moon.  Accretion is driven from the inside-out by gravitational self-stirring and from the outside-in by secular and resonant forcing from Jupiter.  There is a long chaotic phase of growth that involves many giant impacts between planetary embryos (see e.g. \cite{agnor99,kokubo07,quintana16}).  The planets grow on a timescale of 10-100 Myr.  Remnant planetesimals are cleared out on a longer timeframe.  There is sufficient radial mixing during growth for water-rich material from past 2.5 au to have been delivered to the growing Earth (see extensive discussion of the origin of Earth's water in Section 3.4). 

The classical model has a fatal flaw: it systematically produces Mars analogs that are far more massive than the real Mars \cite{wetherill91}.  This can be seen in the simulation from Fig.~\ref{fig:classical}, which formed passable Venus and Earth analog but a Mars analog roughly as massive as Earth.  The true problem is not Mars' absolute mass but the large Earth/Mars mass ratio.  The classical model tends to produce systems in which neighboring planets have comparable masses rather than the $\sim$9:1 Earth/Mars ratio \cite{raymond06b,obrien06,raymond09c,morishima08,morishima10,fischer14b,kaib15}. Classical model simulations tend to have two other problems related to the asteroid belt: large embryos are often stranded in the belt \cite{raymond09c}, and they do not match the belt's eccentricity and inclination distributions \cite{izidoro15c}. 

The building blocks of the terrestrial planets were roughly Mars-mass and thus the inner Solar System may only have contained roughly two dozen embryos \cite{morby12}.  One may wonder how often a small Mars arises because, by chance, it avoided any late giant impacts.  Fischer \& Ciesla \cite{fischer14b} showed that this happens in a few percent of simulations. However, when one takes the asteroid belt constraints into consideration the success rate of the classical model in matching the inner Solar System drops by orders of magnitude \cite{izidoro15c}.

The `small Mars' problem -- first pointed out by Wetherill \cite{wetherill91} and re-emphasized in 2009 \cite{raymond09c} -- is thus the Achilles heel of the classical model. It prompted the development of alternate models whose goal was to explain how two neighboring planets could have such different masses.  

\subsubsection{The Grand Tack model}

\begin{figure}[t]
\centering
\includegraphics[width=12cm]{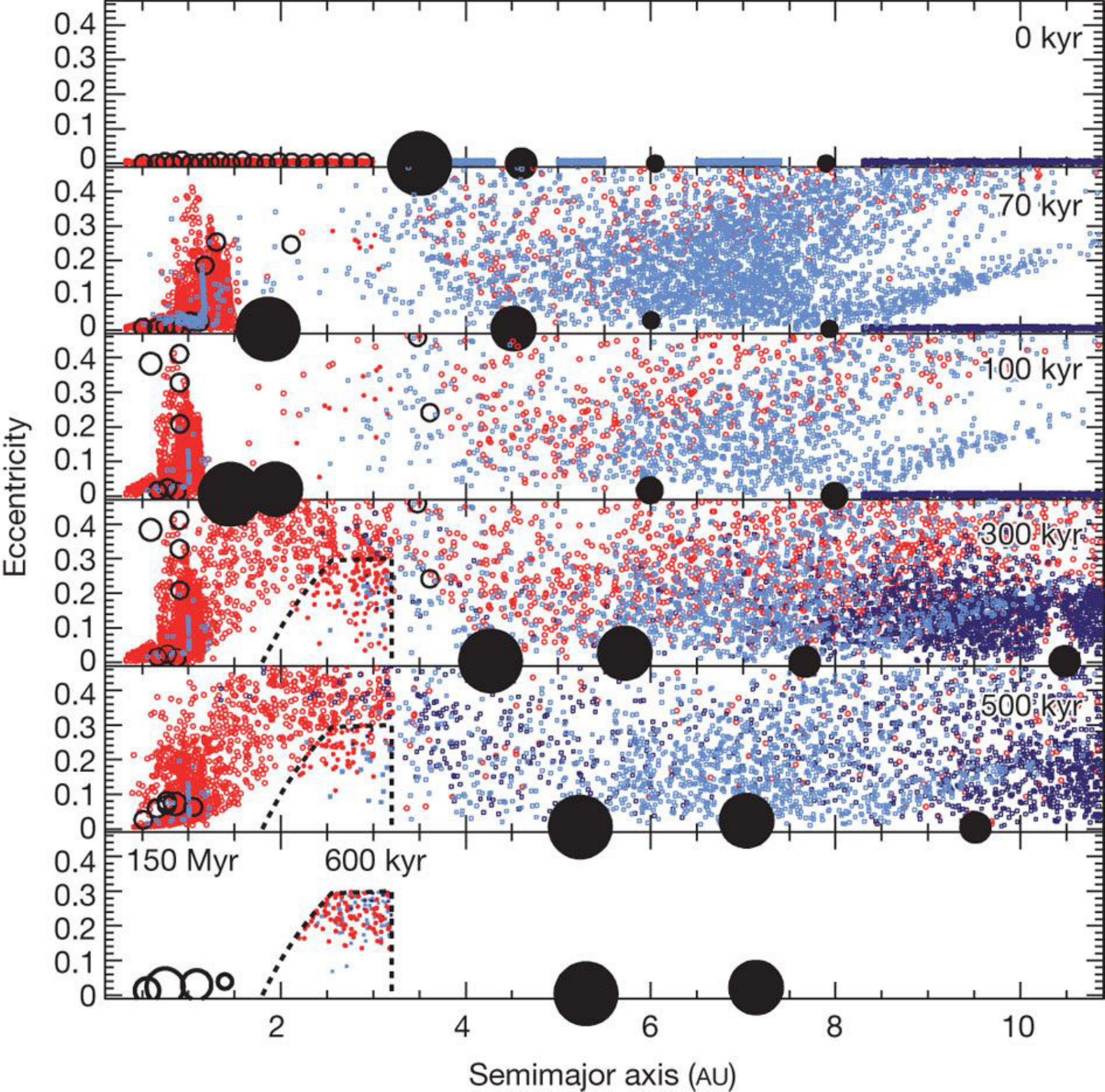}
\caption{Evolution of the Grand Tack model, from \cite{walsh11}.  The giant planets' migration (black symbols) sculpts the distribution of rocky bodies in the inner Solar System.  Here Jupiter migrated inward, then Saturn migrated inward, and then the two planets migrated back out.  The red dots indicate presumably dry, S-type planetesimals that formed interior to Jupiter's orbit, whereas the blue dots represent planetesimals originally between and beyond the giant planets' orbits.  Rocky planetary embryos are shown as empty circles.  The bottom three panels include dashed contours of the present-day main asteroid belt. }
\label{fig:GT}       
\end{figure}

The Grand Tack model \cite{walsh11} proposed that Jupiter's migration during the gaseous disk phase was responsible for depleting Mars' feeding zone.  The Grand Tack is based on hydrodynamical simulations of orbital migration (see Sections 2.4 and 2.5 for more details).  Simulations show the giant planets carve annular gaps in the disk \cite{lin86,ward97,crida06}.  A single planet usually migrates inward on a timescale that is related to the disk's viscous timescale \cite{dangelo03,durmann15}.  Multiple planets often migrate convergently and become trapped in mean-motion resonances \cite{snellgrove01,lee02}, often sharing a common gap.  When the inner planet is two to four times more massive than the outer one and both planets are in the gap-opening regime, gas piles up in the inner disk and a toque imbalance causes the two planets to migrate {\em outward} \cite{masset01,morby07,pierens08,zhang10,pierens11,pierens14}.  

The Grand Tack model proposes that Jupiter formed first and migrated inward.  Saturn formed more slowly, migrated inward and caught up to Jupiter.  The two planets became locked in resonance and migrated outward together until either the disk dissipated or certain conditions slowed their migration (e.g., if the disk was flared such that Saturn dropped below the gap-opening limit in the outer disk).  

The evolution of the Grand Tack model is shown in Figure~\ref{fig:GT}.  If Jupiter's inward migration reached 1.5-2 au then the inner disk would have been truncated at $\sim$1 au, depleting Mars' feeding zone but not Earth's \cite{walsh11,brasser16}.  The terrestrial planets in simulations provide a good match to the real ones~\cite{walsh11,jacobson14b,brasser16}.  

The asteroid belt is severely depleted by Jupiter's migration but is not completely emptied \cite{walsh11,walsh12}.  The S-types were scattered outward by Jupiter's inward migration, then back inward by Jupiter's outward migration, with an efficiency of implantation of $\sim$0.1\%.  The C-types were scattered inward by Jupiter and Saturn's outward migration.  The final belt provides a reasonable match to the real one -- particularly when evolved over Gyr timescales \cite{deienno16}.  It is worth noting, however, that the initial conditions in Fig.~\ref{fig:GT} neglect additional (possibly much more important) phases of implantation that took place during Jupiter and Saturn's rapid gas accretion \cite{raymond17}.  

The Grand Tack's potential Achilles heel is the mechanism of outward migration itself (see discussion in \cite{raymond14c}).  The most stringent constraint on outward migration is that it requires a relatively limited range of mass ratios between Jupiter and Saturn (roughly between a ratio of 2:1 and 4:1; \cite{masset01}). It remains to be seen whether long-range outward migration remains viable when gas accretion is consistently taken into account, as the gas giants' mass ratio should be continuously changing during this phase. There may also be geochemical constraints related to the speed of accretion in the Grand Tack model \cite{zube19}.

\subsubsection{The Low-mass Asteroid belt model}

The Low-Mass Asteroid belt model makes the assumption that Mars' small mass is a consequence of a primordial mass deficit between Earth and Jupiter's present-day orbits.  Gas disks are generally expected to have smooth radial distributions, but this is not the case for dust.  Dust drifts within disks and ALMA images show that dust rings are common in protoplanetary disks \cite{alma15,andrews18}.  Given that planetesimal formation via the streaming instability is sensitive to the local conditions \cite{simon16,yang17}, dust rings may be expected to produce rings of planetesimals.  One model of dust drift and coagulation, combined with the conditions needed to trigger the streaming instability,  showed that planetesimals may indeed form rings \cite{drazkowska16}.

\begin{figure}[t]
\centering
\includegraphics[width=10cm]{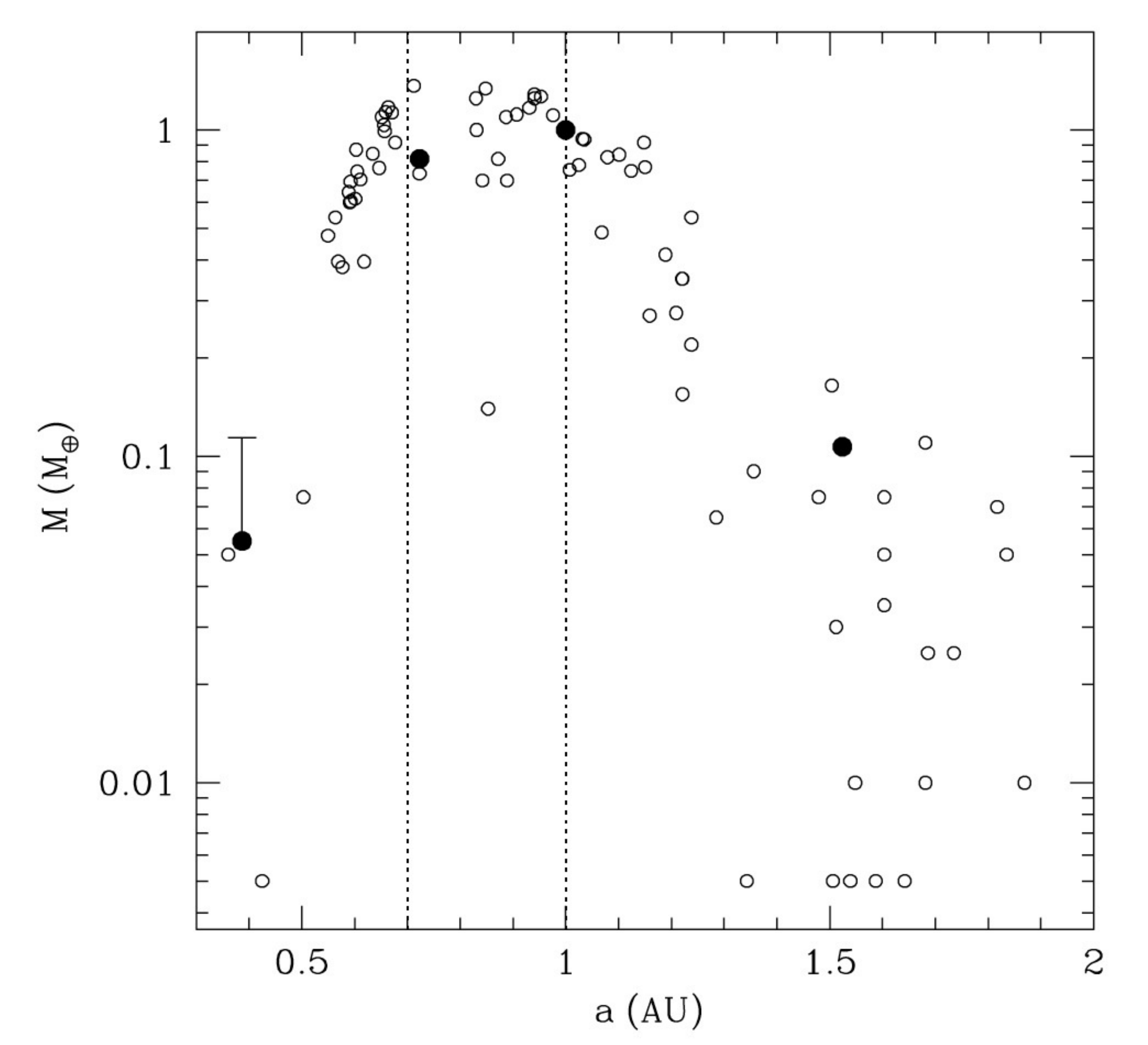}
\caption{Distribution of simulated terrestrial planets formed assuming that all of the terrestrial planets' mass was initially found in a narrow annulus between the orbits of Venus and Earth (denoted by the dashed vertical lines). The open circles show simulated planets and the solid ones are the real planets. From \cite{hansen09}.  }
\label{fig:lmab}       
\end{figure}

Since the 1990s it was known that if an edge existed in the initial distribution of planetary embryos or planetesimals, any planet that formed beyond that edge would be much less massive than the planets that formed within the main disk \cite{wetherill92,wetherill96,chambers98,agnor99,chambers01}.  However, it was Hansen in 2009 \cite{hansen09} who proposed that such an edge might really have existed in the early Solar System.  Indeed, he showed that the terrestrial planets can be matched if they formed from a narrow annulus of rocky material between the orbits of Venus and Earth (see Fig.~\ref{fig:lmab}.  In this model the large Earth/Mars mass ratio is a simple consequence of the depletion of material past Earth's orbit, and the small Mercury/Venus mass ratio is a consequence of the analogous depletion interior to Venus' orbit.  

The most extreme incarnation of the Low-mass asteroid belt model proposes that no planetesimals formed between the orbits of Earth and Jupiter \cite{hansen09,raymond17b}, but the model is also consistent with planetesimal forming in the belt, just at a reduced efficiency. Yet even an empty primordial asteroid belt would have been dynamically re-filled with objects originating across the Solar System. Rapid gas accretion onto Jupiter and later Saturn destabilizes the orbits of nearby planetesimals, many of which are gravitationally scattered in all directions.  Under the action of gas drag, a fraction of planetesimals are trapped on stable, low-eccentricity orbits, preferentially in the asteroid belt \cite{raymond17}, and some are scattered past the asteroid belt to the terrestrial planet region itself (see Fig.~\ref{fig:injection} and Section 3.4).  On a longer timescale, the growing terrestrial planets scatter remnant planetesimals outward, a small fraction of which are trapped in the main belt, preferentially in the inner parts \cite{raymond17b}.  

The main uncertainty in the Low-mass asteroid belt are the initial conditions. When disk models, dust growth and drift, disk observations and interpretation (including studies of ALMA-detected disks), and meteorite constraints (including the broad age distribution of noncarbonaceous chondrites, which indicate many different planetesimal formation events; \cite{kruijer17}) are all accounted for, will it be a reasonable assumption that the terrestrial planets could have formed from a narrow ring of planetesimals?  

\subsubsection{The Early Instability model}

It has been demonstrated numerically that dynamical instabilities among giant planets cause severe damage to the small bodies in their same systems (see Fig.~\ref{fig:unstab}).  The Solar System's giant planets are thought to have undergone an instability, albeit a much less violent one than the typical instability incurred by giant exoplanet systems. Nonetheless, the Solar System's instability is violent enough that, if they were fully-formed at the time, the terrestrial planets would have had a very low probability of survival~\cite{kaib16}. Yet the timing of the instability is uncertain, and it could have happened anytime in the first 100 million years of Solar System evolution~\cite{morby18,mojzsis19}. One may then wonder whether a very early instability could have played a role in shaping the distribution of the terrestrial planets.

\begin{figure}[t]
\centering
\includegraphics[width=11cm]{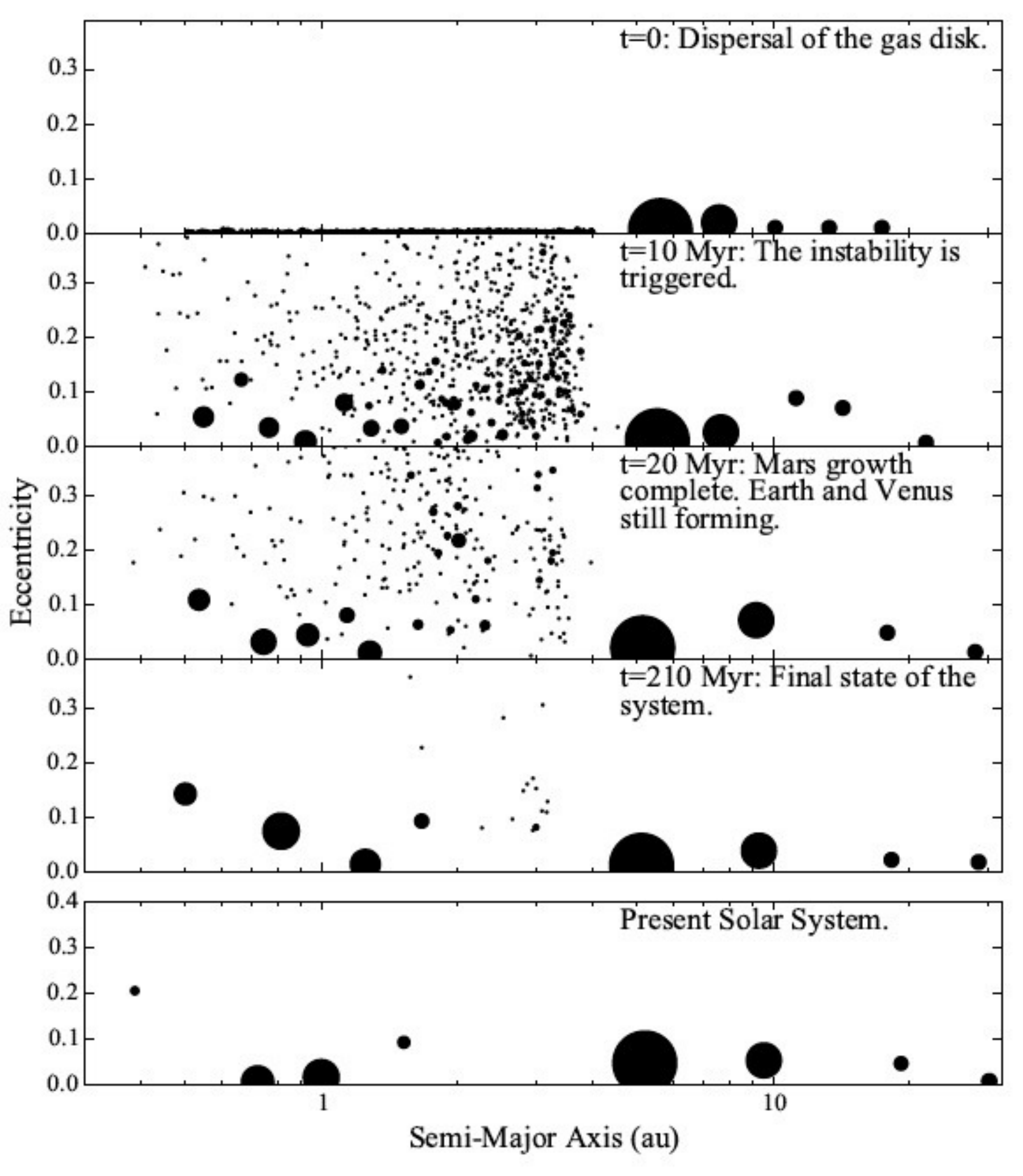}
\caption{Evolution of the Early Instability model (from \cite{clement18}). In this case the giant planets' instability was triggered 10 Myr after the dissipation of the disk. The instability strongly excited the orbits of growing planetesimals and planetary embryos in the asteroid belt and Mars' feeding zone.  Mars' growth was stunted but Earth and Venus continued to accrete for $\sim$100 Myr.  The final system has a large Earth/Mars mass ratio and an overall planetary system architecture similar to the actual Solar System (shown at bottom for comparison). }
\label{fig:Early}       
\end{figure}

The Early Instability model, conceived and developed by Clement et al. \cite{clement18,clement19a,clement19b}, is built on the premise that the giant planets' instability took place within roughly 10 Myr of the dissipation of the gaseous disk.  The evolution of one realization of the model is shown in Figure~\ref{fig:Early}.  The early evolution of the Early Instability model is identical to that in the classical model.  The giant planets' dynamical instability -- triggered after 10 Myr in the simulation from Fig.~\ref{fig:Early} -- strongly excites the orbits of inner Solar System objects extending through the asteroid belt in to Mars' feeding zone~\cite{deienno18,clement19b}.  The belt is strongly depleted and dynamically excited, and Mars' growth is effectively stunted.  In a fraction of simulations no Mars forms at all \cite{clement18}!  The growth of Earth and Venus are largely unperturbed and qualitatively similar to the classical model.  

The instability itself is stochastic in nature.  Matching the instability statistically requires a large number of numerical realizations, which produce a spectrum of Solar Systems with different properties.  Such simulations have more constraints than many because they include all of the planets and not just the terrestrial planets.  One remarkable feature of the Early Instability model is that systems that provide the best match to the outer Solar System are the same ones that provide the best match to the terrestrial planets \cite{clement18,clement19a,clement19b}.  

The main uncertainty in the Early Instability model is simply the timing of the instability.  Mars' formation was largely complete within 5-10 Myr~\cite{nimmo07,dauphas11}, shortly after the disappearance of the gaseous disk.  To affect terrestrial planet formation the giant planets must therefore have gone unstable within perhaps 5 Myr of the disk's dispersal.  This would also imply a cometary bombardment in the inner Solar System that was coincident with the late phases of terrestrial planet growth.  While such a bombardment would deliver only a very small amount of mass to Earth~\cite{morby00,gomes05}, it would provide a large component of Earth's noble gases \cite{marty16}.  The Xe isotopic compositions of the mantle and atmosphere are different~\cite{sujoy12,caracausi16}, and it has been suggested that a comet-delivered component contributed to the atmospheric Xe budget but not to the mantle Xe \cite{marty17}.  It remains to be understood whether or not this implies that the bulk of Earth accreted with little cometary influx, which would constrain the timing of cometary delivery and presumably of the instability itself.

\subsubsection{Other models}

It is worth noting that the three scenarios outlined above as alternatives to the classical model are not the only ones that have been proposed.  For example, it was suggested that sweeping secular resonances during the dissipation of the gaseous disk could have depleted and excited the asteroid belt and generated an edge in the terrestrial planets' mass distribution\cite{nagasawa05,thommes08,bromley17}.  The main uncertainty in that model is whether a non-zero eccentricity of Jupiter can realistically be maintained during the gas dissipation phase.

It has also been suggested that pebble accretion may have played a role in terrestrial planet formation \cite{levison15b,chambers16}.  Isotopic analyses of meteorites may help to constrain the degree to which carbonaceous pebbles from the outer Solar System contributed to the growth of the terrestrial planets~\cite{schiller18,budde19}.  In general, the formation of Earth-mass planets within the disk lifetime poses a problem because, given that the gas disk is required for pebble accretion, such planets should grow fast enough to migrate inward and become super-Earths \cite{lambrechts19}.

\subsection{Origin of water on Earth and rocky exoplanets}

We now turn our attention to the origin of planets' water. Cosmochemical tracers such as isotopic ratios can be used to constrain the potential sources of Earth's water (see, e.g., \cite{morby00,marty12}). While bulk density measurements of solid exoplanets can in principle be used to trace water contents \cite{valencia07,sotin07,fortney07,seager07}, in practice it is extremely challenging \cite{adams08,selsis07b,dorn15}.  There have been a few recent reviews on the origins of Earth's water in dynamical and cosmochemical context \cite{obrien18,mccubbin19,meech19}.

\begin{figure}[t]
\centering
\includegraphics[width=12cm]{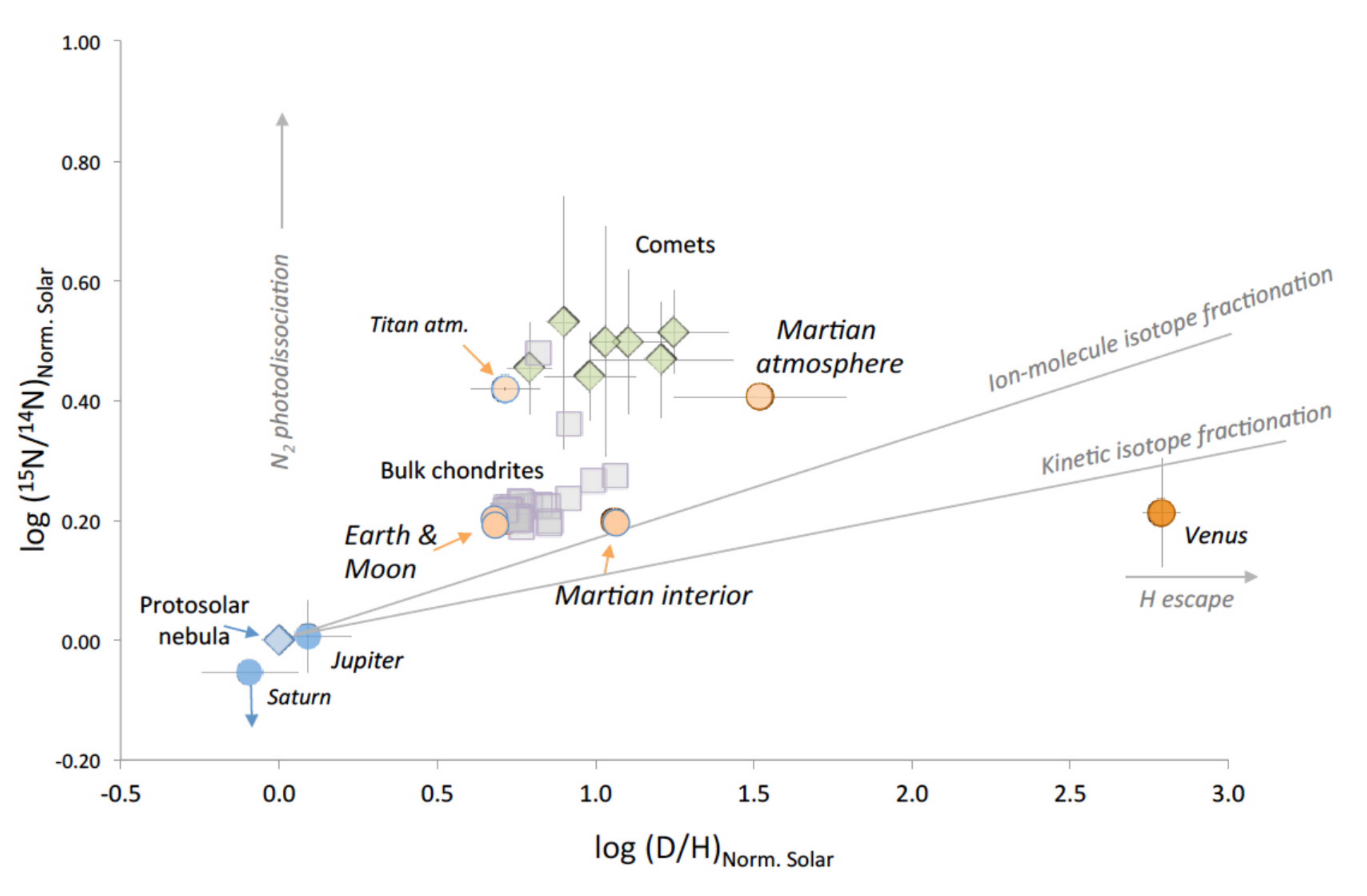}
\caption{Measured $^{15/14}$N isotope ratios vs. D/H ratios of Solar System objects. Arrows indicate how different processes would affect an object's evolution in this isotope space. From Marty et al. \cite{marty16}.}
\label{fig:dh_constraints}       
\end{figure}


Earth is our benchmark for water.  While it appears blue from space, all of Earth's oceans only add up to $\sim 2.5 \times 10^{-4} \mearth$ of water.  This is generally referred to as one ``ocean'' of water.  The water budget of Earth's interior is quite uncertain.  Different studies infer different quantities of water trapped in hydrated silicates, with overall budgets between roughly one and ten oceans \cite{hirschmann06,mottl07,marty12,halliday13}.  Earth's core may be very dry~\cite{badro14} or may contain fifty or more oceans of water~\cite{nomura14}. While papers often quote Earth as being roughly one part in a thousand water by mass, it is important to be aware of the uncertainty.

Figure~\ref{fig:dh_constraints} shows the D/H ratios for a number of Solar System objects (for a compilation of references to the D/H measurements, see \cite{morby00,marty06,alexander12,alexander18}). The Sun and gas-dominated planets have D/H ratios roughly six times lower than Earth's.  This is interpreted as the isotopic composition of the gaseous protoplanetary disk. Carbonaceous chondrite meteorites have similar D/H values to Earth. Most measured comets have higher D/H values than Earth, although two recently-measured Jupiter-family comets \cite{hartogh11,lis13} and one Oort cloud comet \cite{biver16} have Earth-like D/H ratios.  In contrast, ESA's Rosetta mission measured the D/H ratio of comet 67P/C-G to be more than three times the terrestrial value.  A recent result \cite{sosa11,lis19} found that very active comets tended to have Earth-like D/H whereas less active ones have higher D/H. One interpretation is that comets' original water was Earth-like and loss processes during outgassing have fractionated the surviving D/H.  If true, that would mean render the D/H ratio useless as a discriminant between comets and carbonaceous sources of water.  One must then resort to dynamical constraints or perhaps to other isotopic systems. 

Fig.~\ref{fig:dh_constraints} naively would suggest carbonaceous chondrite-like objects as the source of Earth's water and Nitrogen.  Yet even if Earth's water were delivered by carbonaceous objects, that is only a part of the story.  A complete model must explain the full evolution of our planetary system and, in the context of Earth's water, ask: where did the water-bearing objects themselves originate? 

We can break down the various models for water delivery into six rough scenarios that we outline below.  Two of these models invoke local sources of water whereas the others propose that Earth's water was delivered from farther out in the disk. For a comprehensive review of these models, see Meech \& Raymond (2019) \cite{meech19}.

\paragraph{Adsorption of water vapor onto silicate grains}

In a simple picture of the structure of disks, water should exist as a solid past the snow line and as a vapor closer-in. If water vapor was indeed present where silicate grains were coalescing to form the terrestrial planets, then ``in-gassing'' may have attached hydrogen molecules to silicate grains \cite{stimpfl06,muralidharan08,asaduzzaman14,sharp17,dangelo19}.  This process is called adsorption. The mechanism can in principle have seeded Earth with a few oceans of water, albeit without taking any water loss processes into account (such as $^{26}$Al-driven heating \cite{grimm93,monteux18} or impact-related losses \cite{genda05,svetsov07}).  Yet at face value it should have seeded Earth with nebular water, which has a D/H ratio six times smaller than Earth's.  It also cannot explain the abundance of other volatiles such as C, N and the noble gases, which would then require an alternate source containing little hydrogen.  Finally, it begs the question of why the Enstatite chondrites -- meteorites that provide the closest chemical match to Earth's bulk composition \cite{dauphas17} -- appear to have formed without any water.

\paragraph{Oxidation of H-rich primordial envelope}

As planetary embryos grow they gravitationally accrete H-rich envelopes if they are more massive than a few tenths of Earth's mass \cite{ikoma00,lambrechts17}.  This hydrogen could have chemically reacted with Earth's surface magma ocean and generated water by hydrating silicates \cite{ikoma06}.  Given that the D/H ratio of nebular gas was many times smaller than Earth's water, this model, like the previous one, predicts that Earth's initial D/H ratio was small.  In this case, however, it is possible to envision that Earth's D/H changed in time.  The loss of a thick hydrogen envelope would certainly entail mass fractionation, and for a certain range of parameters Earth's final D/H ratio can be matched even though its water was acquired from gas \cite{genda08}.  However, the collateral effects of this presumed atmospheric loss have not been quantified, and appear to be at odds with other constraints, such as the abundance of $^{129}$Xe from the decay of $^{129}$I \cite{marty16}. In addition, it seems quite a coincidence for Earth to match carbonaceous chondrites in multiple isotopic systems as a result of such loss processes, which would affect different molecules differently.

\paragraph{Pebble ``snow''}

Planets beyond the snow line accrete water as a solid.  It is simply a building block.  Yet the snow line is a moving target. As the disk evolves and cools the snow line generally moves inward \cite{sasselov00,lecar06,martin12}.  A planet on a static orbit would see the snow line sweep past it.  Such a planet would start off in the rocky part of the disk but, once the snow line passed interior to its orbit, would find itself outside the snow line, in the presumably icy part of the disk.  

\begin{figure}[t]
\centering
\includegraphics[width=11cm]{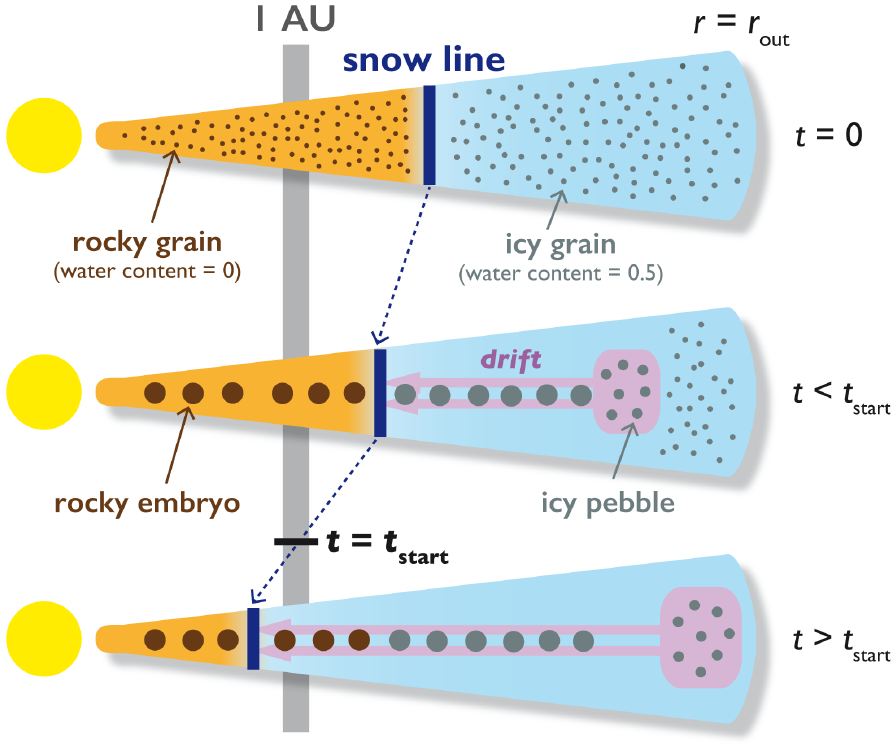}
\caption{The {\em pebble snow} model \cite{oka11,sato16,ida19}. Rocky planetary embryos grow from rocky grains interior to the snow line. As the disk cools, the snow line sweeps inward and ice-rich pebbles from the outer parts of the disk deliver water to the terrestrial planet region. From \cite{sato16}. }
\label{fig:pebblesnow}       
\end{figure}

As the snow line sweeps inward, new ice does not come from the condensation of water vapor.  This is because the speed at which gas moves through the disk is far faster than the motion of the snow line itself \cite{morby16}.  Thus, the gas just interior to the snow line is dry.  Rather, ice at the snow line comes from pebbles and dust that drift inward through the disk.  The source of these drifting pebbles is thought to move outward as an analogous wave of dust coagulation and growth sweeps radially out through the disk\cite{birnstiel12,lambrechts14b,birnstiel16,ida16}.  If anything blocks the inward flow of pebbles -- such as a growing giant planet core \cite{lambrechts14,bitsch18} -- the snow line will continue to move inward but it will not bring any ice along with it \cite{morby16}. The source of water will also drop if the outward-sweeping growth front producing pebbles reaches the outer edge of the disk \cite{ida19}. 

In the Solar System there is evidence that the pebble-sized building blocks carbonaceous and noncarbonaceous were segregated as of 1-1.5 Myr after CAIs \cite{budde16,kruijer17}. This would suggest that carbonaceous pebbles did not deliver Earth's water, as their widespread presence in the inner Solar System would presumably have produced a category of meteorites intermediate in isotopic composition between the carbonaceous and noncarbonaceous.  The Enstatite chondrites, which formed near the end of the disk's lifetime and are the closest chemical match to Earth, are dry \cite{alexander18}.

Nonetheless, pebble snow may be a key mechanism in delivering water to rocky exoplanets (see below).

\paragraph{Wide feeding zone}

A planet's feeding zone is simply the radial distribution of its constituents.  In the classical model the terrestrial planets formed from a broad disk of planetary embryos extending from Venus' orbit out to Jupiter's. For example, in the simulation from Fig.~\ref{fig:classical}, each of the three surviving terrestrial planets accreted material from a broad swath of the disk.  Given that the constituent planetesimals should have a radial gradient in composition based on the local temperature \cite{dodsonrobinson09}, this implies that volatiles are naturally incorporated into growing planets if the planets' feeding zones extend out far enough \cite{morby00,raymond04,raymond07a}. 

The planets' broad feeding zones -- in concert with the isotopic match between Earth's water and that of carbonaceous chondrites -- led to the scenario that Earth's water was delivered by primordial carbonaceous planetesimals or planetary embryos \cite{morby00,raymond04,raymond07a}.  However, this idea was built on the classical model. As the classical model crumbled under the weight of the small Mars problem (see Section 3.3) the idea that Earth's water was a result of its broad feeding zone no longer made sense. The Early Instability model \cite{clement18,clement19a,clement19b} starts from essentially the same initial conditions as the classical model and presents a viable solution to the small Mars problem all the while delivering water to Earth by the same mechanism. However, like the classical model~\cite{raymond09c}, the Early Instability may only adequately deliver water to Earth in simulations that fail to match Mars' small mass (Clement and Rubie, personal communication). The Early Instability model acknowledges that in order to understand in the first place the existence of water-rich asteroids intermixed with water-poor asteroids in the asteroid belt, which may have been the result of the growth and/or migration of the giant planets. That leads us to the next scenario.

\paragraph{External pollution}

The orbits of leftover planetesimals are strongly perturbed by the growth and migration of the giant planets.  The phase of rapid gas accretion is particularly dramatic.  The mass of the giant planets can increase from $\sim 10-20 \mearth$ up to hundreds of Earth masses on a $\sim10^5$ year timescale \cite{pollack96,ida04}.  This rapid gas accretion destabilizes the orbits of any nearby planetesimals that managed to avoid being accreted.  Many planetesimals undergo close gravitational encounters with the growing planet and are scattered in all directions.  Under the action of gas drag, many planetesimals are trapped interior to the giant planet's orbit \cite{raymond17,ronnet18}.  This happens when a planetesimal on an eccentric orbit that crosses the giant planet's orbit at apocenter undergoes sufficient orbital energy loss due to gas drag to drop its apocenter away from the giant planet, releasing it from the gas giant's dynamical clutches.

\begin{figure}[t]
\centering
\includegraphics[width=12cm]{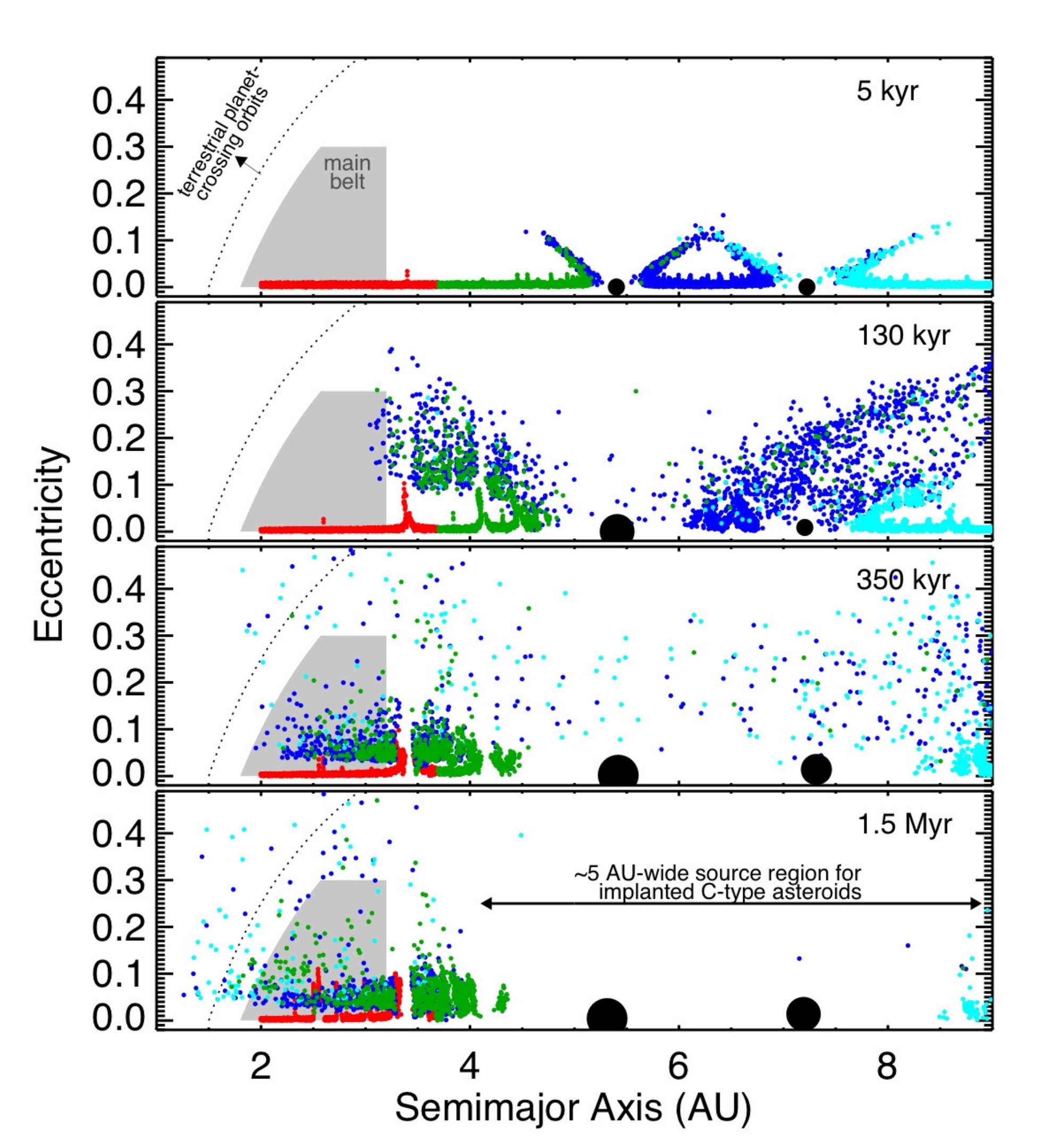}
\caption{Scattering of ten thousand planetesimals driven by the growth of Jupiter (from 100-200 kyr) and Saturn (from 300-400 kyr).  Planetesimals from the Jupiter-Saturn region end up polluting the inner Solar System {\em en masse}.  Some are trapped in the outer parts of the main asteroid belt (shaded) and some are scattered past the belt to the terrestrial planet region (above the dashed line).  And even so, this example shows the minimum expected impact of the giant planets' growth, as it assumed the giant planets to have formed on low-eccentricity, non-migrating orbits in 3:2 resonance.  Planetesimal colors correspond simply to each object's starting location.  There is an underlying gaseous disk in the simulation, whose structure and overall density evolve in time in a consistent way.  In this example planetesimals were assumed to be 100km in diameter for the gas drag calculation.  From \cite{raymond17}. }
\label{fig:injection}       
\end{figure}

Figure~\ref{fig:injection} shows this process in action.  Jupiter's growth triggers a pulse of planetesimal scattering, and a second is triggered when Saturn forms.  The outcomes of these pulses can vary because the disk itself changes in time as it both slowly dissipates and, more importantly, is sculpted by the giant planets. 

The relative number of planetesimals that are trapped in the main asteroid belt (providing an orbital match to the C-types) vs. those that are scattered {\em past} the asteroid belt to the terrestrial planets (to deliver water) varies as a function of the strength of gas drag \cite{raymond17}.  When gas drag is strong -- i.e., for a massive inner gas disk or smaller planetesimals -- scattered planetesimals are rapidly decoupled from Jupiter and are generally trapped in the outer parts of the main belt.  When gas drag is weak -- for a low-density inner disk or large planetesimals -- it takes a large number of orbits for gas drag to decouple planetesimals' orbits from Jupiter.  Planetesimals are more frequently scattered farther inward to pollute the terrestrial planet-forming region and to deliver water.

A similar mechanism is driven by giant planet migration.  In the Grand Tack model, Jupiter and Saturn's outward migration scatters a large number of planetesimals inward (\cite{walsh11,walsh12}; see Fig.~\ref{fig:GT}).  Given that the giant planets are migrating away, scattered planetesimals are more easily decoupled from the giant planets.  The distribution of scattered planetesimals that survive on orbits interior to the giant planets' have a much weaker size dependence than for those that were scattered during the giant planets' growth and for which gas drag played a central role. The dependence on the gas surface density is also much weaker.

Both migration and growth can thus generate populations of terrestrial planet-crossing planetesimals that originated beyond Jupiter.  Given their distant origins these planetesimals tend to have high-eccentricity orbits, but simulations show that they are nonetheless accreted by the terrestrial planets on a geochemically consistent timescale and in sufficient quantity to easily deliver Earth's water budget \cite{obrien14}.  In addition, they naturally match the compositions of carbonaceous meteorites.  

In the wide feeding zone model, the C-type asteroids represented the distant, mother source of Earth's water.  In the pollution model, Earth's water was delivered from the same parent population as the objects that were implanted into the asteroid belt as C-types.  Thus, Earth's water and C-type asteroids are brother and sister.

At present, the pollution scenario is the leading model to explain the origin of Earth's water.  It matches the amount and chemical signature of Earth's water, and naturally fits within different models of Solar System formation.

\paragraph{Inward migration}

Earth-mass planets migrate relatively rapidly in protplanetary disks (see Sections 2.2 and 3.1).  Given that large embryos are thought to form fastest past the snow line \cite{lambrechts14,morby15}, this implies that many inward-migrating planets or cores should be water-rich \cite{bitsch19b}.  Indeed, the migration model for super-Earth formation predicts that most should be water-rich \cite{raymond18b,izidoro19}.  Inward-migrating gas giants can shepherd material interior to their orbits and trigger the formation of very water-rich terrestrial planets \cite{raymond06c,mandell07}.

A number of lines of evidence point to the terrestrial planets having formed from $\sim$Mars-mass planetary embryos \cite{morby12}. This is below the mass threshold for large-scale migration \cite{ward97}, and so we do not think that Earth accreted any water from inward-migrating embryos. 

However, migration may prove a central water delivery mechanism in other systems (see below).

\paragraph{Extrapolation to exoplanets}

Let us now very simply extrapolate these six water delivery (or water production) mechanisms to exoplanet systems.  Most of these mechanisms can account for Earth's water budget with perhaps an order of magnitude variation.  

We propose that the water contents of rocky planets are likely to have a bimodal distribution.  Migration is the key player.  If embryos grew large enough and underwent long-range migration then they must invariably pollute the terrestrial planet-forming region \cite{izidoro14}.  This should lead to elevated water contents of $\sim 10\%$ by mass.  However, if migration was not involved, then the other mechanisms discussed can still deliver an Earth-like or higher water budget.  Gas giants are rare enough around Sun-like and low-mass stars that they probably do not play the central role in water delivery in general terms (although they likely dominate in their own systems).  Pebble snow can in principle deliver tens to hundreds of oceans, but requires a clear path between the outward-sweeping icy pebble front and the rocky planet zone that is unencumbered by pressure bumps, including those produced by growing planets (which are themselves growing by consuming icy pebbles).

If migration is indeed the key, then one might predict an observational marker between the action of migration and the planets' water and other volatile contents.  Such an observational test is beyond current capabilities but may be imaginable in the coming decade or two.

\section{The future of planet formation studies} 

We conclude this broad-sweeping review with our vision for the future of planet formation.  We cannot pretend to have a coherent view of all of the theoretical and empirical challenges that will push the field forward.  Nonetheless, we proceed by highlighting three action items: a key bottleneck for planet formation models, one particularly promising path forward, and a call for connection.

\paragraph{A key bottleneck for planet formation}

We consider the central bottleneck in planet formation to be understanding the underlying structure of protoplanetary disks (see Section 2.1 and discussion in \cite{morbyraymond16}). Disks are the birthplace of planets. Their underlying structure controls how dust grows and drifts, where dust or pebbles pile up to become sufficiently concentrated to form planetesimals, as well as how fast and in what direction planets migrate. Simple viscous disk models do not match observations and have been supplanted by models that include effects such as ambipolar diffusion and wind-driven angular momentum loss.  ALMA observations of disks \cite{andrews18} provide ever more stringent constraints on such models, yet to date there is no underlying model that matches the population of disk observations.  Of course, interactions with other stars during the embedded cluster phase leads to a diversity of disk properties but with statistically described distributions \cite{bate18}.  We encourage future work to develop comprehensive, trustworthy disk models.

\paragraph{Coupling dynamical and chemical models}

One key path forward for understanding the early evolution of the Solar System is coupling dynamical and accretion models with cosmochemical constraints.  Some constraints are already being used by current models.  For example, the D/H ratio -- especially when combined with the $^{15/14}$N ratio \cite{marty06,marty16} -- appears to be a powerful tracer of Earth's water.  Isotope systems such as Hf/W already provide strong constraints on the timing of planetary accretion.  Some formation models also incorporate simple chemical models \cite{rubie15}. Yet there remain many connections to be made between dynamical models and cosmochemistry.

\paragraph{A call to connect different constraints and models}

We conclude this chapter with a call to create as many connections as possible.  This is a large variety of disciplines linked together to create the field of planet formation.  These include observations of protoplanetary disks and debris disks, exoplanet studies, meteorite analysis, planetary surfaces (e.g., crater modeling), orbital dynamics, gasdynamics, small body studies, and a variety of different types of numerical modeling.  We encourage the reader to strive to make connections between their specialty and others.  Many of the most interesting dynamical models come from connections between subdisciplines.  For example, the idea of the Solar System's instability (the {\em Nice model} \cite{tsiganis05,gomes05,nesvorny18b}) was born from a dynamical model to explain the now-defunct terminal lunar cataclysm. The {\em Grand Tack} model of Solar System formation \cite{walsh11} was inspired by numerical studies of planet migration designed to explain the origin of hot Jupiters \cite{lin96}. Giant exoplanets orbiting stars with wide companions have more eccentric orbits than giant exoplanets around single stars; this is a result of Oort cloud comet-like oscillations in the orbits of wide binary stars \cite{kaib13}.  

Connecting the dots in new ways is essential to moving the field of planet formation forward.


%


\begin{acknowledgement}
We are grateful to the Agence Nationale pour la Recherche, who sponsored our research from 2014-2018 (project {\em MOJO}).  We encourage the interested reader to check out the {\em MOJO} videos, which we made to explain some key concepts in planet formation as well as results from the project.  The videos are (poetically) introduced at this URL: https://wp.me/p3BSYQ-1Ys.
\end{acknowledgement}

\clearpage
\newpage

\bibliographystyle{spmpsci}
\bibliography{refs}

\begin{thebibliography}{100}
\providecommand{\url}[1]{{#1}}
\providecommand{\urlprefix}{URL }
\expandafter\ifx\csname urlstyle\endcsname\relax
  \providecommand{\doi}[1]{DOI~\discretionary{}{}{}#1}\else
  \providecommand{\doi}{DOI~\discretionary{}{}{}\begingroup
  \urlstyle{rm}\Url}\fi

\bibitem{abod19}
{Abod}, C.P., {Simon}, J.B., {Li}, R., {Armitage}, P.J., {Youdin}, A.N.,
  {Kretke}, K.A.: {The Mass and Size Distribution of Planetesimals Formed by
  the Streaming Instability. II. The Effect of the Radial Gas Pressure
  Gradient}.
\newblock \apj \textbf{883}(2), 192 (2019).
\newblock \doi{10.3847/1538-4357/ab40a3}

\bibitem{absil13}
{Absil}, O., {Defr{\`e}re}, D., {Coud{\'e} du Foresto}, V., {Di Folco}, E.,
  {M{\'e}rand}, A., {Augereau}, J.C., {Ertel}, S., {Hanot}, C., {Kervella}, P.,
  {Mollier}, B., {Scott}, N., {Che}, X., {Monnier}, J.D., {Thureau}, N.,
  {Tuthill}, P.G., {ten Brummelaar}, T.A., {McAlister}, H.A., {Sturmann}, J.,
  {Sturmann}, L., {Turner}, N.: {A near-infrared interferometric survey of
  debris-disc stars. III. First statistics based on 42 stars observed with
  CHARA/FLUOR}.
\newblock \aap \textbf{555}, A104 (2013).
\newblock \doi{10.1051/0004-6361/201321673}

\bibitem{adams08}
{Adams}, E.R., {Seager}, S., {Elkins-Tanton}, L.: {Ocean Planet or Thick
  Atmosphere: On the Mass-Radius Relationship for Solid Exoplanets with Massive
  Atmospheres}.
\newblock \apj \textbf{673}, 1160--1164 (2008).
\newblock \doi{10.1086/524925}

\bibitem{adams04}
{Adams}, F.C., {Hollenbach}, D., {Laughlin}, G., {Gorti}, U.: {Photoevaporation
  of Circumstellar Disks Due to External Far-Ultraviolet Radiation in Stellar
  Aggregates}.
\newblock \apj \textbf{611}, 360--379 (2004).
\newblock \doi{10.1086/421989}

\bibitem{adams03}
{Adams}, F.C., {Laughlin}, G.: {Migration and dynamical relaxation in crowded
  systems of giant planets}.
\newblock Icarus \textbf{163}, 290--306 (2003).
\newblock \doi{10.1016/S0019-1035(03)00081-2}

\bibitem{agnor99}
{Agnor}, C.B., {Canup}, R.M., {Levison}, H.F.: {On the Character and
  Consequences of Large Impacts in the Late Stage of Terrestrial Planet
  Formation}.
\newblock Icarus \textbf{142}, 219--237 (1999).
\newblock \doi{10.1006/icar.1999.6201}

\bibitem{alexander19a}
{Alexander}, C.M.O.: {Quantitative models for the elemental and isotopic
  fractionations in chondrites: The carbonaceous chondrites}.
\newblock \gca \textbf{254}, 277--309 (2019).
\newblock \doi{10.1016/j.gca.2019.02.008}

\bibitem{alexander19b}
{Alexander}, C.M.O.: {Quantitative models for the elemental and isotopic
  fractionations in the chondrites: The non-carbonaceous chondrites}.
\newblock \gca \textbf{254}, 246--276 (2019).
\newblock \doi{10.1016/j.gca.2019.01.026}

\bibitem{alexander12}
{Alexander}, C.M.O.., {Bowden}, R., {Fogel}, M.L., {Howard}, K.T., {Herd},
  C.D.K., {Nittler}, L.R.: {The Provenances of Asteroids, and Their
  Contributions to the Volatile Inventories of the Terrestrial Planets}.
\newblock Science \textbf{337}, 721 (2012).
\newblock \doi{10.1126/science.1223474}

\bibitem{alexander18}
{Alexander}, C.M.O., {McKeegan}, K.D., {Altwegg}, K.: {Water Reservoirs in
  Small Planetary Bodies: Meteorites, Asteroids, and Comets}.
\newblock \ssr \textbf{214}, 36 (2018).
\newblock \doi{10.1007/s11214-018-0474-9}

\bibitem{alexander14}
{Alexander}, R., {Pascucci}, I., {Andrews}, S., {Armitage}, P., {Cieza}, L.:
  {The Dispersal of Protoplanetary Disks}.
\newblock Protostars and Planets VI pp. 475--496 (2014)

\bibitem{alexander06}
{Alexander}, R.D., {Armitage}, P.J.: {The Stellar Mass-Accretion Rate Relation
  in T Tauri Stars and Brown Dwarfs}.
\newblock \apjl \textbf{639}, L83--L86 (2006).
\newblock \doi{10.1086/503030}

\bibitem{alexander12b}
{Alexander}, R.D., {Pascucci}, I.: {Deserts and pile-ups in the distribution of
  exoplanets due to photoevaporative disc clearing}.
\newblock \mnras \textbf{422}, L82--L86 (2012).
\newblock \doi{10.1111/j.1745-3933.2012.01243.x}

\bibitem{alma15}
{ALMA Partnership}, {Brogan}, C.L., {P{\'e}rez}, L.M., {Hunter}, T.R., {Dent},
  W.R.F., {Hales}, A.S., {Hills}, R.E., {Corder}, S., {Fomalont}, E.B.,
  {Vlahakis}, C., {Asaki}, Y., {Barkats}, D., {Hirota}, A., {Hodge}, J.A.,
  {Impellizzeri}, C.M.V., {Kneissl}, R., {Liuzzo}, E., {Lucas}, R.,
  {Marcelino}, N., {Matsushita}, S., {Nakanishi}, K., {Phillips}, N.,
  {Richards}, A.M.S., {Toledo}, I., {Aladro}, R., {Broguiere}, D., {Cortes},
  J.R., {Cortes}, P.C., {Espada}, D., {Galarza}, F., {Garcia-Appadoo}, D.,
  {Guzman-Ramirez}, L., {Humphreys}, E.M., {Jung}, T., {Kameno}, S., {Laing},
  R.A., {Leon}, S., {Marconi}, G., {Mignano}, A., {Nikolic}, B., {Nyman}, L.A.,
  {Radiszcz}, M., {Remijan}, A., {Rod{\'o}n}, J.A., {Sawada}, T., {Takahashi},
  S., {Tilanus}, R.P.J., {Vila Vilaro}, B., {Watson}, L.C., {Wiklind}, T.,
  {Akiyama}, E., {Chapillon}, E., {de Gregorio-Monsalvo}, I., {Di Francesco},
  J., {Gueth}, F., {Kawamura}, A., {Lee}, C.F., {Nguyen Luong}, Q., {Mangum},
  J., {Pietu}, V., {Sanhueza}, P., {Saigo}, K., {Takakuwa}, S., {Ubach}, C.,
  {van Kempen}, T., {Wootten}, A., {Castro-Carrizo}, A., {Francke}, H.,
  {Gallardo}, J., {Garcia}, J., {Gonzalez}, S., {Hill}, T., {Kaminski}, T.,
  {Kurono}, Y., {Liu}, H.Y., {Lopez}, C., {Morales}, F., {Plarre}, K.,
  {Schieven}, G., {Testi}, L., {Videla}, L., {Villard}, E., {Andreani}, P.,
  {Hibbard}, J.E., {Tatematsu}, K.: {The 2014 ALMA Long Baseline Campaign:
  First Results from High Angular Resolution Observations toward the HL Tau
  Region}.
\newblock \apjl \textbf{808}, L3 (2015).
\newblock \doi{10.1088/2041-8205/808/1/L3}

\bibitem{amelin02}
{Amelin}, Y., {Krot}, A.N., {Hutcheon}, I.D., {Ulyanov}, A.A.: {Lead Isotopic
  Ages of Chondrules and Calcium-Aluminum-Rich Inclusions}.
\newblock Science \textbf{297}(5587), 1678--1683 (2002).
\newblock \doi{10.1126/science.1073950}

\bibitem{andrews18}
{Andrews}, S.M., {Huang}, J., {P{\'e}rez}, L.M., {Isella}, A., {Dullemond},
  C.P., {Kurtovic}, N.T., {Guzm{\'a}n}, V.V., {Carpenter}, J.M., {Wilner},
  D.J., {Zhang}, S., {Zhu}, Z., {Birnstiel}, T., {Bai}, X.N., {Benisty}, M.,
  {Hughes}, A.M., {{\"O}berg}, K.I., {Ricci}, L.: {The Disk Substructures at
  High Angular Resolution Project (DSHARP). I. Motivation, Sample, Calibration,
  and Overview}.
\newblock \apjl \textbf{869}(2), L41 (2018).
\newblock \doi{10.3847/2041-8213/aaf741}

\bibitem{andrews05}
{Andrews}, S.M., {Williams}, J.P.: {Circumstellar Dust Disks in Taurus-Auriga:
  The Submillimeter Perspective}.
\newblock \apj \textbf{631}, 1134--1160 (2005).
\newblock \doi{10.1086/432712}

\bibitem{andrews07a}
{Andrews}, S.M., {Williams}, J.P.: {A Submillimeter View of Circumstellar Dust
  Disks in {$\rho$} Ophiuchi}.
\newblock \apj \textbf{671}, 1800--1812 (2007).
\newblock \doi{10.1086/522885}

\bibitem{andrews07b}
{Andrews}, S.M., {Williams}, J.P.: {High-Resolution Submillimeter Constraints
  on Circumstellar Disk Structure}.
\newblock \apj \textbf{659}, 705--728 (2007).
\newblock \doi{10.1086/511741}

\bibitem{andrews09}
{Andrews}, S.M., {Wilner}, D.J., {Hughes}, A.M., {Qi}, C., {Dullemond}, C.P.:
  {Protoplanetary Disk Structures in Ophiuchus}.
\newblock \apj \textbf{700}, 1502--1523 (2009).
\newblock \doi{10.1088/0004-637X/700/2/1502}

\bibitem{andrews10}
{Andrews}, S.M., {Wilner}, D.J., {Hughes}, A.M., {Qi}, C., {Dullemond}, C.P.:
  {Protoplanetary Disk Structures in Ophiuchus. II. Extension to Fainter
  Sources}.
\newblock \apj \textbf{723}, 1241--1254 (2010).
\newblock \doi{10.1088/0004-637X/723/2/1241}

\bibitem{andrews16}
{Andrews}, S.M., {Wilner}, D.J., {Zhu}, Z., {Birnstiel}, T., {Carpenter}, J.M.,
  {P{\'e}rez}, L.M., {Bai}, X.N., {{\"O}berg}, K.I., {Hughes}, A.M., {Isella},
  A., {Ricci}, L.: {Ringed Substructure and a Gap at 1 au in the Nearest
  Protoplanetary Disk}.
\newblock \apjl \textbf{820}, L40 (2016).
\newblock \doi{10.3847/2041-8205/820/2/L40}

\bibitem{anglada10}
{Anglada-Escud{\'e}}, G., {L{\'o}pez-Morales}, M., {Chambers}, J.E.: {How
  Eccentric Orbital Solutions Can Hide Planetary Systems in 2:1 Resonant
  Orbits}.
\newblock \apj \textbf{709}(1), 168--178 (2010).
\newblock \doi{10.1088/0004-637X/709/1/168}

\bibitem{anglada13}
{Anglada-Escud{\'e}}, G., {Tuomi}, M., {Gerlach}, E., {Barnes}, R., {Heller},
  R., {Jenkins}, J.S., {Wende}, S., {Vogt}, S.S., {Butler}, R.P., {Reiners},
  A., {Jones}, H.R.A.: {A dynamically-packed planetary system around GJ 667C
  with three super-Earths in its habitable zone}.
\newblock \aap \textbf{556}, A126 (2013).
\newblock \doi{10.1051/0004-6361/201321331}

\bibitem{armitage11}
{Armitage}, P.J.: {Dynamics of Protoplanetary Disks}.
\newblock \araa \textbf{49}, 195--236 (2011).
\newblock \doi{10.1146/annurev-astro-081710-102521}

\bibitem{armitage16}
{Armitage}, P.J., {Eisner}, J.A., {Simon}, J.B.: {Prompt Planetesimal Formation
  beyond the Snow Line}.
\newblock \apjl \textbf{828}, L2 (2016).
\newblock \doi{10.3847/2041-8205/828/1/L2}

\bibitem{asaduzzaman14}
{Asaduzzaman}, A.M., {Zega}, T.J., {Laref}, S., {Runge}, K., {Deymier}, P.A.,
  {Muralidharan}, K.: {A computational investigation of adsorption of organics
  on mineral surfaces: Implications for organics delivery in the early solar
  system}.
\newblock Earth and Planetary Science Letters \textbf{408}, 355--361 (2014).
\newblock \doi{10.1016/j.epsl.2014.10.029}

\bibitem{badro14}
{Badro}, J., {Cote}, A., {Brodholt}, J.: {A seismologically consistent
  compositional model of Earth's core}.
\newblock PNAS \textbf{132}, 94 (2014)

\bibitem{bae19}
{Bae}, J., {Zhu}, Z., {Baruteau}, C., {Benisty}, M., {Dullemond}, C.P.,
  {Facchini}, S., {Isella}, A., {Keppler}, M., {P{\'e}rez}, L.M., {Teague}, R.:
  {An Ideal Testbed for Planet-Disk Interaction: Two Giant Protoplanets in
  Resonance Shaping the PDS 70 Protoplanetary Disk}.
\newblock \apjl \textbf{884}(2), L41 (2019).
\newblock \doi{10.3847/2041-8213/ab46b0}

\bibitem{Bai16}
{Bai}, X.N.: {Towards a Global Evolutionary Model of Protoplanetary Disks}.
\newblock \apj \textbf{821}, 80 (2016).
\newblock \doi{10.3847/0004-637X/821/2/80}

\bibitem{bai17}
{Bai}, X.N.: {Global Simulations of the Inner Regions of Protoplanetary Disks
  with Comprehensive Disk Microphysics}.
\newblock \apj \textbf{845}(1), 75 (2017).
\newblock \doi{10.3847/1538-4357/aa7dda}

\bibitem{bai13}
{Bai}, X.N., {Stone}, J.M.: {Wind-driven Accretion in Protoplanetary Disks. I.
  Suppression of the Magnetorotational Instability and Launching of the
  Magnetocentrifugal Wind}.
\newblock \apj \textbf{769}(1), 76 (2013).
\newblock \doi{10.1088/0004-637X/769/1/76}

\bibitem{baillie15}
{Bailli{\'e}}, K., {Charnoz}, S., {Pantin}, E.: {Time evolution of snow regions
  and planet traps in an evolving protoplanetary disk}.
\newblock \aap \textbf{577}, A65 (2015).
\newblock \doi{10.1051/0004-6361/201424987}

\bibitem{balbus98}
{Balbus}, S.A., {Hawley}, J.F.: {Instability, turbulence, and enhanced
  transport in accretion disks}.
\newblock Reviews of Modern Physics \textbf{70}, 1--53 (1998).
\newblock \doi{10.1103/RevModPhys.70.1}

\bibitem{barbato18}
{Barbato}, D., {Sozzetti}, A., {Desidera}, S., {Damasso}, M., {Bonomo}, A.S.,
  {Giacobbe}, P., {Colombo}, L.S., {Lazzoni}, C., {Claudi}, R., {Gratton}, R.,
  {LoCurto}, G., {Marzari}, F., {Mordasini}, C.: {Exploring the realm of scaled
  solar system analogues with HARPS}.
\newblock \aap \textbf{615}, A175 (2018).
\newblock \doi{10.1051/0004-6361/201832791}

\bibitem{barenfeld17}
{Barenfeld}, S.A., {Carpenter}, J.M., {Sargent}, A.I., {Isella}, A., {Ricci},
  L.: {Measurement of Circumstellar Disk Sizes in the Upper Scorpius OB
  Association with ALMA}.
\newblock \apj \textbf{851}(2), 85 (2017).
\newblock \doi{10.3847/1538-4357/aa989d}

\bibitem{baruteau14}
{Baruteau}, C., {Meru}, F., {Paardekooper}, S.J.: {Rapid inward migration of
  planets formed by gravitational instability}.
\newblock \mnras \textbf{416}(3), 1971--1982 (2011).
\newblock \doi{10.1111/j.1365-2966.2011.19172.x}

\bibitem{batalha13}
{Batalha}, N.M., {Rowe}, J.F., {Bryson}, S.T., {Barclay}, T., {Burke}, C.J.,
  {Caldwell}, D.A., {Christiansen}, J.L., {Mullally}, F., {Thompson}, S.E.,
  {Brown}, T.M., {Dupree}, A.K., {Fabrycky}, D.C., {Ford}, E.B., {Fortney},
  J.J., {Gilliland}, R.L., {Isaacson}, H., {Latham}, D.W., {Marcy}, G.W.,
  {Quinn}, S.N., {Ragozzine}, D., {Shporer}, A., {Borucki}, W.J., {Ciardi},
  D.R., {Gautier} III, T.N., {Haas}, M.R., {Jenkins}, J.M., {Koch}, D.G.,
  {Lissauer}, J.J., {Rapin}, W., {Basri}, G.S., {Boss}, A.P., {Buchhave}, L.A.,
  {Carter}, J.A., {Charbonneau}, D., {Christensen-Dalsgaard}, J., {Clarke},
  B.D., {Cochran}, W.D., {Demory}, B.O., {Desert}, J.M., {Devore}, E., {Doyle},
  L.R., {Esquerdo}, G.A., {Everett}, M., {Fressin}, F., {Geary}, J.C.,
  {Girouard}, F.R., {Gould}, A., {Hall}, J.R., {Holman}, M.J., {Howard}, A.W.,
  {Howell}, S.B., {Ibrahim}, K.A., {Kinemuchi}, K., {Kjeldsen}, H., {Klaus},
  T.C., {Li}, J., {Lucas}, P.W., {Meibom}, S., {Morris}, R.L., {Pr{\v s}a}, A.,
  {Quintana}, E., {Sanderfer}, D.T., {Sasselov}, D., {Seader}, S.E., {Smith},
  J.C., {Steffen}, J.H., {Still}, M., {Stumpe}, M.C., {Tarter}, J.C.,
  {Tenenbaum}, P., {Torres}, G., {Twicken}, J.D., {Uddin}, K., {Van Cleve}, J.,
  {Walkowicz}, L., {Welsh}, W.F.: {Planetary Candidates Observed by Kepler.
  III. Analysis of the First 16 Months of Data}.
\newblock \apjs \textbf{204}, 24 (2013).
\newblock \doi{10.1088/0067-0049/204/2/24}

\bibitem{bate18}
{Bate}, M.R.: {On the diversity and statistical properties of protostellar
  discs}.
\newblock \mnras \textbf{475}, 5618--5658 (2018).
\newblock \doi{10.1093/mnras/sty169}

\bibitem{batygin13b}
{Batygin}, K., {Morbidelli}, A.: {Analytical treatment of planetary
  resonances}.
\newblock \aap \textbf{556}, A28 (2013).
\newblock \doi{10.1051/0004-6361/201220907}

\bibitem{batygin15b}
{Batygin}, K., {Morbidelli}, A., {Holman}, M.J.: {Chaotic Disintegration of the
  Inner Solar System}.
\newblock \apj \textbf{799}, 120 (2015).
\newblock \doi{10.1088/0004-637X/799/2/120}

\bibitem{beauge12}
{Beaug{\'e}}, C., {Nesvorn{\'y}}, D.: {Multiple-planet Scattering and the
  Origin of Hot Jupiters}.
\newblock \apj \textbf{751}, 119 (2012).
\newblock \doi{10.1088/0004-637X/751/2/119}

\bibitem{bell94}
{Bell}, K.R., {Lin}, D.N.C.: {Using FU Orionis Outbursts to Constrain
  Self-regulated Protostellar Disk Models}.
\newblock \apj \textbf{427}, 987 (1994).
\newblock \doi{10.1086/174206}

\bibitem{benitez15}
{Ben{\'{\i}}tez-Llambay}, P., {Masset}, F., {Koenigsberger}, G.,
  {Szul{\'a}gyi}, J.: {Planet heating prevents inward migration of planetary
  cores}.
\newblock \nat \textbf{520}, 63--65 (2015).
\newblock \doi{10.1038/nature14277}

\bibitem{benitez18}
{Ben{\'\i}tez-Llambay}, P., {Pessah}, M.E.: {Torques Induced by Scattered
  Pebble-flow in Protoplanetary Disks}.
\newblock \apjl \textbf{855}(2), L28 (2018).
\newblock \doi{10.3847/2041-8213/aab2ae}

\bibitem{bethune17}
{B{\'e}thune}, W., {Lesur}, G., {Ferreira}, J.: {Global simulations of
  protoplanetary disks with net magnetic flux. I. Non-ideal MHD case}.
\newblock \aap \textbf{600}, A75 (2017).
\newblock \doi{10.1051/0004-6361/201630056}

\bibitem{birnstiel16}
{Birnstiel}, T., {Fang}, M., {Johansen}, A.: {Dust Evolution and the Formation
  of Planetesimals}.
\newblock \ssr \textbf{205}, 41--75 (2016).
\newblock \doi{10.1007/s11214-016-0256-1}

\bibitem{birnstiel12}
{Birnstiel}, T., {Klahr}, H., {Ercolano}, B.: {A simple model for the evolution
  of the dust population in protoplanetary disks}.
\newblock \aap \textbf{539}, A148 (2012).
\newblock \doi{10.1051/0004-6361/201118136}

\bibitem{bitsch13}
{Bitsch}, B., {Crida}, A., {Morbidelli}, A., {Kley}, W., {Dobbs-Dixon}, I.:
  {Stellar irradiated discs and implications on migration of embedded planets.
  I. Equilibrium discs}.
\newblock \aap \textbf{549}, A124 (2013).
\newblock \doi{10.1051/0004-6361/201220159}

\bibitem{bitsch19}
{Bitsch}, B., {Izidoro}, A., {Johansen}, A., {Raymond}, S.N., {Morbidelli}, A.,
  {Lambrechts}, M., {Jacobson}, S.A.: {Formation of planetary systems by pebble
  accretion and migration: growth of gas giants}.
\newblock \aap \textbf{623}, A88 (2019).
\newblock \doi{10.1051/0004-6361/201834489}

\bibitem{bitsch15}
{Bitsch}, B., {Johansen}, A., {Lambrechts}, M., {Morbidelli}, A.: {The
  structure of protoplanetary discs around evolving young stars}.
\newblock \aap \textbf{575}, A28 (2015).
\newblock \doi{10.1051/0004-6361/201424964}

\bibitem{bitsch15b}
{Bitsch}, B., {Lambrechts}, M., {Johansen}, A.: {The growth of planets by
  pebble accretion in evolving protoplanetary discs}.
\newblock \aap \textbf{582}, A112 (2015).
\newblock \doi{10.1051/0004-6361/201526463}

\bibitem{bitsch18}
{Bitsch}, B., {Morbidelli}, A., {Johansen}, A., {Lega}, E., {Lambrechts}, M.,
  {Crida}, A.: {Pebble-isolation mass: Scaling law and implications for the
  formation of super-Earths and gas giants}.
\newblock \aap \textbf{612}, A30 (2018).
\newblock \doi{10.1051/0004-6361/201731931}

\bibitem{bitsch14}
{Bitsch}, B., {Morbidelli}, A., {Lega}, E., {Crida}, A.: {Stellar irradiated
  discs and implications on migration of embedded planets. II.
  Accreting-discs}.
\newblock \aap \textbf{564}, A135 (2014).
\newblock \doi{10.1051/0004-6361/201323007}

\bibitem{bitsch19b}
{Bitsch}, B., {Raymond}, S.N., {Izidoro}, A.: {Rocky super-Earths or
  waterworlds: the interplay of planet migration, pebble accretion, and disc
  evolution}.
\newblock \aap \textbf{624}, A109 (2019).
\newblock \doi{10.1051/0004-6361/201935007}

\bibitem{biver16}
{Biver}, N., {Moreno}, R., {Bockel{\'e}e-Morvan}, D., {Sandqvist}, A., {Colom},
  P., {Crovisier}, J., {Lis}, D.C., {Boissier}, J., {Debout}, V., {Paubert},
  G., {Milam}, S., {Hjalmarson}, A., {Lundin}, S., {Karlsson}, T., {Battelino},
  M., {Frisk}, U., {Murtagh}, D., {Odin Team}: {Isotopic ratios of H, C, N, O,
  and S in comets C/2012 F6 (Lemmon) and C/2014 Q2 (Lovejoy)}.
\newblock \aap \textbf{589}, A78 (2016).
\newblock \doi{10.1051/0004-6361/201528041}

\bibitem{blum08}
{Blum}, J., {Wurm}, G.: {The Growth Mechanisms of Macroscopic Bodies in
  Protoplanetary Disks}.
\newblock \araa \textbf{46}, 21--56 (2008).
\newblock \doi{10.1146/annurev.astro.46.060407.145152}

\bibitem{boehnke16}
{Boehnke}, P., {Harrison}, T.M.: {Illusory Late Heavy Bombardments}.
\newblock Proceedings of the National Academy of Science \textbf{113},
  10802--10806 (2016).
\newblock \doi{10.1073/pnas.1611535113}

\bibitem{boisvert18}
{Boisvert}, J.H., {Nelson}, B.E., {Steffen}, J.H.: {Systematic
  mischaracterization of exoplanetary system dynamical histories from a model
  degeneracy near mean-motion resonance}.
\newblock \mnras \textbf{480}, 2846--2852 (2018).
\newblock \doi{10.1093/mnras/sty2023}

\bibitem{boley09}
{Boley}, A.C.: {The Two Modes of Gas Giant Planet Formation}.
\newblock \apjl \textbf{695}(1), L53--L57 (2009).
\newblock \doi{10.1088/0004-637X/695/1/L53}

\bibitem{boley13}
{Boley}, A.C., {Ford}, E.B.: {The Formation of Systems with Tightly-packed
  Inner Planets (STIPs) via Aerodynamic Drift}.
\newblock arXiv:1306.0566  (2013)

\bibitem{boley10}
{Boley}, A.C., {Hayfield}, T., {Mayer}, L., {Durisen}, R.H.: {Clumps in the
  outer disk by disk instability: Why they are initially gas giants and the
  legacy of disruption}.
\newblock \icarus \textbf{207}(2), 509--516 (2010).
\newblock \doi{10.1016/j.icarus.2010.01.015}

\bibitem{bonomo17}
{Bonomo}, A.S., {Desidera}, S., {Benatti}, S., {Borsa}, F., {Crespi}, S.,
  {Damasso}, M., {Lanza}, A.F., {Sozzetti}, A., {Lodato}, G., {Marzari}, F.,
  {Boccato}, C., {Claudi}, R.U., {Cosentino}, R., {Covino}, E., {Gratton}, R.,
  {Maggio}, A., {Micela}, G., {Molinari}, E., {Pagano}, I., {Piotto}, G.,
  {Poretti}, E., {Smareglia}, R., {Affer}, L., {Biazzo}, K., {Bignamini}, A.,
  {Esposito}, M., {Giacobbe}, P., {H{\'e}brard}, G., {Malavolta}, L.,
  {Maldonado}, J., {Mancini}, L., {Martinez Fiorenzano}, A., {Masiero}, S.,
  {Nascimbeni}, V., {Pedani}, M., {Rainer}, M., {Scand ariato}, G.: {The GAPS
  Programme with HARPS-N at TNG . XIV. Investigating giant planet migration
  history via improved eccentricity and mass determination for 231 transiting
  planets}.
\newblock \aap \textbf{602}, A107 (2017).
\newblock \doi{10.1051/0004-6361/201629882}

\bibitem{boss97}
{Boss}, A.P.: {Giant planet formation by gravitational instability.}
\newblock Science \textbf{276}, 1836--1839 (1997).
\newblock \doi{10.1126/science.276.5320.1836}

\bibitem{boss98}
{Boss}, A.P.: {Evolution of the Solar Nebula. IV. Giant Gaseous Protoplanet
  Formation}.
\newblock \apj \textbf{503}(2), 923--937 (1998).
\newblock \doi{10.1086/306036}

\bibitem{boss00}
{Boss}, A.P.: {Possible Rapid Gas Giant Planet Formation in the Solar Nebula
  and Other Protoplanetary Disks}.
\newblock \apjl \textbf{536}(2), L101--L104 (2000).
\newblock \doi{10.1086/312737}

\bibitem{bouvier10}
{Bouvier}, A., {Wadhwa}, M.: {The age of the Solar System redefined by the
  oldest Pb-Pb age of a meteoritic inclusion}.
\newblock Nature Geoscience \textbf{3}, 637--641 (2010).
\newblock \doi{10.1038/ngeo941}

\bibitem{bowler18}
{Bowler}, B.P., {Nielsen}, E.L.: {Occurrence Rates from Direct Imaging
  Surveys}.
\newblock ArXiv e-prints  (2018)

\bibitem{brasser16}
{Brasser}, R., {Matsumura}, S., {Ida}, S., {Mojzsis}, S.J., {Werner}, S.C.:
  {Analysis of Terrestrial Planet Formation by the Grand Tack Model: System
  Architecture and Tack Location}.
\newblock \apj \textbf{821}, 75 (2016).
\newblock \doi{10.3847/0004-637X/821/2/75}

\bibitem{briceno01}
{Brice{\~n}o}, C., {Vivas}, A.K., {Calvet}, N., {Hartmann}, L., {Pacheco}, R.,
  {Herrera}, D., {Romero}, L., {Berlind}, P., {S{\'a}nchez}, G., {Snyder},
  J.A., {Andrews}, P.: {The CIDA-QUEST Large-Scale Survey of Orion OB1:
  Evidence for Rapid Disk Dissipation in a Dispersed Stellar Population}.
\newblock Science \textbf{291}, 93--97 (2001).
\newblock \doi{10.1126/science.291.5501.93}

\bibitem{bromley17}
{Bromley}, B.C., {Kenyon}, S.J.: {Terrestrial Planet Formation: Dynamical
  Shake-up and the Low Mass of Mars}.
\newblock \aj \textbf{153}, 216 (2017).
\newblock \doi{10.3847/1538-3881/aa6aaa}

\bibitem{bryan19}
{Bryan}, M.L., {Knutson}, H.A., {Lee}, E.J., {Fulton}, B.J., {Batygin}, K.,
  {Ngo}, H., {Meshkat}, T.: {An Excess of Jupiter Analogs in Super-Earth
  Systems}.
\newblock \aj \textbf{157}(2), 52 (2019).
\newblock \doi{10.3847/1538-3881/aaf57f}

\bibitem{bryden09}
{Bryden}, G., {Beichman}, C.A., {Carpenter}, J.M., {Rieke}, G.H.,
  {Stapelfeldt}, K.R., {Werner}, M.W., {Tanner}, A.M., {Lawler}, S.M., {Wyatt},
  M.C., {Trilling}, D.E., {Su}, K.Y.L., {Blaylock}, M., {Stansberry}, J.A.:
  {Planets and Debris Disks: Results from a Spitzer/MIPS Search for Infrared
  Excess}.
\newblock \apj \textbf{705}(2), 1226--1236 (2009).
\newblock \doi{10.1088/0004-637X/705/2/1226}

\bibitem{bryden06}
{Bryden}, G., {Beichman}, C.A., {Trilling}, D.E., {Rieke}, G.H., {Holmes},
  E.K., {Lawler}, S.M., {Stapelfeldt}, K.R., {Werner}, M.W., {Gautier}, T.N.,
  {Blaylock}, M., {Gordon}, K.D., {Stansberry}, J.A., {Su}, K.Y.L.: {Frequency
  of Debris Disks around Solar-Type Stars: First Results from a Spitzer MIPS
  Survey}.
\newblock \apj \textbf{636}, 1098--1113 (2006).
\newblock \doi{10.1086/498093}

\bibitem{budde16}
{Budde}, G., {Burkhardt}, C., {Brennecka}, G.A., {Fischer-G{\"o}dde}, M.,
  {Kruijer}, T.S., {Kleine}, T.: {Molybdenum isotopic evidence for the origin
  of chondrules and a distinct genetic heritage of carbonaceous and
  non-carbonaceous meteorites}.
\newblock Earth and Planetary Science Letters \textbf{454}, 293--303 (2016).
\newblock \doi{10.1016/j.epsl.2016.09.020}

\bibitem{budde19}
{Budde}, G., {Burkhardt}, C., {Kleine}, T.: {Molybdenum isotopic evidence for
  the late accretion of outer Solar System material to Earth}.
\newblock Nature Astronomy \textbf{3}, 736--741 (2019).
\newblock \doi{10.1038/s41550-019-0779-y}

\bibitem{butler06}
{Butler}, R.P., {Wright}, J.T., {Marcy}, G.W., {Fischer}, D.A., {Vogt}, S.S.,
  {Tinney}, C.G., {Jones}, H.R.A., {Carter}, B.D., {Johnson}, J.A., {McCarthy},
  C., {Penny}, A.J.: {Catalog of Nearby Exoplanets}.
\newblock \apj \textbf{646}, 505--522 (2006).
\newblock \doi{10.1086/504701}

\bibitem{caracausi16}
{Caracausi}, A., {Avice}, G., {Burnard}, P.G., {F{\"u}ri}, E., {Marty}, B.:
  {Chondritic xenon in the Earth's mantle}.
\newblock \nat \textbf{533}, 82--85 (2016).
\newblock \doi{10.1038/nature17434}

\bibitem{carpenter09}
{Carpenter}, J.M., {Bouwman}, J., {Mamajek}, E.E., {Meyer}, M.R.,
  {Hillenbrand}, L.A., {Backman}, D.E., {Henning}, T., {Hines}, D.C.,
  {Hollenbach}, D., {Kim}, J.S., {Moro-Martin}, A., {Pascucci}, I.,
  {Silverstone}, M.D., {Stauffer}, J.R., {Wolf}, S.: {Formation and Evolution
  of Planetary Systems: Properties of Debris Dust Around Solar-Type Stars}.
\newblock \apjs \textbf{181}, 197--226 (2009).
\newblock \doi{10.1088/0067-0049/181/1/197}

\bibitem{carrera16}
{Carrera}, D., {Davies}, M.B., {Johansen}, A.: {Survival of habitable planets
  in unstable planetary systems}.
\newblock \mnras \textbf{463}, 3226--3238 (2016).
\newblock \doi{10.1093/mnras/stw2218}

\bibitem{carrera18}
{Carrera}, D., {Ford}, E.B., {Izidoro}, A., {Jontof-Hutter}, D., {Raymond},
  S.N., {Wolfgang}, A.: {Identifying Inflated Super-Earths and Photo-evaporated
  Cores}.
\newblock \apj \textbf{866}, 104 (2018).
\newblock \doi{10.3847/1538-4357/aadf8a}

\bibitem{Carrera17}
{Carrera}, D., {Gorti}, U., {Johansen}, A., {Davies}, M.B.: {Planetesimal
  Formation by the Streaming Instability in a Photoevaporating Disk}.
\newblock \apj \textbf{839}, 16 (2017).
\newblock \doi{10.3847/1538-4357/aa6932}

\bibitem{carrera15}
{Carrera}, D., {Johansen}, A., {Davies}, M.B.: {How to form planetesimals from
  mm-sized chondrules and chondrule aggregates}.
\newblock \aap \textbf{579}, A43 (2015).
\newblock \doi{10.1051/0004-6361/201425120}

\bibitem{carrera19}
{Carrera}, D., {Raymond}, S.N., {Davies}, M.B.: {Planet-planet scattering as
  the source of the highest eccentricity exoplanets}.
\newblock \aap \textbf{629}, L7 (2019).
\newblock \doi{10.1051/0004-6361/201935744}

\bibitem{cassan12}
{Cassan}, A., {Kubas}, D., {Beaulieu}, J.P., {Dominik}, M., {Horne}, K.,
  {Greenhill}, J., {Wambsganss}, J., {Menzies}, J., {Williams}, A.,
  {J{\o}rgensen}, U.G., {Udalski}, A., {Bennett}, D.P., {Albrow}, M.D.,
  {Batista}, V., {Brillant}, S., {Caldwell}, J.A.R., {Cole}, A., {Coutures},
  C., {Cook}, K.H., {Dieters}, S., {Dominis Prester}, D., {Donatowicz}, J.,
  {Fouqu{\'e}}, P., {Hill}, K., {Kains}, N., {Kane}, S., {Marquette}, J.B.,
  {Martin}, R., {Pollard}, K.R., {Sahu}, K.C., {Vinter}, C., {Warren}, D.,
  {Watson}, B., {Zub}, M., {Sumi}, T., {Szyma{\'n}ski}, M.K., {Kubiak}, M.,
  {Poleski}, R., {Soszynski}, I., {Ulaczyk}, K., {Pietrzy{\'n}ski}, G.,
  {Wyrzykowski}, {\L}.: {One or more bound planets per Milky Way star from
  microlensing observations}.
\newblock \nat \textbf{481}, 167--169 (2012).
\newblock \doi{10.1038/nature10684}

\bibitem{chambers06}
{Chambers}, J.: {A semi-analytic model for oligarchic growth}.
\newblock Icarus \textbf{180}, 496--513 (2006).
\newblock \doi{10.1016/j.icarus.2005.10.017}

\bibitem{chambers19}
{Chambers}, J.: {An Analytic Model for an Evolving Protoplanetary Disk with a
  Disk Wind}.
\newblock \apj \textbf{879}(2), 98 (2019).
\newblock \doi{10.3847/1538-4357/ab2537}

\bibitem{chambers01}
{Chambers}, J.E.: {Making More Terrestrial Planets}.
\newblock Icarus \textbf{152}, 205--224 (2001).
\newblock \doi{10.1006/icar.2001.6639}

\bibitem{chambers98}
{Chambers}, J.E., {Wetherill}, G.W.: {Making the Terrestrial Planets: N-Body
  Integrations of Planetary Embryos in Three Dimensions}.
\newblock Icarus \textbf{136}, 304--327 (1998).
\newblock \doi{10.1006/icar.1998.6007}

\bibitem{chambers16}
{Chambers}, K.C., {Magnier}, E.A., {Metcalfe}, N., {Flewelling}, H.A., {Huber},
  M.E., {Waters}, C.Z., {Denneau}, L., {Draper}, P.W., {Farrow}, D.,
  {Finkbeiner}, D.P., {Holmberg}, C., {Koppenhoefer}, J., {Price}, P.A.,
  {Saglia}, R.P., {Schlafly}, E.F., {Smartt}, S.J., {Sweeney}, W., {Wainscoat},
  R.J., {Burgett}, W.S., {Grav}, T., {Heasley}, J.N., {Hodapp}, K.W.,
  {Jedicke}, R., {Kaiser}, N., {Kudritzki}, R.P., {Luppino}, G.A., {Lupton},
  R.H., {Monet}, D.G., {Morgan}, J.S., {Onaka}, P.M., {Stubbs}, C.W., {Tonry},
  J.L., {Banados}, E., {Bell}, E.F., {Bender}, R., {Bernard}, E.J.,
  {Botticella}, M.T., {Casertano}, S., {Chastel}, S., {Chen}, W.P., {Chen}, X.,
  {Cole}, S., {Deacon}, N., {Frenk}, C., {Fitzsimmons}, A., {Gezari}, S.,
  {Goessl}, C., {Goggia}, T., {Goldman}, B., {Grebel}, E.K., {Hambly}, N.C.,
  {Hasinger}, G., {Heavens}, A.F., {Heckman}, T.M., {Henderson}, R., {Henning},
  T., {Holman}, M., {Hopp}, U., {Ip}, W.H., {Isani}, S., {Keyes}, C.D.,
  {Koekemoer}, A., {Kotak}, R., {Long}, K.S., {Lucey}, J.R., {Liu}, M.,
  {Martin}, N.F., {McLean}, B., {Morganson}, E., {Murphy}, D.N.A.,
  {Nieto-Santisteban}, M.A., {Norberg}, P., {Peacock}, J.A., {Pier}, E.A.,
  {Postman}, M., {Primak}, N., {Rae}, C., {Rest}, A., {Riess}, A., {Riffeser},
  A., {Rix}, H.W., {Roser}, S., {Schilbach}, E., {Schultz}, A.S.B., {Scolnic},
  D., {Szalay}, A., {Seitz}, S., {Shiao}, B., {Small}, E., {Smith}, K.W.,
  {Soderblom}, D., {Taylor}, A.N., {Thakar}, A.R., {Thiel}, J., {Thilker}, D.,
  {Urata}, Y., {Valenti}, J., {Walter}, F., {Watters}, S.P., {Werner}, S.,
  {White}, R., {Wood-Vasey}, W.M., {Wyse}, R.: {The Pan-STARRS1 Surveys}.
\newblock arxiv:1612.05560  (2016)

\bibitem{chandrasehkar43}
{Chandrasekhar}, S.: {Dynamical Friction. I. General Considerations: the
  Coefficient of Dynamical Friction.}
\newblock \apj \textbf{97}, 255 (1943).
\newblock \doi{10.1086/144517}

\bibitem{chang10}
{Chang}, S.H., {Gu}, P.G., {Bodenheimer}, P.H.: {Tidal and Magnetic
  Interactions Between a Hot Jupiter and its Host Star in the Magnetospheric
  Cavity of a Protoplanetary Disk}.
\newblock \apj \textbf{708}(2), 1692--1702 (2010).
\newblock \doi{10.1088/0004-637X/708/2/1692}

\bibitem{chapman07}
{Chapman}, C.R., {Cohen}, B.A., {Grinspoon}, D.H.: {What are the real
  constraints on the existence and magnitude of the late heavy bombardment?}
\newblock Icarus \textbf{189}, 233--245 (2007).
\newblock \doi{10.1016/j.icarus.2006.12.020}

\bibitem{chatterjee08}
{Chatterjee}, S., {Ford}, E.B., {Matsumura}, S., {Rasio}, F.A.: {Dynamical
  Outcomes of Planet-Planet Scattering}.
\newblock \apj \textbf{686}, 580--602 (2008).
\newblock \doi{10.1086/590227}

\bibitem{chatterjee14}
{Chatterjee}, S., {Tan}, J.C.: {Inside-out Planet Formation}.
\newblock \apj \textbf{780}, 53 (2014).
\newblock \doi{10.1088/0004-637X/780/1/53}

\bibitem{chatterjee15}
{Chatterjee}, S., {Tan}, J.C.: {Vulcan Planets: Inside-out Formation of the
  Innermost Super-Earths}.
\newblock \apjl \textbf{798}, L32 (2015).
\newblock \doi{10.1088/2041-8205/798/2/L32}

\bibitem{chen17}
{Chen}, J., {Kipping}, D.: {Probabilistic Forecasting of the Masses and Radii
  of Other Worlds}.
\newblock \apj \textbf{834}, 17 (2017).
\newblock \doi{10.3847/1538-4357/834/1/17}

\bibitem{chiang13}
{Chiang}, E., {Laughlin}, G.: {The minimum-mass extrasolar nebula: in situ
  formation of close-in super-Earths}.
\newblock \mnras \textbf{431}, 3444--3455 (2013).
\newblock \doi{10.1093/mnras/stt424}

\bibitem{chiang10}
{Chiang}, E., {Youdin}, A.N.: {Forming Planetesimals in Solar and Extrasolar
  Nebulae}.
\newblock Annual Review of Earth and Planetary Sciences \textbf{38}, 493--522
  (2010).
\newblock \doi{10.1146/annurev-earth-040809-152513}

\bibitem{chiang97}
{Chiang}, E.I., {Goldreich}, P.: {Spectral Energy Distributions of T Tauri
  Stars with Passive Circumstellar Disks}.
\newblock \apj \textbf{490}, 368 (1997).
\newblock \doi{10.1086/304869}

\bibitem{clanton14}
{Clanton}, C., {Gaudi}, B.S.: {Synthesizing Exoplanet Demographics from Radial
  Velocity and Microlensing Surveys. II. The Frequency of Planets Orbiting M
  Dwarfs}.
\newblock \apj \textbf{791}, 91 (2014).
\newblock \doi{10.1088/0004-637X/791/2/91}

\bibitem{clanton16}
{Clanton}, C., {Gaudi}, B.S.: {Synthesizing Exoplanet Demographics: A Single
  Population of Long-period Planetary Companions to M Dwarfs Consistent with
  Microlensing, Radial Velocity, and Direct Imaging Surveys}.
\newblock \apj \textbf{819}, 125 (2016).
\newblock \doi{10.3847/0004-637X/819/2/125}

\bibitem{clement19a}
{Clement}, M.S., {Kaib}, N.A., {Raymond}, S.N., {Chambers}, J.E., {Walsh},
  K.J.: {The early instability scenario: Terrestrial planet formation during
  the giant planet instability, and the effect of collisional fragmentation}.
\newblock \icarus \textbf{321}, 778--790 (2019).
\newblock \doi{10.1016/j.icarus.2018.12.033}

\bibitem{clement18}
{Clement}, M.S., {Kaib}, N.A., {Raymond}, S.N., {Walsh}, K.J.: {Mars' growth
  stunted by an early giant planet instability}.
\newblock \icarus \textbf{311}, 340--356 (2018).
\newblock \doi{10.1016/j.icarus.2018.04.008}

\bibitem{clement19b}
{Clement}, M.S., {Raymond}, S.N., {Kaib}, N.A.: {Excitation and Depletion of
  the Asteroid Belt in the Early Instability Scenario}.
\newblock \aj \textbf{157}(1), 38 (2019).
\newblock \doi{10.3847/1538-3881/aaf21e}

\bibitem{coleman17}
{Coleman}, G.A.L., {Papaloizou}, J.C.B., {Nelson}, R.P.: {In situ accretion of
  gaseous envelopes on to planetary cores embedded in evolving protoplanetary
  discs}.
\newblock \mnras \textbf{470}(3), 3206--3219 (2017).
\newblock \doi{10.1093/mnras/stx1297}

\bibitem{connelly08}
{Connelly}, J.N., {Amelin}, Y., {Krot}, A.N., {Bizzarro}, M.: {Chronology of
  the Solar System's Oldest Solids}.
\newblock \apjl \textbf{675}, L121 (2008).
\newblock \doi{10.1086/533586}

\bibitem{connelly12}
{Connelly}, J.N., {Bizzarro}, M., {Krot}, A.N., {Nordlund}, {\AA}., {Wielandt},
  D., {Ivanova}, M.A.: {The Absolute Chronology and Thermal Processing of
  Solids in the Solar Protoplanetary Disk}.
\newblock Science \textbf{338}, 651 (2012).
\newblock \doi{10.1126/science.1226919}

\bibitem{cossou14}
{Cossou}, C., {Raymond}, S.N., {Hersant}, F., {Pierens}, A.: {Hot super-Earths
  and giant planet cores from different migration histories}.
\newblock \aap \textbf{569}, A56 (2014).
\newblock \doi{10.1051/0004-6361/201424157}

\bibitem{cossou13}
{Cossou}, C., {Raymond}, S.N., {Pierens}, A.: {Convergence zones for Type I
  migration: an inward shift for multiple planet systems}.
\newblock \aap \textbf{553}, L2 (2013).
\newblock \doi{10.1051/0004-6361/201220853}

\bibitem{cresswell07}
{Cresswell}, P., {Dirksen}, G., {Kley}, W., {Nelson}, R.P.: {On the evolution
  of eccentric and inclined protoplanets embedded in protoplanetary disks}.
\newblock \aap \textbf{473}, 329--342 (2007).
\newblock \doi{10.1051/0004-6361:20077666}

\bibitem{crida06}
{Crida}, A., {Morbidelli}, A., {Masset}, F.: {On the width and shape of gaps in
  protoplanetary disks}.
\newblock Icarus \textbf{181}, 587--604 (2006).
\newblock \doi{10.1016/j.icarus.2005.10.007}

\bibitem{cumming08}
{Cumming}, A., {Butler}, R.P., {Marcy}, G.W., {Vogt}, S.S., {Wright}, J.T.,
  {Fischer}, D.A.: {The Keck Planet Search: Detectability and the Minimum Mass
  and Orbital Period Distribution of Extrasolar Planets}.
\newblock \pasp \textbf{120}, 531--554 (2008).
\newblock \doi{10.1086/588487}

\bibitem{dangelo03}
{D'Angelo}, G., {Kley}, W., {Henning}, T.: {Orbital Migration and Mass
  Accretion of Protoplanets in Three-dimensional Global Computations with
  Nested Grids}.
\newblock \apj \textbf{586}(1), 540--561 (2003).
\newblock \doi{10.1086/367555}

\bibitem{dangelo19}
{D'Angelo}, M., {Cazaux}, S., {Kamp}, I., {Thi}, W.F., {Woitke}, P.: {Water
  delivery in the inner solar nebula. Monte Carlo simulations of forsterite
  hydration}.
\newblock \aap \textbf{622}, A208 (2019).
\newblock \doi{10.1051/0004-6361/201833715}

\bibitem{dauphas17}
{Dauphas}, N.: {The isotopic nature of the Earth's accreting material through
  time}.
\newblock \nat \textbf{541}, 521--524 (2017).
\newblock \doi{10.1038/nature20830}

\bibitem{dauphas11}
{Dauphas}, N., {Pourmand}, A.: {Hf-W-Th evidence for rapid growth of Mars and
  its status as a planetary embryo}.
\newblock \nat \textbf{473}, 489--492 (2011).
\newblock \doi{10.1038/nature10077}

\bibitem{davis05}
{Davis}, S.S.: {The Surface Density Distribution in the Solar Nebula}.
\newblock \apjl \textbf{627}, L153--L155 (2005).
\newblock \doi{10.1086/432464}

\bibitem{dawson15}
{Dawson}, R.I., {Chiang}, E., {Lee}, E.J.: {A metallicity recipe for rocky
  planets}.
\newblock \mnras \textbf{453}, 1471--1483 (2015).
\newblock \doi{10.1093/mnras/stv1639}

\bibitem{dawson16}
{Dawson}, R.I., {Lee}, E.J., {Chiang}, E.: {Correlations between Compositions
  and Orbits Established by the Giant Impact Era of Planet Formation}.
\newblock \apj \textbf{822}, 54 (2016).
\newblock \doi{10.3847/0004-637X/822/1/54}

\bibitem{dawson13}
{Dawson}, R.I., {Murray-Clay}, R.A.: {Giant Planets Orbiting Metal-rich Stars
  Show Signatures of Planet-Planet Interactions}.
\newblock \apjl \textbf{767}, L24 (2013).
\newblock \doi{10.1088/2041-8205/767/2/L24}

\bibitem{deck15}
{Deck}, K.M., {Batygin}, K.: {Migration of Two Massive Planets into (and out
  of) First Order Mean Motion Resonances}.
\newblock \apj \textbf{810}(2), 119 (2015).
\newblock \doi{10.1088/0004-637X/810/2/119}

\bibitem{deienno16}
{Deienno}, R., {Gomes}, R.S., {Walsh}, K.J., {Morbidelli}, A., {Nesvorn{\'y}},
  D.: {Is the Grand Tack model compatible with the orbital distribution of main
  belt asteroids?}
\newblock \icarus \textbf{272}, 114--124 (2016).
\newblock \doi{10.1016/j.icarus.2016.02.043}

\bibitem{deienno18}
{Deienno}, R., {Izidoro}, A., {Morbidelli}, A., {Gomes}, R.S., {Nesvorn{\'y}},
  D., {Raymond}, S.N.: {Excitation of a Primordial Cold Asteroid Belt as an
  Outcome of Planetary Instability}.
\newblock \apj \textbf{864}, 50 (2018).
\newblock \doi{10.3847/1538-4357/aad55d}

\bibitem{demeo13}
{DeMeo}, F.E., {Carry}, B.: {The taxonomic distribution of asteroids from
  multi-filter all-sky photometric surveys}.
\newblock Icarus \textbf{226}, 723--741 (2013).
\newblock \doi{10.1016/j.icarus.2013.06.027}

\bibitem{demeo14}
{DeMeo}, F.E., {Carry}, B.: {Solar System evolution from compositional mapping
  of the asteroid belt}.
\newblock \nat \textbf{505}, 629--634 (2014).
\newblock \doi{10.1038/nature12908}

\bibitem{demory11}
{Demory}, B.O., {Gillon}, M., {Deming}, D., {Valencia}, D., {Seager}, S.,
  {Benneke}, B., {Lovis}, C., {Cubillos}, P., {Harrington}, J., {Stevenson},
  K.B., {Mayor}, M., {Pepe}, F., {Queloz}, D., {S{\'e}gransan}, D., {Udry}, S.:
  {Detection of a transit of the super-Earth 55 Cancri e with warm Spitzer}.
\newblock \aap \textbf{533}, A114 (2011).
\newblock \doi{10.1051/0004-6361/201117178}

\bibitem{desch15}
{Desch}, S.J., {Turner}, N.J.: {High-temperature Ionization in Protoplanetary
  Disks}.
\newblock \apj \textbf{811}(2), 156 (2015).
\newblock \doi{10.1088/0004-637X/811/2/156}

\bibitem{dodsonrobinson11}
{Dodson-Robinson}, S.E., {Salyk}, C.: {Transitional Disks as Signposts of
  Young, Multiplanet Systems}.
\newblock \apj \textbf{738}(2), 131 (2011).
\newblock \doi{10.1088/0004-637X/738/2/131}

\bibitem{dodsonrobinson09}
{Dodson-Robinson}, S.E., {Willacy}, K., {Bodenheimer}, P., {Turner}, N.J.,
  {Beichman}, C.A.: {Ice lines, planetesimal composition and solid surface
  density in the solar nebula}.
\newblock \icarus \textbf{200}(2), 672--693 (2009).
\newblock \doi{10.1016/j.icarus.2008.11.023}

\bibitem{dong13}
{Dong}, S., {Zhu}, Z.: {Fast Rise of ''Neptune-size'' Planets (4-8 R $_{⊕}$)
  from P \~{} 10 to \~{}250 Days -- Statistics of Kepler Planet Candidates up
  to \~{}0.75 AU}.
\newblock \apj \textbf{778}, 53 (2013).
\newblock \doi{10.1088/0004-637X/778/1/53}

\bibitem{dorn15}
{Dorn}, C., {Khan}, A., {Heng}, K., {Connolly}, J.A.D., {Alibert}, Y., {Benz},
  W., {Tackley}, P.: {Can we constrain the interior structure of rocky
  exoplanets from mass and radius measurements?}
\newblock \aap \textbf{577}, A83 (2015).
\newblock \doi{10.1051/0004-6361/201424915}

\bibitem{dorn18}
{Dorn}, C., {Mosegaard}, K., {Grimm}, S.L., {Alibert}, Y.: {Interior
  Characterization in Multiplanetary Systems: TRAPPIST-1}.
\newblock \apj \textbf{865}(1), 20 (2018).
\newblock \doi{10.3847/1538-4357/aad95d}

\bibitem{drazkowska17}
{Dr{\c a}{\.z}kowska}, J., {Alibert}, Y.: {Planetesimal formation starts at the
  snow line}.
\newblock \aap \textbf{608}, A92 (2017).
\newblock \doi{10.1051/0004-6361/201731491}

\bibitem{drazkowska16}
{Dr{\c a}{\.z}kowska}, J., {Alibert}, Y., {Moore}, B.: {Close-in planetesimal
  formation by pile-up of drifting pebbles}.
\newblock \aap \textbf{594}, A105 (2016).
\newblock \doi{10.1051/0004-6361/201628983}

\bibitem{drazkowska18}
{Dr{\k{a}}{\.z}kowska}, J., {Dullemond}, C.P.: {Planetesimal formation during
  protoplanetary disk buildup}.
\newblock \aap \textbf{614}, A62 (2018).
\newblock \doi{10.1051/0004-6361/201732221}

\bibitem{duffell14}
{Duffell}, P.C., {Haiman}, Z., {MacFadyen}, A.I., {D'Orazio}, D.J., {Farris},
  B.D.: {The Migration of Gap-opening Planets is Not Locked to Viscous Disk
  Evolution}.
\newblock \apjl \textbf{792}(1), L10 (2014).
\newblock \doi{10.1088/2041-8205/792/1/L10}

\bibitem{dullemond18}
{Dullemond}, C.P., {Birnstiel}, T., {Huang}, J., {Kurtovic}, N.T., {Andrews},
  S.M., {Guzm{\'a}n}, V.V., {P{\'e}rez}, L.M., {Isella}, A., {Zhu}, Z.,
  {Benisty}, M., {Wilner}, D.J., {Bai}, X.N., {Carpenter}, J.M., {Zhang}, S.,
  {Ricci}, L.: {The Disk Substructures at High Angular Resolution Project
  (DSHARP). VI. Dust Trapping in Thin-ringed Protoplanetary Disks}.
\newblock \apjl \textbf{869}(2), L46 (2018).
\newblock \doi{10.3847/2041-8213/aaf742}

\bibitem{durmann15}
{D{\"u}rmann}, C., {Kley}, W.: {Migration of massive planets in accreting
  disks}.
\newblock \aap \textbf{574}, A52 (2015).
\newblock \doi{10.1051/0004-6361/201424837}

\bibitem{eisner03}
{Eisner}, J.A., {Carpenter}, J.M.: {Distribution of Circumstellar Disk Masses
  in the Young Cluster NGC 2024}.
\newblock \apj \textbf{598}, 1341--1349 (2003).
\newblock \doi{10.1086/379102}

\bibitem{eklund17}
{Eklund}, H., {Masset}, F.S.: {Evolution of eccentricity and inclination of hot
  protoplanets embedded in radiative discs}.
\newblock \mnras \textbf{469}(1), 206--217 (2017).
\newblock \doi{10.1093/mnras/stx856}

\bibitem{ertel14}
{Ertel}, S., {Absil}, O., {Defr{\`e}re}, D., {Le Bouquin}, J.B., {Augereau},
  J.C., {Marion}, L., {Blind}, N., {Bonsor}, A., {Bryden}, G., {Lebreton}, J.,
  {Milli}, J.: {A near-infrared interferometric survey of debris-disk stars.
  IV. An unbiased sample of 92 southern stars observed in H band with
  VLTI/PIONIER}.
\newblock \aap \textbf{570}, A128 (2014).
\newblock \doi{10.1051/0004-6361/201424438}

\bibitem{fabrycky14}
{Fabrycky}, D.C., {Lissauer}, J.J., {Ragozzine}, D., {Rowe}, J.F., {Steffen},
  J.H., {Agol}, E., {Barclay}, T., {Batalha}, N., {Borucki}, W., {Ciardi},
  D.R., {Ford}, E.B., {Gautier}, T.N., {Geary}, J.C., {Holman}, M.J.,
  {Jenkins}, J.M., {Li}, J., {Morehead}, R.C., {Morris}, R.L., {Shporer}, A.,
  {Smith}, J.C., {Still}, M., {Van Cleve}, J.: {Architecture of Kepler's
  Multi-transiting Systems. II. New Investigations with Twice as Many
  Candidates}.
\newblock \apj \textbf{790}, 146 (2014).
\newblock \doi{10.1088/0004-637X/790/2/146}

\bibitem{fang12}
{Fang}, J., {Margot}, J.L.: {Architecture of Planetary Systems Based on Kepler
  Data: Number of Planets and Coplanarity}.
\newblock \apj \textbf{761}, 92 (2012).
\newblock \doi{10.1088/0004-637X/761/2/92}

\bibitem{fernandes19}
{Fernandes}, R.B., {Mulders}, G.D., {Pascucci}, I., {Mordasini}, C.,
  {Emsenhuber}, A.: {Hints for a Turnover at the Snow Line in the Giant Planet
  Occurrence Rate}.
\newblock \apj \textbf{874}(1), 81 (2019).
\newblock \doi{10.3847/1538-4357/ab0300}

\bibitem{fischer08}
{Fischer}, D.A., {Marcy}, G.W., {Butler}, R.P., {Vogt}, S.S., {Laughlin}, G.,
  {Henry}, G.W., {Abouav}, D., {Peek}, K.M.G., {Wright}, J.T., {Johnson}, J.A.,
  {McCarthy}, C., {Isaacson}, H.: {Five Planets Orbiting 55 Cancri}.
\newblock \apj \textbf{675}, 790--801 (2008).
\newblock \doi{10.1086/525512}

\bibitem{fischer05}
{Fischer}, D.A., {Valenti}, J.: {The Planet-Metallicity Correlation}.
\newblock \apj \textbf{622}, 1102--1117 (2005).
\newblock \doi{10.1086/428383}

\bibitem{fischer14b}
{Fischer}, R.A., {Ciesla}, F.J.: {Dynamics of the terrestrial planets from a
  large number of N-body simulations}.
\newblock Earth and Planetary Science Letters \textbf{392}, 28--38 (2014).
\newblock \doi{10.1016/j.epsl.2014.02.011}

\bibitem{fischer18}
{Fischer}, R.A., {Nimmo}, F.: {Effects of core formation on the Hf-W isotopic
  composition of the Earth and dating of the Moon-forming impact}.
\newblock Earth and Planetary Science Letters \textbf{499}, 257--265 (2018).
\newblock \doi{10.1016/j.epsl.2018.07.030}

\bibitem{flock17}
{Flock}, M., {Fromang}, S., {Turner}, N.J., {Benisty}, M.: {3D Radiation
  Nonideal Magnetohydrodynamical Simulations of the Inner Rim in Protoplanetary
  Disks}.
\newblock \apj \textbf{835}, 230 (2017).
\newblock \doi{10.3847/1538-4357/835/2/230}

\bibitem{flock19}
{Flock}, M., {Turner}, N.J., {Mulders}, G.D., {Hasegawa}, Y., {Nelson}, R.P.,
  {Bitsch}, B.: {Planet formation and migration near the silicate sublimation
  front in protoplanetary disks}.
\newblock \aap \textbf{630}, A147 (2019).
\newblock \doi{10.1051/0004-6361/201935806}

\bibitem{fogg05}
{Fogg}, M.J., {Nelson}, R.P.: {Oligarchic and giant impact growth of
  terrestrial planets in the presence of gas giant planet migration}.
\newblock \aap \textbf{441}, 791--806 (2005).
\newblock \doi{10.1051/0004-6361:20053453}

\bibitem{fogg07}
{Fogg}, M.J., {Nelson}, R.P.: {On the formation of terrestrial planets in
  hot-Jupiter systems}.
\newblock \aap \textbf{461}, 1195--1208 (2007).
\newblock \doi{10.1051/0004-6361:20066171}

\bibitem{ford08}
{Ford}, E.B., {Rasio}, F.A.: {Origins of Eccentric Extrasolar Planets: Testing
  the Planet-Planet Scattering Model}.
\newblock \apj \textbf{686}, 621--636 (2008).
\newblock \doi{10.1086/590926}

\bibitem{foremanmackey16}
{Foreman-Mackey}, D., {Morton}, T.D., {Hogg}, D.W., {Agol}, E.,
  {Sch{\"o}lkopf}, B.: {The Population of Long-period Transiting Exoplanets}.
\newblock \aj \textbf{152}, 206 (2016).
\newblock \doi{10.3847/0004-6256/152/6/206}

\bibitem{fortney07}
{Fortney}, J.J., {Marley}, M.S., {Barnes}, J.W.: {Planetary Radii across Five
  Orders of Magnitude in Mass and Stellar Insolation: Application to Transits}.
\newblock \apj \textbf{659}, 1661--1672 (2007).
\newblock \doi{10.1086/512120}

\bibitem{fressin13}
{Fressin}, F., {Torres}, G., {Charbonneau}, D., {Bryson}, S.T., {Christiansen},
  J., {Dressing}, C.D., {Jenkins}, J.M., {Walkowicz}, L.M., {Batalha}, N.M.:
  {The False Positive Rate of Kepler and the Occurrence of Planets}.
\newblock \apj \textbf{766}, 81 (2013).
\newblock \doi{10.1088/0004-637X/766/2/81}

\bibitem{fulton18}
{Fulton}, B.J., {Petigura}, E.A.: {The California Kepler Survey VII. Precise
  Planet Radii Leveraging Gaia DR2 Reveal the Stellar Mass Dependence of the
  Planet Radius Gap}.
\newblock ArXiv e-prints  (2018)

\bibitem{fulton17}
{Fulton}, B.J., {Petigura}, E.A., {Howard}, A.W., {Isaacson}, H., {Marcy},
  G.W., {Cargile}, P.A., {Hebb}, L., {Weiss}, L.M., {Johnson}, J.A., {Morton},
  T.D., {Sinukoff}, E., {Crossfield}, I.J.M., {Hirsch}, L.A.: {The
  California-Kepler Survey. III. A Gap in the Radius Distribution of Small
  Planets}.
\newblock \aj \textbf{154}, 109 (2017).
\newblock \doi{10.3847/1538-3881/aa80eb}

\bibitem{genda05}
{Genda}, H., {Abe}, Y.: {Enhanced atmospheric loss on protoplanets at the giant
  impact phase in the presence of oceans}.
\newblock \nat \textbf{433}, 842--844 (2005).
\newblock \doi{10.1038/nature03360}

\bibitem{genda08}
{Genda}, H., {Ikoma}, M.: {Origin of the ocean on the Earth: Early evolution of
  water D/H in a hydrogen-rich atmosphere}.
\newblock \icarus \textbf{194}(1), 42--52 (2008).
\newblock \doi{10.1016/j.icarus.2007.09.007}

\bibitem{gillon17}
{Gillon}, M., {Triaud}, A.H.M.J., {Demory}, B.O., {Jehin}, E., {Agol}, E.,
  {Deck}, K.M., {Lederer}, S.M., {de Wit}, J., {Burdanov}, A., {Ingalls}, J.G.,
  {Bolmont}, E., {Leconte}, J., {Raymond}, S.N., {Selsis}, F., {Turbet}, M.,
  {Barkaoui}, K., {Burgasser}, A., {Burleigh}, M.R., {Carey}, S.J., {Chaushev},
  A., {Copperwheat}, C.M., {Delrez}, L., {Fernandes}, C.S., {Holdsworth}, D.L.,
  {Kotze}, E.J., {Van Grootel}, V., {Almleaky}, Y., {Benkhaldoun}, Z.,
  {Magain}, P., {Queloz}, D.: {Seven temperate terrestrial planets around the
  nearby ultracool dwarf star TRAPPIST-1}.
\newblock \nat \textbf{542}, 456--460 (2017).
\newblock \doi{10.1038/nature21360}

\bibitem{ginzburg16}
{Ginzburg}, S., {Schlichting}, H.E., {Sari}, R.: {Super-Earth Atmospheres:
  Self-consistent Gas Accretion and Retention}.
\newblock \apj \textbf{825}, 29 (2016).
\newblock \doi{10.3847/0004-637X/825/1/29}

\bibitem{goldreich14}
{Goldreich}, P., {Schlichting}, H.E.: {Overstable Librations can Account for
  the Paucity of Mean Motion Resonances among Exoplanet Pairs}.
\newblock \aj \textbf{147}(2), 32 (2014).
\newblock \doi{10.1088/0004-6256/147/2/32}

\bibitem{goldreich79}
{Goldreich}, P., {Tremaine}, S.: {The excitation of density waves at the
  Lindblad and corotation resonances by an external potential.}
\newblock \apj \textbf{233}, 857--871 (1979).
\newblock \doi{10.1086/157448}

\bibitem{goldreich80}
{Goldreich}, P., {Tremaine}, S.: {Disk-satellite interactions}.
\newblock \apj \textbf{241}, 425--441 (1980).
\newblock \doi{10.1086/158356}

\bibitem{gomes05}
{Gomes}, R., {Levison}, H.F., {Tsiganis}, K., {Morbidelli}, A.: {Origin of the
  cataclysmic Late Heavy Bombardment period of the terrestrial planets}.
\newblock \nat \textbf{435}, 466--469 (2005).
\newblock \doi{10.1038/nature03676}

\bibitem{gomes04}
{Gomes}, R.S., {Morbidelli}, A., {Levison}, H.F.: {Planetary migration in a
  planetesimal disk: why did Neptune stop at 30 AU?}
\newblock Icarus \textbf{170}, 492--507 (2004).
\newblock \doi{10.1016/j.icarus.2004.03.011}

\bibitem{gonzalez97}
{Gonzalez}, G.: {The stellar metallicity-giant planet connection}.
\newblock \mnras \textbf{285}, 403--412 (1997)

\bibitem{gould10}
{Gould}, A., {Dong}, S., {Gaudi}, B.S., {Udalski}, A., {Bond}, I.A.,
  {Greenhill}, J., {Street}, R.A., {Dominik}, M., {Sumi}, T., {Szyma{\'n}ski},
  M.K., {Han}, C., {Allen}, W., {Bolt}, G., {Bos}, M., {Christie}, G.W.,
  {DePoy}, D.L., {Drummond}, J., {Eastman}, J.D., {Gal-Yam}, A., {Higgins}, D.,
  {Janczak}, J., {Kaspi}, S., {Koz{\l}owski}, S., {Lee}, C., {Mallia}, F.,
  {Maury}, A., {Maoz}, D., {McCormick}, J., {Monard}, L.A.G., {Moorhouse}, D.,
  {Morgan}, N., {Natusch}, T., {Ofek}, E.O., {Park}, B., {Pogge}, R.W.,
  {Polishook}, D., {Santallo}, R., {Shporer}, A., {Spector}, O., {Thornley},
  G., {Yee}, J.C., {{$\mu$}FUN Collaboration}, {Kubiak}, M., {Pietrzy{\'n}ski},
  G., {Soszy{\'n}ski}, I., {Szewczyk}, O., {Wyrzykowski}, {\L}., {Ulaczyk}, K.,
  {Poleski}, R., {OGLE Collaboration}, {Abe}, F., {Bennett}, D.P., {Botzler},
  C.S., {Douchin}, D., {Freeman}, M., {Fukui}, A., {Furusawa}, K., {Hearnshaw},
  J.B., {Hosaka}, S., {Itow}, Y., {Kamiya}, K., {Kilmartin}, P.M., {Korpela},
  A., {Lin}, W., {Ling}, C.H., {Makita}, S., {Masuda}, K., {Matsubara}, Y.,
  {Miyake}, N., {Muraki}, Y., {Nagaya}, M., {Nishimoto}, K., {Ohnishi}, K.,
  {Okumura}, T., {Perrott}, Y.C., {Philpott}, L., {Rattenbury}, N., {Saito},
  T., {Sako}, T., {Sullivan}, D.J., {Sweatman}, W.L., {Tristram}, P.J., {von
  Seggern}, E., {Yock}, P.C.M., {MOA Collaboration}, {Albrow}, M., {Batista},
  V., {Beaulieu}, J.P., {Brillant}, S., {Caldwell}, J., {Calitz}, J.J.,
  {Cassan}, A., {Cole}, A., {Cook}, K., {Coutures}, C., {Dieters}, S., {Dominis
  Prester}, D., {Donatowicz}, J., {Fouqu{\'e}}, P., {Hill}, K., {Hoffman}, M.,
  {Jablonski}, F., {Kane}, S.R., {Kains}, N., {Kubas}, D., {Marquette}, J.,
  {Martin}, R., {Martioli}, E., {Meintjes}, P., {Menzies}, J., {Pedretti}, E.,
  {Pollard}, K., {Sahu}, K.C., {Vinter}, C., {Wambsganss}, J., {Watson}, R.,
  {Williams}, A., {Zub}, M., {PLANET Collaboration}, {Allan}, A., {Bode}, M.F.,
  {Bramich}, D.M., {Burgdorf}, M.J., {Clay}, N., {Fraser}, S., {Hawkins}, E.,
  {Horne}, K., {Kerins}, E., {Lister}, T.A., {Mottram}, C., {Saunders}, E.S.,
  {Snodgrass}, C., {Steele}, I.A., {Tsapras}, Y., {RoboNet Collaboration},
  {J{\o}rgensen}, U.G., {Anguita}, T., {Bozza}, V., {Calchi Novati}, S.,
  {Harps{\o}e}, K., {Hinse}, T.C., {Hundertmark}, M., {Kj{\ae}rgaard}, P.,
  {Liebig}, C., {Mancini}, L., {Masi}, G., {Mathiasen}, M., {Rahvar}, S.,
  {Ricci}, D., {Scarpetta}, G., {Southworth}, J., {Surdej}, J., {Th{\"o}ne},
  C.C., {MiNDSTEp Consortium}: {Frequency of Solar-like Systems and of Ice and
  Gas Giants Beyond the Snow Line from High-magnification Microlensing Events
  in 2005-2008}.
\newblock \apj \textbf{720}, 1073--1089 (2010).
\newblock \doi{10.1088/0004-637X/720/2/1073}

\bibitem{gradie82}
{Gradie}, J., {Tedesco}, E.: {Compositional structure of the asteroid belt}.
\newblock Science \textbf{216}, 1405--1407 (1982).
\newblock \doi{10.1126/science.216.4553.1405}

\bibitem{greaves11}
{Greaves}, J.S., {Rice}, W.K.M.: {Do all Sun-like stars have planets?
  Inferences from the disc mass reservoirs of Class 0 protostars}.
\newblock \mnras \textbf{412}(1), L88--L92 (2011).
\newblock \doi{10.1111/j.1745-3933.2011.01011.x}

\bibitem{greenberg78}
{Greenberg}, R., {Hartmann}, W.K., {Chapman}, C.R., {Wacker}, J.F.:
  {Planetesimals to planets - Numerical simulation of collisional evolution}.
\newblock Icarus \textbf{35}, 1--26 (1978).
\newblock \doi{10.1016/0019-1035(78)90057-X}

\bibitem{gressel15}
{Gressel}, O., {Turner}, N.J., {Nelson}, R.P., {McNally}, C.P.: {Global
  Simulations of Protoplanetary Disks With Ohmic Resistivity and Ambipolar
  Diffusion}.
\newblock \apj \textbf{801}(2), 84 (2015).
\newblock \doi{10.1088/0004-637X/801/2/84}

\bibitem{grimm93}
{Grimm}, R.E., {McSween}, H.Y.: {Heliocentric zoning of the asteroid belt by
  aluminum-26 heating}.
\newblock Science \textbf{259}, 653--655 (1993)

\bibitem{grishin15}
{Grishin}, E., {Perets}, H.B.: {Application of Gas Dynamical Friction for
  Planetesimals. I. Evolution of Single Planetesimals}.
\newblock \apj \textbf{811}(1), 54 (2015).
\newblock \doi{10.1088/0004-637X/811/1/54}

\bibitem{gupta19}
{Gupta}, A., {Schlichting}, H.E.: {Sculpting the valley in the radius
  distribution of small exoplanets as a by-product of planet formation: the
  core-powered mass-loss mechanism}.
\newblock \mnras \textbf{487}(1), 24--33 (2019).
\newblock \doi{10.1093/mnras/stz1230}

\bibitem{hahn09}
Hahn, J.M.: The Dynamics of Planetary Systems and Astrophysical Disks.
\newblock WILEY-VCH (2009)

\bibitem{haisch01}
{Haisch} Jr., K.E., {Lada}, E.A., {Lada}, C.J.: {Disk Frequencies and Lifetimes
  in Young Clusters}.
\newblock \apjl \textbf{553}, L153--L156 (2001).
\newblock \doi{10.1086/320685}

\bibitem{halliday13}
{Halliday}, A.N.: {The origins of volatiles in the terrestrial planets}.
\newblock \gca \textbf{105}, 146--171 (2013).
\newblock \doi{10.1016/j.gca.2012.11.015}

\bibitem{halliday06}
{Halliday}, A.N., {Kleine}, T.: {Meteorites and the Timing, Mechanisms, and
  Conditions of Terrestrial Planet Accretion and Early Differentiation}, pp.
  775--801 (2006)

\bibitem{hansen09}
{Hansen}, B.M.S.: {Formation of the Terrestrial Planets from a Narrow Annulus}.
\newblock \apj \textbf{703}, 1131--1140 (2009).
\newblock \doi{10.1088/0004-637X/703/1/1131}

\bibitem{hansen12}
{Hansen}, B.M.S., {Murray}, N.: {Migration Then Assembly: Formation of
  Neptune-mass Planets inside 1 AU}.
\newblock \apj \textbf{751}, 158 (2012).
\newblock \doi{10.1088/0004-637X/751/2/158}

\bibitem{hansen13}
{Hansen}, B.M.S., {Murray}, N.: {Testing in Situ Assembly with the Kepler
  Planet Candidate Sample}.
\newblock \apj \textbf{775}, 53 (2013).
\newblock \doi{10.1088/0004-637X/775/1/53}

\bibitem{hartmann98}
{Hartmann}, L., {Calvet}, N., {Gullbring}, E., {D'Alessio}, P.: {Accretion and
  the Evolution of T Tauri Disks}.
\newblock \apj \textbf{495}, 385--400 (1998).
\newblock \doi{10.1086/305277}

\bibitem{hartmann19}
{Hartmann}, W.K.: {The Collapse of the Terminal Cataclysm Paradigm...and Where
  We Go from Here}.
\newblock In: Lunar and Planetary Science Conference, Lunar and Planetary
  Science Conference, p. 1064 (2019)

\bibitem{hartogh11}
{Hartogh}, P., {Lis}, D.C., {Bockel{\'e}e-Morvan}, D., {de Val-Borro}, M.,
  {Biver}, N., {K{\"u}ppers}, M., {Emprechtinger}, M., {Bergin}, E.A.,
  {Crovisier}, J., {Rengel}, M., {Moreno}, R., {Szutowicz}, S., {Blake}, G.A.:
  {Ocean-like water in the Jupiter-family comet 103P/Hartley 2}.
\newblock \nat \textbf{478}, 218--220 (2011).
\newblock \doi{10.1038/nature10519}

\bibitem{hasegawa12}
{Hasegawa}, Y., {Pudritz}, R.E.: {Evolutionary Tracks of Trapped, Accreting
  Protoplanets: The Origin of the Observed Mass-Period Relation}.
\newblock \apj \textbf{760}, 117 (2012).
\newblock \doi{10.1088/0004-637X/760/2/117}

\bibitem{hayashi81}
{Hayashi}, C.: {Structure of the Solar Nebula, Growth and Decay of Magnetic
  Fields and Effects of Magnetic and Turbulent Viscosities on the Nebula}.
\newblock Progress of Theoretical Physics Supplement \textbf{70}, 35--53
  (1981).
\newblock \doi{10.1143/PTPS.70.35}

\bibitem{helled14}
{Helled}, R., {Bodenheimer}, P., {Podolak}, M., {Boley}, A., {Meru}, F.,
  {Nayakshin}, S., {Fortney}, J.J., {Mayer}, L., {Alibert}, Y., {Boss}, A.P.:
  {Giant Planet Formation, Evolution, and Internal Structure}.
\newblock Protostars and Planets VI pp. 643--665 (2014)

\bibitem{henon64}
{Henon}, M., {Heiles}, C.: {The applicability of the third integral of motion:
  Some numerical experiments}.
\newblock \aj \textbf{69}, 73 (1964).
\newblock \doi{10.1086/109234}

\bibitem{hillenbrand08}
{Hillenbrand}, L.A., {Carpenter}, J.M., {Kim}, J.S., {Meyer}, M.R., {Backman},
  D.E., {Moro-Mart{\'{\i}}n}, A., {Hollenbach}, D.J., {Hines}, D.C.,
  {Pascucci}, I., {Bouwman}, J.: {The Complete Census of 70 {$\mu$}m-bright
  Debris Disks within ``the Formation and Evolution of Planetary Systems''
  Spitzer Legacy Survey of Sun-like Stars}.
\newblock \apj \textbf{677}, 630--656 (2008).
\newblock \doi{10.1086/529027}

\bibitem{hirschmann06}
{Hirschmann}, M.M.: {Water, Melting, and the Deep Earth H2O Cycle}.
\newblock Annual Review of Earth and Planetary Sciences \textbf{34}, 629--653
  (2006).
\newblock \doi{10.1146/annurev.earth.34.031405.125211}

\bibitem{hollenbach94}
{Hollenbach}, D., {Johnstone}, D., {Lizano}, S., {Shu}, F.: {Photoevaporation
  of disks around massive stars and application to ultracompact H II regions}.
\newblock \apj \textbf{428}, 654--669 (1994).
\newblock \doi{10.1086/174276}

\bibitem{howard12}
{Howard}, A.W., {Marcy}, G.W., {Bryson}, S.T., {Jenkins}, J.M., {Rowe}, J.F.,
  {Batalha}, N.M., {Borucki}, W.J., {Koch}, D.G., {Dunham}, E.W., {Gautier}
  III, T.N., {Van Cleve}, J., {Cochran}, W.D., {Latham}, D.W., {Lissauer},
  J.J., {Torres}, G., {Brown}, T.M., {Gilliland}, R.L., {Buchhave}, L.A.,
  {Caldwell}, D.A., {Christensen-Dalsgaard}, J., {Ciardi}, D., {Fressin}, F.,
  {Haas}, M.R., {Howell}, S.B., {Kjeldsen}, H., {Seager}, S., {Rogers}, L.,
  {Sasselov}, D.D., {Steffen}, J.H., {Basri}, G.S., {Charbonneau}, D.,
  {Christiansen}, J., {Clarke}, B., {Dupree}, A., {Fabrycky}, D.C., {Fischer},
  D.A., {Ford}, E.B., {Fortney}, J.J., {Tarter}, J., {Girouard}, F.R.,
  {Holman}, M.J., {Johnson}, J.A., {Klaus}, T.C., {Machalek}, P., {Moorhead},
  A.V., {Morehead}, R.C., {Ragozzine}, D., {Tenenbaum}, P., {Twicken}, J.D.,
  {Quinn}, S.N., {Isaacson}, H., {Shporer}, A., {Lucas}, P.W., {Walkowicz},
  L.M., {Welsh}, W.F., {Boss}, A., {Devore}, E., {Gould}, A., {Smith}, J.C.,
  {Morris}, R.L., {Prsa}, A., {Morton}, T.D., {Still}, M., {Thompson}, S.E.,
  {Mullally}, F., {Endl}, M., {MacQueen}, P.J.: {Planet Occurrence within 0.25
  AU of Solar-type Stars from Kepler}.
\newblock \apjs \textbf{201}, 15 (2012).
\newblock \doi{10.1088/0067-0049/201/2/15}

\bibitem{howard10}
{Howard}, A.W., {Marcy}, G.W., {Johnson}, J.A., {Fischer}, D.A., {Wright},
  J.T., {Isaacson}, H., {Valenti}, J.A., {Anderson}, J., {Lin}, D.N.C., {Ida},
  S.: {The Occurrence and Mass Distribution of Close-in Super-Earths, Neptunes,
  and Jupiters}.
\newblock Science \textbf{330}, 653-- (2010).
\newblock \doi{10.1126/science.1194854}

\bibitem{hu17}
{Hu}, X., {Tan}, J.C., {Zhu}, Z., {Chatterjee}, S., {Birnstiel}, T., {Youdin},
  A.N., {Mohanty}, S.: {Inside-Out Planet Formation. IV. Pebble Evolution and
  Planet Formation Timescales}.
\newblock ArXiv e-prints  (2017)

\bibitem{hu16}
{Hu}, X., {Zhu}, Z., {Tan}, J.C., {Chatterjee}, S.: {Inside-out Planet
  Formation. III. Planet-Disk Interaction at the Dead Zone Inner Boundary}.
\newblock \apj \textbf{816}, 19 (2016).
\newblock \doi{10.3847/0004-637X/816/1/19}

\bibitem{hubickyj05}
{Hubickyj}, O., {Bodenheimer}, P., {Lissauer}, J.J.: {Accretion of the gaseous
  envelope of Jupiter around a 5 10 Earth-mass core}.
\newblock Icarus \textbf{179}, 415--431 (2005).
\newblock \doi{10.1016/j.icarus.2005.06.021}

\bibitem{hughes18}
{Hughes}, A.M., {Duch{\^e}ne}, G., {Matthews}, B.C.: {Debris Disks: Structure,
  Composition, and Variability}.
\newblock \araa \textbf{56}, 541--591 (2018).
\newblock \doi{10.1146/annurev-astro-081817-052035}

\bibitem{ida16b}
{Ida}, S., {Guillot}, T.: {Formation of dust-rich planetesimals from sublimated
  pebbles inside of the snow line}.
\newblock \aap \textbf{596}, L3 (2016).
\newblock \doi{10.1051/0004-6361/201629680}

\bibitem{ida16}
{Ida}, S., {Guillot}, T., {Morbidelli}, A.: {The radial dependence of pebble
  accretion rates: A source of diversity in planetary systems. I. Analytical
  formulation}.
\newblock \aap \textbf{591}, A72 (2016).
\newblock \doi{10.1051/0004-6361/201628099}

\bibitem{ida04}
{Ida}, S., {Lin}, D.N.C.: {Toward a Deterministic Model of Planetary Formation.
  I. A Desert in the Mass and Semimajor Axis Distributions of Extrasolar
  Planets}.
\newblock \apj \textbf{604}, 388--413 (2004).
\newblock \doi{10.1086/381724}

\bibitem{ida08}
{Ida}, S., {Lin}, D.N.C.: {Toward a Deterministic Model of Planetary Formation.
  IV. Effects of Type I Migration}.
\newblock \apj \textbf{673}, 487--501 (2008).
\newblock \doi{10.1086/523754}

\bibitem{ida10}
{Ida}, S., {Lin}, D.N.C.: {Toward a Deterministic Model of Planetary Formation.
  VI. Dynamical Interaction and Coagulation of Multiple Rocky Embryos and
  Super-Earth Systems around Solar-type Stars}.
\newblock \apj \textbf{719}, 810--830 (2010).
\newblock \doi{10.1088/0004-637X/719/1/810}

\bibitem{ida13}
{Ida}, S., {Lin}, D.N.C., {Nagasawa}, M.: {Toward a Deterministic Model of
  Planetary Formation. VII. Eccentricity Distribution of Gas Giants}.
\newblock \apj \textbf{775}, 42 (2013).
\newblock \doi{10.1088/0004-637X/775/1/42}

\bibitem{ida19}
{Ida}, S., {Yamamura}, T., {Okuzumi}, S.: {Water delivery by pebble accretion
  to rocky planets in habitable zones in evolving disks}.
\newblock \aap \textbf{624}, A28 (2019).
\newblock \doi{10.1051/0004-6361/201834556}

\bibitem{ikoma06}
{Ikoma}, M., {Genda}, H.: {Constraints on the Mass of a Habitable Planet with
  Water of Nebular Origin}.
\newblock \apj \textbf{648}(1), 696--706 (2006).
\newblock \doi{10.1086/505780}

\bibitem{ikoma00}
{Ikoma}, M., {Nakazawa}, K., {Emori}, H.: {Formation of Giant Planets:
  Dependences on Core Accretion Rate and Grain Opacity}.
\newblock \apj \textbf{537}, 1013--1025 (2000).
\newblock \doi{10.1086/309050}

\bibitem{inamdar15}
{Inamdar}, N.K., {Schlichting}, H.E.: {The formation of super-Earths and
  mini-Neptunes with giant impacts}.
\newblock \mnras \textbf{448}, 1751--1760 (2015).
\newblock \doi{10.1093/mnras/stv030}

\bibitem{izidoro19}
{Izidoro}, A., {Bitsch}, B., {Raymond}, S.N., {Johansen}, A., {Morbidelli}, A.,
  {Lambrechts}, M., {Jacobson}, S.A.: {Formation of planetary systems by pebble
  accretion and migration: Hot super-Earth systems from breaking compact
  resonant chains}.
\newblock arXiv e-prints arXiv:1902.08772 (2019)

\bibitem{izidoro14}
{Izidoro}, A., {Morbidelli}, A., {Raymond}, S.N.: {Terrestrial Planet Formation
  in the Presence of Migrating Super-Earths}.
\newblock \apj \textbf{794}, 11 (2014).
\newblock \doi{10.1088/0004-637X/794/1/11}

\bibitem{izidoro15b}
{Izidoro}, A., {Morbidelli}, A., {Raymond}, S.N., {Hersant}, F., {Pierens}, A.:
  {Accretion of Uranus and Neptune from inward-migrating planetary embryos
  blocked by Jupiter and Saturn}.
\newblock \aap \textbf{582}, A99 (2015).
\newblock \doi{10.1051/0004-6361/201425525}

\bibitem{izidoro17}
{Izidoro}, A., {Ogihara}, M., {Raymond}, S.N., {Morbidelli}, A., {Pierens}, A.,
  {Bitsch}, B., {Cossou}, C., {Hersant}, F.: {Breaking the chains: hot
  super-Earth systems from migration and disruption of compact resonant
  chains}.
\newblock \mnras \textbf{470}, 1750--1770 (2017).
\newblock \doi{10.1093/mnras/stx1232}

\bibitem{izidoro15a}
{Izidoro}, A., {Raymond}, S.N., {Morbidelli}, A., {Hersant}, F., {Pierens}, A.:
  {Gas Giant Planets as Dynamical Barriers to Inward-Migrating Super-Earths}.
\newblock \apjl \textbf{800}, L22 (2015).
\newblock \doi{10.1088/2041-8205/800/2/L22}

\bibitem{izidoro15c}
{Izidoro}, A., {Raymond}, S.N., {Morbidelli}, A., {Winter}, O.C.: {Terrestrial
  planet formation constrained by Mars and the structure of the asteroid belt}.
\newblock \mnras \textbf{453}, 3619--3634 (2015).
\newblock \doi{10.1093/mnras/stv1835}

\bibitem{jacobson14b}
{Jacobson}, S.A., {Morbidelli}, A.: {Lunar and terrestrial planet formation in
  the Grand Tack scenario}.
\newblock Philosophical Transactions of the Royal Society of London Series A
  \textbf{372}, 0174 (2014).
\newblock \doi{10.1098/rsta.2013.0174}

\bibitem{jacobson14}
{Jacobson}, S.A., {Morbidelli}, A., {Raymond}, S.N., {O'Brien}, D.P., {Walsh},
  K.J., {Rubie}, D.C.: {Highly siderophile elements in Earth's mantle as a
  clock for the Moon-forming impact}.
\newblock \nat \textbf{508}, 84--87 (2014).
\newblock \doi{10.1038/nature13172}

\bibitem{jacquet11}
{Jacquet}, E., {Balbus}, S., {Latter}, H.: {On linear dust-gas streaming
  instabilities in protoplanetary discs}.
\newblock \mnras \textbf{415}(4), 3591--3598 (2011).
\newblock \doi{10.1111/j.1365-2966.2011.18971.x}

\bibitem{jin18}
{Jin}, S., {Mordasini}, C.: {Compositional Imprints in Density--Distance--Time:
  A Rocky Composition for Close-in Low-mass Exoplanets from the Location of the
  Valley of Evaporation}.
\newblock \apj \textbf{853}, 163 (2018).
\newblock \doi{10.3847/1538-4357/aa9f1e}

\bibitem{johansen14}
{Johansen}, A., {Blum}, J., {Tanaka}, H., {Ormel}, C., {Bizzarro}, M.,
  {Rickman}, H.: {The Multifaceted Planetesimal Formation Process}.
\newblock Protostars and Planets VI pp. 547--570 (2014)

\bibitem{johansen12}
{Johansen}, A., {Davies}, M.B., {Church}, R.P., {Holmelin}, V.: {Can Planetary
  Instability Explain the Kepler Dichotomy?}
\newblock \apj \textbf{758}, 39 (2012).
\newblock \doi{10.1088/0004-637X/758/1/39}

\bibitem{johansen17}
{Johansen}, A., {Lambrechts}, M.: {Forming Planets via Pebble Accretion}.
\newblock Annual Review of Earth and Planetary Sciences \textbf{45}, 359--387
  (2017).
\newblock \doi{10.1146/annurev-earth-063016-020226}

\bibitem{johansen15}
{Johansen}, A., {Mac Low}, M.M., {Lacerda}, P., {Bizzarro}, M.: {Growth of
  asteroids, planetary embryos, and Kuiper belt objects by chondrule
  accretion}.
\newblock Science Advances \textbf{1}, 1500109 (2015).
\newblock \doi{10.1126/sciadv.1500109}

\bibitem{johansen09}
{Johansen}, A., {Youdin}, A., {Klahr}, H.: {Zonal Flows and Long-lived
  Axisymmetric Pressure Bumps in Magnetorotational Turbulence}.
\newblock \apj \textbf{697}, 1269--1289 (2009).
\newblock \doi{10.1088/0004-637X/697/2/1269}

\bibitem{johnson16}
Johnson, B.C., Walsh, K.J., Minton, D.A., Krot, A.N., Levison, H.F.: Timing of
  the formation and migration of giant planets as constrained by cb chondrites.
\newblock Science Advances \textbf{2}(12) (2016).
\newblock \doi{10.1126/sciadv.1601658}.
\newblock \urlprefix\url{http://advances.sciencemag.org/content/2/12/e1601658}

\bibitem{johnson07}
{Johnson}, J.A., {Butler}, R.P., {Marcy}, G.W., {Fischer}, D.A., {Vogt}, S.S.,
  {Wright}, J.T., {Peek}, K.M.G.: {A New Planet around an M Dwarf: Revealing a
  Correlation between Exoplanets and Stellar Mass}.
\newblock \apj \textbf{670}, 833--840 (2007).
\newblock \doi{10.1086/521720}

\bibitem{jones06}
{Jones}, H.R.A., {Butler}, R.P., {Tinney}, C.G., {Marcy}, G.W., {Carter}, B.D.,
  {Penny}, A.J., {McCarthy}, C., {Bailey}, J.: {High-eccentricity planets from
  the Anglo-Australian Planet Search}.
\newblock \mnras \textbf{369}, 249--256 (2006).
\newblock \doi{10.1111/j.1365-2966.2006.10298.x}

\bibitem{juric08}
{Juri{\'c}}, M., {Tremaine}, S.: {Dynamical Origin of Extrasolar Planet
  Eccentricity Distribution}.
\newblock \apj \textbf{686}, 603--620 (2008).
\newblock \doi{10.1086/590047}

\bibitem{kaib16}
{Kaib}, N.A., {Chambers}, J.E.: {The fragility of the terrestrial planets
  during a giant-planet instability}.
\newblock \mnras \textbf{455}, 3561--3569 (2016).
\newblock \doi{10.1093/mnras/stv2554}

\bibitem{kaib15}
{Kaib}, N.A., {Cowan}, N.B.: {The feeding zones of terrestrial planets and
  insights into Moon formation}.
\newblock \icarus \textbf{252}, 161--174 (2015).
\newblock \doi{10.1016/j.icarus.2015.01.013}

\bibitem{kaib13}
{Kaib}, N.A., {Raymond}, S.N., {Duncan}, M.: {Planetary system disruption by
  Galactic perturbations to wide binary stars}.
\newblock \nat \textbf{493}, 381--384 (2013).
\newblock \doi{10.1038/nature11780}

\bibitem{kanagawa15}
{Kanagawa}, K.D., {Tanaka}, H., {Muto}, T., {Tanigawa}, T., {Takeuchi}, T.:
  {Formation of a disc gap induced by a planet: effect of the deviation from
  Keplerian disc rotation}.
\newblock \mnras \textbf{448}(1), 994--1006 (2015).
\newblock \doi{10.1093/mnras/stv025}

\bibitem{kanagawa18}
{Kanagawa}, K.D., {Tanaka}, H., {Szuszkiewicz}, E.: {Radial Migration of
  Gap-opening Planets in Protoplanetary Disks. I. The Case of a Single Planet}.
\newblock \apj \textbf{861}(2), 140 (2018).
\newblock \doi{10.3847/1538-4357/aac8d9}

\bibitem{kataoka13}
{Kataoka}, A., {Tanaka}, H., {Okuzumi}, S., {Wada}, K.: {Fluffy dust forms icy
  planetesimals by static compression}.
\newblock \aap \textbf{557}, L4 (2013).
\newblock \doi{10.1051/0004-6361/201322151}

\bibitem{kerridge85}
{Kerridge}, J.F.: {Carbon, hydrogen and nitrogen in carbonaceous chondrites
  Abundances and isotopic compositions in bulk samples}.
\newblock \gca \textbf{49}, 1707--1714 (1985).
\newblock \doi{10.1016/0016-7037(85)90141-3}

\bibitem{kimura74}
{Kimura}, K., {Lewis}, R.S., {Anders}, E.: {Distribution of gold and rhenium
  between nickel-iron and silicate melts: implications for the abundance of
  siderophile elements on the Earth and Moon}.
\newblock \gca \textbf{38}, 683--701 (1974).
\newblock \doi{10.1016/0016-7037(74)90144-6}

\bibitem{kimura12}
{Kimura}, S.S., {Tsuribe}, T.: {Conditions of Gravitational Instability in
  Protoplanetary Disks}.
\newblock \pasj \textbf{64}, 116 (2012).
\newblock \doi{10.1093/pasj/64.5.116}

\bibitem{kita05}
{Kita}, N.T., {Huss}, G.R., {Tachibana}, S., {Amelin}, Y., {Nyquist}, L.E.,
  {Hutcheon}, I.D.: {Constraints on the Origin of Chondrules and CAIs from
  Short-lived and Long-Lived Radionuclides}.
\newblock In: A.N. {Krot}, E.R.D. {Scott}, B.~{Reipurth} (eds.) Chondrites and
  the Protoplanetary Disk, \emph{Astronomical Society of the Pacific Conference
  Series}, vol. 341, p. 558 (2005)

\bibitem{kleine09}
{Kleine}, T., {Touboul}, M., {Bourdon}, B., {Nimmo}, F., {Mezger}, K., {Palme},
  H., {Jacobsen}, S.B., {Yin}, Q.Z., {Halliday}, A.N.: {Hf-W chronology of the
  accretion and early evolution of asteroids and terrestrial planets}.
\newblock \gca \textbf{73}, 5150--5188 (2009)

\bibitem{kokubo96}
{Kokubo}, E., {Ida}, S.: {On Runaway Growth of Planetesimals}.
\newblock Icarus \textbf{123}, 180--191 (1996).
\newblock \doi{10.1006/icar.1996.0148}

\bibitem{kokubo98}
{Kokubo}, E., {Ida}, S.: {Oligarchic Growth of Protoplanets}.
\newblock Icarus \textbf{131}, 171--178 (1998).
\newblock \doi{10.1006/icar.1997.5840}

\bibitem{kokubo00}
{Kokubo}, E., {Ida}, S.: {Formation of Protoplanets from Planetesimals in the
  Solar Nebula}.
\newblock Icarus \textbf{143}, 15--27 (2000).
\newblock \doi{10.1006/icar.1999.6237}

\bibitem{kokubo07}
{Kokubo}, E., {Ida}, S.: {Formation of Terrestrial Planets from Protoplanets.
  II. Statistics of Planetary Spin}.
\newblock \apj \textbf{671}, 2082--2090 (2007).
\newblock \doi{10.1086/522364}

\bibitem{kral17}
{Kral}, Q., {Krivov}, A.V., {Defr{\`e}re}, D., {van Lieshout}, R., {Bonsor},
  A., {Augereau}, J.C., {Th{\'e}bault}, P., {Ertel}, S., {Lebreton}, J.,
  {Absil}, O.: {Exozodiacal clouds: hot and warm dust around main sequence
  stars}.
\newblock The Astronomical Review \textbf{13}(2), 69--111 (2017).
\newblock \doi{10.1080/21672857.2017.1353202}

\bibitem{krasinsky02}
{Krasinsky}, G.A., {Pitjeva}, E.V., {Vasilyev}, M.V., {Yagudina}, E.I.: {Hidden
  Mass in the Asteroid Belt}.
\newblock \icarus \textbf{158}, 98--105 (2002).
\newblock \doi{10.1006/icar.2002.6837}

\bibitem{kretke12}
{Kretke}, K.A., {Lin}, D.N.C.: {The Importance of Disk Structure in Stalling
  Type I Migration}.
\newblock \apj \textbf{755}, 74 (2012).
\newblock \doi{10.1088/0004-637X/755/1/74}

\bibitem{krivov10}
{Krivov}, A.V.: {Debris disks: seeing dust, thinking of planetesimals and
  planets}.
\newblock Research in Astronomy and Astrophysics \textbf{10}, 383--414 (2010).
\newblock \doi{10.1088/1674-4527/10/5/001}

\bibitem{krot05}
{Krot}, A.N., {Amelin}, Y., {Cassen}, P., {Meibom}, A.: {Young chondrules in CB
  chondrites from a giant impact in the early Solar System}.
\newblock \nat \textbf{436}, 989--992 (2005).
\newblock \doi{10.1038/nature03830}

\bibitem{kruijer17}
{Kruijer}, T.S., {Burkhardt}, C., {Budde}, C., {Kleine}, T.: {Age of Jupiter
  inferred from the distinct genetics and formation times of meteorites}.
\newblock PNAS  (2017)

\bibitem{kruijer14}
{Kruijer}, T.S., {Touboul}, M., {Fischer-G{\"o}dde}, M., {Bermingham}, K.R.,
  {Walker}, R.J., {Kleine}, T.: {Protracted core formation and rapid accretion
  of protoplanets}.
\newblock Science \textbf{344}, 1150--1154 (2014).
\newblock \doi{10.1126/science.1251766}

\bibitem{kuchynka13}
{Kuchynka}, P., {Folkner}, W.M.: {A new approach to determining asteroid masses
  from planetary range measurements}.
\newblock \icarus \textbf{222}, 243--253 (2013).
\newblock \doi{10.1016/j.icarus.2012.11.003}

\bibitem{lambrechts12}
{Lambrechts}, M., {Johansen}, A.: {Rapid growth of gas-giant cores by pebble
  accretion}.
\newblock \aap \textbf{544}, A32 (2012).
\newblock \doi{10.1051/0004-6361/201219127}

\bibitem{lambrechts14}
{Lambrechts}, M., {Johansen}, A.: {Forming the cores of giant planets from the
  radial pebble flux in protoplanetary discs}.
\newblock \aap \textbf{572}, A107 (2014).
\newblock \doi{10.1051/0004-6361/201424343}

\bibitem{lambrechts14b}
{Lambrechts}, M., {Johansen}, A., {Morbidelli}, A.: {Separating gas-giant and
  ice-giant planets by halting pebble accretion}.
\newblock \aap \textbf{572}, A35 (2014).
\newblock \doi{10.1051/0004-6361/201423814}

\bibitem{lambrechts17}
{Lambrechts}, M., {Lega}, E.: {Reduced gas accretion on super-Earths and ice
  giants}.
\newblock \aap \textbf{606}, A146 (2017).
\newblock \doi{10.1051/0004-6361/201731014}

\bibitem{lambrechts19b}
{Lambrechts}, M., {Lega}, E., {Nelson}, R.P., {Crida}, A., {Morbidelli}, A.:
  {Quasi-static contraction during runaway gas accretion onto giant planets}.
\newblock \aap \textbf{630}, A82 (2019).
\newblock \doi{10.1051/0004-6361/201834413}

\bibitem{lambrechts19}
{Lambrechts}, M., {Morbidelli}, A., {Jacobson}, S.A., {Johansen}, A., {Bitsch},
  B., {Izidoro}, A., {Raymond}, S.N.: {Formation of planetary systems by pebble
  accretion and migration. How the radial pebble flux determines a
  terrestrial-planet or super-Earth growth mode}.
\newblock \aap \textbf{627}, A83 (2019).
\newblock \doi{10.1051/0004-6361/201834229}

\bibitem{Laplace}
{Laplace}, P.S., {Bowditch}, N., {Bowditch}, N.I.: {M{\'e}canique c{\'e}leste}
  (1829)

\bibitem{laskar97}
{Laskar}, J.: {Large scale chaos and the spacing of the inner planets.}
\newblock \aap \textbf{317}, L75--L78 (1997)

\bibitem{laws03}
{Laws}, C., {Gonzalez}, G., {Walker}, K.M., {Tyagi}, S., {Dodsworth}, J.,
  {Snider}, K., {Suntzeff}, N.B.: {Parent Stars of Extrasolar Planets. VII. New
  Abundance Analyses of 30 Systems}.
\newblock \aj \textbf{125}, 2664--2677 (2003).
\newblock \doi{10.1086/374626}

\bibitem{lecar06}
{Lecar}, M., {Podolak}, M., {Sasselov}, D., {Chiang}, E.: {On the Location of
  the Snow Line in a Protoplanetary Disk}.
\newblock \apj \textbf{640}, 1115--1118 (2006).
\newblock \doi{10.1086/500287}

\bibitem{lee16}
{Lee}, E.J., {Chiang}, E.: {Breeding Super-Earths and Birthing Super-puffs in
  Transitional Disks}.
\newblock \apj \textbf{817}, 90 (2016).
\newblock \doi{10.3847/0004-637X/817/2/90}

\bibitem{lee17}
{Lee}, E.J., {Chiang}, E.: {Magnetospheric Truncation, Tidal Inspiral, and the
  Creation of Short-period and Ultra-short-period Planets}.
\newblock \apj \textbf{842}, 40 (2017).
\newblock \doi{10.3847/1538-4357/aa6fb3}

\bibitem{lee14}
{Lee}, E.J., {Chiang}, E., {Ormel}, C.W.: {Make Super-Earths, Not Jupiters:
  Accreting Nebular Gas onto Solid Cores at 0.1 AU and Beyond}.
\newblock \apj \textbf{797}, 95 (2014).
\newblock \doi{10.1088/0004-637X/797/2/95}

\bibitem{lee02}
{Lee}, M.H., {Peale}, S.J.: {Dynamics and Origin of the 2:1 Orbital Resonances
  of the GJ 876 Planets}.
\newblock \apj \textbf{567}, 596--609 (2002).
\newblock \doi{10.1086/338504}

\bibitem{lega14}
{Lega}, E., {Crida}, A., {Bitsch}, B., {Morbidelli}, A.: {Migration of
  Earth-sized planets in 3D radiative discs}.
\newblock \mnras \textbf{440}, 683--695 (2014).
\newblock \doi{10.1093/mnras/stu304}

\bibitem{lestrade12}
{Lestrade}, J.F., {Matthews}, B.C., {Sibthorpe}, B., {Kennedy}, G.M., {Wyatt},
  M.C., {Bryden}, G., {Greaves}, J.S., {Thilliez}, E., {Moro-Mart{\'\i}n}, A.,
  {Booth}, M., {Dent}, W.R.F., {Duch{\^e}ne}, G., {Harvey}, P.M., {Horner}, J.,
  {Kalas}, P., {Kavelaars}, J.J., {Phillips}, N.M., {Rodriguez}, D.R., {Su},
  K.Y.L., {Wilner}, D.J.: {A DEBRIS disk around the planet hosting M-star GJ
  581 spatially resolved with Herschel}.
\newblock \aap \textbf{548}, A86 (2012).
\newblock \doi{10.1051/0004-6361/201220325}

\bibitem{levison15b}
{Levison}, H.F., {Kretke}, K.A., {Walsh}, K.J., {Bottke}, W.F.: {Growing the
  terrestrial planets from the gradual accumulation of sub-meter sized
  objects}.
\newblock Proceedings of the National Academy of Science \textbf{112},
  14180--14185 (2015).
\newblock \doi{10.1073/pnas.1513364112}

\bibitem{levison11}
{Levison}, H.F., {Morbidelli}, A., {Tsiganis}, K., {Nesvorn{\'y}}, D., {Gomes},
  R.: {Late Orbital Instabilities in the Outer Planets Induced by Interaction
  with a Self-gravitating Planetesimal Disk}.
\newblock \aj \textbf{142}, 152 (2011).
\newblock \doi{10.1088/0004-6256/142/5/152}

\bibitem{levison08}
{Levison}, H.F., {Morbidelli}, A., {Vanlaerhoven}, C., {Gomes}, R., {Tsiganis},
  K.: {Origin of the structure of the Kuiper belt during a dynamical
  instability in the orbits of Uranus and Neptune}.
\newblock Icarus \textbf{196}, 258--273 (2008).
\newblock \doi{10.1016/j.icarus.2007.11.035}

\bibitem{levison01}
{Levison}, H.F., {Stewart}, G.R.: {Remarks on Modeling the Formation of Uranus
  and Neptune}.
\newblock Icarus \textbf{153}, 224--228 (2001).
\newblock \doi{10.1006/icar.2001.6672}

\bibitem{levison10}
{Levison}, H.F., {Thommes}, E., {Duncan}, M.J.: {Modeling the Formation of
  Giant Planet Cores. I. Evaluating Key Processes}.
\newblock \aj \textbf{139}, 1297--1314 (2010).
\newblock \doi{10.1088/0004-6256/139/4/1297}

\bibitem{lichtenberg19}
{Lichtenberg}, T., {Golabek}, G.J., {Burn}, R., {Meyer}, M.R., {Alibert}, Y.,
  {Gerya}, T.V., {Mordasini}, C.: {A water budget dichotomy of rocky
  protoplanets from $^{26}$Al-heating}.
\newblock Nature Astronomy \textbf{3}, 307--313 (2019).
\newblock \doi{10.1038/s41550-018-0688-5}

\bibitem{lin96}
{Lin}, D.N.C., {Bodenheimer}, P., {Richardson}, D.C.: {Orbital migration of the
  planetary companion of 51 Pegasi to its present location}.
\newblock \nat \textbf{380}, 606--607 (1996).
\newblock \doi{10.1038/380606a0}

\bibitem{lin97}
{Lin}, D.N.C., {Ida}, S.: {On the Origin of Massive Eccentric Planets}.
\newblock \apj \textbf{477}, 781--+ (1997).
\newblock \doi{10.1086/303738}

\bibitem{lin79}
{Lin}, D.N.C., {Papaloizou}, J.: {Tidal torques on accretion discs in binary
  systems with extreme mass ratios.}
\newblock \mnras \textbf{186}, 799--812 (1979).
\newblock \doi{10.1093/mnras/186.4.799}

\bibitem{lin86}
{Lin}, D.N.C., {Papaloizou}, J.: {On the tidal interaction between protoplanets
  and the protoplanetary disk. III - Orbital migration of protoplanets}.
\newblock \apj \textbf{309}, 846--857 (1986).
\newblock \doi{10.1086/164653}

\bibitem{lis13}
{Lis}, D.C., {Biver}, N., {Bockel{\'e}e-Morvan}, D., {Hartogh}, P., {Bergin},
  E.A., {Blake}, G.A., {Crovisier}, J., {de Val-Borro}, M., {Jehin}, E.,
  {K{\"u}ppers}, M., {Manfroid}, J., {Moreno}, R., {Rengel}, M., {Szutowicz},
  S.: {A Herschel Study of D/H in Water in the Jupiter-family Comet
  45P/Honda-Mrkos-Pajdu{\v s}{\'a}kov{\'a} and Prospects for D/H Measurements
  with CCAT}.
\newblock \apjl \textbf{774}, L3 (2013).
\newblock \doi{10.1088/2041-8205/774/1/L3}

\bibitem{lis19}
{Lis}, D.C., {Bockel{\'e}e-Morvan}, D., {G{\"u}sten}, R., {Biver}, N.,
  {Stutzki}, J., {Delorme}, Y., {Dur{\'a}n}, C., {Wiesemeyer}, H., {Okada}, Y.:
  {Terrestrial deuterium-to-hydrogen ratio in water in hyperactive comets}.
\newblock \aap \textbf{625}, L5 (2019).
\newblock \doi{10.1051/0004-6361/201935554}

\bibitem{lissauer87}
{Lissauer}, J.J.: {Timescales for planetary accretion and the structure of the
  protoplanetary disk}.
\newblock Icarus \textbf{69}, 249--265 (1987).
\newblock \doi{10.1016/0019-1035(87)90104-7}

\bibitem{lissauer93}
{Lissauer}, J.J.: {Planet formation}.
\newblock \araa \textbf{31}, 129--174 (1993).
\newblock \doi{10.1146/annurev.aa.31.090193.001021}

\bibitem{lissauer09}
{Lissauer}, J.J., {Hubickyj}, O., {D'Angelo}, G., {Bodenheimer}, P.: {Models of
  Jupiter's growth incorporating thermal and hydrodynamic constraints}.
\newblock Icarus \textbf{199}, 338--350 (2009).
\newblock \doi{10.1016/j.icarus.2008.10.004}

\bibitem{lissauer11b}
{Lissauer}, J.J., {Ragozzine}, D., {Fabrycky}, D.C., {Steffen}, J.H., {Ford},
  E.B., {Jenkins}, J.M., {Shporer}, A., {Holman}, M.J., {Rowe}, J.F.,
  {Quintana}, E.V., {Batalha}, N.M., {Borucki}, W.J., {Bryson}, S.T.,
  {Caldwell}, D.A., {Carter}, J.A., {Ciardi}, D., {Dunham}, E.W., {Fortney},
  J.J., {Gautier} III, T.N., {Howell}, S.B., {Koch}, D.G., {Latham}, D.W.,
  {Marcy}, G.W., {Morehead}, R.C., {Sasselov}, D.: {Architecture and Dynamics
  of Kepler's Candidate Multiple Transiting Planet Systems}.
\newblock \apjs \textbf{197}, 8 (2011).
\newblock \doi{10.1088/0067-0049/197/1/8}

\bibitem{lissauer07}
{Lissauer}, J.J., {Stevenson}, D.J.: {Formation of Giant Planets}.
\newblock Protostars and Planets V pp. 591--606 (2007)

\bibitem{lodders03}
{Lodders}, K.: {Solar System Abundances and Condensation Temperatures of the
  Elements}.
\newblock \apj \textbf{591}, 1220--1247 (2003).
\newblock \doi{10.1086/375492}

\bibitem{looney03}
{Looney}, L.W., {Mundy}, L.G., {Welch}, W.J.: {Envelope Emission in Young
  Stellar Systems: A Subarcsecond Survey of Circumstellar Structure}.
\newblock \apj \textbf{592}, 255--265 (2003).
\newblock \doi{10.1086/375582}

\bibitem{lopez17}
{Lopez}, E.D.: {Born dry in the photoevaporation desert: Kepler's
  ultra-short-period planets formed water-poor}.
\newblock \mnras \textbf{472}, 245--253 (2017).
\newblock \doi{10.1093/mnras/stx1558}

\bibitem{lovis07}
{Lovis}, C., {Mayor}, M.: {Planets around evolved intermediate-mass stars. I.
  Two substellar companions in the open clusters NGC 2423 and NGC 4349}.
\newblock \aap \textbf{472}, 657--664 (2007).
\newblock \doi{10.1051/0004-6361:20077375}

\bibitem{luger17}
{Luger}, R., {Sestovic}, M., {Kruse}, E., {Grimm}, S.L., {Demory}, B.O.,
  {Agol}, E., {Bolmont}, E., {Fabrycky}, D., {Fernandes}, C.S., {Van Grootel},
  V., {Burgasser}, A., {Gillon}, M., {Ingalls}, J.G., {Jehin}, E., {Raymond},
  S.N., {Selsis}, F., {Triaud}, A.H.M.J., {Barclay}, T., {Barentsen}, G.,
  {Howell}, S.B., {Delrez}, L., {de Wit}, J., {Foreman-Mackey}, D.,
  {Holdsworth}, D.L., {Leconte}, J., {Lederer}, S., {Turbet}, M., {Almleaky},
  Y., {Benkhaldoun}, Z., {Magain}, P., {Morris}, B.M., {Heng}, K., {Queloz},
  D.: {A seven-planet resonant chain in TRAPPIST-1}.
\newblock Nature Astronomy \textbf{1}, 0129 (2017).
\newblock \doi{10.1038/s41550-017-0129}

\bibitem{lykawka19}
{Lykawka}, P.S., {Ito}, T.: {Constraining the Formation of the Four Terrestrial
  Planets in the Solar System}.
\newblock arXiv e-prints arXiv:1908.04934 (2019)

\bibitem{lyndenbell74}
{Lynden-Bell}, D., {Pringle}, J.E.: {The evolution of viscous discs and the
  origin of the nebular variables.}
\newblock \mnras \textbf{168}, 603--637 (1974)

\bibitem{lyra10}
{Lyra}, W., {Paardekooper}, S.J., {Mac Low}, M.M.: {Orbital Migration of
  Low-mass Planets in Evolutionary Radiative Models: Avoiding Catastrophic
  Infall}.
\newblock \apjl \textbf{715}, L68--L73 (2010).
\newblock \doi{10.1088/2041-8205/715/2/L68}

\bibitem{mamajek09}
{Mamajek}, E.E.: {Initial Conditions of Planet Formation: Lifetimes of
  Primordial Disks}.
\newblock In: T.~{Usuda}, M.~{Tamura}, M.~{Ishii} (eds.) American Institute of
  Physics Conference Series, \emph{American Institute of Physics Conference
  Series}, vol. 1158, pp. 3--10 (2009).
\newblock \doi{10.1063/1.3215910}

\bibitem{manara18}
{Manara}, C.F., {Morbidelli}, A., {Guillot}, T.: {Why do protoplanetary disks
  appear not massive enough to form the known exoplanet population?}
\newblock \aap \textbf{618}, L3 (2018).
\newblock \doi{10.1051/0004-6361/201834076}

\bibitem{mandell07}
{Mandell}, A.M., {Raymond}, S.N., {Sigurdsson}, S.: {Formation of Earth-like
  Planets During and After Giant Planet Migration}.
\newblock \apj \textbf{660}, 823--844 (2007).
\newblock \doi{10.1086/512759}

\bibitem{marcy14}
{Marcy}, G.W., {Isaacson}, H., {Howard}, A.W., {Rowe}, J.F., {Jenkins}, J.M.,
  {Bryson}, S.T., {Latham}, D.W., {Howell}, S.B., {Gautier} III, T.N.,
  {Batalha}, N.M., {Rogers}, L., {Ciardi}, D., {Fischer}, D.A., {Gilliland},
  R.L., {Kjeldsen}, H., {Christensen-Dalsgaard}, J., {Huber}, D., {Chaplin},
  W.J., {Basu}, S., {Buchhave}, L.A., {Quinn}, S.N., {Borucki}, W.J., {Koch},
  D.G., {Hunter}, R., {Caldwell}, D.A., {Van Cleve}, J., {Kolbl}, R., {Weiss},
  L.M., {Petigura}, E., {Seager}, S., {Morton}, T., {Johnson}, J.A., {Ballard},
  S., {Burke}, C., {Cochran}, W.D., {Endl}, M., {MacQueen}, P., {Everett},
  M.E., {Lissauer}, J.J., {Ford}, E.B., {Torres}, G., {Fressin}, F., {Brown},
  T.M., {Steffen}, J.H., {Charbonneau}, D., {Basri}, G.S., {Sasselov}, D.D.,
  {Winn}, J., {Sanchis-Ojeda}, R., {Christiansen}, J., {Adams}, E., {Henze},
  C., {Dupree}, A., {Fabrycky}, D.C., {Fortney}, J.J., {Tarter}, J., {Holman},
  M.J., {Tenenbaum}, P., {Shporer}, A., {Lucas}, P.W., {Welsh}, W.F., {Orosz},
  J.A., {Bedding}, T.R., {Campante}, T.L., {Davies}, G.R., {Elsworth}, Y.,
  {Handberg}, R., {Hekker}, S., {Karoff}, C., {Kawaler}, S.D., {Lund}, M.N.,
  {Lundkvist}, M., {Metcalfe}, T.S., {Miglio}, A., {Silva Aguirre}, V.,
  {Stello}, D., {White}, T.R., {Boss}, A., {Devore}, E., {Gould}, A., {Prsa},
  A., {Agol}, E., {Barclay}, T., {Coughlin}, J., {Brugamyer}, E., {Mullally},
  F., {Quintana}, E.V., {Still}, M., {Thompson}, S.E., {Morrison}, D.,
  {Twicken}, J.D., {D{\'e}sert}, J.M., {Carter}, J., {Crepp}, J.R.,
  {H{\'e}brard}, G., {Santerne}, A., {Moutou}, C., {Sobeck}, C., {Hudgins}, D.,
  {Haas}, M.R., {Robertson}, P., {Lillo-Box}, J., {Barrado}, D.: {Masses,
  Radii, and Orbits of Small Kepler Planets: The Transition from Gaseous to
  Rocky Planets}.
\newblock \apjs \textbf{210}, 20 (2014).
\newblock \doi{10.1088/0067-0049/210/2/20}

\bibitem{marois08}
{Marois}, C., {Macintosh}, B., {Barman}, T., {Zuckerman}, B., {Song}, I.,
  {Patience}, J., {Lafreni{\`e}re}, D., {Doyon}, R.: {Direct Imaging of
  Multiple Planets Orbiting the Star HR 8799}.
\newblock Science \textbf{322}, 1348-- (2008).
\newblock \doi{10.1126/science.1166585}

\bibitem{marois10}
{Marois}, C., {Zuckerman}, B., {Konopacky}, Q.M., {Macintosh}, B., {Barman},
  T.: {Images of a fourth planet orbiting HR 8799}.
\newblock \nat \textbf{468}, 1080--1083 (2010).
\newblock \doi{10.1038/nature09684}

\bibitem{marshall14}
{Marshall}, J.P., {Moro-Mart{\'{\i}}n}, A., {Eiroa}, C., {Kennedy}, G., {Mora},
  A., {Sibthorpe}, B., {Lestrade}, J.F., {Maldonado}, J., {Sanz-Forcada}, J.,
  {Wyatt}, M.C., {Matthews}, B., {Horner}, J., {Montesinos}, B., {Bryden}, G.,
  {del Burgo}, C., {Greaves}, J.S., {Ivison}, R.J., {Meeus}, G., {Olofsson},
  G., {Pilbratt}, G.L., {White}, G.J.: {Correlations between the stellar,
  planetary, and debris components of exoplanet systems observed by Herschel}.
\newblock \aap \textbf{565}, A15 (2014).
\newblock \doi{10.1051/0004-6361/201323058}

\bibitem{martin12}
{Martin}, R.G., {Livio}, M.: {On the evolution of the snow line in
  protoplanetary discs}.
\newblock \mnras \textbf{425}, L6--L9 (2012).
\newblock \doi{10.1111/j.1745-3933.2012.01290.x}

\bibitem{martin15}
{Martin}, R.G., {Livio}, M.: {The Solar System as an Exoplanetary System}.
\newblock \apj \textbf{810}, 105 (2015).
\newblock \doi{10.1088/0004-637X/810/2/105}

\bibitem{marty12}
{Marty}, B.: {The origins and concentrations of water, carbon, nitrogen and
  noble gases on Earth}.
\newblock Earth and Planetary Science Letters \textbf{313}, 56--66 (2012).
\newblock \doi{10.1016/j.epsl.2011.10.040}

\bibitem{marty17}
{Marty}, B., {Altwegg}, K., {Balsiger}, H., {Bar-Nun}, A., {Bekaert}, D.V.,
  {Berthelier}, J.J., {Bieler}, A., {Briois}, C., {Calmonte}, U., {Combi}, M.,
  {De Keyser}, J., {Fiethe}, B., {Fuselier}, S.A., {Gasc}, S., {Gombosi}, T.I.,
  {Hansen}, K.C., {H{\"a}ssig}, M., {J{\"a}ckel}, A., {Kopp}, E., {Korth}, A.,
  {Le Roy}, L., {Mall}, U., {Mousis}, O., {Owen}, T., {R{\`e}me}, H., {Rubin},
  M., {S{\'e}mon}, T., {Tzou}, C.Y., {Waite}, J.H., {Wurz}, P.: {Xenon isotopes
  in 67P/Churyumov-Gerasimenko show that comets contributed to Earth's
  atmosphere}.
\newblock Science \textbf{356}, 1069--1072 (2017).
\newblock \doi{10.1126/science.aal3496}

\bibitem{marty16}
{Marty}, B., {Avice}, G., {Sano}, Y., {Altwegg}, K., {Balsiger}, H.,
  {H{\"a}ssig}, M., {Morbidelli}, A., {Mousis}, O., {Rubin}, M.: {Origins of
  volatile elements (H, C, N, noble gases) on Earth and Mars in light of recent
  results from the ROSETTA cometary mission}.
\newblock Earth and Planetary Science Letters \textbf{441}, 91--102 (2016).
\newblock \doi{10.1016/j.epsl.2016.02.031}

\bibitem{marty06}
{Marty}, B., {Yokochi}, R.: {Water in the Early Earth}.
\newblock Rev. Mineral Geophys. \textbf{62}, 421--450 (2006)

\bibitem{marzari14}
{Marzari}, F.: {Impact of planet-planet scattering on the formation and
  survival of debris discs}.
\newblock \mnras \textbf{444}, 1419--1424 (2014).
\newblock \doi{10.1093/mnras/stu1544}

\bibitem{marzari02}
{Marzari}, F., {Weidenschilling}, S.J.: {Eccentric Extrasolar Planets: The
  Jumping Jupiter Model}.
\newblock Icarus \textbf{156}, 570--579 (2002).
\newblock \doi{10.1006/icar.2001.6786}

\bibitem{masset01}
{Masset}, F., {Snellgrove}, M.: {Reversing type II migration: resonance
  trapping of a lighter giant protoplanet}.
\newblock \mnras \textbf{320}, L55--L59 (2001).
\newblock \doi{10.1046/j.1365-8711.2001.04159.x}

\bibitem{masset17}
{Masset}, F.S.: {Coorbital thermal torques on low-mass protoplanets}.
\newblock \mnras \textbf{472}(4), 4204--4219 (2017).
\newblock \doi{10.1093/mnras/stx2271}

\bibitem{masset06}
{Masset}, F.S., {Morbidelli}, A., {Crida}, A., {Ferreira}, J.: {Disk Surface
  Density Transitions as Protoplanet Traps}.
\newblock \apj \textbf{642}, 478--487 (2006).
\newblock \doi{10.1086/500967}

\bibitem{matsumoto12}
{Matsumoto}, Y., {Nagasawa}, M., {Ida}, S.: {The orbital stability of planets
  trapped in the first-order mean-motion resonances}.
\newblock \icarus \textbf{221}(2), 624--631 (2012).
\newblock \doi{10.1016/j.icarus.2012.08.032}

\bibitem{matsumura13}
{Matsumura}, S., {Ida}, S., {Nagasawa}, M.: {Effects of Dynamical Evolution of
  Giant Planets on Survival of Terrestrial Planets}.
\newblock \apj \textbf{767}, 129 (2013).
\newblock \doi{10.1088/0004-637X/767/2/129}

\bibitem{matsumura10}
{Matsumura}, S., {Thommes}, E.W., {Chatterjee}, S., {Rasio}, F.A.: {Unstable
  Planetary Systems Emerging Out of Gas Disks}.
\newblock \apj \textbf{714}, 194--206 (2010).
\newblock \doi{10.1088/0004-637X/714/1/194}

\bibitem{matthews14}
{Matthews}, B.C., {Krivov}, A.V., {Wyatt}, M.C., {Bryden}, G., {Eiroa}, C.:
  {Observations, Modeling, and Theory of Debris Disks}.
\newblock In: H.~{Beuther}, R.S. {Klessen}, C.P. {Dullemond}, T.~{Henning}
  (eds.) Protostars and Planets VI, p. 521 (2014)

\bibitem{mayer07}
{Mayer}, L., {Lufkin}, G., {Quinn}, T., {Wadsley}, J.: {Fragmentation of
  Gravitationally Unstable Gaseous Protoplanetary Disks with Radiative
  Transfer}.
\newblock \apjl \textbf{661}(1), L77--L80 (2007).
\newblock \doi{10.1086/518433}

\bibitem{mayer02}
{Mayer}, L., {Quinn}, T., {Wadsley}, J., {Stadel}, J.: {Formation of Giant
  Planets by Fragmentation of Protoplanetary Disks}.
\newblock Science \textbf{298}(5599), 1756--1759 (2002).
\newblock \doi{10.1126/science.1077635}

\bibitem{mayor11}
{Mayor}, M., {Marmier}, M., {Lovis}, C., {Udry}, S., {S{\'e}gransan}, D.,
  {Pepe}, F., {Benz}, W., {Bertaux}, J.., {Bouchy}, F., {Dumusque}, X., {Lo
  Curto}, G., {Mordasini}, C., {Queloz}, D., {Santos}, N.C.: {The HARPS search
  for southern extra-solar planets XXXIV. Occurrence, mass distribution and
  orbital properties of super-Earths and Neptune-mass planets}.
\newblock arXiv:1109.2497  (2011)

\bibitem{mccubbin19}
{McCubbin}, F.M., {Barnes}, J.J.: {Origin and abundances of H$_{2}$O in the
  terrestrial planets, Moon, and asteroids}.
\newblock Earth and Planetary Science Letters \textbf{526}, 115771 (2019).
\newblock \doi{10.1016/j.epsl.2019.115771}

\bibitem{mcnally19}
{McNally}, C.P., {Nelson}, R.P., {Paardekooper}, S.J., {Ben{\'\i}tez-Llambay},
  P.: {Migrating super-Earths in low-viscosity discs: unveiling the roles of
  feedback, vortices, and laminar accretion flows}.
\newblock \mnras \textbf{484}(1), 728--748 (2019).
\newblock \doi{10.1093/mnras/stz023}

\bibitem{mcneil10}
{McNeil}, D.S., {Nelson}, R.P.: {On the formation of hot Neptunes and
  super-Earths}.
\newblock \mnras \textbf{401}, 1691--1708 (2010).
\newblock \doi{10.1111/j.1365-2966.2009.15805.x}

\bibitem{meech19}
{Meech}, K., {Raymond}, S.N.: {Origin of Earth's water: sources and
  constraints}.
\newblock arXiv e-prints arXiv:1912.04361 (2019)

\bibitem{meru11}
{Meru}, F., {Bate}, M.R.: {On the fragmentation criteria of self-gravitating
  protoplanetary discs}.
\newblock \mnras \textbf{410}(1), 559--572 (2011).
\newblock \doi{10.1111/j.1365-2966.2010.17465.x}

\bibitem{meyer08}
{Meyer}, M.R., {Carpenter}, J.M., {Mamajek}, E.E., {Hillenbrand}, L.A.,
  {Hollenbach}, D., {Moro-Martin}, A., {Kim}, J.S., {Silverstone}, M.D.,
  {Najita}, J., {Hines}, D.C., {Pascucci}, I., {Stauffer}, J.R., {Bouwman}, J.,
  {Backman}, D.E.: {Evolution of Mid-Infrared Excess around Sun-like Stars:
  Constraints on Models of Terrestrial Planet Formation}.
\newblock \apjl \textbf{673}, L181--L184 (2008).
\newblock \doi{10.1086/527470}

\bibitem{millholland17}
{Millholland}, S., {Wang}, S., {Laughlin}, G.: {Kepler Multi-planet Systems
  Exhibit Unexpected Intra-system Uniformity in Mass and Radius}.
\newblock \apjl \textbf{849}(2), L33 (2017).
\newblock \doi{10.3847/2041-8213/aa9714}

\bibitem{mills16}
{Mills}, S.M., {Fabrycky}, D.C., {Migaszewski}, C., {Ford}, E.B., {Petigura},
  E., {Isaacson}, H.: {A resonant chain of four transiting, sub-Neptune
  planets}.
\newblock \nat \textbf{533}, 509--512 (2016).
\newblock \doi{10.1038/nature17445}

\bibitem{moeckel09}
{Moeckel}, N., {Throop}, H.B.: {Bondi-Hoyle-Lyttleton Accretion Onto a
  Protoplanetary Disk}.
\newblock \apj \textbf{707}(1), 268--277 (2009).
\newblock \doi{10.1088/0004-637X/707/1/268}

\bibitem{mojzsis19}
{Mojzsis}, S.J., {Brasser}, R., {Kelly}, N.M., {Abramov}, O., {Werner}, S.C.:
  {Onset of Giant Planet Migration before 4480 Million Years Ago}.
\newblock \apj \textbf{881}(1), 44 (2019).
\newblock \doi{10.3847/1538-4357/ab2c03}

\bibitem{montesinos16}
{Montesinos}, B., {Eiroa}, C., {Krivov}, A.V., {Marshall}, J.P., {Pilbratt},
  G.L., {Liseau}, R., {Mora}, A., {Maldonado}, J., {Wolf}, S., {Ertel}, S.,
  {Bayo}, A., {Augereau}, J.C., {Heras}, A.M., {Fridlund}, M., {Danchi}, W.C.,
  {Solano}, E., {Kirchschlager}, F., {del Burgo}, C., {Montes}, D.: {Incidence
  of debris discs around FGK stars in the solar neighbourhood}.
\newblock \aap \textbf{593}, A51 (2016).
\newblock \doi{10.1051/0004-6361/201628329}

\bibitem{monteux18}
{Monteux}, J., {Golabek}, G.J., {Rubie}, D.C., {Tobie}, G., {Young}, E.D.:
  {Water and the Interior Structure of Terrestrial Planets and Icy Bodies}.
\newblock \ssr \textbf{214}(1), 39 (2018).
\newblock \doi{10.1007/s11214-018-0473-x}

\bibitem{moorhead05}
{Moorhead}, A.V., {Adams}, F.C.: {Giant planet migration through the action of
  disk torques and planet planet scattering}.
\newblock Icarus \textbf{178}, 517--539 (2005).
\newblock \doi{10.1016/j.icarus.2005.05.005}

\bibitem{morales19}
{Morales}, J.C., {Mustill}, A.J., {Ribas}, I., {Davies}, M.B., {Reiners}, A.,
  {Bauer}, F.F., {Kossakowski}, D., {Herrero}, E., {Rodr{\'\i}guez}, E.,
  {L{\'o}pez-Gonz{\'a}lez}, M.J., {Rodr{\'\i}guez-L{\'o}pez}, C., {B{\'e}jar},
  V.J.S., {Gonz{\'a}lez-Cuesta}, L., {Luque}, R., {Pall{\'e}}, E., {Perger},
  M., {Baroch}, D., {Johansen}, A., {Klahr}, H., {Mordasini}, C.,
  {Anglada-Escud{\'e}}, G., {Caballero}, J.A., {Cort{\'e}s-Contreras}, M.,
  {Dreizler}, S., {Lafarga}, M., {Nagel}, E., {Passegger}, V.M., {Reffert}, S.,
  {Rosich}, A., {Schweitzer}, A., {Tal-Or}, L., {Trifonov}, T., {Zechmeister},
  M., {Quirrenbach}, A., {Amado}, P.J., {Guenther}, E.W., {Hagen}, H.J.,
  {Henning}, T., {Jeffers}, S.V., {Kaminski}, A., {K{\"u}rster}, M., {Montes},
  D., {Seifert}, W., {Abell{\'a}n}, F.J., {Abril}, M., {Aceituno}, J.,
  {Aceituno}, F.J., {Alonso-Floriano}, F.J., {Ammler-von Eiff}, M., {Antona},
  R., {Arroyo-Torres}, B., {Azzaro}, M., {Barrado}, D., {Becerril-Jarque}, S.,
  {Ben{\'\i}tez}, D., {Berdi{\~n}as}, Z.M., {Bergond}, G., {Brinkm{\"o}ller},
  M., {del Burgo}, C., {Burn}, R., {Calvo-Ortega}, R., {Cano}, J.,
  {C{\'a}rdenas}, M.C., {Guill{\'e}n}, C.C., {Carro}, J., {Casal}, E.,
  {Casanova}, V., {Casasayas-Barris}, N., {Chaturvedi}, P., {Cifuentes}, C.,
  {Claret}, A., {Colom{\'e}}, J., {Czesla}, S., {D{\'\i}ez-Alonso}, E.,
  {Dorda}, R., {Emsenhuber}, A., {Fern{\'a}ndez}, M.,
  {Fern{\'a}ndez-Mart{\'\i}n}, A., {Ferro}, I.M., {Fuhrmeister}, B.,
  {Galad{\'\i}-Enr{\'\i}quez}, D., {Cava}, I.G., {Vargas}, M.L.G.,
  {Garcia-Piquer}, A., {Gesa}, L., {Gonz{\'a}lez-{\'A}lvarez}, E.,
  {Hern{\'a}ndez}, J.I.G., {Gonz{\'a}lez-Peinado}, R., {Gu{\`a}rdia}, J.,
  {Guijarro}, A., {de Guindos}, E., {Hatzes}, A.P., {Hauschildt}, P.H.,
  {Hedrosa}, R.P., {Hermelo}, I., {Arabi}, R.H., {Otero}, F.H., {Hintz}, D.,
  {Holgado}, G., {Huber}, A., {Huke}, P., {Johnson}, E.N., {de Juan}, E.,
  {Kehr}, M., {Kemmer}, J., {Kim}, M., {Kl{\"u}ter}, J., {Klutsch}, A.,
  {Labarga}, F., {Labiche}, N., {Lalitha}, S., {Lamp{\'o}n}, M., {Lara}, L.M.,
  {Launhardt}, R., {L{\'a}zaro}, F.J., {Lizon}, J.L., {Llamas}, M., {Lodieu},
  N., {L{\'o}pez del Fresno}, M., {Salas}, J.F.L., {L{\'o}pez-Santiago}, J.,
  {Madinabeitia}, H.M., {Mall}, U., {Mancini}, L., {Mand el}, H., {Marfil}, E.,
  {Molina}, J.A.M., {Mart{\'\i}n}, E.L., {Mart{\'\i}n-Fern{\'a}ndez}, P.,
  {Mart{\'\i}n-Ruiz}, S., {Mart{\'\i}nez-Rodr{\'\i}guez}, H., {Marvin}, C.J.,
  {Mirabet}, E., {Moya}, A., {Naranjo}, V., {Nelson}, R.P., {Nortmann}, L.,
  {Nowak}, G., {Ofir}, A., {Pascual}, J., {Pavlov}, A., {Pedraz}, S.,
  {Medialdea}, D.P., {P{\'e}rez-Calpena}, A., {Perryman}, M.A.C., {Rabaza}, O.,
  {Ballesta}, A.R., {Rebolo}, R., {Redondo}, P., {Rix}, H.W., {Rodler}, F.,
  {Trinidad}, A.R., {Sabotta}, S., {Sadegi}, S., {Salz}, M.,
  {S{\'a}nchez-Blanco}, E., {Carrasco}, M.A.S., {S{\'a}nchez-L{\'o}pez}, A.,
  {Sanz-Forcada}, J., {Sarkis}, P., {Sarmiento}, L.F., {Sch{\"a}fer}, S.,
  {Schlecker}, M., {Schmitt}, J.H.M.M., {Sch{\"o}fer}, P., {Solano}, E.,
  {Sota}, A., {Stahl}, O., {Stock}, S., {Stuber}, T., {St{\"u}rmer}, J.,
  {Su{\'a}rez}, J.C., {Tabernero}, H.M., {Tulloch}, S.M., {Veredas}, G.,
  {Vico-Linares}, J.I., {Vilardell}, F., {Wagner}, K., {Winkler}, J.,
  {Wolthoff}, V., {Yan}, F., {Osorio}, M.R.Z.: {A giant exoplanet orbiting a
  very-low-mass star challenges planet formation models}.
\newblock Science \textbf{365}(6460), 1441--1445 (2019).
\newblock \doi{10.1126/science.aax3198}

\bibitem{morby16}
{Morbidelli}, A., {Bitsch}, B., {Crida}, A., {Gounelle}, M., {Guillot}, T.,
  {Jacobson}, S., {Johansen}, A., {Lambrechts}, M., {Lega}, E.: {Fossilized
  condensation lines in the Solar System protoplanetary disk}.
\newblock Icarus \textbf{267}, 368--376 (2016).
\newblock \doi{10.1016/j.icarus.2015.11.027}

\bibitem{morby00}
{Morbidelli}, A., {Chambers}, J., {Lunine}, J.I., {Petit}, J.M., {Robert}, F.,
  {Valsecchi}, G.B., {Cyr}, K.E.: {Source regions and time scales for the
  delivery of water to Earth}.
\newblock Meteoritics and Planetary Science \textbf{35}, 1309--1320 (2000).
\newblock \doi{10.1111/j.1945-5100.2000.tb01518.x}

\bibitem{morby15}
{Morbidelli}, A., {Lambrechts}, M., {Jacobson}, S., {Bitsch}, B.: {The great
  dichotomy of the Solar System: Small terrestrial embryos and massive giant
  planet cores}.
\newblock Icarus \textbf{258}, 418--429 (2015).
\newblock \doi{10.1016/j.icarus.2015.06.003}

\bibitem{morby05}
{Morbidelli}, A., {Levison}, H.F., {Tsiganis}, K., {Gomes}, R.: {Chaotic
  capture of Jupiter's Trojan asteroids in the early Solar System}.
\newblock \nat \textbf{435}, 462--465 (2005).
\newblock \doi{10.1038/nature03540}

\bibitem{morby12}
{Morbidelli}, A., {Nesvorny}, D.: {Dynamics of pebbles in the vicinity of a
  growing planetary embryo: hydro-dynamical simulations}.
\newblock \aap \textbf{546}, A18 (2012).
\newblock \doi{10.1051/0004-6361/201219824}

\bibitem{morby18}
{Morbidelli}, A., {Nesvorny}, D., {Laurenz}, V., {Marchi}, S., {Rubie}, D.C.,
  {Elkins-Tanton}, L., {Wieczorek}, M., {Jacobson}, S.: {The timeline of the
  lunar bombardment: Revisited}.
\newblock \icarus \textbf{305}, 262--276 (2018).
\newblock \doi{10.1016/j.icarus.2017.12.046}

\bibitem{morbyraymond16}
{Morbidelli}, A., {Raymond}, S.N.: {Challenges in planet formation}.
\newblock Journal of Geophysical Research (Planets) \textbf{121}, 1962--1980
  (2016).
\newblock \doi{10.1002/2016JE005088}

\bibitem{morby07}
{Morbidelli}, A., {Tsiganis}, K., {Crida}, A., {Levison}, H.F., {Gomes}, R.:
  {Dynamics of the Giant Planets of the Solar System in the Gaseous
  Protoplanetary Disk and Their Relationship to the Current Orbital
  Architecture}.
\newblock \aj \textbf{134}, 1790--1798 (2007).
\newblock \doi{10.1086/521705}

\bibitem{morbywood15}
{Morbidelli}, A., {Wood}, B.J.: {Late Accretion and the Late Veneer}.
\newblock Washington DC American Geophysical Union Geophysical Monograph Series
  \textbf{212}, 71--82 (2015).
\newblock \doi{10.1002/9781118860359.ch4}

\bibitem{mordasini09}
{Mordasini}, C., {Alibert}, Y., {Benz}, W.: {Extrasolar planet population
  synthesis. I. Method, formation tracks, and mass-distance distribution}.
\newblock \aap \textbf{501}, 1139--1160 (2009).
\newblock \doi{10.1051/0004-6361/200810301}

\bibitem{mori19}
{Mori}, S., {Bai}, X.N., {Okuzumi}, S.: {Temperature Structure in the Inner
  Regions of Protoplanetary Disks: Inefficient Accretion Heating Controlled by
  Nonideal Magnetohydrodynamics}.
\newblock \apj \textbf{872}(1), 98 (2019).
\newblock \doi{10.3847/1538-4357/ab0022}

\bibitem{moriarty16}
{Moriarty}, J., {Ballard}, S.: {The Kepler Dichotomy in Planetary Disks:
  Linking Kepler Observables to Simulations of Late-stage Planet Formation}.
\newblock \apj \textbf{832}, 34 (2016).
\newblock \doi{10.3847/0004-637X/832/1/34}

\bibitem{morishima08}
{Morishima}, R., {Schmidt}, M.W., {Stadel}, J., {Moore}, B.: {Formation and
  Accretion History of Terrestrial Planets from Runaway Growth through to Late
  Time: Implications for Orbital Eccentricity}.
\newblock \apj \textbf{685}, 1247--1261 (2008).
\newblock \doi{10.1086/590948}

\bibitem{morishima10}
{Morishima}, R., {Stadel}, J., {Moore}, B.: {From planetesimals to terrestrial
  planets: N-body simulations including the effects of nebular gas and giant
  planets}.
\newblock Icarus \textbf{207}, 517--535 (2010).
\newblock \doi{10.1016/j.icarus.2009.11.038}

\bibitem{moromartin07}
{Moro-Mart{\'{\i}}n}, A., {Carpenter}, J.M., {Meyer}, M.R., {Hillenbrand},
  L.A., {Malhotra}, R., {Hollenbach}, D., {Najita}, J., {Henning}, T., {Kim},
  J.S., {Bouwman}, J., {Silverstone}, M.D., {Hines}, D.C., {Wolf}, S.,
  {Pascucci}, I., {Mamajek}, E.E., {Lunine}, J.: {Are Debris Disks and Massive
  Planets Correlated?}
\newblock \apj \textbf{658}, 1312--1321 (2007).
\newblock \doi{10.1086/511746}

\bibitem{moromartin15}
{Moro-Mart{\'{\i}}n}, A., {Marshall}, J.P., {Kennedy}, G., {Sibthorpe}, B.,
  {Matthews}, B.C., {Eiroa}, C., {Wyatt}, M.C., {Lestrade}, J.F., {Maldonado},
  J., {Rodriguez}, D., {Greaves}, J.S., {Montesinos}, B., {Mora}, A., {Booth},
  M., {Duch{\^e}ne}, G., {Wilner}, D., {Horner}, J.: {Does the Presence of
  Planets Affect the Frequency and Properties of Extrasolar Kuiper Belts?
  Results from the Herschel Debris and Dunes Surveys}.
\newblock \apj \textbf{801}, 143 (2015).
\newblock \doi{10.1088/0004-637X/801/2/143}

\bibitem{mottl07}
{Mottl}, M., {Glazer}, B., {Kaiser}, R., {Meech}, K.: {Water and astrobiology}.
\newblock Chemie der Erde / Geochemistry \textbf{67}(4), 253--282 (2007).
\newblock \doi{10.1016/j.chemer.2007.09.002}

\bibitem{sujoy12}
{Mukhopadhyay}, S.: {Early differentiation and volatile accretion recorded in
  deep-mantle neon and xenon}.
\newblock \nat \textbf{486}, 101--104 (2012).
\newblock \doi{10.1038/nature11141}

\bibitem{mulders19}
{Mulders}, G.D., {Mordasini}, C., {Pascucci}, I., {Ciesla}, F.J., {Emsenhuber},
  A., {Apai}, D.: {The Exoplanet Population Observation Simulator. II --
  Population Synthesis in the Era of Kepler}.
\newblock arXiv e-prints arXiv:1905.08804 (2019)

\bibitem{mulders18}
{Mulders}, G.D., {Pascucci}, I., {Apai}, D., {Ciesla}, F.J.: {The Exoplanet
  Population Observation Simulator. I. The Inner Edges of Planetary Systems}.
\newblock \aj \textbf{156}, 24 (2018).
\newblock \doi{10.3847/1538-3881/aac5ea}

\bibitem{mundy00}
{Mundy}, L.G., {Looney}, L.W., {Welch}, W.J.: {The Structure and Evolution of
  Envelopes and Disks in Young Stellar Systems}.
\newblock Protostars and Planets IV pp. 355--+ (2000)

\bibitem{muralidharan08}
{Muralidharan}, K., {Deymier}, P., {Stimpfl}, M., {de Leeuw}, N.H., {Drake},
  M.J.: {Origin of water in the inner Solar System: A kinetic Monte Carlo study
  of water adsorption on forsterite}.
\newblock \icarus \textbf{198}(2), 400--407 (2008).
\newblock \doi{10.1016/j.icarus.2008.07.017}

\bibitem{nagasawa08}
{Nagasawa}, M., {Ida}, S., {Bessho}, T.: {Formation of Hot Planets by a
  Combination of Planet Scattering, Tidal Circularization, and the Kozai
  Mechanism}.
\newblock \apj \textbf{678}, 498--508 (2008).
\newblock \doi{10.1086/529369}

\bibitem{nagasawa05}
{Nagasawa}, M., {Lin}, D.N.C., {Thommes}, E.: {Dynamical Shake-up of Planetary
  Systems. I. Embryo Trapping and Induced Collisions by the Sweeping Secular
  Resonance and Embryo-Disk Tidal Interaction}.
\newblock \apj \textbf{635}, 578--598 (2005).
\newblock \doi{10.1086/497386}

\bibitem{nayakshin15}
{Nayakshin}, S.: {Tidal downsizing model - I. Numerical methods: saving giant
  planets from tidal disruptions}.
\newblock \mnras \textbf{454}(1), 64--82 (2015).
\newblock \doi{10.1093/mnras/stv1915}

\bibitem{ndugu18}
{Ndugu}, N., {Bitsch}, B., {Jurua}, E.: {Planet population synthesis driven by
  pebble accretion in cluster environments}.
\newblock \mnras \textbf{474}(1), 886--897 (2018).
\newblock \doi{10.1093/mnras/stx2815}

\bibitem{neishtadt84}
{Neishtadt}, A.: {The separation of motions in systems with rapidly rotating
  phase}.
\newblock Journal of Applied Mathematics and Mechanics \textbf{48}(2), 133--139
  (1984).
\newblock \doi{10.1016/0021-8928(84)90078-9}

\bibitem{nesvorny15}
{Nesvorn{\'y}}, D.: {Evidence for Slow Migration of Neptune from the
  Inclination Distribution of Kuiper Belt Objects}.
\newblock \aj \textbf{150}, 73 (2015).
\newblock \doi{10.1088/0004-6256/150/3/73}

\bibitem{nesvorny18b}
{Nesvorn{\'y}}, D.: {Dynamical Evolution of the Early Solar System}.
\newblock \araa \textbf{56}, 137--174 (2018).
\newblock \doi{10.1146/annurev-astro-081817-052028}

\bibitem{Nesvorny19}
{Nesvorn{\'y}}, D., {Li}, R., {Youdin}, A.N., {Simon}, J.B., {Grundy}, W.M.:
  {Trans-Neptunian binaries as evidence for planetesimal formation by the
  streaming instability}.
\newblock Nature Astronomy \textbf{3}, 808--812 (2019).
\newblock \doi{10.1038/s41550-019-0806-z}

\bibitem{nesvorny12}
{Nesvorn{\'y}}, D., {Morbidelli}, A.: {Statistical Study of the Early Solar
  System's Instability with Four, Five, and Six Giant Planets}.
\newblock \aj \textbf{144}, 117 (2012).
\newblock \doi{10.1088/0004-6256/144/4/117}

\bibitem{nesvorny18}
{Nesvorn{\'y}}, D., {Vokrouhlick{\'y}}, D., {Bottke}, W.F., {Levison}, H.F.:
  {Evidence for very early migration of the Solar System planets from the
  Patroclus-Menoetius binary Jupiter Trojan}.
\newblock Nature Astronomy \textbf{2}, 878--882 (2018).
\newblock \doi{10.1038/s41550-018-0564-3}

\bibitem{nesvorny14}
{Nesvorn{\'y}}, D., {Vokrouhlick{\'y}}, D., {Deienno}, R.: {Capture of
  Irregular Satellites at Jupiter}.
\newblock \apj \textbf{784}(1), 22 (2014).
\newblock \doi{10.1088/0004-637X/784/1/22}

\bibitem{nesvorny13}
{Nesvorn{\'y}}, D., {Vokrouhlick{\'y}}, D., {Morbidelli}, A.: {Capture of
  Trojans by Jumping Jupiter}.
\newblock \apj \textbf{768}, 45 (2013).
\newblock \doi{10.1088/0004-637X/768/1/45}

\bibitem{nesvorny10b}
{Nesvorn{\'y}}, D., {Youdin}, A.N., {Richardson}, D.C.: {Formation of Kuiper
  Belt Binaries by Gravitational Collapse}.
\newblock \aj \textbf{140}(3), 785--793 (2010).
\newblock \doi{10.1088/0004-6256/140/3/785}

\bibitem{nimmo07}
{Nimmo}, F., {Kleine}, T.: {How rapidly did Mars accrete? Uncertainties in the
  Hf W timing of core formation}.
\newblock Icarus \textbf{191}, 497--504 (2007).
\newblock \doi{10.1016/j.icarus.2007.05.002}

\bibitem{noll08}
{Noll}, K.S., {Grundy}, W.M., {Chiang}, E.I., {Margot}, J.L., {Kern}, S.D.:
  {Binaries in the Kuiper Belt}, p. 345 (2008)

\bibitem{nomura14}
{Nomura}, R., {Hirose}, K., {Uesegi}, K., {Ohishi}, Y., {Tsuchiyama}, A.,
  {Miyake}, A., {Ueno}, Y.: {Low Core-Mantle Boundary Temperature Inferred from
  the Solidus of Pyrolite}.
\newblock Science \textbf{343}, 522--525 (2014)

\bibitem{nyquist09}
{Nyquist}, L.E., {Kleine}, T., {Shih}, C.Y., {Reese}, Y.D.: {The distribution
  of short-lived radioisotopes in the early solar system and the chronology of
  asteroid accretion, differentiation, and secondary mineralization}.
\newblock \gca \textbf{73}, 5115--5136 (2009).
\newblock \doi{10.1016/j.gca.2008.12.031}

\bibitem{obrien18}
{O'Brien}, D.P., {Izidoro}, A., {Jacobson}, S.A., {Raymond}, S.N., {Rubie},
  D.C.: {The Delivery of Water During Terrestrial Planet Formation}.
\newblock \ssr \textbf{214}(1), 47 (2018).
\newblock \doi{10.1007/s11214-018-0475-8}

\bibitem{obrien06}
{O'Brien}, D.P., {Morbidelli}, A., {Levison}, H.F.: {Terrestrial planet
  formation with strong dynamical friction}.
\newblock Icarus \textbf{184}, 39--58 (2006).
\newblock \doi{10.1016/j.icarus.2006.04.005}

\bibitem{obrien14}
{O'Brien}, D.P., {Walsh}, K.J., {Morbidelli}, A., {Raymond}, S.N., {Mandell},
  A.M.: {Water delivery and giant impacts in the Grand Tack scenario}.
\newblock Icarus \textbf{239}, 74--84 (2014).
\newblock \doi{10.1016/j.icarus.2014.05.009}

\bibitem{ogihara09}
{Ogihara}, M., {Ida}, S.: {N-Body Simulations of Planetary Accretion Around M
  Dwarf Stars}.
\newblock \apj \textbf{699}, 824--838 (2009).
\newblock \doi{10.1088/0004-637X/699/1/824}

\bibitem{ogihara18}
{Ogihara}, M., {Kokubo}, E., {Suzuki}, T.K., {Morbidelli}, A.: {Formation of
  close-in super-Earths in evolving protoplanetary disks due to disk winds}.
\newblock \aap \textbf{615}, A63 (2018).
\newblock \doi{10.1051/0004-6361/201832720}

\bibitem{ogihara18b}
{Ogihara}, M., {Kokubo}, E., {Suzuki}, T.K., {Morbidelli}, A.: {Formation of
  the terrestrial planets in the solar system around 1 au via radial
  concentration of planetesimals}.
\newblock \aap \textbf{612}, L5 (2018).
\newblock \doi{10.1051/0004-6361/201832654}

\bibitem{ogihara15}
{Ogihara}, M., {Morbidelli}, A., {Guillot}, T.: {A reassessment of the in situ
  formation of close-in super-Earths}.
\newblock \aap \textbf{578}, A36 (2015).
\newblock \doi{10.1051/0004-6361/201525884}

\bibitem{oka11}
{Oka}, A., {Nakamoto}, T., {Ida}, S.: {Evolution of Snow Line in Optically
  Thick Protoplanetary Disks: Effects of Water Ice Opacity and Dust Grain
  Size}.
\newblock \apj \textbf{738}(2), 141 (2011).
\newblock \doi{10.1088/0004-637X/738/2/141}

\bibitem{okuzumi12}
{Okuzumi}, S., {Tanaka}, H., {Kobayashi}, H., {Wada}, K.: {Rapid Coagulation of
  Porous Dust Aggregates outside the Snow Line: A Pathway to Successful Icy
  Planetesimal Formation}.
\newblock \apj \textbf{752}(2), 106 (2012).
\newblock \doi{10.1088/0004-637X/752/2/106}

\bibitem{ormel10}
{Ormel}, C.W., {Klahr}, H.H.: {The effect of gas drag on the growth of
  protoplanets. Analytical expressions for the accretion of small bodies in
  laminar disks}.
\newblock \aap \textbf{520}, A43 (2010).
\newblock \doi{10.1051/0004-6361/201014903}

\bibitem{owen11}
{Owen}, J.E., {Ercolano}, B., {Clarke}, C.J.: {Protoplanetary disc evolution
  and dispersal: the implications of X-ray photoevaporation}.
\newblock \mnras \textbf{412}(1), 13--25 (2011).
\newblock \doi{10.1111/j.1365-2966.2010.17818.x}

\bibitem{owen13}
{Owen}, J.E., {Wu}, Y.: {Kepler Planets: A Tale of Evaporation}.
\newblock \apj \textbf{775}, 105 (2013).
\newblock \doi{10.1088/0004-637X/775/2/105}

\bibitem{owen17}
{Owen}, J.E., {Wu}, Y.: {The Evaporation Valley in the Kepler Planets}.
\newblock \apj \textbf{847}, 29 (2017).
\newblock \doi{10.3847/1538-4357/aa890a}

\bibitem{paardekooper14}
{Paardekooper}, S.J.: {Dynamical corotation torques on low-mass planets}.
\newblock \mnras \textbf{444}, 2031--2042 (2014).
\newblock \doi{10.1093/mnras/stu1542}

\bibitem{paardekooper10}
{Paardekooper}, S.J., {Baruteau}, C., {Crida}, A., {Kley}, W.: {A torque
  formula for non-isothermal type I planetary migration - I. Unsaturated
  horseshoe drag}.
\newblock \mnras \textbf{401}, 1950--1964 (2010).
\newblock \doi{10.1111/j.1365-2966.2009.15782.x}

\bibitem{paardekooper11}
{Paardekooper}, S.J., {Baruteau}, C., {Kley}, W.: {A torque formula for
  non-isothermal Type I planetary migration - II. Effects of diffusion}.
\newblock \mnras \textbf{410}, 293--303 (2011).
\newblock \doi{10.1111/j.1365-2966.2010.17442.x}

\bibitem{paardekooper06}
{Paardekooper}, S.J., {Mellema}, G.: {Halting type I planet migration in
  non-isothermal disks}.
\newblock \aap \textbf{459}, L17--L20 (2006).
\newblock \doi{10.1051/0004-6361:20066304}

\bibitem{paetzold19}
{P{\"a}tzold}, M., {Andert}, T.P., {Hahn}, M., {Barriot}, J.P., {Asmar}, S.W.,
  {H{\"a}usler}, B., {Bird}, M.K., {Tellmann}, S., {Oschlisniok}, J., {Peter},
  K.: {The Nucleus of comet 67P/Churyumov-Gerasimenko - Part I: The global view
  - nucleus mass, mass-loss, porosity, and implications}.
\newblock \mnras \textbf{483}(2), 2337--2346 (2019).
\newblock \doi{10.1093/mnras/sty3171}

\bibitem{pecaut16}
{Pecaut}, M.J., {Mamajek}, E.E.: {The star formation history and accretion-disc
  fraction among the K-type members of the Scorpius-Centaurus OB association}.
\newblock \mnras \textbf{461}(1), 794--815 (2016).
\newblock \doi{10.1093/mnras/stw1300}

\bibitem{petigura13}
{Petigura}, E.A., {Howard}, A.W., {Marcy}, G.W.: {Prevalence of Earth-size
  planets orbiting Sun-like stars}.
\newblock Proceedings of the National Academy of Science \textbf{110},
  19273--19278 (2013)

\bibitem{pfalzner14}
{Pfalzner}, S., {Steinhausen}, M., {Menten}, K.: {Short Dissipation Times of
  Proto-planetary Disks: An Artifact of Selection Effects?}
\newblock \apjl \textbf{793}, L34 (2014).
\newblock \doi{10.1088/2041-8205/793/2/L34}

\bibitem{pichierri18}
{Pichierri}, G., {Morbidelli}, A., {Crida}, A.: {Capture into first-order
  resonances and long-term stability of pairs of equal-mass planets}.
\newblock Celestial Mechanics and Dynamical Astronomy \textbf{130}(8), 54
  (2018).
\newblock \doi{10.1007/s10569-018-9848-2}

\bibitem{pierens15}
{Pierens}, A.: {Fast migration of low-mass planets in radiative discs}.
\newblock \mnras \textbf{454}, 2003--2014 (2015).
\newblock \doi{10.1093/mnras/stv2024}

\bibitem{pierens08}
{Pierens}, A., {Nelson}, R.P.: {Constraints on resonant-trapping for two
  planets embedded in a protoplanetary disc}.
\newblock \aap \textbf{482}, 333--340 (2008).
\newblock \doi{10.1051/0004-6361:20079062}

\bibitem{pierens11}
{Pierens}, A., {Raymond}, S.N.: {Two phase, inward-then-outward migration of
  Jupiter and Saturn in the gaseous solar nebula}.
\newblock \aap \textbf{533}, A131 (2011).
\newblock \doi{10.1051/0004-6361/201117451}

\bibitem{pierens16}
{Pierens}, A., {Raymond}, S.N.: {Migration of accreting planets in radiative
  discs from dynamical torques}.
\newblock \mnras \textbf{462}(4), 4130--4140 (2016).
\newblock \doi{10.1093/mnras/stw1904}

\bibitem{pierens14}
{Pierens}, A., {Raymond}, S.N., {Nesvorny}, D., {Morbidelli}, A.: {Outward
  Migration of Jupiter and Saturn in 3:2 or 2:1 Resonance in Radiative Disks:
  Implications for the Grand Tack and Nice models}.
\newblock \apjl \textbf{795}, L11 (2014).
\newblock \doi{10.1088/2041-8205/795/1/L11}

\bibitem{pinte16}
{Pinte}, C., {Dent}, W.R.F., {M{\'e}nard}, F., {Hales}, A., {Hill}, T.,
  {Cortes}, P., {de Gregorio-Monsalvo}, I.: {Dust and Gas in the Disk of HL
  Tauri: Surface Density, Dust Settling, and Dust-to-gas Ratio}.
\newblock \apj \textbf{816}(1), 25 (2016).
\newblock \doi{10.3847/0004-637X/816/1/25}

\bibitem{pirani19}
{Pirani}, S., {Johansen}, A., {Bitsch}, B., {Mustill}, A.J., {Turrini}, D.:
  {Consequences of planetary migration on the minor bodies of the early solar
  system}.
\newblock \aap \textbf{623}, A169 (2019).
\newblock \doi{10.1051/0004-6361/201833713}

\bibitem{piso14}
{Piso}, A.M.A., {Youdin}, A.N.: {On the Minimum Core Mass for Giant Planet
  Formation at Wide Separations}.
\newblock \apj \textbf{786}(1), 21 (2014).
\newblock \doi{10.1088/0004-637X/786/1/21}

\bibitem{poincare1892}
{Poincare}, H.: {Les methodes nouvelles de la mecanique celeste} (1892)

\bibitem{poincare1897}
{Poincare}, H.: {Sur une forme Nouvelle des Equations du Probleme des Trois
  Corps}.
\newblock Bulletin Astronomique, Serie I \textbf{14}, 53--67 (1897)

\bibitem{pollack96}
{Pollack}, J.B., {Hubickyj}, O., {Bodenheimer}, P., {Lissauer}, J.J.,
  {Podolak}, M., {Greenzweig}, Y.: {Formation of the Giant Planets by
  Concurrent Accretion of Solids and Gas}.
\newblock Icarus \textbf{124}, 62--85 (1996).
\newblock \doi{10.1006/icar.1996.0190}

\bibitem{quarles19}
{Quarles}, B., {Kaib}, N.: {Instabilities in the Early Solar System Due to a
  Self-gravitating Disk}.
\newblock \aj \textbf{157}(2), 67 (2019).
\newblock \doi{10.3847/1538-3881/aafa71}

\bibitem{quintana16}
{Quintana}, E.V., {Barclay}, T., {Borucki}, W.J., {Rowe}, J.F., {Chambers},
  J.E.: {The Frequency of Giant Impacts on Earth-like Worlds}.
\newblock \apj \textbf{821}, 126 (2016).
\newblock \doi{10.3847/0004-637X/821/2/126}

\bibitem{quintana14}
{Quintana}, E.V., {Barclay}, T., {Raymond}, S.N., {Rowe}, J.F., {Bolmont}, E.,
  {Caldwell}, D.A., {Howell}, S.B., {Kane}, S.R., {Huber}, D., {Crepp}, J.R.,
  {Lissauer}, J.J., {Ciardi}, D.R., {Coughlin}, J.L., {Everett}, M.E., {Henze},
  C.E., {Horch}, E., {Isaacson}, H., {Ford}, E.B., {Adams}, F.C., {Still}, M.,
  {Hunter}, R.C., {Quarles}, B., {Selsis}, F.: {An Earth-Sized Planet in the
  Habitable Zone of a Cool Star}.
\newblock Science \textbf{344}, 277--280 (2014).
\newblock \doi{10.1126/science.1249403}

\bibitem{rafikov04}
{Rafikov}, R.R.: {Fast Accretion of Small Planetesimals by Protoplanetary
  Cores}.
\newblock \aj \textbf{128}, 1348--1363 (2004).
\newblock \doi{10.1086/423216}

\bibitem{rasio96}
{Rasio}, F.A., {Ford}, E.B.: {Dynamical instabilities and the formation of
  extrasolar planetary systems}.
\newblock Science \textbf{274}, 954--956 (1996).
\newblock \doi{10.1126/science.274.5289.954}

\bibitem{raymond10}
{Raymond}, S.N., {Armitage}, P.J., {Gorelick}, N.: {Planet-Planet Scattering in
  Planetesimal Disks. II. Predictions for Outer Extrasolar Planetary Systems}.
\newblock \apj \textbf{711}, 772--795 (2010).
\newblock \doi{10.1088/0004-637X/711/2/772}

\bibitem{raymond11}
{Raymond}, S.N., {Armitage}, P.J., {Moro-Mart{\'{\i}}n}, A., {Booth}, M.,
  {Wyatt}, M.C., {Armstrong}, J.C., {Mandell}, A.M., {Selsis}, F., {West},
  A.A.: {Debris disks as signposts of terrestrial planet formation}.
\newblock \aap \textbf{530}, A62 (2011).
\newblock \doi{10.1051/0004-6361/201116456}

\bibitem{raymond12}
{Raymond}, S.N., {Armitage}, P.J., {Moro-Mart{\'{\i}}n}, A., {Booth}, M.,
  {Wyatt}, M.C., {Armstrong}, J.C., {Mandell}, A.M., {Selsis}, F., {West},
  A.A.: {Debris disks as signposts of terrestrial planet formation. II.
  Dependence of exoplanet architectures on giant planet and disk properties}.
\newblock \aap \textbf{541}, A11 (2012).
\newblock \doi{10.1051/0004-6361/201117049}

\bibitem{raymond18c}
{Raymond}, S.N., {Armitage}, P.J., {Veras}, D.: {Interstellar Object
  {\textquoteright}Oumuamua as an Extinct Fragment of an Ejected Cometary
  Planetesimal}.
\newblock \apjl \textbf{856}(1), L7 (2018).
\newblock \doi{10.3847/2041-8213/aab4f6}

\bibitem{raymond18}
{Raymond}, S.N., {Armitage}, P.J., {Veras}, D., {Quintana}, E.V., {Barclay},
  T.: {Implications of the interstellar object 1I/'Oumuamua for planetary
  dynamics and planetesimal formation}.
\newblock \mnras \textbf{476}, 3031--3038 (2018).
\newblock \doi{10.1093/mnras/sty468}

\bibitem{raymond08b}
{Raymond}, S.N., {Barnes}, R., {Armitage}, P.J., {Gorelick}, N.: {Mean Motion
  Resonances from Planet-Planet Scattering}.
\newblock \apjl \textbf{687}, L107--L110 (2008).
\newblock \doi{10.1086/593301}

\bibitem{raymond08a}
{Raymond}, S.N., {Barnes}, R., {Mandell}, A.M.: {Observable consequences of
  planet formation models in systems with close-in terrestrial planets}.
\newblock \mnras \textbf{384}, 663--674 (2008).
\newblock \doi{10.1111/j.1365-2966.2007.12712.x}

\bibitem{raymond09a}
{Raymond}, S.N., {Barnes}, R., {Veras}, D., {Armitage}, P.J., {Gorelick}, N.,
  {Greenberg}, R.: {Planet-Planet Scattering Leads to Tightly Packed Planetary
  Systems}.
\newblock \apjl \textbf{696}, L98--L101 (2009).
\newblock \doi{10.1088/0004-637X/696/1/L98}

\bibitem{raymond18b}
{Raymond}, S.N., {Boulet}, T., {Izidoro}, A., {Esteves}, L., {Bitsch}, B.:
  {Migration-driven diversity of super-Earth compositions}.
\newblock \mnras \textbf{479}, L81--L85 (2018).
\newblock \doi{10.1093/mnrasl/sly100}

\bibitem{raymond14b}
{Raymond}, S.N., {Cossou}, C.: {No universal minimum-mass extrasolar nebula:
  evidence against in situ accretion of systems of hot super-Earths}.
\newblock \mnras \textbf{440}, L11--L15 (2014).
\newblock \doi{10.1093/mnrasl/slu011}

\bibitem{raymond17}
{Raymond}, S.N., {Izidoro}, A.: {Origin of water in the inner Solar System:
  Planetesimals scattered inward during Jupiter and Saturn's rapid gas
  accretion}.
\newblock Icarus \textbf{297}, 134--148 (2017).
\newblock \doi{10.1016/j.icarus.2017.06.030}

\bibitem{raymond17b}
{Raymond}, S.N., {Izidoro}, A.: {The empty primordial asteroid belt}.
\newblock Science Advances \textbf{3}, e1701138 (2017).
\newblock \doi{10.1126/sciadv.1701138}

\bibitem{raymond18d}
{Raymond}, S.N., {Izidoro}, A., {Morbidelli}, A.: {Solar System Formation in
  the Context of Extra-Solar Planets}.
\newblock arXiv e-prints arXiv:1812.01033 (2018)

\bibitem{raymond14}
{Raymond}, S.N., {Kokubo}, E., {Morbidelli}, A., {Morishima}, R., {Walsh},
  K.J.: {Terrestrial Planet Formation at Home and Abroad}.
\newblock Protostars and Planets VI pp. 595--618 (2014)

\bibitem{raymond06c}
{Raymond}, S.N., {Mandell}, A.M., {Sigurdsson}, S.: {Exotic Earths: Forming
  Habitable Worlds with Giant Planet Migration}.
\newblock Science \textbf{313}, 1413--1416 (2006).
\newblock \doi{10.1126/science.1130461}

\bibitem{raymond14c}
{Raymond}, S.N., {Morbidelli}, A.: {The Grand Tack model: a critical review}.
\newblock In: Complex Planetary Systems, Proceedings of the International
  Astronomical Union, \emph{IAU Symposium}, vol. 310, pp. 194--203 (2014).
\newblock \doi{10.1017/S1743921314008254}

\bibitem{raymond09c}
{Raymond}, S.N., {O'Brien}, D.P., {Morbidelli}, A., {Kaib}, N.A.: {Building the
  terrestrial planets: Constrained accretion in the inner Solar System}.
\newblock Icarus \textbf{203}, 644--662 (2009).
\newblock \doi{10.1016/j.icarus.2009.05.016}

\bibitem{raymond04}
{Raymond}, S.N., {Quinn}, T., {Lunine}, J.I.: {Making other earths: dynamical
  simulations of terrestrial planet formation and water delivery}.
\newblock Icarus \textbf{168}, 1--17 (2004).
\newblock \doi{10.1016/j.icarus.2003.11.019}

\bibitem{raymond06b}
{Raymond}, S.N., {Quinn}, T., {Lunine}, J.I.: {High-resolution simulations of
  the final assembly of Earth-like planets I. Terrestrial accretion and
  dynamics}.
\newblock Icarus \textbf{183}, 265--282 (2006).
\newblock \doi{10.1016/j.icarus.2006.03.011}

\bibitem{raymond07a}
{Raymond}, S.N., {Quinn}, T., {Lunine}, J.I.: {High-Resolution Simulations of
  The Final Assembly of Earth-Like Planets. 2. Water Delivery And Planetary
  Habitability}.
\newblock Astrobiology \textbf{7}, 66--84 (2007).
\newblock \doi{10.1089/ast.2006.06-0126}

\bibitem{raymond07b}
{Raymond}, S.N., {Scalo}, J., {Meadows}, V.S.: {A Decreased Probability of
  Habitable Planet Formation around Low-Mass Stars}.
\newblock \apj \textbf{669}, 606--614 (2007).
\newblock \doi{10.1086/521587}

\bibitem{raymond13}
{Raymond}, S.N., {Schlichting}, H.E., {Hersant}, F., {Selsis}, F.: {Dynamical
  and collisional constraints on a stochastic late veneer on the terrestrial
  planets}.
\newblock \icarus \textbf{226}(1), 671--681 (2013).
\newblock \doi{10.1016/j.icarus.2013.06.019}

\bibitem{ribas07}
{Ribas}, I., {Miralda-Escud{\'e}}, J.: {The eccentricity-mass distribution of
  exoplanets: signatures of different formation mechanisms?}
\newblock \aap \textbf{464}, 779--785 (2007).
\newblock \doi{10.1051/0004-6361:20065726}

\bibitem{ribeiro20}
{Ribeiro}, R., {Morbidelli}, A., {Raymond}, S., {Izidoro}, A., {Gomes}, R.:
  {Dynamical evidence for an early giant planet instability}.
\newblock Submitted to Icarus  (2020)

\bibitem{rivera10}
{Rivera}, E.J., {Laughlin}, G., {Butler}, R.P., {Vogt}, S.S., {Haghighipour},
  N., {Meschiari}, S.: {The Lick-Carnegie Exoplanet Survey: a Uranus-Mass
  Fourth Planet for GJ 876 in an Extrasolar Laplace Configuration}.
\newblock \apj \textbf{719}, 890--899 (2010).
\newblock \doi{10.1088/0004-637X/719/1/890}

\bibitem{robert18}
{Robert}, C.M.T., {Crida}, A., {Lega}, E., {M{\'e}heut}, H., {Morbidelli}, A.:
  {Toward a new paradigm for Type II migration}.
\newblock \aap \textbf{617}, A98 (2018).
\newblock \doi{10.1051/0004-6361/201833539}

\bibitem{robert77}
{Robert}, F., {Merlivat}, L., {Javoy}, M.: {Water and Deuterium Content in
  Eight Condrites}.
\newblock Meteoritics \textbf{12}, 349 (1977)

\bibitem{rogers15}
{Rogers}, L.A.: {Most 1.6 Earth-radius Planets are Not Rocky}.
\newblock \apj \textbf{801}, 41 (2015).
\newblock \doi{10.1088/0004-637X/801/1/41}

\bibitem{ronnet18}
{Ronnet}, T., {Mousis}, O., {Vernazza}, P., {Lunine}, J.I., {Crida}, A.:
  {Saturn{\textquoteright}s Formation and Early Evolution at the Origin of
  Jupiter{\textquoteright}s Massive Moons}.
\newblock \aj \textbf{155}(5), 224 (2018).
\newblock \doi{10.3847/1538-3881/aabcc7}

\bibitem{rowan16}
{Rowan}, D., {Meschiari}, S., {Laughlin}, G., {Vogt}, S.S., {Butler}, R.P.,
  {Burt}, J., {Wang}, S., {Holden}, B., {Hanson}, R., {Arriagada}, P.,
  {Keiser}, S., {Teske}, J., {Diaz}, M.: {The Lick-Carnegie Exoplanet Survey:
  HD 32963 -- A New Jupiter Analog Orbiting a Sun-like Star}.
\newblock \apj \textbf{817}, 104 (2016).
\newblock \doi{10.3847/0004-637X/817/2/104}

\bibitem{rowe14}
{Rowe}, J.F., {Bryson}, S.T., {Marcy}, G.W., {Lissauer}, J.J., {Jontof-Hutter},
  D., {Mullally}, F., {Gilliland}, R.L., {Issacson}, H., {Ford}, E., {Howell},
  S.B., {Borucki}, W.J., {Haas}, M., {Huber}, D., {Steffen}, J.H., {Thompson},
  S.E., {Quintana}, E., {Barclay}, T., {Still}, M., {Fortney}, J., {Gautier}
  III, T.N., {Hunter}, R., {Caldwell}, D.A., {Ciardi}, D.R., {Devore}, E.,
  {Cochran}, W., {Jenkins}, J., {Agol}, E., {Carter}, J.A., {Geary}, J.:
  {Validation of Kepler's Multiple Planet Candidates. III. Light Curve Analysis
  and Announcement of Hundreds of New Multi-planet Systems}.
\newblock \apj \textbf{784}, 45 (2014).
\newblock \doi{10.1088/0004-637X/784/1/45}

\bibitem{rubie15}
{Rubie}, D.C., {Jacobson}, S.A., {Morbidelli}, A., {O'Brien}, D.P., {Young},
  E.D., {de Vries}, J., {Nimmo}, F., {Palme}, H., {Frost}, D.J.: {Accretion and
  differentiation of the terrestrial planets with implications for the
  compositions of early-formed Solar System bodies and accretion of water}.
\newblock \icarus \textbf{248}, 89--108 (2015).
\newblock \doi{10.1016/j.icarus.2014.10.015}

\bibitem{safronov72}
{Safronov}, V.S.: {Evolution of the protoplanetary cloud and formation of the
  earth and planets.} (1972)

\bibitem{santos01}
{Santos}, N.C., {Israelian}, G., {Mayor}, M.: {The metal-rich nature of stars
  with planets}.
\newblock \aap \textbf{373}, 1019--1031 (2001).
\newblock \doi{10.1051/0004-6361:20010648}

\bibitem{sasselov00}
{Sasselov}, D.D., {Lecar}, M.: {On the Snow Line in Dusty Protoplanetary
  Disks}.
\newblock \apj \textbf{528}, 995--998 (2000).
\newblock \doi{10.1086/308209}

\bibitem{sato16}
{Sato}, T., {Okuzumi}, S., {Ida}, S.: {On the water delivery to terrestrial
  embryos by ice pebble accretion}.
\newblock \aap \textbf{589}, A15 (2016).
\newblock \doi{10.1051/0004-6361/201527069}

\bibitem{schaefer17}
{Sch{\"a}fer}, U., {Yang}, C.C., {Johansen}, A.: {Initial mass function of
  planetesimals formed by the streaming instability}.
\newblock \aap \textbf{597}, A69 (2017).
\newblock \doi{10.1051/0004-6361/201629561}

\bibitem{schiller18}
{Schiller}, M., {Bizzarro}, M., {Fernandes}, V.A.: {Isotopic evolution of the
  protoplanetary disk and the building blocks of Earth and the Moon}.
\newblock \nat \textbf{555}(7697), 507--510 (2018).
\newblock \doi{10.1038/nature25990}

\bibitem{schiller15}
{Schiller}, M., {Connelly}, J.N., {Glad}, A.C., {Mikouchi}, T., {Bizzarro}, M.:
  {Early accretion of protoplanets inferred from a reduced inner solar system
  $^{26}$Al inventory}.
\newblock Earth and Planetary Science Letters \textbf{420}, 45--54 (2015).
\newblock \doi{10.1016/j.epsl.2015.03.028}

\bibitem{schlichting14}
{Schlichting}, H.E.: {Formation of Close in Super-Earths and Mini-Neptunes:
  Required Disk Masses and their Implications}.
\newblock \apjl \textbf{795}, L15 (2014).
\newblock \doi{10.1088/2041-8205/795/1/L15}

\bibitem{schoonenberg17}
{Schoonenberg}, D., {Ormel}, C.W.: {Planetesimal formation near the snowline:
  in or out?}
\newblock \aap \textbf{602}, A21 (2017).
\newblock \doi{10.1051/0004-6361/201630013}

\bibitem{seager07}
{Seager}, S., {Kuchner}, M., {Hier-Majumder}, C.A., {Militzer}, B.:
  {Mass-Radius Relationships for Solid Exoplanets}.
\newblock \apj \textbf{669}(2), 1279--1297 (2007).
\newblock \doi{10.1086/521346}

\bibitem{selsis07b}
{Selsis}, F., {Chazelas}, B., {Bord{\'e}}, P., {Ollivier}, M., {Brachet}, F.,
  {Decaudin}, M., {Bouchy}, F., {Ehrenreich}, D., {Grie{\ss}meier}, J.M.,
  {Lammer}, H., {Sotin}, C., {Grasset}, O., {Moutou}, C., {Barge}, P.,
  {Deleuil}, M., {Mawet}, D., {Despois}, D., {Kasting}, J.F., {L{\'e}ger}, A.:
  {Could we identify hot ocean-planets with CoRoT, Kepler and Doppler
  velocimetry?}
\newblock Icarus \textbf{191}, 453--468 (2007).
\newblock \doi{10.1016/j.icarus.2007.04.010}

\bibitem{sessin84}
{Sessin}, W., {Ferraz-Mello}, S.: {Motion of two planets with periods
  commensurable in the ratio 2{\ensuremath{:}}1 solutions of the hori auxiliary
  system}.
\newblock Celestial Mechanics \textbf{32}(4), 307--332 (1984).
\newblock \doi{10.1007/BF01229087}

\bibitem{shakura73}
{Shakura}, N.I., {Sunyaev}, R.A.: {Black holes in binary systems. Observational
  appearance.}
\newblock \aap \textbf{24}, 337--355 (1973)

\bibitem{sharp17}
{Sharp}, Z.D.: {Nebular ingassing as a source of volatiles to the Terrestrial
  planets}.
\newblock Chemical Geology \textbf{448}, 137--150 (2017).
\newblock \doi{10.1016/j.chemgeo.2016.11.018}

\bibitem{simon16}
{Simon}, J.B., {Armitage}, P.J., {Li}, R., {Youdin}, A.N.: {The Mass and Size
  Distribution of Planetesimals Formed by the Streaming Instability. I. The
  Role of Self-gravity}.
\newblock \apj \textbf{822}, 55 (2016).
\newblock \doi{10.3847/0004-637X/822/1/55}

\bibitem{simon17}
{Simon}, J.B., {Armitage}, P.J., {Youdin}, A.N., {Li}, R.: {Evidence for
  Universality in the Initial Planetesimal Mass Function}.
\newblock \apjl \textbf{847}, L12 (2017).
\newblock \doi{10.3847/2041-8213/aa8c79}

\bibitem{snellgrove01}
{Snellgrove}, M.D., {Papaloizou}, J.C.B., {Nelson}, R.P.: {On disc driven
  inward migration of resonantly coupled planets with application to the system
  around GJ876}.
\newblock \aap \textbf{374}, 1092--1099 (2001).
\newblock \doi{10.1051/0004-6361:20010779}

\bibitem{sosa11}
{Sosa}, A., {Fern{\'a}ndez}, J.A.: {Masses of long-period comets derived from
  non-gravitational effects - analysis of the computed results and the
  consistency and reliability of the non-gravitational parameters}.
\newblock \mnras \textbf{416}(1), 767--782 (2011).
\newblock \doi{10.1111/j.1365-2966.2011.19111.x}

\bibitem{sotin07}
{Sotin}, C., {Grasset}, O., {Mocquet}, A.: {Mass radius curve for extrasolar
  Earth-like planets and ocean planets}.
\newblock \icarus \textbf{191}(1), 337--351 (2007).
\newblock \doi{10.1016/j.icarus.2007.04.006}

\bibitem{stimpfl06}
{Stimpfl}, M., {Walker}, A.M., {Drake}, M.J., {de Leeuw}, N.H., {Deymier}, P.:
  {An {\r{a}}ngstr{\"o}m-sized window on the origin of water in the inner solar
  system: Atomistic simulation of adsorption of water on olivine}.
\newblock Journal of Crystal Growth \textbf{294}(1), 83--95 (2006).
\newblock \doi{10.1016/j.jcrysgro.2006.05.057}

\bibitem{stone98}
{Stone}, J.M., {Ostriker}, E.C., {Gammie}, C.F.: {Dissipation in Compressible
  Magnetohydrodynamic Turbulence}.
\newblock \apjl \textbf{508}(1), L99--L102 (1998).
\newblock \doi{10.1086/311718}

\bibitem{su06}
{Su}, K.Y.L., {Rieke}, G.H., {Stansberry}, J.A., {Bryden}, G., {Stapelfeldt},
  K.R., {Trilling}, D.E., {Muzerolle}, J., {Beichman}, C.A., {Moro-Martin}, A.,
  {Hines}, D.C., {Werner}, M.W.: {Debris Disk Evolution around A Stars}.
\newblock \apj \textbf{653}, 675--689 (2006).
\newblock \doi{10.1086/508649}

\bibitem{suzuki16b}
{Suzuki}, D., {Bennett}, D.P., {Sumi}, T., {Bond}, I.A., {Rogers}, L.A., {Abe},
  F., {Asakura}, Y., {Bhattacharya}, A., {Donachie}, M., {Freeman}, M.,
  {Fukui}, A., {Hirao}, Y., {Itow}, Y., {Koshimoto}, N., {Li}, M.C.A., {Ling},
  C.H., {Masuda}, K., {Matsubara}, Y., {Muraki}, Y., {Nagakane}, M., {Onishi},
  K., {Oyokawa}, H., {Rattenbury}, N., {Saito}, T., {Sharan}, A., {Shibai}, H.,
  {Sullivan}, D.J., {Tristram}, P.J., {Yonehara}, A., {MOA Collaboration}: {The
  Exoplanet Mass-ratio Function from the MOA-II Survey: Discovery of a Break
  and Likely Peak at a Neptune Mass}.
\newblock \apj \textbf{833}, 145 (2016).
\newblock \doi{10.3847/1538-4357/833/2/145}

\bibitem{suzuki16}
{Suzuki}, T.K., {Ogihara}, M., {Morbidelli}, A., {Crida}, A., {Guillot}, T.:
  {Evolution of protoplanetary discs with magnetically driven disc winds}.
\newblock \aap \textbf{596}, A74 (2016).
\newblock \doi{10.1051/0004-6361/201628955}

\bibitem{svetsov07}
{Svetsov}, V.V.: {Atmospheric erosion and replenishment induced by impacts of
  cosmic bodies upon the Earth and Mars}.
\newblock Solar System Research \textbf{41}(1), 28--41 (2007).
\newblock \doi{10.1134/S0038094607010030}

\bibitem{tanaka02}
{Tanaka}, H., {Takeuchi}, T., {Ward}, W.R.: {Three-Dimensional Interaction
  between a Planet and an Isothermal Gaseous Disk. I. Corotation and Lindblad
  Torques and Planet Migration}.
\newblock \apj \textbf{565}, 1257--1274 (2002).
\newblock \doi{10.1086/324713}

\bibitem{tanaka04}
{Tanaka}, H., {Ward}, W.R.: {Three-dimensional Interaction between a Planet and
  an Isothermal Gaseous Disk. II. Eccentricity Waves and Bending Waves}.
\newblock \apj \textbf{602}, 388--395 (2004).
\newblock \doi{10.1086/380992}

\bibitem{tera74}
{Tera}, F., {Papanastassiou}, D.A., {Wasserburg}, G.J.: {Isotopic evidence for
  a terminal lunar cataclysm}.
\newblock Earth and Planetary Science Letters \textbf{22}, 1 (1974).
\newblock \doi{10.1016/0012-821X(74)90059-4}

\bibitem{terquem07}
{Terquem}, C., {Papaloizou}, J.C.B.: {Migration and the Formation of Systems of
  Hot Super-Earths and Neptunes}.
\newblock \apj \textbf{654}, 1110--1120 (2007).
\newblock \doi{10.1086/509497}

\bibitem{thommes03}
{Thommes}, E.W., {Duncan}, M.J., {Levison}, H.F.: {Oligarchic growth of giant
  planets}.
\newblock Icarus \textbf{161}, 431--455 (2003).
\newblock \doi{10.1016/S0019-1035(02)00043-X}

\bibitem{thommes08}
{Thommes}, E.W., {Matsumura}, S., {Rasio}, F.A.: {Gas Disks to Gas Giants:
  Simulating the Birth of Planetary Systems}.
\newblock Science \textbf{321}, 814-- (2008).
\newblock \doi{10.1126/science.1159723}

\bibitem{throop08}
{Throop}, H.B., {Bally}, J.: {Tail-End Bondi-Hoyle Accretion in Young Star
  Clusters: Implications for Disks, Planets, and Stars}.
\newblock \aj \textbf{135}(6), 2380--2397 (2008).
\newblock \doi{10.1088/0004-6256/135/6/2380}

\bibitem{tominaga19}
{Tominaga}, R.T., {Takahashi}, S.Z., {Inutsuka}, S.i.: {Revised Description of
  Dust Diffusion and a New Instability Creating Multiple Rings in
  Protoplanetary Disks}.
\newblock \apj \textbf{881}(1), 53 (2019).
\newblock \doi{10.3847/1538-4357/ab25ea}

\bibitem{toomre64}
{Toomre}, A.: {On the gravitational stability of a disk of stars.}
\newblock \apj \textbf{139}, 1217--1238 (1964).
\newblock \doi{10.1086/147861}

\bibitem{touboul07}
{Touboul}, M., {Kleine}, T., {Bourdon}, B., {Palme}, H., {Wieler}, R.: {Late
  formation and prolonged differentiation of the Moon inferred from W isotopes
  in lunar metals}.
\newblock \nat \textbf{450}, 1206--1209 (2007).
\newblock \doi{10.1038/nature06428}

\bibitem{tremaine12}
{Tremaine}, S., {Dong}, S.: {The Statistics of Multi-planet Systems}.
\newblock \aj \textbf{143}, 94 (2012).
\newblock \doi{10.1088/0004-6256/143/4/94}

\bibitem{trilling08}
{Trilling}, D.E., {Bryden}, G., {Beichman}, C.A., {Rieke}, G.H., {Su}, K.Y.L.,
  {Stansberry}, J.A., {Blaylock}, M., {Stapelfeldt}, K.R., {Beeman}, J.W.,
  {Haller}, E.E.: {Debris Disks around Sun-like Stars}.
\newblock \apj \textbf{674}, 1086--1105 (2008).
\newblock \doi{10.1086/525514}

\bibitem{tsiganis05}
{Tsiganis}, K., {Gomes}, R., {Morbidelli}, A., {Levison}, H.F.: {Origin of the
  orbital architecture of the giant planets of the Solar System}.
\newblock \nat \textbf{435}, 459--461 (2005).
\newblock \doi{10.1038/nature03539}

\bibitem{turner14}
{Turner}, N.J., {Fromang}, S., {Gammie}, C., {Klahr}, H., {Lesur}, G.,
  {Wardle}, M., {Bai}, X.N.: {Transport and Accretion in Planet-Forming Disks}.
\newblock Protostars and Planets VI pp. 411--432 (2014)

\bibitem{udry07b}
{Udry}, S., {Santos}, N.C.: {Statistical Properties of Exoplanets}.
\newblock \araa \textbf{45}, 397--439 (2007).
\newblock \doi{10.1146/annurev.astro.45.051806.110529}

\bibitem{uribe13}
{Uribe}, A.L., {Klahr}, H., {Henning}, T.: {Accretion of Gas onto Gap-opening
  Planets and Circumplanetary Flow Structure in Magnetized Turbulent Disks}.
\newblock \apj \textbf{769}(2), 97 (2013).
\newblock \doi{10.1088/0004-637X/769/2/97}

\bibitem{valencia07}
{Valencia}, D., {Sasselov}, D.D., {O'Connell}, R.J.: {Detailed Models of
  Super-Earths: How Well Can We Infer Bulk Properties?}
\newblock \apj \textbf{665}, 1413--1420 (2007).
\newblock \doi{10.1086/519554}

\bibitem{varniere04}
{Varni{\`e}re}, P., {Quillen}, A.C., {Frank}, A.: {The Evolution of
  Protoplanetary Disk Edges}.
\newblock \apj \textbf{612}(2), 1152--1162 (2004).
\newblock \doi{10.1086/422542}

\bibitem{veras05}
{Veras}, D., {Armitage}, P.J.: {The Influence of Massive Planet Scattering on
  Nascent Terrestrial Planets}.
\newblock \apjl \textbf{620}, L111--L114 (2005).
\newblock \doi{10.1086/428831}

\bibitem{veras06}
{Veras}, D., {Armitage}, P.J.: {Predictions for the Correlation between Giant
  and Terrestrial Extrasolar Planets in Dynamically Evolved Systems}.
\newblock \apj \textbf{645}, 1509--1515 (2006).
\newblock \doi{10.1086/504582}

\bibitem{villeneuve09}
{Villeneuve}, J., {Chaussidon}, M., {Libourel}, G.: {Homogeneous Distribution
  of $^{26}$Al in the Solar System from the Mg Isotopic Composition of
  Chondrules}.
\newblock Science \textbf{325}(5943), 985 (2009).
\newblock \doi{10.1126/science.1173907}

\bibitem{walsh11}
{Walsh}, K.J., {Morbidelli}, A., {Raymond}, S.N., {O'Brien}, D.P., {Mandell},
  A.M.: {A low mass for Mars from Jupiter's early gas-driven migration}.
\newblock \nat \textbf{475}, 206--209 (2011).
\newblock \doi{10.1038/nature10201}

\bibitem{walsh12}
{Walsh}, K.J., {Morbidelli}, A., {Raymond}, S.N., {O'Brien}, D.P., {Mandell},
  A.M.: {Populating the asteroid belt from two parent source regions due to the
  migration of giant planets--``The Grand Tack''}.
\newblock Meteoritics and Planetary Science \textbf{47}, 1941--1947 (2012).
\newblock \doi{10.1111/j.1945-5100.2012.01418.x}

\bibitem{ward86}
{Ward}, W.R.: {Density waves in the solar nebula - Differential Lindblad
  torque}.
\newblock Icarus \textbf{67}, 164--180 (1986).
\newblock \doi{10.1016/0019-1035(86)90182-X}

\bibitem{ward97}
{Ward}, W.R.: {Protoplanet Migration by Nebula Tides}.
\newblock Icarus \textbf{126}, 261--281 (1997).
\newblock \doi{10.1006/icar.1996.5647}

\bibitem{warren11}
{Warren}, P.H.: {Stable-isotopic anomalies and the accretionary assemblage of
  the Earth and Mars: A subordinate role for carbonaceous chondrites}.
\newblock Earth and Planetary Science Letters \textbf{311}, 93--100 (2011).
\newblock \doi{10.1016/j.epsl.2011.08.047}

\bibitem{weber18}
{Weber}, P., {Ben{\'\i}tez-Llambay}, P., {Gressel}, O., {Krapp}, L., {Pessah},
  M.E.: {Characterizing the Variable Dust Permeability of Planet-induced Gaps}.
\newblock \apj \textbf{854}(2), 153 (2018).
\newblock \doi{10.3847/1538-4357/aaab63}

\bibitem{weidenschilling77b}
{Weidenschilling}, S.J.: {Aerodynamics of solid bodies in the solar nebula}.
\newblock \mnras \textbf{180}, 57--70 (1977)

\bibitem{weidenschilling77}
{Weidenschilling}, S.J.: {The distribution of mass in the planetary system and
  solar nebula}.
\newblock \apss \textbf{51}, 153--158 (1977).
\newblock \doi{10.1007/BF00642464}

\bibitem{weidenschilling96}
{Weidenschilling}, S.J., {Marzari}, F.: {Gravitational scattering as a possible
  origin for giant planets at small stellar distances}.
\newblock \nat \textbf{384}, 619--621 (1996).
\newblock \doi{10.1038/384619a0}

\bibitem{weiss14}
{Weiss}, L.M., {Marcy}, G.W.: {The Mass-Radius Relation for 65 Exoplanets
  Smaller than 4 Earth Radii}.
\newblock \apjl \textbf{783}, L6 (2014).
\newblock \doi{10.1088/2041-8205/783/1/L6}

\bibitem{weiss18}
{Weiss}, L.M., {Marcy}, G.W., {Petigura}, E.A., {Fulton}, B.J., {Howard}, A.W.,
  {Winn}, J.N., {Isaacson}, H.T., {Morton}, T.D., {Hirsch}, L.A., {Sinukoff},
  E.J., {Cumming}, A., {Hebb}, L., {Cargile}, P.A.: {The California-Kepler
  Survey. V. Peas in a Pod: Planets in a Kepler Multi-planet System Are Similar
  in Size and Regularly Spaced}.
\newblock \aj \textbf{155}, 48 (2018).
\newblock \doi{10.3847/1538-3881/aa9ff6}

\bibitem{weiss13}
{Weiss}, L.M., {Marcy}, G.W., {Rowe}, J.F., {Howard}, A.W., {Isaacson}, H.,
  {Fortney}, J.J., {Miller}, N., {Demory}, B.O., {Fischer}, D.A., {Adams},
  E.R., {Dupree}, A.K., {Howell}, S.B., {Kolbl}, R., {Johnson}, J.A., {Horch},
  E.P., {Everett}, M.E., {Fabrycky}, D.C., {Seager}, S.: {The Mass of KOI-94d
  and a Relation for Planet Radius, Mass, and Incident Flux}.
\newblock \apj \textbf{768}, 14 (2013).
\newblock \doi{10.1088/0004-637X/768/1/14}

\bibitem{wetherill78}
{Wetherill}, G.W.: {Accumulation of the terrestrial planets}.
\newblock In: T.~{Gehrels} (ed.) IAU Colloq. 52: Protostars and Planets, pp.
  565--598 (1978)

\bibitem{wetherill91}
{Wetherill}, G.W.: {Why Isn't Mars as Big as Earth?}
\newblock In: Lunar and Planetary Institute Science Conference Abstracts,
  \emph{Lunar and Planetary Inst. Technical Report}, vol.~22, p. 1495 (1991)

\bibitem{wetherill92}
{Wetherill}, G.W.: {An alternative model for the formation of the asteroids}.
\newblock \icarus \textbf{100}, 307--325 (1992).
\newblock \doi{10.1016/0019-1035(92)90103-E}

\bibitem{wetherill96}
{Wetherill}, G.W.: {The Formation and Habitability of Extra-Solar Planets}.
\newblock Icarus \textbf{119}, 219--238 (1996).
\newblock \doi{10.1006/icar.1996.0015}

\bibitem{wetherill89}
{Wetherill}, G.W., {Stewart}, G.R.: {Accumulation of a swarm of small
  planetesimals}.
\newblock Icarus \textbf{77}, 330--357 (1989).
\newblock \doi{10.1016/0019-1035(89)90093-6}

\bibitem{wetherill93}
{Wetherill}, G.W., {Stewart}, G.R.: {Formation of planetary embryos - Effects
  of fragmentation, low relative velocity, and independent variation of
  eccentricity and inclination}.
\newblock Icarus \textbf{106}, 190 (1993).
\newblock \doi{10.1006/icar.1993.1166}

\bibitem{williams11}
{Williams}, J.P., {Cieza}, L.A.: {Protoplanetary Disks and Their Evolution}.
\newblock \araa \textbf{49}, 67--117 (2011).
\newblock \doi{10.1146/annurev-astro-081710-102548}

\bibitem{windmark12}
{Windmark}, F., {Birnstiel}, T., {G{\"u}ttler}, C., {Blum}, J., {Dullemond},
  C.P., {Henning}, T.: {Planetesimal formation by sweep-up: how the bouncing
  barrier can be beneficial to growth}.
\newblock \aap \textbf{540}, A73 (2012).
\newblock \doi{10.1051/0004-6361/201118475}

\bibitem{winn15}
{Winn}, J.N., {Fabrycky}, D.C.: {The Occurrence and Architecture of
  Exoplanetary Systems}.
\newblock \araa \textbf{53}, 409--447 (2015).
\newblock \doi{10.1146/annurev-astro-082214-122246}

\bibitem{wittenmyer16}
{Wittenmyer}, R.A., {Butler}, R.P., {Tinney}, C.G., {Horner}, J., {Carter},
  B.D., {Wright}, D.J., {Jones}, H.R.A., {Bailey}, J., {O'Toole}, S.J.: {The
  Anglo-Australian Planet Search XXIV: The Frequency of Jupiter Analogs}.
\newblock \apj \textbf{819}, 28 (2016).
\newblock \doi{10.3847/0004-637X/819/1/28}

\bibitem{wittenmyer20}
{Wittenmyer}, R.A., {Wang}, S., {Horner}, J., {Butler}, R.P., {Tinney}, C.G.,
  {Carter}, B.D., {Wright}, D.J., {Jones}, H.R.A., {Bailey}, J., {O'Toole},
  S.J., {Johns}, D.: {Cool Jupiters greatly outnumber their toasty siblings:
  occurrence rates from the Anglo-Australian Planet Search}.
\newblock \mnras \textbf{492}(1), 377--383 (2020).
\newblock \doi{10.1093/mnras/stz3436}

\bibitem{wolfgang16}
{Wolfgang}, A., {Rogers}, L.A., {Ford}, E.B.: {Probabilistic Mass-Radius
  Relationship for Sub-Neptune-Sized Planets}.
\newblock \apj \textbf{825}, 19 (2016).
\newblock \doi{10.3847/0004-637X/825/1/19}

\bibitem{wright11}
{Wright}, J.T., {Fakhouri}, O., {Marcy}, G.W., {Han}, E., {Feng}, Y.,
  {Johnson}, J.A., {Howard}, A.W., {Fischer}, D.A., {Valenti}, J.A.,
  {Anderson}, J., {Piskunov}, N.: {The Exoplanet Orbit Database}.
\newblock ArXiv e-prints  (2010)

\bibitem{wright08}
{Wright}, J.T., {Marcy}, G.W., {Butler}, R.P., {Vogt}, S.S., {Henry}, G.W.,
  {Isaacson}, H., {Howard}, A.W.: {The Jupiter Twin HD 154345b}.
\newblock \apjl \textbf{683}, L63--L66 (2008).
\newblock \doi{10.1086/587461}

\bibitem{wright09}
{Wright}, J.T., {Upadhyay}, S., {Marcy}, G.W., {Fischer}, D.A., {Ford}, E.B.,
  {Johnson}, J.A.: {Ten New and Updated Multiplanet Systems and a Survey of
  Exoplanetary Systems}.
\newblock \apj \textbf{693}, 1084--1099 (2009).
\newblock \doi{10.1088/0004-637X/693/2/1084}

\bibitem{wyatt08}
{Wyatt}, M.C.: {Evolution of Debris Disks}.
\newblock \araa \textbf{46}, 339--383 (2008).
\newblock \doi{10.1146/annurev.astro.45.051806.110525}

\bibitem{wyatt12}
{Wyatt}, M.C., {Kennedy}, G., {Sibthorpe}, B., {Moro-Mart{\'\i}n}, A.,
  {Lestrade}, J.F., {Ivison}, R.J., {Matthews}, B., {Udry}, S., {Greaves},
  J.S., {Kalas}, P., {Lawler}, S., {Su}, K.Y.L., {Rieke}, G.H., {Booth}, M.,
  {Bryden}, G., {Horner}, J., {Kavelaars}, J.J., {Wilner}, D.: {Herschel
  imaging of 61 Vir: implications for the prevalence of debris in low-mass
  planetary systems}.
\newblock \mnras \textbf{424}(2), 1206--1223 (2012).
\newblock \doi{10.1111/j.1365-2966.2012.21298.x}

\bibitem{yang17}
{Yang}, C.C., {Johansen}, A., {Carrera}, D.: {Concentrating small particles in
  protoplanetary disks through the streaming instability}.
\newblock \aap \textbf{606}, A80 (2017).
\newblock \doi{10.1051/0004-6361/201630106}

\bibitem{youdin05}
{Youdin}, A.N., {Goodman}, J.: {Streaming Instabilities in Protoplanetary
  Disks}.
\newblock \apj \textbf{620}, 459--469 (2005).
\newblock \doi{10.1086/426895}

\bibitem{zellner17}
{Zellner}, N.E.B.: {Cataclysm No More: New Views on the Timing and Delivery of
  Lunar Impactors}.
\newblock Origins of Life and Evolution of the Biosphere \textbf{47}, 261--280
  (2017).
\newblock \doi{10.1007/s11084-017-9536-3}

\bibitem{zhang10}
{Zhang}, H., {Zhou}, J.L.: {On the Orbital Evolution of a Giant Planet Pair
  Embedded in a Gaseous Disk. I. Jupiter-Saturn Configuration}.
\newblock \apj \textbf{714}, 532--548 (2010).
\newblock \doi{10.1088/0004-637X/714/1/532}

\bibitem{Zhang18}
{Zhang}, Q.: {Prospects for Backtracing 1I/{\lsquo}Oumuamua and Future
  Interstellar Objects}.
\newblock \apjl \textbf{852}, L13 (2018).
\newblock \doi{10.3847/2041-8213/aaa2f7}

\bibitem{zhou05}
{Zhou}, J.L., {Aarseth}, S.J., {Lin}, D.N.C., {Nagasawa}, M.: {Origin and
  Ubiquity of Short-Period Earth-like Planets: Evidence for the Sequential
  Accretion Theory of Planet Formation}.
\newblock \apjl \textbf{631}, L85--L88 (2005).
\newblock \doi{10.1086/497094}

\bibitem{zhu18}
{Zhu}, W., {Wu}, Y.: {The Super Earth-Cold Jupiter Relations}.
\newblock \aj \textbf{156}(3), 92 (2018).
\newblock \doi{10.3847/1538-3881/aad22a}

\bibitem{zube19}
{Zube}, N.G., {Nimmo}, F., {Fischer}, R.A., {Jacobson}, S.A.: {Constraints on
  terrestrial planet formation timescales and equilibration processes in the
  Grand Tack scenario from Hf-W isotopic evolution}.
\newblock Earth and Planetary Science Letters \textbf{522}, 210--218 (2019).
\newblock \doi{10.1016/j.epsl.2019.07.001}

\end{thebibliography}

\end{document}